\documentclass[10pt]{iopart}
\usepackage{lipsum,appendix,enumerate,graphicx,caption,algpseudocode,algorithm, footnote}
\usepackage{hyperref}
\usepackage{iopams,soul,dirtytalk}
\usepackage[table,xcdraw]{xcolor}
\usepackage{titlesec}

\titleformat{\section}[block]{\normalfont\Large\bfseries}{\thesection}{1em}{} 
\titleformat{name=\section,numberless}[block]{\normalfont\Large\bfseries}{}{0em}{}

\expandafter\let\csname equation*\endcsname\relax
\expandafter\let\csname endequation*\endcsname\relax
\usepackage{amsmath,mathrsfs,bm,scalerel}
\usepackage{soul}
\usepackage{etoolbox}
\makeatletter
\def\@mkboth#1#2{}
\newlength\appendixwidth
\preto\appendix{\addtocontents{toc}{\protect\patchl@section}}
\newcommand{\patchl@section}{%
  \settowidth{\appendixwidth}{\textbf{Appendix }}%
  \addtolength{\appendixwidth}{1.5em}%
  \patchcmd{\l@section}{1.5em}{\appendixwidth}{}{\ddt}%
}
\makeatother

\definecolor{mDarkRed}{HTML}{CF0A0A}
\definecolor{mLightBrown}{HTML}{FF7000}

\usepackage{array}
\usepackage{multirow}
\usepackage{orcidlink}
\usepackage{siunitx} 
\usepackage{adjustbox}
\hypersetup{
    colorlinks=true,
   breaklinks=true,
    bookmarksopen=true,
    linkcolor=purple,  
    citecolor=blue, 
    urlcolor=purple     
}
\begin{document}

\title{Magnetic Skyrmion: From Fundamental Physics to Pioneering Applications}

\author{Kishan K. Mishra$^{\ast}$\orcidlink{https://orcid.org/0000-0002-5369-0880}}

\address{Department of Physics \& Astrophysics, University of Delhi, India}
\ead{Physics.kishan@gmail.com}

\author{Aijaz H. Lone$^{\ast}$\orcidlink{https://orcid.org/0000-0002-1687-2917}, Hossein Fariborzi, Gianluca Setti}%
        \address{Division of Computer, Electrical and Mathematical Sciences and Engineering (CEMSE), King Abdullah University of Science and Technology (KAUST), Saudi Arabia}
        \ead{aijaz.lone@kaust.edu.sa}

\author{Srikant Srinivasan}%
        \address{Plaksha University, India}
\vspace{10pt}
\begin{indented}
\item[5th]~September 2023
\end{indented}

\begin{abstract}
Skyrmionic devices exhibit energy-efficient and high-integration data storage and computing capabilities due to their small size, topological protection, and low drive current requirements. So, to realize these devices, an extensive study, from fundamental physics to practical applications, becomes essential. In this article, we present an exhaustive review of the advancements in understanding the fundamental physics behind magnetic skyrmions and the novel data storage and computing technologies based on them. We begin with an in-depth discussion of fundamental concepts such as topological protection, stability, statics and dynamics essential for understanding skyrmions, henceforth the foundation of skyrmion technologies. For the realization of CMOS-compatible skyrmion functional devices, the writing and reading of the skyrmions are crucial. We discuss the developments in different writing schemes such as STT, SOT, and VCMA. The reading of skyrmions is predominantly achieved via two mechanisms: the Magnetoresistive Tunnel Junction (MTJ) TMR effect and topological resistivity (THE). So, a thorough investigation into the Skyrmion Hall Effect, topological properties, and emergent fields is also provided, concluding the discussion on skyrmion reading developments. Based on the writing and reading schemes, we discuss the applications of the skyrmions in conventional logic, unconventional logic, memory applications, and neuromorphic computing in particular. Subsequently, we present an overview of the potential of skyrmion-hosting Majorana Zero Modes (MZMs) in the emerging Topological Quantum Computation and helicity-dependent skyrmion qubits.
Finally, we address the future prospects and challenges for integrating skyrmion-based devices into CMOS-integrated circuits and systems. Principally, this comprehensive review aims to contribute to a broader understanding and realization of the potential of skyrmionic devices in modern technologies.
\end{abstract}
\vspace{2pc}

\maketitle
%
\ioptwocol
\tableofcontents
\iopamstrue
\markboth{Magnetic Skyrmion: From Fundamental Physics to Pioneering Applications}{Magnetic Skyrmion: From Fundamental Physics to Pioneering Applications}
\section{\label{sec:1}Literature and Historical Advances}
    For several decades, the progress of digital electronics has been fueled by continuous hardware scaling and Moore's law, primarily measured by the increasing number of transistors per microprocessor chip. However, as transistors shrink and billions are packed into a single chip, challenges related to the dissipation of Joule heat and quantum mechanical behavior of electrons at the atomic level arise. When we scale down further to the characteristic length of the system, quantum mechanical concerns related to carrier mean free path, spin discussion length and magnetic exchange length become prominent results in the emergence of novel properties that the shift from classical to quantum domain becomes apparent.
    To address these challenges, alternative approaches have been proposed, including information processing using light in optical fiber (based on charge) and spintronic devices that utilize both the charge and spin of electrons to carry and process information. These approaches offer potential solutions to overcome the limitations imposed by shrinking transistor sizes and increasing heat dissipation.
    The growing disparity between memory and processor performance, known as the memory wall problem, is a major obstacle to improving computer performance. In the traditional Von Neumann computer architecture, the processor must access data stored in distinct memory chips to perform logical operations. Nonetheless, the memory scaling rate, increasing by only $1.1\times$ every two years, has not kept pace with the rapid $2 \times$ increase in processor speed over the same period. To address this issue, leveraging the spin properties of electrons for data storage has emerged as a promising approach that minimizes computational energy consumption.
    Magnetic RAM, such as Spin Transfer Torque (STT)-RAM and Spin-Orbit Torque (SOT)-RAM, has gained attention as commercially viable options for on-chip cache memory. These technologies provide strong compatibility with CMOS processes and voltages, no standby leakage (non-volatility), scalability, excellent endurance, long retention time, and overall high reliability. They demonstrate comparable performance to SRAM and hold promise for improving on-chip memory capabilities. In 2009, experimentally observed chiral magnetization structures, such as Skyrmions, which were similar to magnetic domains but possessed the advantage of being smaller in size and operating at faster speeds. This makes them promising candidates for replacing conventional devices.\\
\begin{figure*}
        \centering
        \includegraphics[width = \linewidth]{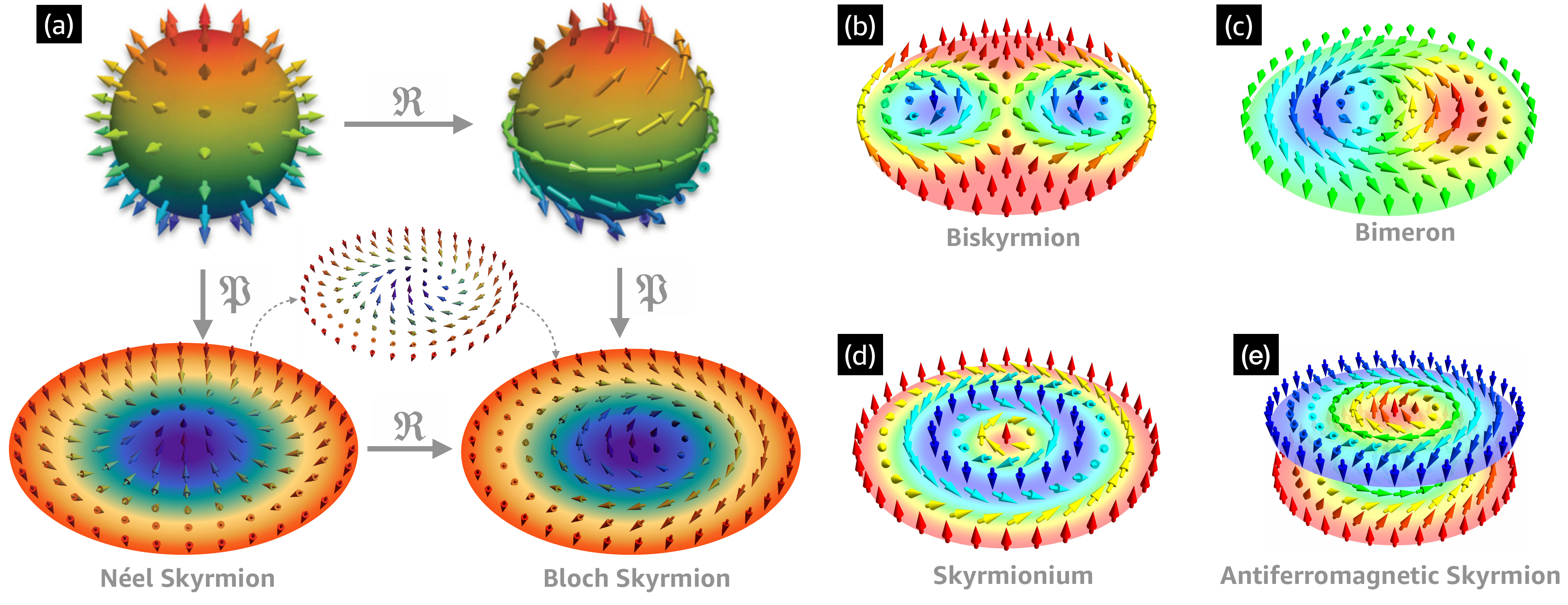}
        \caption{\textbf{Class of Topological Solitons.} \textbf{(a)}~The stereographic projection (denoted $\mathfrak{P}$) squashes the hedgehog into $D = 2$ skyrmion. N\'{e}el (left) and a combed over spins (denoted $\mathfrak{R}$) is a Bloch skyrmion; \textbf{(b)}~Biskyrmion; \textbf{(c)}~Bimeron; \textbf{(d)}~Skyrmionium; \textbf{(e)}~Antiferromagnetic Skyrmion.}
        \label{topo-soliton}
    \end{figure*}
    Since their theoretical proposals and discovery in 2009-10, magnetic textures have become a focal point of research and technology exploration, leading to fundamental understanding during the years 2010-15. The search for materials hosting skyrmions intensified between 2012-17, resulting in the discovery of new compounds and heterostructures exhibiting skyrmionic behaviour, broadening the scope of potential applications. During the years 2015 to 2018, techniques for controlling and manipulating skyrmions were developed, utilizing spin currents and electric fields. This led to their significant presence in spintronics, where skyrmions are recognized for their potential in low-power, high-density data storage and logic devices, with concepts like skyrmion-based racetrack memory and logic gates being proposed. Moreover, researchers explored skyrmions in nanoscale materials such as nanoparticles and thin films, creating artificial skyrmion lattices with new possibilities for manipulation. The practical applications of skyrmions, including memory devices, spin logic circuits and neuromorphic computing, started to emerge, while ongoing research continues to optimize their properties for real-world implementations. As a result, these advancements hold the potential to shape the future landscape of digital electronics.
\section{\label{sec:level2} Skyrmion: A Topological Soliton}
    In the early 1960s, when quantum chromodynamics (QCD) was in its nascent stages, T.H.R. Skyrme introduced the Skyrme model~\cite{skyrme1962unified, skyrme1961non} to describe baryons as soliton solutions using a non-linear sigma model (NLSM) with a triplet of pion fields. While the model was originally developed to understand the interior of nuclei, it later inspired the concept of magnetic skyrmions as topological soliton solutions in the magnetization vector field. These topologically protected soliton structures were initially proposed by Bogdanov and Ivanov~\cite{bogdanov1989thermodynamically28c,stefanovich1986two29c,ivanov1990magnetic30c} in the late 1980s, initially referred to as \textit{magnetic vortices} and \textit{topological solitons}, and later \textit{Magnetic Skyrmion}(\emph{Skyrmion} hereafter). The researchers extensively studied the mean-field theory of ferromagnets with chiral spin-orbit interactions. They specifically pinpointed certain parameter ranges where a mixed state, resembling the vortex lattice found in type II superconductors, emerges as the thermodynamically stable phase, featuring a finite density of skyrmions \cite{81c}. Although soliton solutions are mostly dynamic cases, the Hobart-Derrick theorem~\cite{derrick} suggests that a stable localized soliton structure is impossible in the most condensed matter systems. But magnetic skyrmions, in contrast, are stabilized by the Dzyaloinshii-Moriya Interaction (DMI)~\cite{29,30}, an anti-symmetric exchange interaction which prevents the collapse of the magnetization vector field into a magnetic singularity. Magnetic skyrmions can exist in both two-dimensional (2D)\footnote{also known as 'baby skyrmions'.} and three-dimensional (3D) systems, and its roots are spread in diverse sub-fields of many-body physics, including 2D electron gases~\cite{58c,59c}, superconductors~\cite{60c,61c,62c,63c,64c,65c,yokoyama2015josephson,68}, spinor Bose-Einstein condensates~\cite{ho1998spinor69c, al2001skyrmions70c,marzlin2000creation71c,battye2002stable72c,savage2003energetically73c,leslie2009creation74c,choi2012observation75c}, Particle physics \cite{58c}, string theory \cite{vilenkin1994cosmic} and quasiclassical magnetic systems.
    Skyrmions, in addition to their intriguing fundamental properties, have garnered significant attention for their potential applications in spintronics~\cite{fert2013skyrmions, kiselev2011chiral68kk}. 
    So, understanding skyrmion dynamics is crucial for harnessing their potential. They induce electric and magnetic fields within conductive materials and exhibit a phenomenon referred to as the Skyrmion Hall Effect (SHE). This effect causes skyrmions to deflect from the applied force direction, enabling their motion under external drives. They have shown promising potential for ultra-low current densities, thanks to techniques such as Spin-Transfer (STT)~\cite{jonietz2010spin80c, yu2012skyrmion}, Spin-orbit Torques (SOT)~\cite{sampaio2013nucleation88c} and Spin hall torque magnetometry of Dzyaloshinskii
    domain walls. They are considered as potential next-generation information carriers due to their small size~\cite{buttner2015dynamics}, topological stability~\cite{luo2021skyrmion}, nucleation and detection methods~\cite{jiang2015blowing, sampaio2013nucleation88c, iwasaki2013current85c, heinonen2016generation73kk,crum2015perpendicular74kk,hanneken2015electrical75kk,parkin2008magnetic,yu2014biskyrmion47c}, and sensitivity to external forces~\cite{nagaosa2013topological36}. They can be utilized as bits in storage devices and logic processing units~\cite{yan2021skyrmion,zhang2015magnetic,sisodia2022programmable,luo2018reconfigurable}. The concept of skyrmion racetracks~\cite{zhang2015topological, zhang2018manipulation, litzius2017skyrmion47k, yu2017room}, akin to domain wall racetracks~\cite{parkin2008magnetic}, has been proposed, where magnetic skyrmions serve as mobile bits within shift-register lines allowing for denser memory bit integration, unlike DWs. Numerous device designs have been suggested, but practical implementation faces two main obstacles.
    The first challenge is finding a material that fulfils the necessary criteria, including a diameter on the order of less than 10 nm, stability at room temperature and stability at remanence. While several materials, significantly, those lacking inversion symmetry, such as the B20-class ``chiral magnet'' [see Table~\ref{tab:table-sky}] systems host skyrmions, none meet all three requirements. Some materials, such as FeGe and $\mathrm{Co_xZn_yMn_z}$, satisfy the first two criteria, while certain magnetic multilayers \cite{woo2016observation46k} fulfil the latter two.
    The second obstacle lies in understanding the structure and behavior of magnetic skyrmions, which is the primary focus of the ongoing research. Addressing these challenges is crucial for realizing the practical applications of skyrmions in various technological domains. Furthermore, Thiele's equation is a model for skyrmion dynamics and addresses its limitations by deriving a general interaction potential. The equation is extended to incorporate external driving forces, and numerical simulations provide insights into the behavior of skyrmion crystals.\\
    The distinctive magnetization topology of skyrmions has an effect on electron transport, leading to the emergence of the Topological Hall Effect (THE)~\cite{neubauer2009topological,nagaosa2013topological36, kanazawa2011large,kanazawa2015discretized87k}. Understanding and harnessing the properties of magnetic skyrmions hold promise for advancing various technological applications in Computing, data processing and storage applications~\cite{87}.
\section{\label{sec:level3}Magnetic Skyrmion}
    \textit{Magnetic Skyrmions} is a particle-like (localized), swirling spin texture found in some magnetically ordered materials with a distinct topological nature. Typically, they arise due to the chiral interaction among atomic spin present in magnetic materials without inversion symmetry, such as non-centrosymmetric (NCS) compounds or thin films with interfaces that break inversion symmetry. Additionally, they can also occur in heavy metals employing certain mechanisms. Initially, magnetic skyrmions were identified in single crystals of magnetic compounds having a non-centrosymmetric lattice and explained by the existence of Dzyaloshinskii-Moriya Interaction (DMI) introduced by spin-orbit coupling in the absence of inversion symmetry in the lattice of the crystal.
    In the classical micromagnetic model~\cite{aharoni2000introduction32}, magnetic skyrmions are visualized as swirling or hedgehog-like structures [see FIG.~\ref{topo-soliton}(a)], depending on the specific type of exchange interaction that stabilizes them. The first observation of skyrmions in magnetic materials with DMI was reported in 2009~\cite{muhlbauer2009skyrmion38c}.
    
\begin{table*}
    \centering
    \resizebox{\linewidth}{!}{%
        \begin{tabular}{cccccc}
            \hline \hline
            \rowcolor[HTML]{FAFCCE}
            \textbf{Materials} & \textbf{Category} & \textbf{Temp.} & \textbf{Topological Structure} & \textbf{Skyrmion Size$^b$} & \textbf{Detection Methods} \\
            \hline
    \rowcolor[HTML]{F6CDCD}
    MnSi (bulk) & NCS-phase system &28-29.5 & Bloch~\cite{muhlbauer2009skyrmion38c} & 18 & SANS \\
    \rowcolor[HTML]{FAFCCE} 
    $\mathrm{FeGe}$ (thin film)$^c$ & NCS-phase system & 60-280 & Bloch~\cite{yu2011near42c} & 90 & LTEM \\
    \rowcolor[HTML]{F6CDCD} 
    $\mathrm{Cu_OSeO_3}$ (thin film) & NCS-phase system & 5-57 & Bloch \cite{seki2012magnetoelectric44c} & 50 & LTEM \\
    \rowcolor[HTML]{FAFCCE} 
    $\mathrm{Fe_{0.5}Co_{0.5}Si}$ (thin film) & NCS-phase system & 5-40 & Bloch~\cite{yu2010real41c} & 90 & LTEM \\
    \rowcolor[HTML]{F6CDCD} 
    $\mathrm{Co_8Zn_9Mn_3}$(thin film) & NCS-phase system & 295-320 & Bloch \cite{tokunaga2015new} & $\sim $125 & LTEM \\
    \rowcolor[HTML]{FAFCCE} 
    $\mathrm{Gd_2PdSi_3}$ (bulk) & CS-phase system & 2-20 & Bloch \cite{kurumaji2019skyrmion} & 2.5 & TM, RXS \\
    \rowcolor[HTML]{F6CDCD} 
    $\mathrm{Gd_3Ru_4Al_{12}}$ (bulk) & CS-phase system & 5-8 & Bloch \cite{hirschberger2019skyrmion} & 2.8 & TM, RXS, LTEM, SANS \\
    \rowcolor[HTML]{FAFCCE} 
    $\mathrm{GdRu_2Si_2}$ (bulk) & CS-phase system & 0-20 & Bloch \cite{khanh2020nanometric}& 1.9 & TM, RXS, LTEM \\
    \rowcolor[HTML]{F6CDCD}
    $\mathrm{EuAl_4}$ & CS-phase system & 5 & Bloch \cite{takagi2022square} & 3.5 & SANS, RXS \\
    \rowcolor[HTML]{FAFCCE} 
    $\mathrm{GaV_4S_8}$ (bulk) & NCS-phase system & 9-13 & N\'{e}el \cite{kezsmarki2015neel} & 22.2 & MFM, SANS \\
    \rowcolor[HTML]{F6CDCD}
    $\mathrm{(Fe_{0.5}Co_{0.5})_{5}GeT_2}$(nanoflake) & NCS-phase system & RT & N\'{e}el \cite{zhang2022room} & -$^d$ & LTEM, MFM, TM \\
    \rowcolor[HTML]{FAFCCE} 
    $\mathrm{Fe/Ir(111)}$ & FM/HM interface & 11 & N\'{e}el \cite{heinze2011spontaneous47} & 1 & SP-STM \\
    \rowcolor[HTML]{F6CDCD}
    $\mathrm{PdFe/Ir(111)}$ & FM/HM interface & 4.2-8 & N\'{e}el \cite{romming2013writing48, romming2015field60k} & 6-7$^e$ & SP-STM \\
    \rowcolor[HTML]{FAFCCE} 
    $\mathrm{Pt/Co/MgO}$ &  Magnetic multilayer & RT & N\'{e}el \cite{boulle2016room} & $\sim 130$ & XMCD-PEEM \\
    \rowcolor[HTML]{F6CDCD}
    $\mathrm{[Ir/Co/Pt]_{10}}$ & Magnetic multilayer & RT & N\'{e}el~\cite{moreau2016additive} & 30-90 & STXM\\
    \rowcolor[HTML]{FAFCCE} 
    $\mathrm{[Ir/Fe/Co/Pt]_n}$ & Magnetic multilayer & RT & N\'{e}el~\cite{soumyanarayanan2016emergent} & $<50$ & MTXM, TM, MFM \\
    \rowcolor[HTML]{F6CDCD}
    $\mathrm{[Pt/Co/Ta]_{15}}$ & Magnetic multilayer & RT & N\'{e}el~\cite{woo2016observation46k} & 200-250$^f$ & STXM \\
    \rowcolor[HTML]{FAFCCE} 
    $\mathrm{Fe/Ni/Cu/Ni/Cu(001)}$ & Magnetic multilayer & RT & N\'{e}el~\cite{chen2015room} & $\sim 200$ & SPLEEM \\
    \rowcolor[HTML]{F6CDCD}
    $\mathrm{WTe_2/Fe_3/GeTe_2}$ & vdW heterostructures & 94, 198 & N\'{e}el~\cite{wu2020neel} & 150, 80 & LTEM, TM\\
    \rowcolor[HTML]{FAFCCE} 
    $\mathrm{Fe_3TeGe_2/Cr_2Te_2Ge_6}$ & vdW heterostructures & 20-100 & N\'{e}el~\cite{wu2022van} & $\sim 100$ & MFM, TM\\
    \rowcolor[HTML]{F6CDCD} 
    $\mathrm{Fe_GeTe_2/[Co/Pd]_{10}}$ & vdW heterostructures & Liquid $\mathrm{N_2}$ & N\'{e}el~\cite{yang2020creation} &$\sim$ 50 & XMCD-PEEM, LTEM \\
    \rowcolor[HTML]{FAFCCE} 
    $\mathrm{Mn_{1.4}Pt_{0.9}Pd_{0.1}Sn}$ (thin film) & NCS-phase system & 100-400 & Anti-skyrmion \cite{nayak2017magnetic, peng2020controlled} & 135 & LTEM \\
    \rowcolor[HTML]{F6CDCD}
    $\mathrm{Fe_{1.9}Ni_{0.9}Pd_{0.2}P}$ (thin film) & NC-phase system & 100-400 & Anti-skyrmion \cite{karube2021room} & -$^g$ & LTEM, MFM \\
    \rowcolor[HTML]{FAFCCE} 
    $\mathrm{Pt/Gd_{44}Co_{56}/TaO_x}$ & FM multilayer & RT & AFSK~\cite{caretta2018fast} & 10-35 & X-ray holography \\
    \rowcolor[HTML]{F6CDCD}
    $\mathrm{[Pt/Gd_{25}Fe_{65.6}Co_{9.4}/MgO]_{20}}$ & FM multilayer & RT & AFSK~\cite{woo2018current} & $\sim 180$ & STXM \\
    \rowcolor[HTML]{FAFCCE} 
    $\mathrm{Pt/Co/CoFeB/Ir/Co/CoFeB/W}$ & SAF multilayer & RT & AFSK~\cite{dohi2019formation} & $\sim 1600$ & MOKE \\
    \rowcolor[HTML]{F6CDCD} 
    $\mathrm{Pt/Ru/Co/Pt/Ru/Co/Pt/BL/Pt}$ & SAF & RT & AFSK~\cite{legrand2018hybrid} & 20-30 & MFM \\
    \rowcolor[HTML]{FAFCCE} 
    $\mathrm{[Co/Pd]/Ru/[Co/Pd]}$ & 
    SAF multilayer & RT & AFSK~\cite{chen2020realization} & $\sim 80$ & LTEM \\
    \rowcolor[HTML]{F6CDCD}
    $\mathrm{\alpha-Fe_2O_3/Pt}$ & Antiferromagnet & $>240$ &  AF half-SK~\cite{jani2021antiferromagnetic} & $\sim$ 100 & XLMD-PEEM \\
    \hline \hline
    \end{tabular}%
    }
    \caption[Short version of the caption for the list of tables]{%
        \textbf{Details$^a$ of skyrmion hosting materials.}\\Note: Refer to (\ref{SnA}) for detailing of symbols and abbreviations, with permission from Ref.~\cite{li2023magnetic}
        }
    \scriptsize{$^a$A concise summary of temperature dependence, skyrmion size, and detection method provided,
    {$^b$For lattice-stabilized skyrmions, the skyrmion size corresponds to the lattice constant,}
    {$^c$The study explores multiple films with different thicknesses and the presented data refers to the thickness 15 nm,}
    {$^d$The thickness of the sample determines its magnetic periodicity},
    {$^e$Magnetic field controlled skyrmion size,}
    {$^f$Magnetic field controlled skyrmion size,}
    {$^g$The thickness of the sample determines its magnetic periodicity}}
        \label{tab:table-sky}
\end{table*}
    
The DMI
    promotes perpendicular orientations of adjacent magnetizations, while the ferromagnetic exchange interaction promotes parallel orientations~\cite{beg2015ground16, 89, bogdanov2002magnetic79c}. The interplay between these interactions results in different skyrmion textures. Skyrmions can be classified in two prominent textures Bloch and N\'{e}el types [see FIG.~\ref{topo-soliton}(a)], depending on the type of DM interaction that stabilizes them. Bloch skyrmions are stabilized by bulk DMI ($b-\text{DMI}$) and feature helicoidal magnetic structures, whereas N\'{e}el skyrmions are stabilized interfacial DMI ($i-\text{DMI}$) and showcase cycloidal magnetic textures. The sign of DM interaction dictates the chirality of these textures. As a topologically protected soliton skyrmion exhibits a non-trivial topology and topological properties. Topological protection ensures the stability of skyrmions in magnetic systems by creating energy barriers, as discussed by Braun \text{et al.} in Ref.~\cite{braun2012topological}. Nevertheless, the presence of discrete magnetic moments and thermal fluctuation can disrupt and overcome these barriers. Skyrmions are stable~\cite{hagemeister2015stability} under applied fields and at low temperatures, but their practical use in devices is still limited. They have been found stabilized in Bose-Einstein condensates~\cite{al2001skyrmions70c}, superconductors~\cite{mascot2021topological}, liquid crystals~\cite{fukuda2011quasi}, acoustic systems~\cite{ge2021observation}, photonic systems~\cite{du2019deep} and other magnetic materials.\\
   Aside from skyrmions, several other particle-like nontrivial magnetic textures with distinct topological numbers have been identified, as illustrated in FIG.~\ref{topo-soliton}(b-e). A class of soliton recently been observed in thin plate~\cite{yu2018transformation} is \textit{Merons} and \textit{Antimerons} ($N_{\text{sk}} = \pm 1/2$) that are half-skyrmion like magnetic textures with in-plane magnetization~\cite{hayami2021meron,lin2015skyrmion}. \textit{Biskyrmion} ($N_{\text{sk}} = \pm 2$) with a pair of skyrmions of opposite helicities~\cite{capic2019stabilty} in another kind. In three-dimensional space, hedgehogs and anti-hedgehogs ($N_{\text{sk}} = \pm 1$) behave like emergent \textit{magnetic monopoles} and \textit{antimonopoles}~\cite{kanazawa2017topological}. An intriguing non-topological soliton, known as \textit{skyrmionium}, is created when two skyrmions with opposite topological numbers come together at their center, merge, annihilate and form the texture. This unique structure exhibits a complete absence of the parasitic Skyrmion Hall effect. Most of these solitons are computationally realised but thin ferrimagnetic films subjected to ultrafast laser pulses have experimentally achieved many such structures~\cite{finazzi2013laser}. Similar to Skyrmion (Sk) in a Ferromagnetic (FM) background, an Antiferromagnetic Skyrmion (AFMSk) is a specific type of skyrmion that occurs in Antiferromagnetic Materials (AFM) possessing significant advantages over other solitons which we will discuss in detail in a coming section.
    \begin{figure*}
        \centering
        \includegraphics[width = \linewidth]{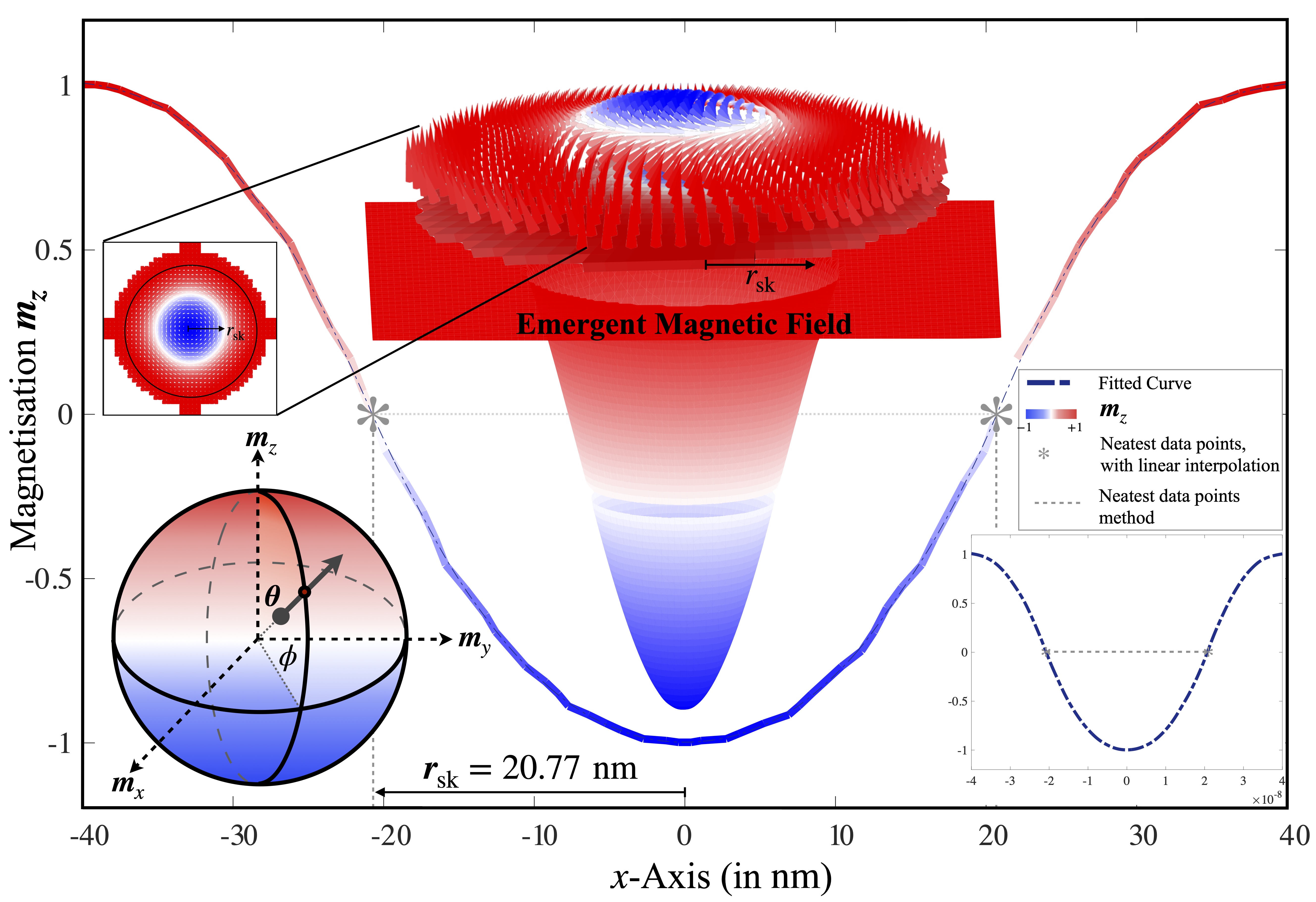}
        \caption{\textbf{Magnetic Profile of Skyrmion.} simulated in a 80 nm $\times$ 80 nm $\times$ 2 nm thin film. Nearest interpolation was performed to calculate the radius, $r_{\text{sk}} = 20.77$ nm. The emergent magnetic field was calculated using $\theta(r)$ profile.}
        \label{fig:profile-sky}
    \end{figure*}
\subsection{\label{sec:level4} Numerical Profile of Skyrmion}
    In order to manipulate the characteristics of skyrmions, such as their size, chirality, or stability, it is essential to develop a predictive model for their profiles. It's been observed that stable ferromagnetic skyrmions in magnetic materials possess cylindrical symmetry around a vertical axis aligned with $\hat{z}$, which also serves as their center. Consequently, it is advantageous to focus solely on their radial profiles, as they are invariant upon rotation around this vertical symmetry axis. Thus, it is generally convenient to represent a skyrmion field using spherical coordinates ($r, \theta, \phi$) in spin space, as this configuration can be likened to a ``hedgehog'' spin configuration on a 3D sphere [refer to FIG.~\ref{topo-soliton}(a)]. Thus we can express for unit-length $\boldsymbol{m}$ vector
    \begin{equation}
    \boldsymbol{m}(r, \theta, \phi) = {\begin{pmatrix}
    \sin{\theta} \cos{\phi} \\
     \sin{\theta} \sin{\phi} \\ 
     \sin{\theta}
    \end{pmatrix}}
    \end{equation} 
    where the angles are introduced in FIG.~\ref{fig:profile-sky} (bottom left). Considering the simplicity of theoretical and computational studies, it's convenient to define a model skyrmion as $\theta (0) = \pi$ (spin down $\uparrow$ at the center) and $\theta (\infty) = 0$ (spin up $\downarrow$ on the periphery), if not mentioned otherwise.
    To significantly simplify calculations,  one can introduce a cylindrical frame denoted as  $\boldsymbol{r} = (x, y, z) = (r, \varphi, z)$. With this approach, the spin at each point in the magnetization space can be described using just two spherical angles
    $$\begin{aligned}
        \theta &= \theta (r, \varphi) \Longrightarrow \text{the angle w.r.t the positive $z$ axis} \\    
        \phi & = \phi (r, \varphi) \Longrightarrow \text{the angle measured counterclockwise from} \\
        & \text{the positive $x$ axis on the $xy$ plane}
    \end{aligned}$$ 
    Here, the $z$ axis specifies the $xy$ plane where the Skyrmion is located, and the magnetic field is oriented along the $z$ direction. \\
    Employing above definitions 
    \begin{equation}
    \boldsymbol{m}(\theta, \phi) = 
    \begin{pmatrix}
    \sin{\theta(r,\varphi )} \cos{\phi(r,\varphi)} \\
    \sin{\theta(r,\varphi )} \sin{\phi(r,\varphi)} \\ 
    \sin{\theta(r,\varphi )}
    \end{pmatrix}
    \label{eqn:9}
    \end{equation}
    For the perfectly \textit{axially symmetrical} and undistorted skyrmion, the spherical angles take values
    \begin{center}
        $\theta = \theta(r)$ \& $\phi = \phi(\varphi)$
    \end{center}
    \begin{figure*}[htbp]
        \centering
        \includegraphics[width = \linewidth]{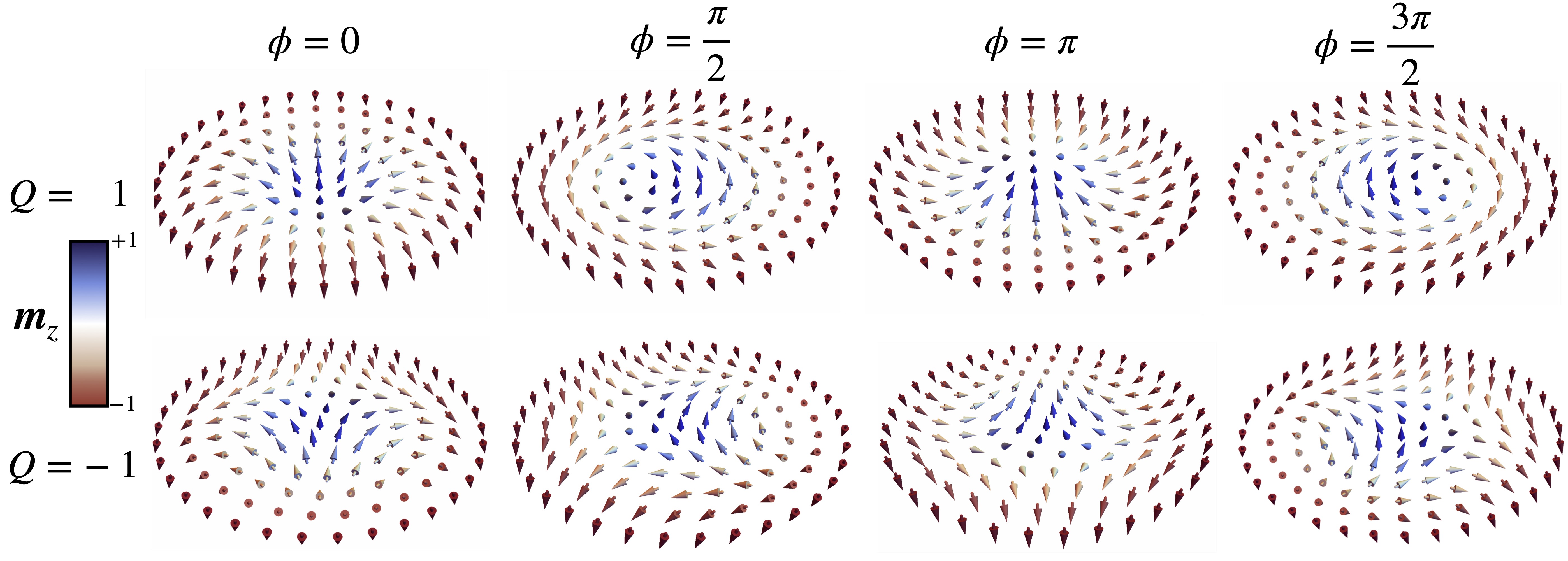}
        \caption{\textbf{Class of skyrmions.} (Top) Magnetization configuration of skyrmions ($Q = 1$) and (Bottom) antiskyrmions ($Q = -1$) with fixed polarity ($p = 1$) and varying helicity. Here, the arrows represent the in-plane component of $\boldsymbol{m}$, and the color indicates $\boldsymbol{m_z}$.}
        \label{fig:topo-class}
    \end{figure*}
    The straightforward algebra gives,
    \begin{equation}
    \frac{\partial \theta}{\partial x}=\frac{d \theta}{d r} \cos \varphi,\quad\frac{\partial \phi}{\partial x}=-\frac{1}{r} \frac{d \phi}{d \varphi} \sin \varphi
    \end{equation}
    \begin{equation}
        \frac{\partial \theta}{\partial y}=\frac{d \theta}{d r} \sin \varphi,\quad\frac{\partial \phi}{\partial y}=\frac{1}{r} \frac{d \phi}{d \varphi} \cos \varphi
    \end{equation}
    In the case of an axially-symmetric skyrmion, the function $\phi(\varphi)$ is a smooth continuous function defined modulo $2\pi$ has the description
    \begin{equation}
        \phi(\varphi, \gamma) = w \varphi+ \gamma
    \end{equation}
    where $\varphi$ represents the azimuthal angle (ranging from $0$ to $2\pi$), $w$ is an integer-valued \textit{Vorticity} that determines the number of twists of a vector field around a closed loop, giving the skyrmion number $Q = w$ [see FIG.~\ref{fig:topo-class}].
    The parameter $\gamma$ corresponds to the \textit{Helicity}, an arbitrary phase that may or may not be dictated by the underlying crystallographic symmetry. Special values of the helicity ($\gamma = 0, \pi/2, \pi, 3\pi/2$) play important roles~\cite{zhang2017skyrmion} in determining the chirality of the skyrmion. The helicity of N'{e}el skyrmions can be expressed as $\gamma = 0$ or $\gamma = \pi$, determining the orientation of the in-plane spins as inward or outward, respectively. On the other hand, Bloch skyrmions exhibit a helicity of $\gamma = \pm \pi/2$, signifying a clockwise or counterclockwise rotation of the in-plane spins [see FIG.~\ref{fig:sky-fig}(a,b)]. Different helicities correspond to different DMIs. To simplify, we can label a skyrmion as $(Q, \gamma)$ using its topological number and helicity. It's important to note that $Q$ changes sign when all magnetic moment directions are reversed, $\boldsymbol{m} \rightarrow - \boldsymbol{m}$, leading to a configuration known as an \textit{Anti-skyrmion} [see FIG.~\ref{fig:sky-fig}(c)].
\subsection{\label{sec:level5}Topological Invariants}
    Having established the parameterization of the profile, we can now delve into defining topological invariants that characterize a skyrmion. These invariants are integer numbers that exhibit relative robustness against moderate perturbations unless they are compelled to overcome the topological barrier by using a collapsing mechanism~\cite{derras2022dynamics}. The primary topological descriptors are the \textit{topological charge}, representing the number of spin windings around an order-parameter sphere, and the \textit{helicity}, governing the spin's twisting from spin-down at the skyrmion's core to spin-up at its boundaries. [refer to FIG.~\ref{fig:topo-class} and \ref{fig:windno}]
    Magnetic skyrmions are observed in both two-dimensional systems, like magnetic thin films, and in bulk materials, where they can be regarded as stacked two-dimensional copies. As a result, the homotopy group characterizing magnetic skyrmions in two dimensions is denoted by $\pi_2(\mathbb{S}^2)$, while the original Skyrme model in three dimensions corresponds to the homotopy group $\pi_3(\mathbb{S}^3)$. Due to the reduced dimensionality of the magnetization field's domain and target space, magnetic skyrmions are often called ``baby skyrmions''. The mathematical expression for the topological charge $Q$ of a $N=2$-dimensional vector field $\hat{\boldsymbol{n}}=n^1\hat{\mathbf{e_1}}+n^2\mathbf{e_2}$ can be written as
    \begin{equation}
        Q=\frac{1}{2 \pi} \int d x \epsilon_{a b} n^{a} \frac{d n^{b}}{d x}
    \end{equation}
    By employing the parametrization given in Eqn.~(\ref{eqn:9}), we establish the below results for the magnetization vector field ${\boldsymbol{m}}(\boldsymbol{r})$. Subsequently, the skyrmion's topological charge\footnote{also known as the topological number, winding number, chern number and skyrmion number} $Q$ is
    \begin{equation}
        Q =  \frac{1}{4\pi} \int   \boldsymbol{m} \cdot \left(\frac{\partial \boldsymbol{m}}{\partial x} \times \frac{\partial \boldsymbol{m}}{\partial y}  \right)d x d y
        \label{Eqn:topo}
    \end{equation}
    The expression above yields a topological charge, an integer value ($\in \mathbb{Z}$), which takes on values like $\pm1, \pm2, \pm3$, and so on, for spin-field configurations that smoothly vary and possess topological nontriviality. We are particularly interested in magnetic skyrmions, which have a topological number of $\pm 1$. \\
    In the forthcoming discussion, we will find that we can define an emergent magnetic (or berry) field as,
    \begin{equation}
        \mathscr{B} = \frac{1}{2} \boldsymbol{m} \cdot \left(\frac{\partial \boldsymbol{m}}{\partial x} \times \frac{\partial \boldsymbol{m}}{\partial y}  \right)
    \end{equation}
    and that the skyrmion possesses an emergent magnetic flux denoted by $\Phi$ which corresponds to $2\pi Q$. With the considered parametrization, the resulting magnetic field simplifies into
    \begin{equation}
        \mathscr{B} = \frac{1}{2} \boldsymbol{m} \cdot \left(\frac{\partial \boldsymbol{m}}{\partial x} \times \frac{\partial \boldsymbol{m}}{\partial y}  \right) = \frac{\sin \theta}{2r} \frac{d \theta}{dr} \frac{d \phi}{d \varphi}
    \end{equation}
    which corresponds to a two-dimensional magnetic monopole with a topological charge
    \begin{equation}
    \begin{aligned}
        Q=\frac{1}{2 \pi} \int \mathscr{B} d^{2} \mathbf{r} & =\frac{1}{4 \pi} \int_{0}^{2 \pi} d \varphi \frac{d \phi}{d \varphi} \int_{0}^{\infty} d r \frac{d \theta}{d r} \sin \theta \\
         & =-\left.\left.\frac{1}{4 \pi} \phi(\varphi)\right|_{0} ^{2 \pi} \cos \theta(r)\right|_{0} ^{\infty}
    \end{aligned}
    \end{equation}
        \begin{figure*}[t]
    	\centering
             \includegraphics[width = \linewidth]{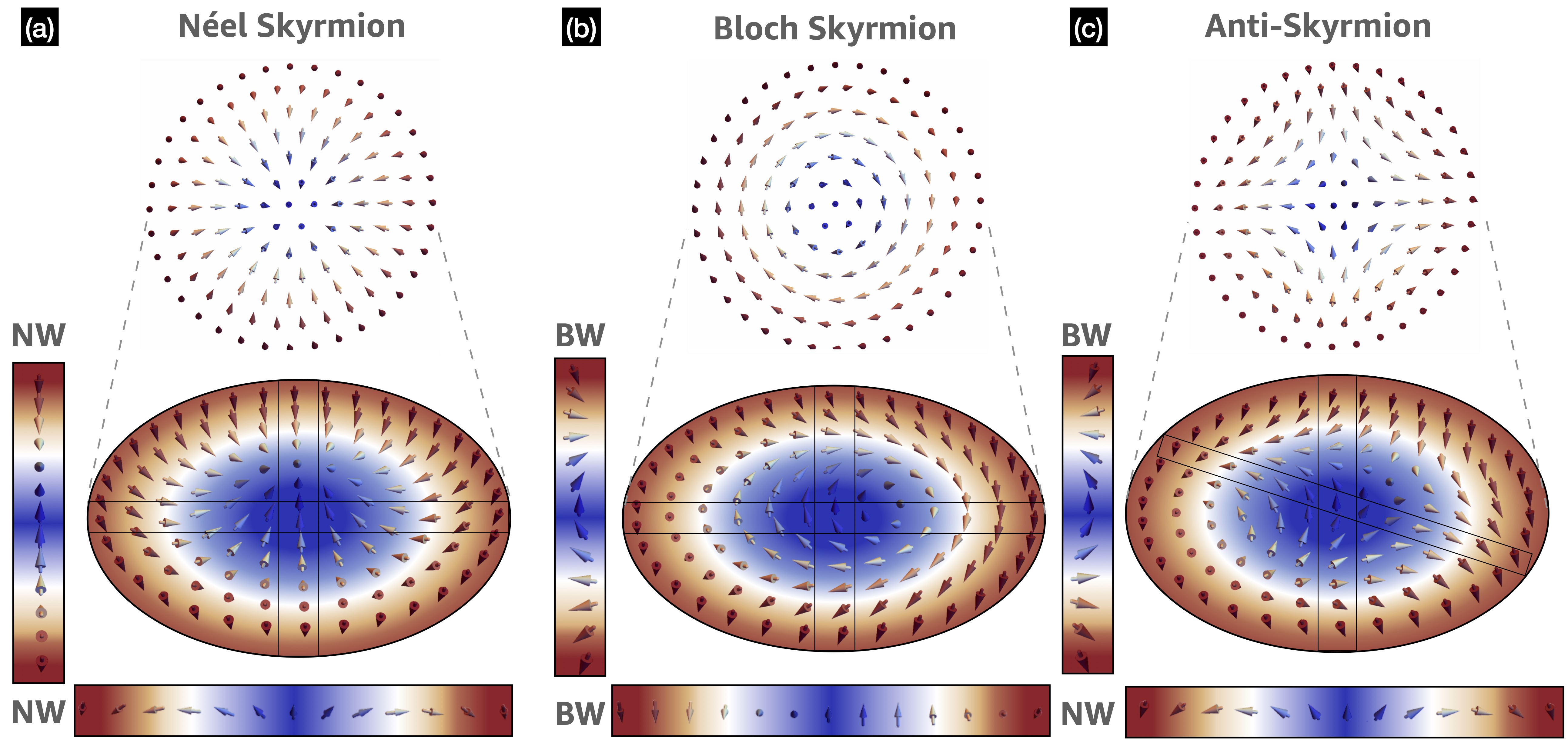}
    	\caption{\textbf{N\'{e}el, Bloch and Antiskyrmions.}   Diagonal Cuts through a skyrmion give a domain wall: (N\'{e}el Skyrmion, N\'{e}el Wall (NW); Bloch Skyrmion, Bloch Wall (BW) and Antiskyrmion with both NW and BW at $45^{\circ}$ to each other.)}
    	\label{fig:sky-fig}
    \end{figure*}
    Given a ferromagnetic background with specific boundary conditions ($\theta(0) = \pi, \theta(\infty) = 0$), the topological charge of a skyrmion can be determined as
    \begin{equation}
        Q = -w, ~~w\in \mathbb{Z}
    \end{equation}
    where $w$ is an integer-valued vorticity of a vector field [refer to FIG.~\ref{fig:windno}]. Thus, even minimally charged ($Q = \pm 1$) skyrmions represent an extensive family of topological textures with $Q = 0$ as the trivial spin configuration field. Several of the most important of them are discussed in the next subsections. he magnetization field characterizing a magnetic skyrmion with $w=1$ gives rise to a topological charge of $-1$. Conversely, an antiskyrmion is represented by a spin configuration with $w=-1$ and $Q=1$, as depicted in FIG.~\ref{fig:topo-class}. It should be noted that the idealized symmetry used in the ansatz [Eqn.~(\ref{eqn:9})] for a skyrmion may not accurately represent its true shape in reality. In many cases, skyrmions are observed to be distorted due to various factors such as their dynamics or imperfections in the sample. Experimentally, only the cases $\gamma = 0, \pi$; a N\'{e}el skyrmion in FIG.~\ref{fig:sky-fig}(a) and $\gamma =\pi/2, -\pi/2$; a Bloch skyrmion in FIG.~\ref{fig:sky-fig}(b) have been observed so far, while intermediate skyrmions are theoretically predicted. As anticipated, both magnetic skyrmion types are topologically identical, as they can be smoothly interconverted by rotating the unit vector field at every point [depicted in FIG.~\ref{topo-soliton}(a)].\\
    Additionally, for simplicity, another parameter \textit{polarity} is defined as
    \begin{equation}
        Q = p\cdot w, \quad p = \pm 1 , \quad w = \pm 0, \pm 1, \pm 2 , \dots
    \end{equation}
    which indicates whether the central spin of the skyrmion points into $+z$ direction $(p = +1)$ or $-z$ direction $(p = -1)$.
\subsection{\label{sec:level6}Skyrmion Zoology}
\begin{figure*}[htbp]
    	\centering
            \includegraphics[width=\linewidth]{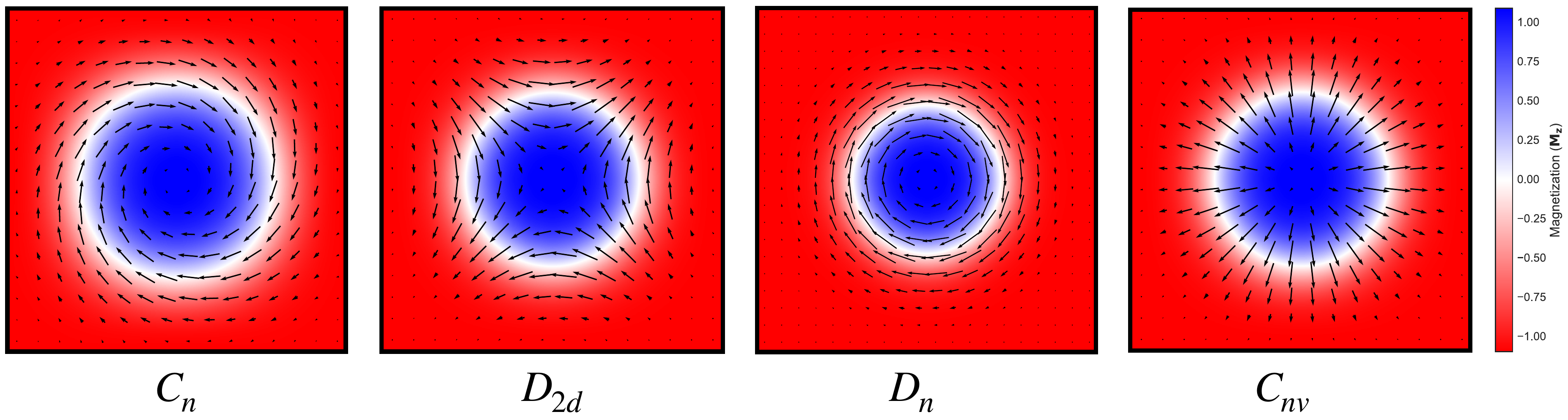}
    	\caption{\textbf{Skyrmion zoology.} Skyrmions exhibit different appearances based on the crystal symmetry promoting the chiral (DM) interaction. The figure shows spin projections of various skyrmions on the basal plane: (a) $C_n$ skyrmion, (b) $D_{2d}$ skyrmion, (c) $D_n$ skyrmion (Bloch skyrmion with $\gamma = 0$), and (d) $C_{nv}$ skyrmion (N\'{e}el skyrmion with $\gamma = \pi/2$).}
    	\label{fig:12}
    \end{figure*}

    In most cases, skyrmions can be characterized based on their topological charge, which often corresponds to the lowest-energy field configurations. Generally, the lowest-energy field configurations correspond to a minimum topological charge of $Q = \pm1$ which depends on a specific form of a Hamiltonian. The helicity $\gamma$ of the system is determined by underlying crystallographic symmetry. Among the various 2D skyrmion patterns, the N\'{e}el-type skyrmion and the Bloch-type skyrmion [as shown in FIG.~\ref{fig:sky-fig}] are the two most significant ones. 
    In order to observe the distinct chirality and topological charge of skyrmions, it is typically necessary to incorporate an inversion-breaking mechanism in the Hamiltonian. This mechanism, known as the DMI\footnote{Referred to as \textit{antisymmetric exchange} analogous to symmetric Heisenberg exchange.}, is present in non-centrosymmetric magnets and magnetic systems like thin films with complex surface interactions and interfaces. The DMI effectively breaks inversion symmetry, allowing the manifestation of the unique properties of skyrmions. The physics of chiral (\textit{twisted}) spin structures can be traced back to the 1960s to the works of Dzyaloshinskii and Moriya~\cite{29,30}. Theoretically, the DMI's ability to stabilize two-dimensional topological solitons, initially known as ``(chiral) magnetic vortices'', was proposed back in 1989 and became evident when the presence of a skyrmion lattice was first observed in the bulk of the B20 chiral magnet MnSi through neutron scattering experiment in reciprocal space~\cite{muhlbauer2009skyrmion38c}.\\
    The presence of DMI introduces certain symmetries that give rise to a distinct category of solutions characterized by fixed chirality and helicity. For instance, N'{e}el skyrmions [FIG.~\ref{fig:sky-fig}(a)] are favored in cases of interfacial DMI, while Bloch skyrmions [FIG.~\ref{fig:sky-fig}(b)] are preferred when bulk DMI is present. Additionally, antiskyrmions [FIG.~\ref{fig:sky-fig}(c)] can be stabilized through anisotropic DMI as reported in tetragonal Heusler compounds, where their existence was observed using Lorentz transmission electron microscopy (TEM)~\cite{hoffmann2017antiskyrmions, jena2019observation}.\\
    In the most general case, the DMI can be expressed as a microscopic Hamiltonian \begin{equation}
        \mathcal{H}_{\mathrm{DMI}} = \sum_{i,j} \mathscr{D}_{ij} \cdot (\boldsymbol{S}_i \times \boldsymbol{S}_j)
    \end{equation} where $\mathscr{D}_{ij}$ is the Dzyaloshinkskii vector\footnote{The DMI can exist in centrosymmetric crystal structures along noncentrosymmetric bonds, but its overall effect cancels out within the unit cell, except in the case of noncentrosymmetric structures.}. Depending on the lattice's underlying symmetry, the expression for the two skyrmions mentioned above can be further simplified [refer to FIG.~\ref{fig:12}] as
    \begin{equation}
    \begin{aligned} 
    \mathcal{H}_{\mathrm{DMI}}^{\mathrm {Bloch }}& =-D \sum_{i} \boldsymbol{S}_{i} \times \boldsymbol{S}_{i+\hat{x}} \cdot \hat{x}+\boldsymbol{S}_{i} \times \boldsymbol{S}_{i+\hat{y}} \cdot \hat{y} \\ 
    \mathcal{H}_{\mathrm{DMI}}^{\mathrm{N\acute{e}el}}&=-D \sum_{i} \boldsymbol{S}_{i} \times \boldsymbol{S}_{i+\hat{x}} \cdot \hat{y}-\boldsymbol{S}_{i} \times \boldsymbol{S}_{i+\hat{y}} \cdot \hat{x}
    \end{aligned}
    \end{equation}
    Initially, Dzyaloshinskii analysed the anisotropic interaction in terms of Lifshitz invariants - the lowest-order inversion-breaking terms that contain spatial derivatives
    \begin{equation}
        \mathscr{L}_{\alpha \beta}^{\gamma} = {m}_{\alpha} \frac{\partial {m}_{\beta}}{\partial x_{\gamma}} -{m}_{\beta} \frac{\partial {m}_{\alpha}}{\partial x_{\gamma} }
        \label{17}
    \end{equation}
    Here, for the sake of convenience, we use $\alpha$ and $\beta$ as covariant coordinates (representing the directions $x, y, z$) and $\gamma$ as contra-variant coordinates. Note that $\mathscr{L}_{\alpha \beta}^{\gamma}$is an antisymmetric tensor w.r.t the exchange of indices $\alpha \longleftrightarrow \beta$.\\
    The expression of DM energy density $\mathcal{W}_D$ relies on the combination of Lifshitz invariants, governed by the crystallographic point group symmetry as studied Refs.~\cite{dzyaloshinskii1964theory,bogdanov1994thermodynamically16,kataoka1981helical}
    \begin{align*}
        \mathcal{W}^{(\pm)}_1 &= \mathscr{L}^{(x)}_{zx} \pm \mathscr{L}^{(y)}_{zy}, \quad\mathcal{W}^{(\pm)}_2 = \mathscr{L}^{(y)}_{zx} \pm \mathscr{L}^{(x)}_{zy} \\
        C_{nv} & : [\mathcal{W}^{(+)}_1], ~~~\qquad D_{2d} : [\mathcal{W}^{(-)}_1] \\
        D_n & :[\mathcal{W}^{(-)}_2 \mathscr{L}^{(z)}_{xy}], \quad C_n : [\mathcal{W}^{(+)}_1, \mathcal{W}^{(+)}_2\mathscr{L}^{(z)}_{xy}]
    \end{align*}
    As demonstrated by Bogdanov and Hubert, the energy densities of skyrmions in all these structures can be reduced into a common functional form~\cite{bogdanov1994thermodynamically16}. Specifically, in uniaxial crystallographic classes, such as $C_{nv}$ (resulting in N'{e}el-type skyrmion) and $D_{n}$ (resulting in Bloch-type skyrmion), the continuous-limit DMI energies are given by the following expressions
    \begin{equation}
    \begin{aligned}
      \mathcal{W}_{\mathrm{DMI}}^{\mathrm{N\acute{e}el}} &= \langle D (\mathscr{L}_{x z}^{x}-\mathscr{L}_{y z}^{y}) \rangle \\
        \mathcal{W}_{\mathrm{DMI}}^{\mathrm{Bloch}} & = \langle D_1 (\mathscr{L}_{y z}^{x}+\mathscr{L}_{x z}^{y}) + D_2 \mathscr{L}_{x y}^{z}\rangle
    \end{aligned}
    \label{19}
    \end{equation} 
    With the energetic considerations on the basis of Eqn.~(\ref{eqn:9}), for a minimally topologically charged skyrmion $(|Q| = 1)$, the Eqn.~(\ref{19}) of both N\'{e}el and Bloch skyrmion can be written as
    \begin{equation}
        \mathcal{W}_{\mathrm{DMI}}^{\mathrm{N\acute{e}el}}= D \left( \frac{d \theta}{dr}+ \frac{\sin 2 \theta}{2r}\right);~~\mathcal{W}_{\mathrm{DMI}}^{\mathrm{Bloch}}= D \left( \frac{d \theta}{dr}+ \frac{\sin 2 \theta}{2r}\right)
    \end{equation} 
\section{\label{sec:level8}Energy Profile of Magnetic Skyrmion}
   In the theory of magnetically ordered materials within the continuum approximation, the internal energy of the system is expressed as a functional dependent on the magnetization $\mathcal{E}\left[\boldsymbol{m(r)}\right]$, rather than individual magnetic moments. This approach, introduced by W.F. Brown~\cite{brown1963micromagnefics}, is based on the variational principle\footnote{As per this principle, when the total free energy is minimized absolutely or relatively while adhering to the constraint $\boldsymbol{m^2} = 1$, we obtain the magnetization vector field $\boldsymbol{m(r)}$ (Hubert and Sch\"{a}fer 1998).} and is widely studied. The total Gibbs free energy of a magnetic system is expressed as
   \begin{equation}
        \mathcal{E}_{\mathrm{total}} = \mathrm{U}- \mu_0 \int \boldsymbol{m} \cdot \mathbf{H}_{\mathrm{ext}} \mathrm{dV}
    \end{equation}
    where $\mathrm{U}$, $\boldsymbol{m}$ and $\mathbf{H}_{\mathrm{ext}}$ denote the internal energy, local magnetization and an external magnetic field, respectively.  The internal energy includes contributions from the exchange, magnetocrystalline, surface anisotropy, asymmetric exchange, and stray field energies. In real space, the magnetization is described by a continuous vector field, where distinct components along the orthogonal directions ($x, y$, and $z$) correspond to the layers' plane and the films' normal direction.
    \begin{equation}
    \mathcal{E}_{\text{total}} = \mathcal{E}_{\text{exch}} + \mathcal{E}_{\text{dmi}} + \mathcal{E}_{\text{zee}} + \mathcal{E}_{\text{anis}} + \mathcal{E}_{\text{stray}} + \mathcal{E}_{\text{demag}}
    \end{equation}
    where $\mathcal{E}_{\text{exch}},~ \mathcal{E}_{\text{dmi}},~ \mathcal{E}_{\text{zee}},~ \mathcal{E}_{\text{anis}},~ \mathcal{E}_{\text{stray}}~ \text{and}~ \mathcal{E}_{\text{demag}}$ are the Isotropic exchange interaction energy, Dzyaloshinskii-Moriya (anisotropic) Exchange interaction energy, Zeeman energy, Magnetocrystalline anisotropy energy, Stray field energy and Demagnetisation energy, respectively. The ground state of the system is determined by minimizing the energy, subject to the equilibrium equation,
    \begin{equation}
         \delta \mathcal{E}[\boldsymbol{m(r)}] = 0
     \end{equation}
    Within the micromagnetic approach, it is important to identify the different energy contributions to the system. These magnetic energy densities can be expressed as functions of the components of the magnetization $\boldsymbol{m}$ and their derivatives, derived from the initial Hamiltonian descriptions of the interactions and considering all magnetic moments and pairs of moments.\\

    \noindent \textbf{Exchange Energy ($\mathcal{E}_{\mathbf{exch}}$)}: Heisenberg introduced the concept of exchange interaction in 1928 to explain the strong magnetic field observed in ferromagnetic materials. Dirac later formulated the atomistic expression for the interaction energy between neighboring sites as $-2\mathcal{J}_{ij}{m}_i\cdot {m}_j$, where $\mathcal{J}_{ij}$ represents the exchange integral between sites $i$ and $j$. This interaction decays rapidly with distance, accounting for its short-range nature.
    \begin{figure}[b]
        \centering
        \includegraphics[width = \linewidth]{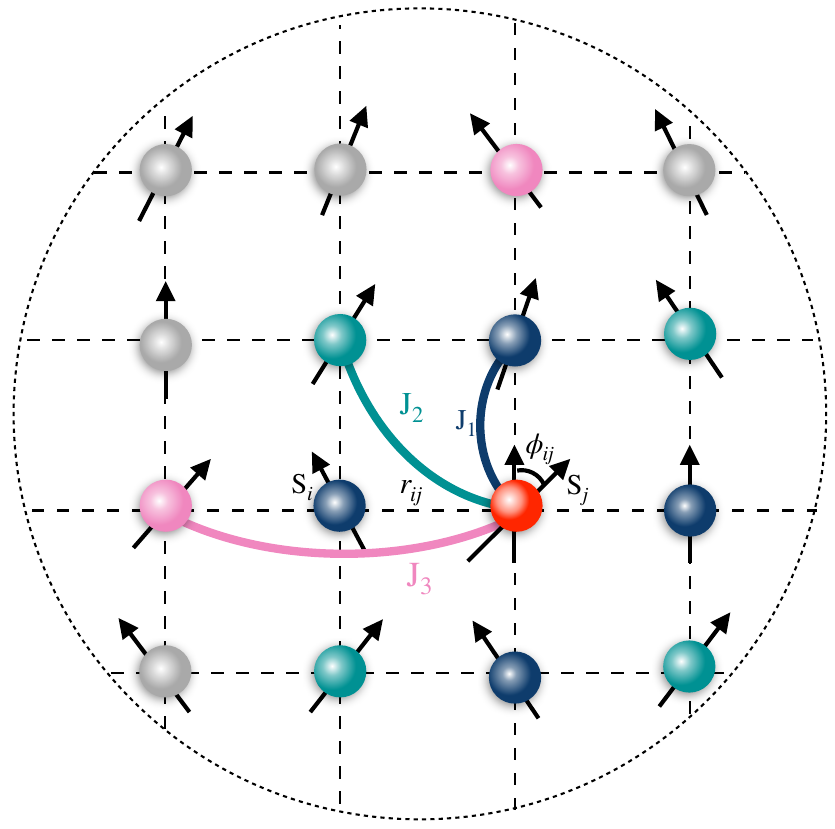}
        \caption{\textbf{A simple cubic lattice}, with nearest neighbours interactions.}
        \label{fig:5}
    \end{figure}
    The energy arising from the exchange interaction in a system of $\mathrm{N}$ spins can be expressed as follows
    \begin{equation}
        {\mathcal{H}}_{\mathrm{exch}} = -\sum_{ij} \mathcal{J}_{ij} \mathrm{S}_i \cdot \mathrm{S}_j
        \label{25}
    \end{equation}
    where $\mathcal{J}_{ij}$ is the tensorial exchange coupling between atomic sites $i$ and $j$ arises from the spatial overlap of the electron wave functions at these sites
    \begin{equation}
        \mathcal{J}_{ij} = \begin{bmatrix}
     \mathrm{J}^{xx}_{ij} & \mathrm{J}^{xy}_{ij} & \mathrm{J}^{xz}_{ij} \\
    \mathrm{J}^{yx}_{ij} & \mathrm{J}^{yy}_{ij} & \mathrm{J}^{yz}_{ij} \\
    \mathrm{J}^{zx}_{ij} & \mathrm{J}^{zy}_{ij} & \mathrm{J}^{zz}_{ij}
    \end{bmatrix}
    \end{equation}
    The matrix, represented as $\mathrm{J}^{\alpha \beta}_{ij}$ with $\alpha, \beta = x, y, z$, can be decomposed into three distinct interactions~\cite{ex3}
    \begin{equation}
        \mathcal{J}_{ij} = \mathrm{J}_{ij} \mathcal{I}+\mathcal{J}^S_{ij}+\mathcal{J}^A_{ij}
    \end{equation}
    where $\mathcal{I}$ is the identity matrix and $\mathrm{J}_{ij}$ represents the \textit{isotropic exchange interaction or classical Heisenberg exchange}
    \begin{equation}
        \mathrm{J}_{ij} = \frac{1}{3} \mathrm{Tr}(\mathcal{J}_{ij})
    \end{equation}
    The isotropic exchange can be referred to as ferromagnetic (FM) if it favours parallel spins ($\mathrm{J}_{ij} > 0$) or antiferromagnetic (AFM) if it favours anti-parallel spins ($\mathrm{J}_{ij} < 0$).
    For a small misalignment between neighbouring spins, Eqn.~(\ref{25}) can be re-written as
    \begin{equation}
        {\mathcal{E}}_{\mathrm{exch}} = -\sum_{ij} \mathcal{J}_{ij} \mathrm{S}^2 \cos \phi_{ij}
    \end{equation}
    where $\phi_{ij}$ is the angle between spins at sites $i$ and $j$ and $r_{ij}$ represents the distance between them, as sketched in FIG.~\ref{fig:5}.\\
    The exchange energy can be simplified by considering a basic cubic lattice and using a uniform coupling constant $\mathcal{J}_{ij}$ for all neighboring interactions as
    \begin{equation}
        \mathcal{E}_{\mathrm{exch}} = \mathcal{A} \int \left[(\nabla {m}_x)^2 +(\nabla {m}_y)^2+(\nabla {m}_z)^2\right]{dV}
    \end{equation}
    The exchange stiffness, $\mathcal{A}$, is given by
    \[\mathcal{A} = \frac{2\mathcal{J}\mathrm{S}^2n}{a}\]
    where $\mathcal{J}$ is the exchange constant, $S$ is the spin quantum number, $n$ is the number of sites in a unit cell, and $a$ is the nearest-neighbor distance.
    In transition metal ferromagnets (FMs), the typical range of $\mathcal{A}$ is $5-30~\si{\pico\joule\per\meter}$ while in the bulk limit, $\mathcal{A}$ is approximately $30~\si{\pico\joule\per\meter}$ for Co, $20~\si{\pico\joule\per\meter}$ for Fe, and $10~\si{\pico\joule\per\meter}$ for Ni~\cite{100,101}.\\
    
    \noindent \textbf{RKKY Interaction} ($\mathcal{E}_{\mathbf{RKKY}}$):
        \begin{figure}[b]
        \centering
        \includegraphics[width = \linewidth]{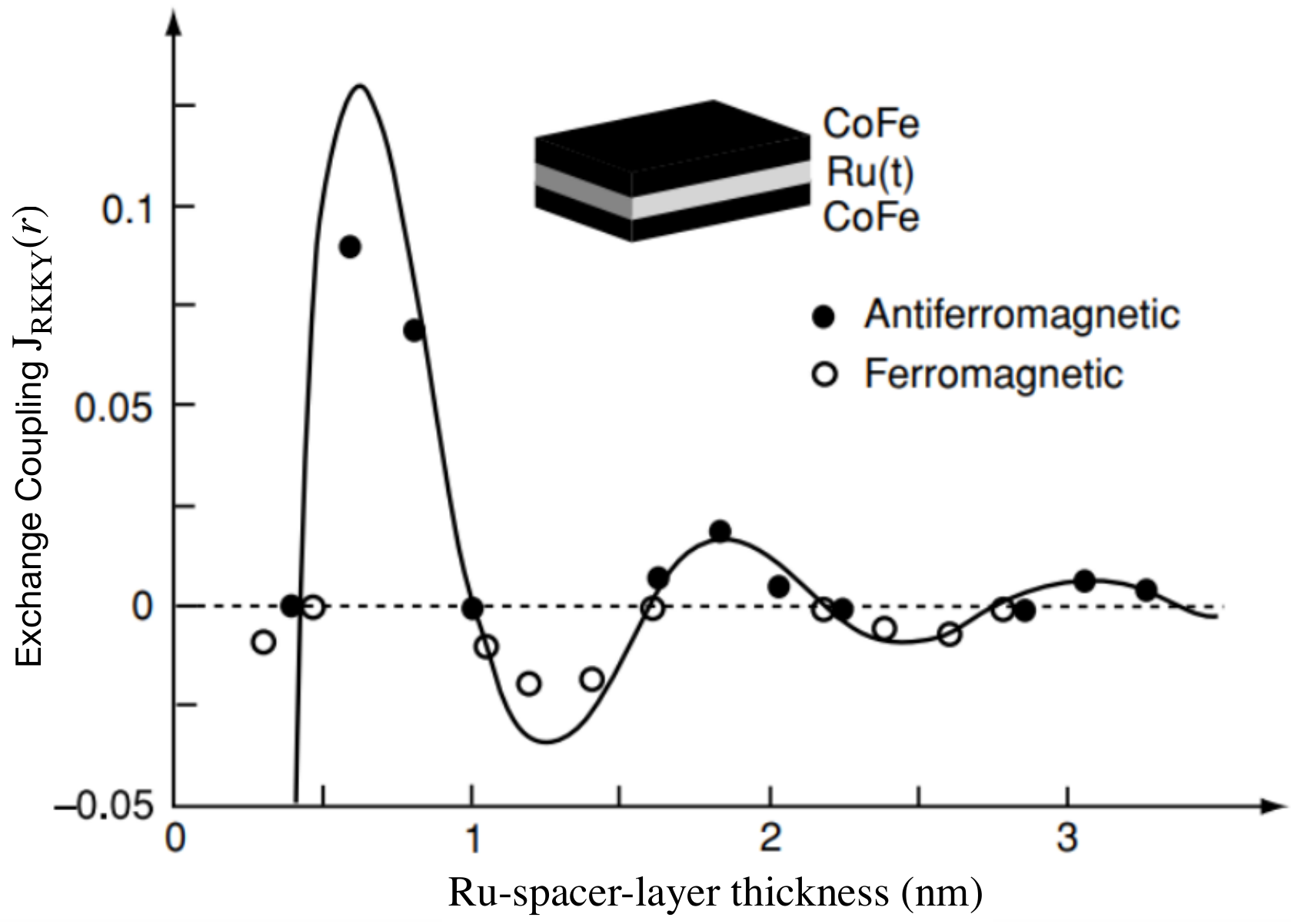}
        \caption{\textbf{RKKY Interaction.} Oscillating exchange coupling in a CoFe-Ru-CoFe trilayer, from Ref.~\cite{ie}.}
        \label{fig:6}
    \end{figure}In metallic ferromagnets, in addition to the direct exchange interaction between nearest-neighbour atoms, there exists an important indirect exchange interaction known as the Ruderman-Kittel-Kasuya-Yosida (RKKY) interaction. This interaction was first proposed by Ruderman and Kittel in 1954~\cite{12} and later expanded upon by Kasuya and Yosida~\cite{13,14}. The RKKY interaction involves localized electrons ( $d$ or $f$) with magnetic moments influencing the magnetization of delocalized conduction electrons ($s$).
    This polarization of conduction electrons, in turn, couples to neighboring localized spin moments. The sign and approximate magnitude of the RKKY interaction is given in Ref.~\cite{10} as
    \begin{equation}
        \mathcal{J}_{\mathrm{RKKY}}(r)  = \frac{\sin x- x \cos x}{x^4} \rightarrow \frac{1}{x^3}(x \rightarrow \infty)
    \end{equation}
    where $x = 2k_F r$, $k_F$ is the Fermi sphere radius and $\mathcal{J}_{\mathrm{RKKY}}(r)$ is the separation dependent RKKY exchange constant. The oscillating exchange coupling variation is shown in FIG.~\ref{fig:6} for the multilayer stack observed in~\cite{ie}. \\
    \noindent \textbf{Dzyaloshinskii-Moriya Interaction (DMI)} ($\mathcal{E}_{\mathbf{dmi}}$): The Dzyaloshinskii-Moriya interaction (DMI), is an antisymmetric, anisotropic exchange term which arises in low symmetry crystals due to SOC, often leading to tilting of adjacent spins by a small angle. This represents a portion of the overall magnetic exchange interaction between two adjacent magnetic spins, denoted as $\mathrm{S}_1$ and $\mathrm{S}_2$. The system's Hamiltonian is provided as follows
    \begin{equation}
        \mathcal{H}_{\mathrm{DMI}} = \mathscr{D}_{12} \cdot \mathrm{S}_1 \times \mathrm{S}_2
    \end{equation}
    where $\mathscr{D}_{12}$ is the resulting DMI vector, which is perpendicular to the plane of the triangle (as illustrated in the FIG.~\ref{fig:8}). \\
    In the Heisenberg model, the total DMI energy of a system of ${N}$ interacting spins is expressed as
    \begin{equation}
        \mathcal{E}_{\mathrm{DM}} = -\sum_{<ij>}^N \mathscr{D}_{ij} \cdot ({m}_i \times {m}_j)
        \label{34}
    \end{equation}
    The DMI described by the Dzyaloshinskii vector, denoted as $\mathscr{D}_{ij}$ is typically much weaker than the isotropic exchange interaction. The orientation of this vector depends on the crystal lattice symmetries, and it becomes zero when there is a center of inversion between two sites, labelled as $i$ and $j$. According to Eqn.~(\ref{34}), the energy contribution associated with the DMI is minimized when neighboring spins are perpendicular to each other, such as the spin configurations $\uparrow \rightarrow \downarrow$ or $\uparrow \leftarrow \downarrow$, with the rotation direction determined by the orientation of $\mathscr{D}_{ij}$. Consequently, the DMI competes with other interactions, particularly the Heisenberg Exchange, to induce spin rotations, leading to various non-collinear magnetic configurations characterized by unique magnetic textures and chirality. These configurations include helicoids, conical and skyrmions, which are the main focus of this study. Furthermore, the DMI also removes degeneracy by favouring a specific chirality in existing spin textures, such as those observed in domain walls and magnetic vortices.\\
        \begin{figure*}[t]
        \centering
        \includegraphics[width = \linewidth]{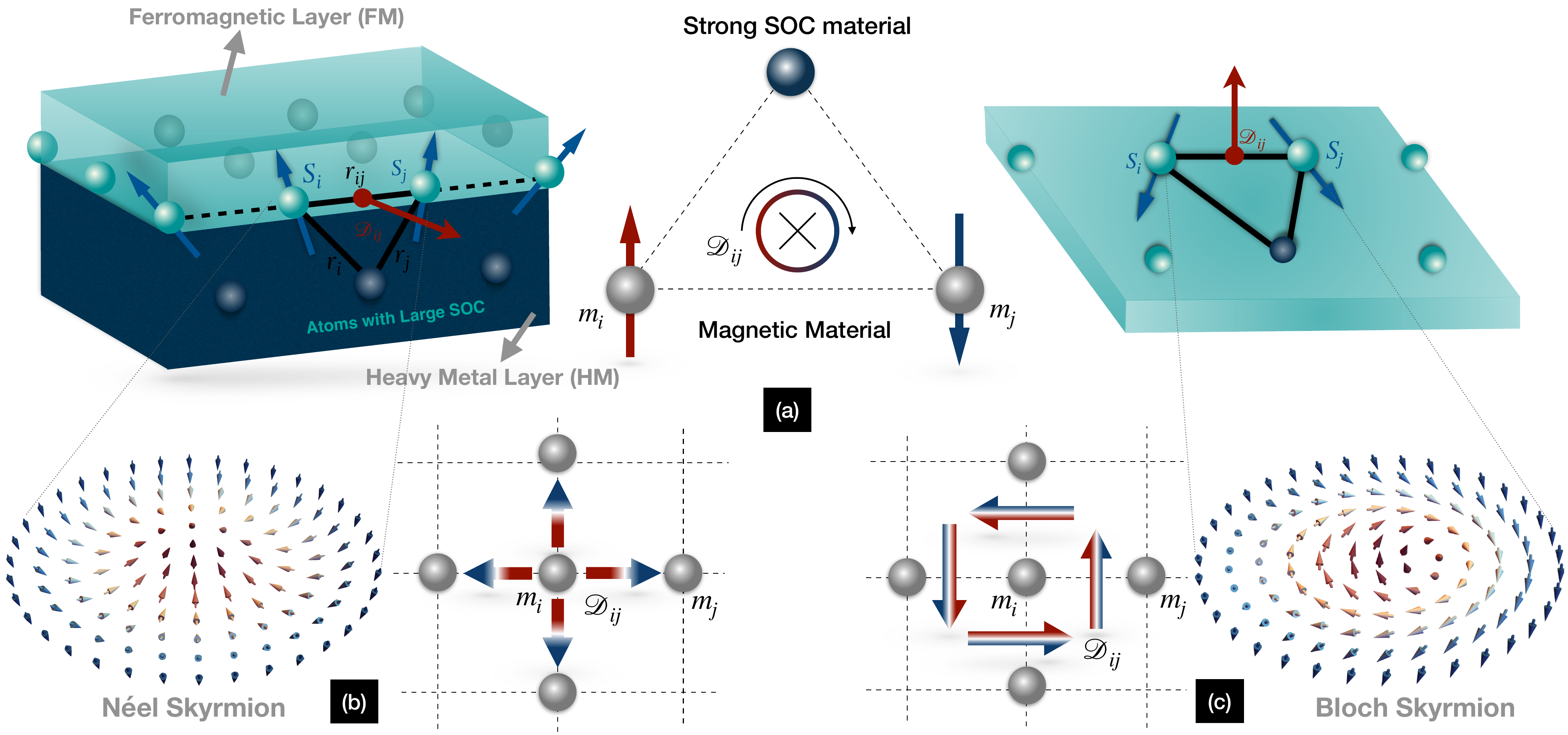}
        \caption{\textbf{Interfacial and Bulk DMI.} (a)~DMI induced by the vertical breaking of the inversion symmetry at an interface between a magnetic material and a material exhibiting strong SOC, 
        (b)~Interfacial DMI between a ferromagnetic layer (top) and a heavy layer (bottom) (c) Bulk DMI in Bulk material with horizontal inversion breaking. Inspired from~\cite{fert2013skyrmions}}
        \label{fig:8}
    \end{figure*}
   DMI was originally introduced by Dzyaloshinskii~\cite{29}, who showed that the interplay of spin-orbit coupling and low symmetry results in an antisymmetric exchange contribution. Subsequently, Moriya~\cite{30} proposed a microscopically derived mechanism that accounts for the existence of this term in systems featuring spin-orbit coupling. The presence of DMI is notably observed in crystals with reduced symmetry, especially in non-centrosymmetric magnetic crystals. Its effects were initially observed in weak ferromagnetism in antiferromagnetic compounds $\alpha-\mathrm{Fe_2O_3}$ and $\mathrm{CrF_3}$ and later in chiral bulk magnets like MnSi~\cite{87,88}. An external mechanism can induce the Dzyaloshinskii-Moriya interaction (DMI) even in crystals with centrosymmetric symmetry. The absence of inversion symmetry at the surface of thin films results in interfacial DMI ($i-\mathrm{DMI}$)~\cite{14, 32, 34, 89}. This interfacial DMI becomes particularly notable in small magnetic structures like thin films, multilayers, nanowires, and nanodots.\\
    In contrast, the robust Dzyaloshinskii-Moriya interaction (DMI) in such systems arises from the combination of a thin ferromagnetic film (e.g., Co, Fe) with a non-magnetic heavy material displaying a substantial spin-orbit coupling (SOC) (e.g., Ir, Pt). The direction of the $\mathscr{D-}$vector relies on the system's symmetries. In thin films with broken vertical inversion symmetry, the $\mathscr{D-}$vector aligns perpendicular to the displacement vector between atomic sites, thereby stabilizing N'{e}el-type skyrmions [refer to FIG.~\ref{fig:8}(a)]~\cite{89}. On the other hand, in bulk materials where the horizontal inversion symmetry is broken, the $\mathscr{D-}$vector aligns with the displacement vector, resulting in the stabilization of Bloch-type skyrmions [see FIG.~\ref{fig:8}(b)]~\cite{89}. The strength, sign, and direction of the DM interaction determine the possibility of stabilizing chiral skyrmions and their characteristics~\cite{wang2018theory30, nagaosa2013topological36}. Vector plots of Bloch and N\'{e}el-type skyrmions show distinct spin distributions, but they can be mathematically mapped onto a unit sphere, indicating their topological equivalence [see FIG.~\ref{fig:sky-fig} and \ref{topo-soliton}].
    The expressions of dimensionless DMI fields with symmetry breaking along $z$ for bulk and interfacial can be quantified respectively,
    \begin{equation}
    \begin{aligned}
    \mathcal{H}_{b \text{-DMI }}&=-\frac{1}{\mu_{0} M_{\mathrm{s}}^{2}} \frac{\delta \mathcal{E}_{\mathrm{BulkDMI}}}{\partial \boldsymbol{m}}=-\frac{2 D}{\mu_{0} M_{\mathrm{s}}^{2}} [\nabla \times \boldsymbol{m}] \\
    \mathcal{H}_{i \text{-DMI }}&=-\frac{1}{\mu_{0} M_{\mathrm{s}}^{2}} \frac{\delta \mathcal{E}_{i-\mathrm{DMI}}}{\partial \boldsymbol{m}}=\frac{2 D}{\mu_{0} M_{s}}\left[\nabla {m}_{z}-(\nabla \cdot \boldsymbol{m}) \widehat{z}\right]
    \end{aligned}
    \end{equation}
     where the DMI coupling constant is expressed in units of energy per unit area $\mathrm{(J/m^2)}$, with $\mathscr{D} = 2\frac{D}{a^2}$ on the simple cubic lattice. Nevertheless, the full variational calculus shows that DMI also affects the system boundary conditions.
    Experimentally, researchers have observed various effects of the DMI including phenomena such as chiral domain walls~\cite{34}, asymmetric bubble expansion~\cite{35,36}, asymmetric propagation of spin waves~\cite{37}, and the emergence of new types of magnetization patterns like Skyrmions and helices etc. When the DMI is sufficiently strong, a spin spiral state with the spiral axis lying in the plane becomes favorable. On the other hand, for lower values of the DMI strength (${D}$), the ground state corresponds to a lattice of Skyrmions~\cite{rohart2013skyrmion}. Although we will see that the DMI is not a necessary requirement for obtaining skyrmion spin structures with a specific topology, as they can be stabilized by dipolar interactions in thin films, the presence of the DMI does modify the spin structure.\\
    
    \noindent \textbf{External Field (Zeeman) Energy} ($\mathcal{E}_{\textbf{Zee}}$): The magnetic field energy can be separated into two parts: the external and the stray field energies. The first part is the interaction energy between the magnetization vector field and an external field $\mathbf{B} = \mu_0 \mathbf{H}$, which can be expressed as
    \begin{equation}
         \mathcal{E}_{\mathrm{Zee}} = - \mu_0 \mathrm{M}_s \sum_{i = 1}^{N} {m}_i \cdot \mathbf{H}
    \label{38}
    \end{equation}
    where $\mu_0 = 4\pi \times 10^{-7}$ $\mathrm{N/A^2}$ is the permeability of free space, $\mathbf{H} \in \mathbb{R}^3$ \si{A/m} is the magnetic $\mathbf{H-}$field with strength equal to the magnitude of the vector. The energy is minimized when the magnetic moments align with the strong external magnetic field $\mathbf{H}$, effectively disrupting the spin-orbit coupling (SOC) of electrons. In this case, the energy contribution from the magnetic field dominates over other energy terms.\\
    In experiments, the magnetic moment field can be controlled by adjusting the strength and direction of an external magnetic field. When the external field is uniform, the energy of the system is determined solely by the average magnetization and is not affected by the shape or domain structure of the sample. Exchange interactions among magnetic moments cause them to twist relative to each other but without a specific preferred direction. However, when combined with an external magnetic field, these exchange effects not only lead to the formation of chiral micromagnetic structures but also influence their type and alignment.\\
            \begin{figure*}[t]
        \centering
        \includegraphics[width = \linewidth]{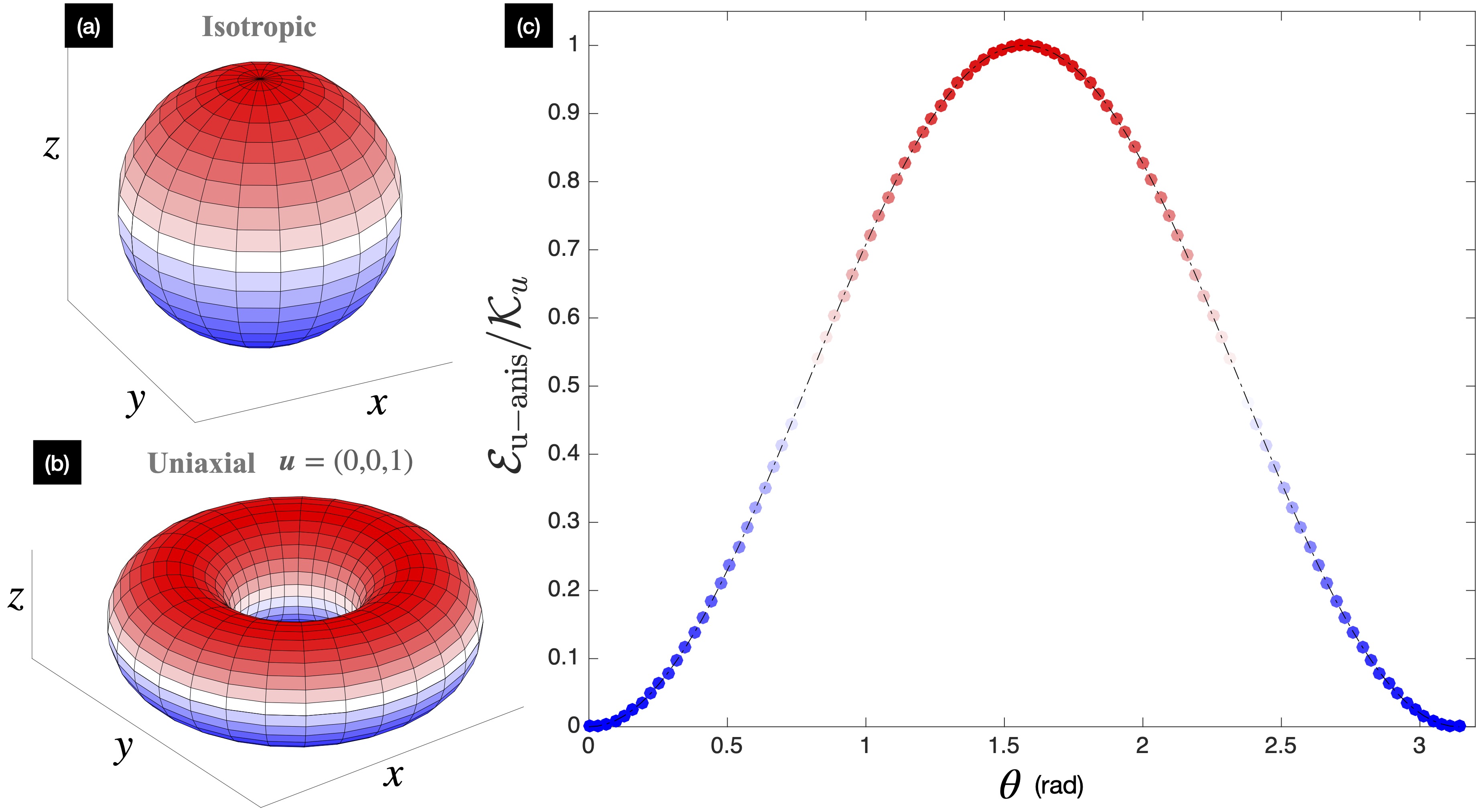}
        \caption{\textbf{3D representation and Energy Plot.} (a) Isotropic energy, any direction of the magnetization has the same energy. (b)~Uniaxial anisotropy energy minimum along the $-z$ direction. (c)~Uniaxial anisotropy energy for $\mathcal{K}_u = (0, 0, 1) $. }
        \label{fig:9}
    \end{figure*}
    
    \noindent \textbf{Dipolar or Demagnetisation Energy}($\mathcal{E}_{\textbf{demag}}$): The magnetostatic energy term in a material arises from the magnetic field generated by the sample itself, rather than an external field. Just like how a bar magnet generates a magnetic field around itself, the magnetic moments within a material lattice are affected by the magnetic fields ($\mathbf{H}_\mathrm{d}$) produced by all other neighboring magnetic moments. The interaction potential between any two magnetic moments, denoted as $m_i$ and $m_j$, can be described by
    \begin{equation} \mathcal{E}_{ij}^{\mathrm{dip}}=\frac{\mathrm{I}^{\mathrm{dip}}}{\left|\mathbf{r}_{{ij}}\right|^3}\left[{m}_{i}\cdot{m}_{j}-3\left({m}_{i}\cdot\hat{{r}}_{ij}\right)\left({m}_{j}\cdot\hat{{r}}_{ij}\right)\right]
    \end{equation} where $\mathbf{r}_{ij}=\mathbf{r}_j-\mathbf{r}_i,\hat{{r}}_{ij}=\mathbf{r}_{ij}/\left|\mathbf{r}_{ij}\right|$, and $\mathrm{I}^{\mathrm{dip}}$ is the dipolar coupling strength. In the limit of continuous systems, the coupling between the magnetization and the demag field, generated by the material itself, contributes to the magnetostatic energy. 
    A more convenient approach is to use Maxwell's equations to calculate the magnetic field generated by a certain magnetic configuration. The field, $\mathbf{H}_\mathrm{d}$, in absence of any currents, or displacement currents~\cite{brown1962magnetostatic45}
    \begin{equation}
    \begin{aligned}
         \nabla \times \mathbf{H}_\mathrm{d} & = 0  \\
         \nabla \cdot (\mathbf{H}_\mathrm{d}+\boldsymbol{m}) & = 0
    \end{aligned}
    \end{equation}
    As the curl of $\mathbf{H}_\mathrm{d}$ is zero the field is the gradient of a scalar potential $\Phi$,
    \begin{equation}
        \mathbf{H}_\mathrm{d} = -\nabla \Phi
        \label{42}
    \end{equation}
    The magnetic field induction ($\mathbf{B-}$field) is given by $\mathbf{B} = \mu_0(\mathbf{H}_\mathrm{d}+\boldsymbol{m})$. Combining these equations yields \begin{equation}
    \begin{aligned}
         \nabla^2 \Phi_{\mathrm{in}} & = \nabla \cdot \boldsymbol{m} \\
          \nabla^2 \Phi_{\mathrm{out}} & = 0
    \end{aligned}
    \end{equation} 
    Maxwell's equations require that both the component of $\mathbf{H}_\mathrm{d}$ parallel to the surface of the boundary and the component of $\mathbf{B}$ perpendicular to the surface are continuous over the boundary of a material. With required boundary conditions for the scalar potential, the energy can then be evaluated as\begin{equation}
        \mathcal{H}_{\mathrm{demag}} = -\frac{1}{2} \int (\mathbf{H}_\mathrm{d} \cdot \boldsymbol{m}) \mathrm{d}V
    \end{equation}The magnetic scalar potential as expressed in Eqn.~(\ref{42}) gives
    \begin{equation}
        \mathbf{H}_\mathrm{d} = -(\nabla \cdot \boldsymbol{m}) = -\rho_{\boldsymbol{m}}
    \end{equation}
    So, in terms of volume charge density $(\boldsymbol{\rho}_{\boldsymbol{m}} = -\nabla \cdot \boldsymbol{m})$ and surface $(\boldsymbol{\sigma}_{\boldsymbol{m}} = \boldsymbol{n} \cdot \boldsymbol{m})$ poles, analogous to their electric counterparts:
    \begin{equation}
    \Phi(\boldsymbol{r})=\frac{1}{4 \pi}\left[\int_{V} \frac{\nabla \cdot \boldsymbol{m}\left(\boldsymbol{r}^{\prime}\right)}{\left|\boldsymbol{r}-\boldsymbol{r}^{\prime}\right|} d V^{\prime}+\int_{S} \frac{\boldsymbol{n} \cdot \boldsymbol{m}\left(\boldsymbol{r}^{\prime}\right)}{\left|\boldsymbol{r}-\boldsymbol{r}^{\prime}\right|} d S^{\prime}\right]
    \end{equation}
    where $\boldsymbol{r}^{\prime}$ is the variable to be integrated over the volume and surface, respectively.\\
    
\begin{figure}[t]
        \centering
        \includegraphics[width = \linewidth]{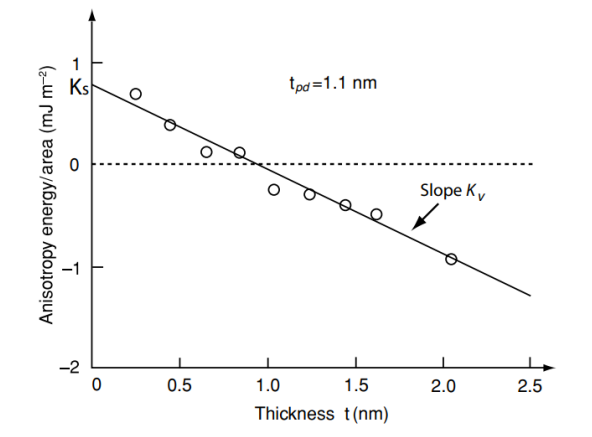}
        \caption{\textbf{Anistropy Energy/Area.} Plot to determine the surface and volume anisotropy terms for a Co-Pd multilayer, from Ref.~\cite{den1991magnetic5}.}
        \label{fig:122}
    \end{figure}

    \noindent \textbf{Magnetic Anisotropy Energy} ($\mathcal{E}_{\mathbf{anis}}$): Magnetocrystalline Anisotropy (MA) is a characteristic behavior observed in materials where the arrangement of atoms in a crystal structure leads to variations in magnetization. It arises from the preferential alignment of magnetic moments along specific directions due to the influence of electron orbitals. The primary cause of magnetocrystalline anisotropy is the presence of SOC, which arises from the relativistic interaction between spin and orbital motion. At the micromagnetic level, the exchange energy is uniform and directionless, although the atomistic exchange coupling can exhibit anisotropic behavior depending on the crystal lattice directions.\\
    In micromagnetic studies, the anisotropy energy is characterised as a function expressed below that depends on the orientation of the magnetization and the direction of the east axis
    \begin{equation}
        \mathcal{E}_{u-\mathrm{Anis}} = f(\boldsymbol{m}(r), \hat{\boldsymbol{u}}) \Delta V
    \end{equation}
    where $\boldsymbol{m}(r) = \mathbf{M}(r)/\mathbf{M}_s$ and $\hat{\boldsymbol{u}}$ represents the anisotropy direction.\\
    \textit{Uniaxial anisotropy} aligns the magnetization along a preferred axis in which the energy is minimized.
    \begin{equation}
        \mathcal{E}_{u-\mathrm{Anis}} = \mathcal{K}_u\left(1-(\boldsymbol{m}(r) \cdot \hat{\mathbf{u}} )^2\right)\Delta V
        \label{Eq:48}
    \end{equation}
    It is characterized by the uniaxial anisotropy constant $\mathcal{K}_u$, which quantifies the strength of this anisotropy and is measured in units of $\mathrm{J/m^3}$. FIG.~\ref{fig:9}(b) illustrates a 3D depiction of the anisotropy energy, specifically for $\mathcal{K}_u = (0, 0, 1)$, while FIG.~\ref{fig:9}(a) displays the isotropic energy. Notably, the energy exhibits a minimum along the $z$ direction. Employing polar coordinates, we can express Eqn.~(\ref{eqn:9}) due to the unitary nature of $\boldsymbol{m}(r)$. Thus
    \begin{equation}
        \mathcal{E}_{u-\mathrm{Anis}} = \mathcal{K}_u \sin^2 \theta
    \end{equation}
    The energy plot in FIG~\ref{fig:9}(c) exhibits two minima at $\theta = 0$ and $\theta = \pi$, representing full alignment of the magnetization along the $z$ direction, which corresponds to the direction of the anisotropy.\\
    \textit{Shape anisotropy} is a type of magnetic anisotropy that arises from the shape of a magnetic crystal or structure, rather than its crystal structure in magnetocrystalline anisotropy. It is caused by the anisotropic dipolar interactions among magnetic poles, including demagnetization and stray fields. Shape anisotropy tends to align magnetic elements in a way that minimizes the magnetostatic energy, favoring magnetization directions parallel to the surfaces and avoiding perpendicular orientations that would generate surface magnetic charges. It is particularly relevant in magnetic structures with different shapes, such as magnetic disks, ellipses, or stripes. Shape anisotropy plays a significant role in determining the preferred direction of magnetization in these structures, influencing phenomena such as the vortex state in disks or the alignment of magnetization along the axis of magnetic stripes.\\
    Similarly, \textit{Surface} and \textit{Interface anisotropy} in magnetic materials refer to the alteration of magnetocrystalline anisotropy near surfaces or interfaces. 
    This effect is a direct consequence of electron spin and orbital motion i.e., SO interaction, leading to disruptions in the crystal lattice. Surface anisotropy was initially proposed by N\'{e}el in 1953 and arises from the interaction between surface atoms and the anisotropic crystal field. The lack of symmetry at surfaces and interfaces in thin films and magnetic multilayers is the primary factor contributing to this phenomenon. Surface atoms, particularly those in the first few layers, undergo structural relaxation perpendicular to the surface, leading to modifications in the magnetic properties. This disruption in the crystal structure affects the magnetocrystalline anisotropy, influencing the preferred magnetization direction near the surface or interface. Understanding and controlling surface and interface anisotropy is essential for manipulating the magnetic behaviour and stability of thin films and multilayers.\\
    In ultrathin films, it becomes crucial to write magnetic anisotropy in surface and volume terms with their anisotropy constants represented by $\mathcal{K}_s$ and $\mathcal{K}_v$ respectively.
    \begin{equation}
        \mathcal{K}_u = \mathcal{K}_v+\frac{\mathcal{K}_s}{t}
    \end{equation}
    The interplay between these terms leads to a dependence of magnetization on the film thickness $t$. As thickness increases to a critical thickness ($t_c = -\frac{2\mathcal{K}_s}{\mathcal{K}_v}$), the magnetocrystalline anisotropy term dominates, causing the magnetization to switch to a perpendicular direction. On the other hand, in thicker films, the primary contribution comes from shape anisotropy which governs the preferred orientations of the film's magnetization~\cite{coey2010magnetism43,johnson1996magnetic54,parlak2015thickness53}.
    Furthermore, by plotting the anisotropy energy per unit area against the thickness $t$ for multiple films with shared easy axes, both  $\mathcal{K}_v$ and $\mathcal{K}_s$ can be inferred [refer to FIG.~\ref{fig:122}]. \\
    Clearly, as $t$ is reduced the surface term starts to dominate and can explain why a rotation from in-plane to out-of-plane magnetisation is seen as the film thickness is reduced.\\
    For a cubic crystal, surface anisotropy in an isotropic case
    \begin{equation}
        \mathcal{E}_s = \mathcal{K}_s\left[1- (\boldsymbol{m}\cdot \boldsymbol{n})^2\right] = \mathcal{K}_s \sin^2{(\theta_{\text{n}})},
    \end{equation}
    where $\mathcal{K}_{s}$ is anisotripic constant and $\text{n}$ is the surface normal. If $\mathcal{K}_{s} > 0$, $\mathcal{E}_s$ is the minimum for the magnetization perpendicular to the surface. This kind of energy can also exist at interfaces between ferromagnetic and nonmagnetic materials.
\section{Magnetic Domain Walls (DWs)}
\label{sec:level22}
\begin{figure}[b]
        \centering
        \includegraphics[width = \linewidth]{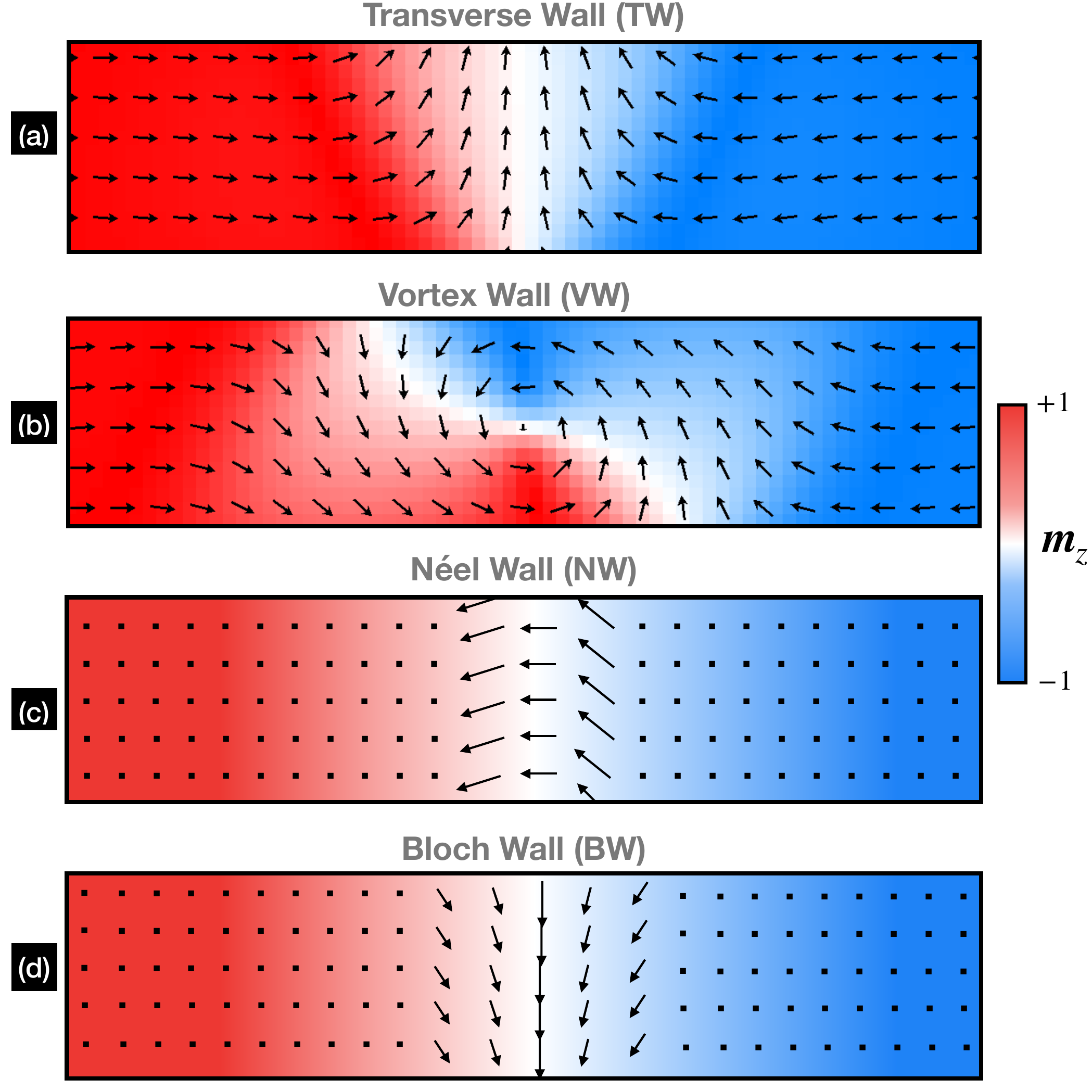}
        \caption{\textbf{Domain Walls.}~(a) Transverse wall (b) Vortex wall (c) Bloch Wall and (d) N\'{e}el Wall. The $z-$ component of magnetization is depicted in the colour scheme.}
        \label{fig:13}
    \end{figure}

    In this paper, the focus will revolve around magnetic Domain Walls(DWs), which are specific configurations found in ferromagnetic crystals. Magnetic domain walls originally proposed by Weiss can be seen as static topological solitons~\cite{yoshimura2016soliton}. These walls minimize the magneto-static energy of the sample by reducing magnetic charges on the surface, although they increase the exchange energy by creating non-parallel spin regions. The formation of domains is favoured in larger systems where the decrease in magnetostatic energy dominates over the increase in exchange energy, extensively analysed in our previous paper Ref.~\cite{mishra2020cross}. \\
    Different types of DWs exist in ferromagnetic materials, including Bloch walls~\cite{bloch1932theorie93}, N\'{e}el walls~\cite{103}, transverse walls~\cite{kravchuk2014influence}, vortex walls~\cite{landeros2009equilibrium}, chiral wall~\cite{ryu2013chiral} etc. The orientation of magnetization in ferromagnetic materials can be either in-plane or out-of-plane, depending on the thickness of the magnetic layer and the surrounding layers. Each structure exhibits specific types of domain walls.
    When considering structures like nano-strips,  with in-plane magnetization, two types of domain walls can be observed- transverse wall (TW) and vortex wall (VW), see FIG.~\ref{fig:13}(a-b). And in an out-of-plane system, N\'{e}el wall (NW) and Bloch wall (BW), see FIG.~\ref{fig:13}(c-d). N\'{e}el walls have magnetization pointing in the $x-$direction within the wall, while Bloch walls have magnetization pointing in the $y-$direction within the wall. The type of wall that is stabilized depends on the cross-section of the magnetic strip. In the case of wider magnetic stripes, the presence of more volume charges leads to increased energy of N\'{e}el walls and favours the stabilization of Bloch walls. In what follows, we highlight the characteristic quantities in a one-dimensional chain of magnetic moments along the $x-$axis, with an easy axis along the $z-$direction -i.e., wall profile, wall width, and wall energy based on Ref.~\cite{craik1995magnetism}.\\
    The total energy density per unit area of the wall is given by the sum of the exchange energy contribution $\mathcal{E}_{\mathrm{exch}}$, and the anisotropy energy contribution $\mathcal{E}_{\mathrm{anis}}$
    \begin{equation}
        \mathcal{E} = \mathcal{E}_{\mathrm{exch}}+\mathcal{E}_{\mathrm{anis}} = \int_{-\infty}^{\infty}\left[\mathcal{A} \left(\frac{\partial \theta}{\partial x}\right)^2 + f(\theta)\right] dx
        \label{TWWAlls}
    \end{equation}
    Doing the variational study of the profile, the energy expression can be quantified for both Bloch Wall(BW) and N\'{e}el Wall (NW). The energy of a Bloch ($\pi-$)Wall
    \begin{equation}
        \mathcal{E}_{\mathrm{BW}} = 2 \sqrt{\mathcal{A} \mathcal{K}} \int_{0}^{\pi} \sin \theta d\theta = 4 \sqrt{\mathcal{A} \mathcal{K}} 
        \label{Eqn. DW}
    \end{equation}
    The wall profile
    \begin{equation}
        x(\theta) = \sqrt{\frac{\mathcal{A}}{\mathcal{K}}} \int_{\pi/2}^{0}\frac{d \theta'}{\sin \theta'} = \sqrt{\frac{\mathcal{A}}{\mathcal{K}}} \ln{\left(\tan \frac{\theta}{2}\right)}
    \end{equation}
    The characteristic (asymptotic) width of the wall $\delta$
    \begin{equation}
        \delta_{\mathrm{BW}}  = \sqrt{\frac{\mathcal{A}}{\mathcal{K}}}
    \end{equation}
    and \begin{equation}
        \theta (x) = 2 \arctan \left(e^{x/{\delta_{\mathrm{BW}}}}\right)
    \end{equation}
    Similarly, for N\'{e}el Wall (NW),
    the energy of the wall, \begin{equation}
        \mathcal{E}_{\mathrm{NW}} = 4 \sqrt{\mathcal{A}\left(\mathcal{K} + \frac{1}{2} \mu_0 M^2_s\right)}
        \label{Eqn.NW}
    \end{equation}
    and wall width, \begin{equation}
            \delta_{\mathrm{NW}} = \frac{1}{\left(\frac{\mathcal{K}}{\mathcal{A}}+\frac{\mu_0 M^2_s}{2\mathcal{A}}\right)^{1/2}}
    \end{equation}
    Eqns.~(\ref{Eqn. DW}) and (\ref{Eqn.NW}) show that the N\'{e}el wall has a larger wall energy than the Bloch wall in bulk materials.\\
    
    \noindent \textbf{\label{sec:level24}Effect of DMI or Chiral DWs:}
    The addition of DMI (Dzyaloshinskii-Moriya interaction) in the Eqn.~(\ref{TWWAlls}) can stabilize N\'{e}el domain walls (DWs) with \textit{a fixed chirality}\footnote{Fixed chirality means that the magnetization rotates consistently in the same direction when transitioning between up and down domains. The conventional right-handed chirality with positive DMI $(\mathcal{D} > 0)$ and left-handed chirality with negative DMI $(\mathcal{D} < 0)$.}, even in cases where Bloch DWs would be favoured, [see FIG.~\ref{fig:sky-fig}]. \\
    In the presence of DMI, the total micromagnetic energy of the system
    \begin{equation}
        \mathcal{E}[\theta(x)]= \left[ \mathcal{A} \left(\frac{\partial \theta}{\partial x}\right)^2 +\mathcal{D}\frac{\partial \theta}{\partial x}  +\mathcal{K}_{\mathrm{eff}}  \sin^2\theta  \right] dx
    \end{equation}
    where $\mathcal{K}_{\mathrm{eff}}  = \mathcal{K} -1/2 \mu_0 M^2_s$ is an effective anisotropy which takes into account the shape anisotropy. Solving the Euler equation yields the domain wall width ($\delta$) and the domain wall energy with DMI ($\sigma$)~\cite{heide2008dzyaloshinskii95, thiaville2012dynamics96}, respectively
    \begin{equation}
        \delta  = \pi \Delta = \pi\sqrt{\mathcal{A}/\mathcal{K}_{\mathrm{eff}}}
    \end{equation}
    \begin{equation}
        \sigma= 4 \sqrt{\mathcal{A}\mathcal{K}_{\mathrm{eff}}} \mp \pi \mathcal{D}
        \label{eq:chiral}
    \end{equation}
    where $\Delta = \mathcal{A}/\mathcal{K}_{\mathrm{eff}}$ is the Bloch wall width parameter. 
    Notably, the Dzyaloshinskii-Moriya interaction (DMI) does not change the shape of the one-dimensional (1D) domain wall, but it does introduce chirality, with its sign determined by that of $\mathcal{D}$. When the chirality is most favorable, it results in lowered energy, giving rise to intriguing dynamic properties of Dzyaloshinskii domain wall (D-walls)~\cite{dzyaloshinskii1964theory}. Thiaville \textit{et al.} defined the critical DMI energy constant $\mathcal{D}_c$ as the limit when $sigma$ approaches zero and is given by $\mathcal{D}_c = 4\sqrt{\mathcal{A}\mathcal{K}_{\mathrm{eff}}}/\pi$~\cite{thiaville2012dynamics96}. Above this critical value, the domain wall energy becomes negative, leading to the proliferation of domain walls throughout the sample.
\section{\label{sec:level25}Skyrmion Energetic: A Variational Approach}
    \begin{figure*}[htbp]
        \centering
        \includegraphics[width=\linewidth]{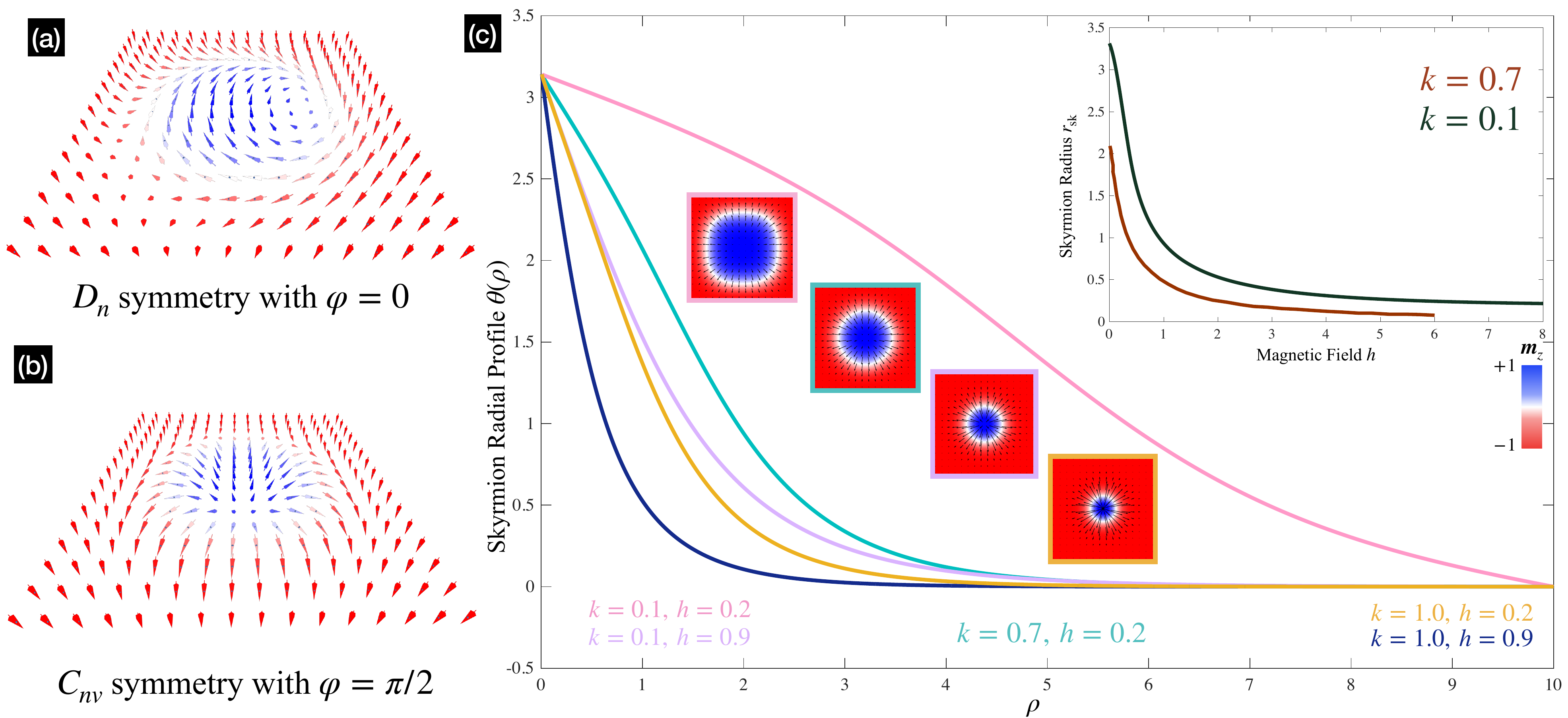}
        \caption{\textbf{Skyrmion radial profile.} (a-b)~Two different configurations of Axisymmetric isolated skyrmions illustrating the angle $\varphi$; (c)~Typical localized solutions of the boundary value problem with magnetization profile $\theta(\rho)$.}
        \label{fig:numer-sky}
    \end{figure*}
    In this section, we consider a variational calculation for a 2D axisymmetric skyrmion in a ferromagnetic background. In a magnetic material, skyrmions are regions where the magnetization vectors are standing in a spiral-like configuration. So we want to know how exactly the magnetization vector depends on the position in space. The unit magnetization vector at a point $\boldsymbol{r}$ in space is given by the vector $\boldsymbol{m(r)}$. Considering the parametrization defined in Eqn.~(\ref{eqn:9})
    \begin{equation}
        \boldsymbol{m(r)} = \sin \theta \cos \phi \hat{r} +\sin \theta \sin \phi \hat{\varphi} + \cos \theta \hat{z}
        \label{Eq: 66}
    \end{equation}
    where $\theta$ and $\phi$ depend on $\boldsymbol{r}$, which we consider in cylindrical coordinates, so that $\boldsymbol{r} = (r, \varphi, z)$. Given a certain $\boldsymbol{m(r)}$, the energy in terms of functional of this configuration is given by:
   \begin{equation}
    \begin{aligned}
    \mathrm{E}[{\boldsymbol{m}}(\boldsymbol{r})]&=\int \Biggl\{-\frac{\mathcal{J}}{2} {\boldsymbol{m}} \cdot \nabla^{2} {\boldsymbol{m}}  +\frac{\mathcal{D}}{2} {\boldsymbol{m}} \cdot(\nabla \times {\boldsymbol{m}})\\
    &+\mathcal{K}\left(1-{m}_{z}^2\right)+\mu_{0} \mathbf{H}_{\text{ext}} \mathbf{M}\left(1-m_{z}\right)\\
    &-\mu_{0} \mathbf{M} ({\boldsymbol{m}} \cdot \mathbf{H}_{{d}}) \Biggl\} {d} \boldsymbol{r}
    \label{Eqn. 67}
    \end{aligned}
    \end{equation}
    with $\mathcal{J}$ being the spin stiffness, $\mathcal{D}$ the Dzyaloshinskii-Moriya interaction constant, $\mathcal{K}$ the anisotropy constant and $\mu_0$ the vacuum permeability. We want to find the configuration that minimizes the energy in Eqn.~(\ref{Eqn. 67}). We only look for solutions with $\phi = \phi_0$ constant that has rotational symmetry in the $\varphi$ direction and translational symmetry in the $z$ direction, so that $\theta$ only depends on $r$ [see inset of FIG.~\ref{fig:profile-sky}].\\
    Two main skyrmionic states Bloch-type modulations with $\mathrm{D}_n$ symmetry [FIG.~\ref{fig:numer-sky}(a)] observed in free-standing nanolayers of cubic helimagnets [see e.g.~\cite{yu2010real41c, yu2011near42c, yu2012skyrmion, romming2013writing48}], and skyrmion lattices with$\mathrm{C_{nv}}$ symmetry in N\'{e}el-type modulations [FIG.~\ref{fig:numer-sky}(b)] observed in Fe/Ir(111) and PdFe/Ir(111) nanolayers~\cite{yu2010real41c, romming2013writing48} and in the rhombohedral ferromagnet $\mathrm{GaV_4O_8}$ with $\mathrm{C}_{nv}$ symmetry~\cite{kezsmarki2015neel} by direct experimental observations.
    With our suitable Eqn.~(\ref{Eq: 66}), the minimised skyrmion energy functional with respect to a field-polarized (ferromagnetic) background therefore reads
    \begin{equation}
    \begin{aligned}
         E[\theta(r)]_{\text{sk}} & = \int r dz d\varphi \int dr \Biggr[ \frac{\mathcal{J}}{2} \biggl\{\left(\frac{d \theta}{d r}\right)^2 +\frac{\sin^2 \theta}{r^2}\biggl\} \\
         & + \frac{\mathcal{D}}{2} \biggl\{ \frac{d \theta}{d r}+ \frac{\sin \theta \cos \theta}{r}\biggl\} + \mathcal{K} \sin^2 \theta \\
         & + \mu_0 \mathbf{H}_{\text{ext}}  \mathbf{M} (1-\cos \theta)\Biggr] \\
         & = {2\pi t_c} \int_0^{\infty} f(\theta, r) rdr
    \end{aligned}
    \label{Eqn: enermin}
    \end{equation}
    where the $2\pi$ and the $t_c$ account for integration in the $\phi$ and $z$ direction respectively, $t_c$ is the thickness of the ferromagnetic layer. $f(\theta, r ) = E\left[\theta(r)\right]_{\mathrm{sk}}-E(0)_{\mathrm{Sk}}$ is the difference between the skyrmion energy density and that of the saturated state, $E(0)_{\mathrm{sk}} = -\mathcal{K} - \mu_0 \mathbf{M}H_{\mathrm{ext}}$.\\
    After minimisation, the Euler equation for energy functional Eqn.~(\ref{Eqn: enermin}) turns out to be
    \begin{equation}
    \begin{aligned}
        \mathcal{J} \left(\theta_{rr}+ \frac{1}{r}\theta_r-\frac{\sin \theta \cos \theta}{r^2}\right) &+ \mathcal{D}~ \frac{\sin^2 \theta}{r} - 2\mathcal{K} \sin \theta \cos \theta \\
       & - \mu_0 \mathbf{M} \mathbf{H}_{\mathrm{ext}} \sin \theta = 0 
       \end{aligned}
    \label{eqn:ansat}
    \end{equation}
    with boundary conditions:
    \begin{equation}
        \theta(0) = \pi; \quad \theta(\infty) = 0
        \label{Eq:bc}
    \end{equation}
    yields the equilibrium structure of isolated axisymmetric skyrmions \cite{bogdanov1999stability, bogdanov1989thermodynamically28c}. The above equation is a non-linear differential equation which can be solved using Numerical methods with dimensionless parameters which results in the equation for axisymmetric skyrmions [see FIG.~\ref{fig:numer-sky}(a,b)] of the form
    \begin{equation}
        \theta_{\rho\rho} + \frac{\theta_{\rho}}{\rho} - \frac{1}{\rho^2} \sin \theta \cos \theta + \frac{2 \sin^2 \theta}{\rho} - k{\sin \theta \cos \theta} - h \sin \theta = 0
        \label{eq77}
    \end{equation}
     \begin{figure*}[htbp]
        \centering
        \includegraphics[width=\linewidth]{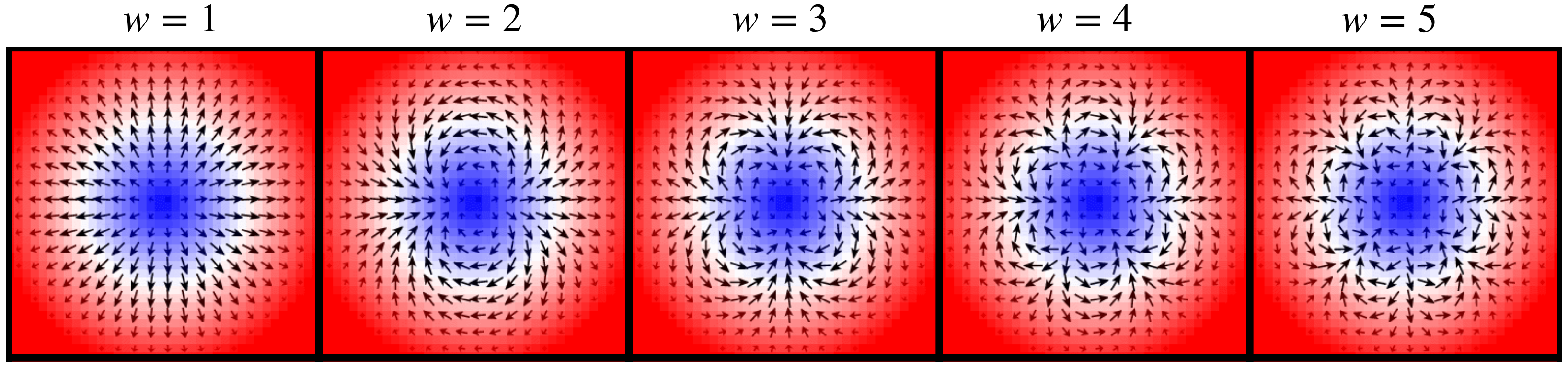}
        \caption{\textbf{Magnetic skyrmion solutions} specified by Eqn.~(\ref{Eqn:77}). The field solution $\boldsymbol{m(r)}$ is classified by the winding number $w$. The field configurations for $1 \leq w \leq 5$ are plotted.}
        \label{fig:windno}
    \end{figure*}
    with boundary conditions (\ref{Eq:bc}). Note that these boundary conditions depend on the sign of $\mathbf{H}_{\mathrm{ext}}$. If $\mathbf{H}_{\mathrm{ext}}$ would be negative, the boundary conditions would be interchanged. We can show that for a generalised case of $\phi = w\varphi+\gamma$, a solution can indeed carry non-zero topological charge $Q$, by plugging Eqn.~(\ref{Eq: 66}) into Eqn.~(\ref{Eqn:topo}) yields
    \begin{equation}
    \begin{aligned}
            Q & = \frac{1}{4\pi} \int \left(\partial_i \boldsymbol{m} \times \partial_j \boldsymbol{m} \right) \cdot \boldsymbol{m} d^2r \\
            &\quad = \frac{1}{4\pi} \int_0^{\infty} d\rho \frac{d \theta(\rho)}{d\rho} \sin \theta(\rho) \int_0^{2\pi} d \varphi \frac{d[w(\varphi + \gamma)]}{d \varphi} \\
            &\quad = w
    \end{aligned}
    \label{Eqn:77}
    \end{equation}
    Therefore, the ansatz in Eqn.~(\ref{Eq: 66}) indeed describes a `topologically protected' soliton solution with integer winding number $w$. Different values of the winding number will represent the different magnetic structures [see FIG.~\ref{fig:windno}].
    The equilibrium skyrmion profiles $\theta(\rho)$ of axisymmetric skyrmions are derived by solving the boundary value problem (\ref{eq77}) and (\ref{Eq:bc}) with a finite-difference method~\cite{bogdanov1999stability} using MATLAB software, also investigated in~\cite{wilson2014chiral217, keesman2015degeneracies219, rybakov2013three235, rybakov2015new236,bub1, butenko2010stabilization242,lin2013particle243,rohart2013skyrmion,kim2014breathing245,zhang2015skyrmion250}. Typical solutions of Eqn.~(\ref{eq77}) are plotted in FIG.~\ref{fig:numer-sky}(c) and the existing areas for isolated skyrmions are indicated in the phase diagram of the solutions [Illustrated in FIG.~\ref{fig:phasediag}, recreated with permission from Ref.~\cite{wilson2014chiral217}].
    We see a rich spectrum of magnetic states characteristic for chiral skyrmions and various scenarios of their evolution under the influence of applied fields, also in~\cite{sampaio2013nucleation88c, zhang2015skyrmion250, iwasaki2013current85c}. In micromagnetism, the characteristic size of a localized magnetization profile $\theta(\rho)$ is defined as~\cite{hubert2008magnetic}
    \begin{equation}
        r_{\text{sk}} = r_0-{\theta_0}{\left(\frac{d\theta}{dr}\right)_{r = r_0}}
    \end{equation}
    where $(r_0, \theta_0)$ is the inflection point of the profile $\theta(r)$ which we have calculated using nearest data-interpolation on magnetization profile [See inset of FIG.~\ref{fig:profile-sky}].
    \begin{figure}[b]
        \centering
        \includegraphics[width = \linewidth]{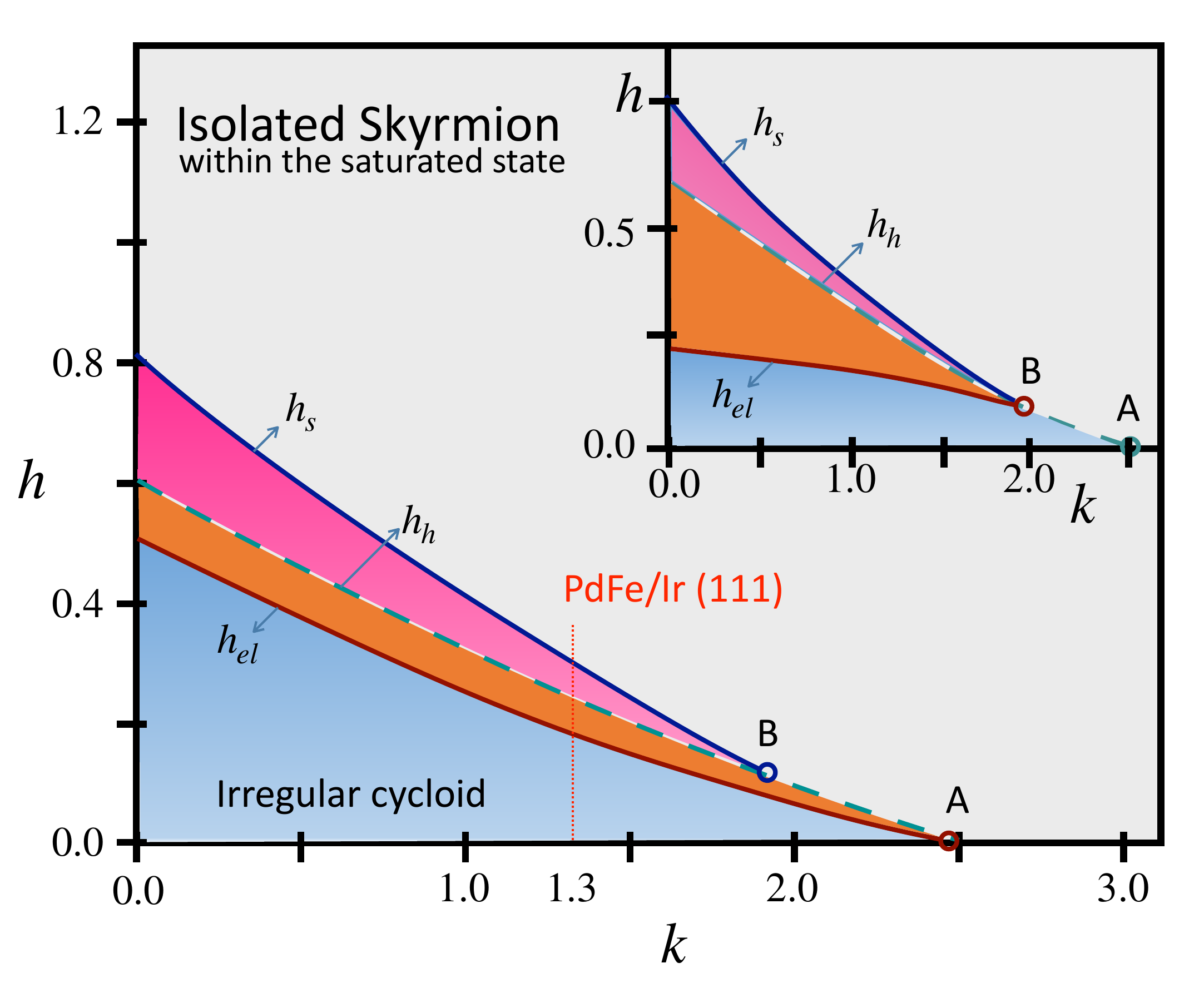}
        \caption{\textbf{$k-h$ phase diagram.} Recreated with permission from Ref.~\cite{leonov2016properties, wilson2014chiral217}}
        \label{fig:phasediag}
    \end{figure}
    We also have plotted the emergent magnetic field of skyrmion mapping using an ansatz for $\theta(r)$ profile $\theta(r)=\pi[1-r/r_{\text{sk}}]$~$(r<r_{\text{sk}})$ and $\theta(r) = 0~(r>r_{\text{sk}})$. Furthermore, the ansatz, $\theta(r/r_{\text{sk}}) = 4\arctan[\exp{(-r/r_{\text{sk}})}]$ based on solutions for isolated $360^{\circ}$ Bloch walls provides a good fit to the solutions of Eqn.~(\ref{eqn:ansat}). $D_c = 4\sqrt{\mathcal{J} K}\pi$ is the critical DM interaction for the onset of spin spirals at zero fields in a single thin layer. Two ranges have to be considered, according to the value of $D$ compared to $D_c$. For example, skyrmion confined in nanodots, using the 1D solution for $\theta(r)$, the skyrmion energy is
    \begin{equation}
        E_s \sim 2\pi R_s t \sigma(D)+\frac{4\pi t \Delta}{R_s}
    \end{equation}
    where the first term is the domain wall energy [Eqn.~(\ref{eq:chiral})] cost and the second one is the curvature energy cost. The minimisation of this equation gives the skyrmion equilibrium size
    \begin{equation}
        R_s \sim \frac{\Delta}{\sqrt{1-D/D_c}}
    \end{equation}
    Various other ansatzs have been extensively analysed to map the numerical and experimental analysis~\cite{braun1994fluctuations,romming2015field232,wang2018theory30} findings on the size of skyrmions~\cite{rohart2013skyrmion, zhang2015magnetic, castro2016skyrmion,iwasaki2013current85c,vidal2017stability}. In
    comparison with previous studies However, a most general theory for skyrmion's size was proposed by Wang \textit{et al.}, Ref.~\cite{wang2018theory30} with the radius of skyrmion [see Eqn.~(\ref{eq:radi})] probed exhaustively with DMI, Anisotropy, exchange and field variations.
    \begin{equation}
    R=\pi D \sqrt{\frac{A}{16 A K^{2}-\pi^{2} D^{2} K}}
    \label{eq:radi}
    \end{equation}
     The formulas and corresponding relations agree very well with simulations and experiments.\\
    Pertaining to our Eqn.~(\ref{Eqn:77}), a $k-h$ phase diagram is analysed in Ref.~\cite{wilson2014chiral217}. These findings highlight the significance of anisotropy symmetry in both bulk and confined cubic helimagnets when it comes to the development of chiral patterns. Moreover, these results offer further proof regarding the physical characteristics of `A-phase' states in various B20 compounds.
\subsection{\label{sec:level29}Theoretical Caveats on Stability}
    \begin{figure*}[htbp]
        \centering
        \includegraphics[width = \linewidth]{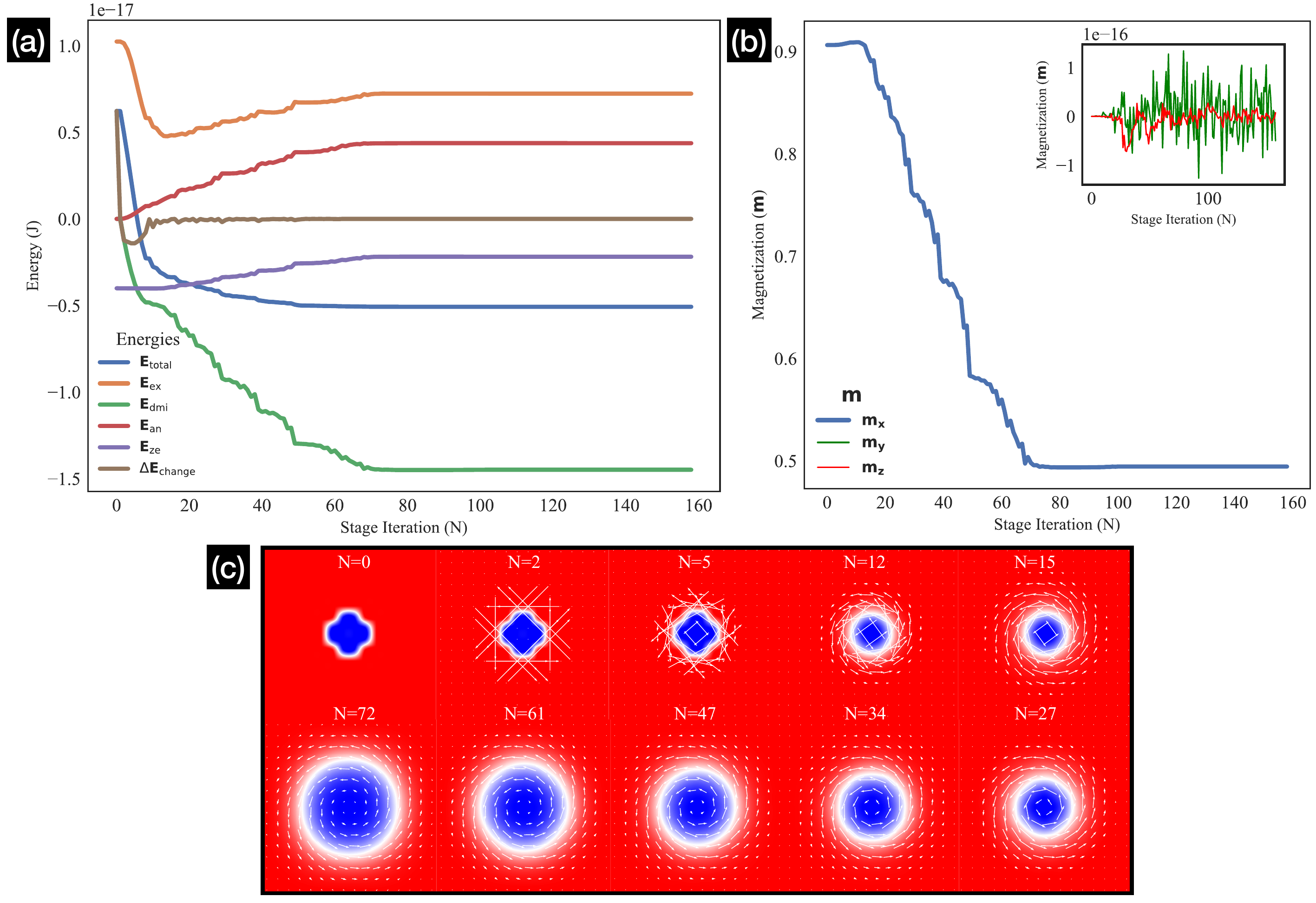}
        \caption{\textbf{Skyrmion nucleation and the corresponding energetics.} (a)~Different Energy Variation with nucleation, (b)~magnetization $\boldsymbol{m}$ with nucleation and (c)~Nucleation of Skyrmion in FM in $N$ no of steps.}
        \label{fig:energy-sk}
    \end{figure*}
    Compared to the domain wall, the vortex and hedgehog bring a new aspect - they require an infinite amount of energy. Well, this can be understood by looking at the swirly nature of the vortex, where the fields never become uniform. Hobart-Derrick \cite{derrick} and the Mermin-Wagner theorems, explain that if an object requires an infinite amount of energy to create, it cannot exist. The instability of the vortex is a result of the lack of any stationary point in its energy as its size increases, and the core of the vortex or monopole also costs a lot of energy due to the swirliness near it. So, initially, it appears that magnetic skyrmions cannot exist. Additionally, because these spin textures are situated on a lattice, it is difficult to utilize topology, which makes any topological protection invalid. Nonetheless, experiments have shown that magnetic skyrmions can bypass the constraints imposed by these theorems, and they do acquire some level of protection from the nontrivial topology of their continuum limit. Therefore, it is crucial to consider the energetic cost when judging whether these objects can exist.
    Under scaling transformation, for example, a rescaled function
    $\theta(r) = \theta(r/\eta)$, the skyrmion energy Eqn.~(\ref{Eqn: enermin}) can be expressed as a function of $\eta$
    \begin{equation}
    \widetilde{E}[\theta(\eta)]_{\text{sk}}=\mathcal{E}_{e}-\mathcal{E}_{D} \eta+\mathcal{E}_{0} \eta^{2}
\end{equation}
    which shows that the DM energy plays a crucial role in stabilizing skyrmions~\cite{bogdanov1994thermodynamically16}. The FIG.~\ref{fig:energy-sk} illustrates the results obtained from a micromagnetic simulation focused on the nucleation of skyrmion in FM material. It presents the variation of both the total energy and individual energies. Evidently, the DMI energy plays a vital role in effectively minimizing the total energy. This minimal energy state corresponds to the stability, or sometimes meta-stability, of these unique magnetic skyrmionic textures. Furthermore, the size of the skyrmion system is controlled by an external magnetic field, as demonstrated in the inset plot and texture plot of the skyrmion in FIG.~\ref{fig:numer-sky}(c) for different values of $h$ and $k$. Moreover, FIG.~\ref{fig:phasediag} showcases the phase transitions between stripe domains, skyrmion states, and saturated states, all of which can be induced by variations in the external magnetic field.
\section{\label{sec:level31} Creation, Deletion and Detection}
    The study of controllable creation, detection, and deletion of nm-scale skyrmions systems offers a large degree of control over the skyrmion states~\cite{wiesendanger2016nanoscale} and will be necessary for any future applications of skyrmionics. The study constitutes, in some sense, a proof of concept demonstrating that beyond the fundamental properties of skyrmions, they may perfectly find applications in electronic devices that could harness their rich behaviours. This section gives an overview of these elementary functionalities. A novel device implementation based on these textures requires their stability at room temperature (RT) which remains a challenge due to the limitations of their energy compared to thermal agitation in ultra-thin systems. Recently, RT stabilization of skyrmions in magnetic multilayers has been achieved~\cite{sampaio2013nucleation88c, fert2013skyrmions}, which opens up possibilities for skyrmion-based technologies. Several extensive reviews on the issue have been recently published ~\cite{jiang2017skyrmions,kang2016skyrmion,fert2017magnetic}.
\subsection{\label{sec:level32} Creation \& Deletion}
    \begin{figure*}[htbp]
        \centering
        \includegraphics[width = \linewidth]{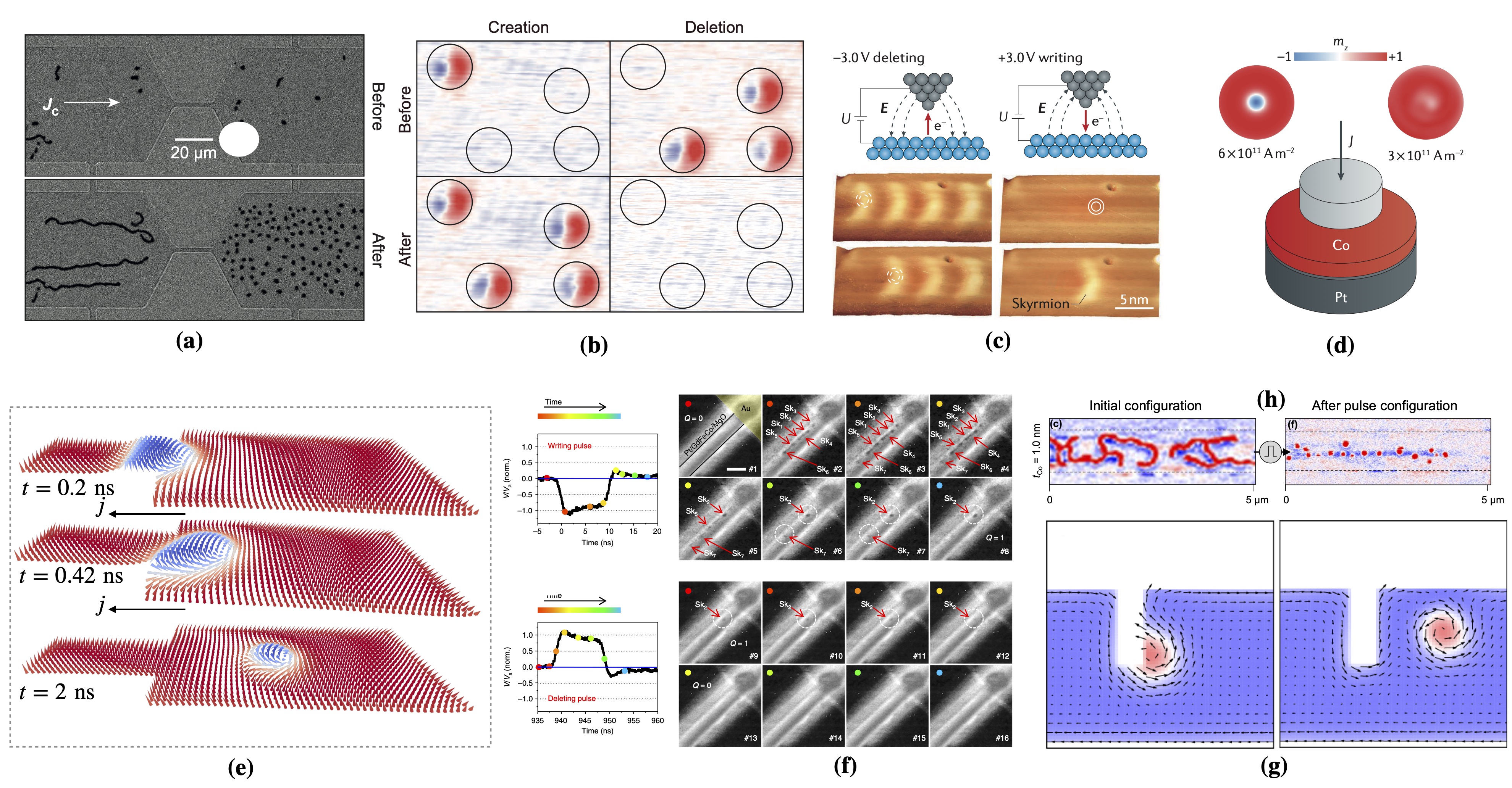}
        \caption{\textbf{Nucleation and Destruction.} (a) Experimental Nucleation of skyrmionic bubbles by driving stripe domains with current ($j_e = \pm 5\times 10^5 \mathrm{A/cm^2}$) through a geometric constriction, with permission from Ref.~\cite{jiang2015blowing}.
        (b) Writing and Deleting of individual skyrmions in Fe/Pd/Ir(111) before and after SP-STM manipulation at specific locations, with permission from Ref.~\cite{romming2013writing48}).
        (c) Deleting (left) and writing (right) skyrmions by local electric fields, with permission from Ref.~\cite{hsu2017electric}.
        (d) Illustration of Skyrmion nucleation by vertical spin injection device utilizing a magnetic tunnel junction with current pulses $j = 3\times 10^{11}$ and $6 \times 10^{11}$, with permission from Ref.~\cite{sampaio2013nucleation88c}.
        (e) Nucleation of Skyrmion from Domain Wall-pair by applying a spin-polarized current, recreated with permission from Ref.~\cite{zhou2014reversible}.
        (f) Magnetic skyrmion configuration for writing and deleting. Scale bar, 500 nm, with permission from Ref.~\cite{woo2018deterministic}.
        (g) Snapshots of the creation of a skyrmion around a notch by current-induced spin transfer torque, with permission from Ref.~\cite{iwasaki2013current85c}.
        (h) MFM images of the $1~\mathrm{\mu m}$ wide tracks before any pulses and after the injection of 1000 current pulses of 200 ns in 1.0 nm thick Co multilayers, with permission from Ref.~\cite{legrand2018hybrid}.}
        \label{fig:nucdel}
    \end{figure*}
    Skyrmions have several important implications due to their unique topology. One of these is that they are protected from creation or annihilation by an energy barrier, as long as they remain away from the edges of the sample. This topological stability allows for a wide region in the phase diagram where skyrmions and the ferromagnetic state can coexist in a hysteresis, with the number of skyrmions in a given area being a variable parameter. This is particularly important for applications in data storage technologies. However, it has been demonstrated that the artificial creation and annihilation of skyrmions is possible within this region of the phase diagram.
    Numerous techniques have been suggested for generating skyrmions, with the magnetic field being a vital and easily controllable stimulus commonly achievable in most laboratories. In the initial phase of skyrmion research, extensive theoretical and experimental investigations were conducted on the magnetic phase diagram. Typically, under specific out-of-the plane magnetic fields and temperature ranges, chiral magnets tend to exhibit the formation of skyrmion crystal and isolated skyrmions~\cite{bogdanov1994thermodynamically16, muhlbauer2009skyrmion38c, jonietz2010spin80c, yu2010real41c,yu2011near42c,milde2013unwinding59k,seki2012magnetoelectric44c, huang2012extended226,mochizuki2015writing94c}. That is to say, by applying an external magnetic field perpendicularly to the magnetic film with DM interactions, it is possible to control the formation of skyrmion texture~\cite{romming2013writing48}.
    On the other hand, the size of a nucleated skyrmion in a magnetic film depends on several factors, including the strength of the external out-of-the plane magnetic field, which interacts with other energetic contributions like DMI, magnetic anisotropy, demag energy and exchange interaction~\cite{rohart2013skyrmion,tomasello2014strategy108c,wang2018theory30}. Typically, the formation of skyrmion requires a small out-of-the plane magnetic field, while a large field can lead to their collapse and annihilation. In 2016,  J M\"{u}ller demonstrated that magnetic field pulse can nucleate skyrmions near the edge of chiral magnet due to edge instability~\cite{muller2016edge0018} and a year later, M. Mochizuki in Ref.~\cite{mochizuki2015writing94c} showed that individual skyrmions can be created in a magnetic film with a fabricated hole or notch using an external magnetic field in a controlled manner. Recently, Garanin \textit{et al.} proposed a method for writing skyrmions in a uniform magnetized film using a magnetic dipole dip~\cite{garanin2018writing0020} and subsequent experimental work using Magnetic Force Microscopy (MFM) tips confirmed that localised magnetic field can create skyrmions~\cite{zhang2018direct0021} by Zhang \textit{et al}. Additionally, S. Woo demonstrated the creation of skyrmions at room temperature without an external magnetic field by using dipolar voltage pulses~\cite{woo2017spin0019}. Micromagnetic simulations have depicted the magnetic field-assisted nucleation stages of a skyrmion on a track~[see FIG.~\ref{fig:energy-sk}(c)], and further dynamics of this process will be discussed in the following section. \\
    Another technique involves employing spin-polarized electric currents [refer to Section~\ref{SkyrmionDynamics}] to induce STTs or SOTs on magnetic moments, which plays a crucial role in driving magnetization dynamics in modern spintronic devices~\cite{ralph2008spin, sinova2015spin}. In 2011, Everschor \textit{et al.} theoretically explored the interaction between STTs and skyrmion lattices in chiral magnets~\cite{everschor2011current}. Subsequently, Tchoe \textit{et al.} made theoretical predictions of the writing of a magnetic skyrmion using spin-polarized electric currents in 2012, demonstrating that skyrmions can be nucleated from a ferromagnetic background with an estimated vertical current of $j \sim 10^{10}-10^{11} \mathrm{Am^{-2}}$\cite{tchoe2012skyrmion}. A year later, Iwasaki \textit{et al.} [see FIG.~\ref{fig:nucdel}(g)] and Sampio \textit{et al.} [see FIG.~\ref{fig:nucdel}(d)] independently investigated current-induced skyrmion creation in confined geometries through theoretical and numerical apporaches~\cite{iwasaki2013current85c, sampaio2013nucleation88c}. In particular, Iwasaki demonstrated through numerical simulation that skyrmion can be nucleated and created from the notch in a nanotrack, offering an effective method for writing skyrmions at specific locations. Theoretical studies have also discussed various mechanisms for the nucleation of skyrmions by electric current pulses in nanodisks \cite{lin2013particle243}, circulating spin current~\cite{tchoe2012skyrmion}, local heating~\cite{koshibae2014creation}, thermal fluctuations~\cite{hagemeister2015stability}, and time-dependent magnetic fields~\cite{heo2016switching}. One such breakthrough was made by N. Romming \textit{et al.} in 2013~\cite{romming2013writing48}, they used a scanning tunneling microscope (STM) to inject spin-polarized current($\sim 1$ nA) into a Pd/Fe bilayer to write a skyrmion [FIG.~\ref{fig:nucdel}(b)]. They also found that a skyrmion could be deleted simply by raising the applied field above 2 Tesla. In this particular case, the externally applied magnetic field is chosen such that the skyrmion and the ferromagnetic states are energetically equivalent and the temperature has been decreased to 4.2 K so that thermal switching is extremely improbable. By increasing or decreasing the field, either the ferromagnetic or the skyrmionic state becomes more favorable, respectively, allowing for a deterministic switching of bits.
    Later in 2017, it was demonstrated that nanoscale skyrmions can also be created and deleted using local electric fields generated from a non-magnetic tip [see FIG.~\ref{fig:nucdel}(c)], where the polarity of the applied bias voltage determines the switching direction, offering an all-electric alternative to spin-transfer-torque-based switching via local spin-polarized current injection~\cite{hsu2017electric}.\\
    In MTJ-based nucleation~\cite{sampaio2013nucleation88c}, by increasing the tunneling current and applied bias voltage, the SP-STM can be used to write and delete nanoscale skyrmions in a reproducible manner via local injection of spin-polarized electrons. In 2014, Zhou and Ezawa theoretically predicted that a magnetic skyrmion can be converted from a domain-wall pair in a junction geometry ($j \sim 1.8\times 10{12}~\mathrm{A m^{-2}}$)~\cite{zhou2014reversible}. Snapshot of which is shown in FIG.~\ref{fig:nucdel}(e).
    In 2015, W. Jian \textit{et al.} demonstrated in room-temperature MOKE experiments that magnetic stripe domains pushed through a geometric constriction with spin current ($j \sim \pm 5\times 10^5 \mathrm{A/cm^2}$) could blow skyrmion bubbles similar to soap films~\cite{jiang2015blowing} [see FIG.~\ref{fig:nucdel}(a)]. In the presence of geometric constriction, such as a corner, a notch, or a narrow electrode, the current density is locally increased. This locally higher current density at the constriction can generate large torques in sandwiched HM and FM layer devices, which are prone to excite magnetization dynamics and induce skyrmion nucleation. Numerical simulations in chiral magnets have demonstrated that corners are effective at generating skyrmions~\cite{wang2022electrical}. The creation of skyrmion bubbles in such a junction constriction driven by inhomogeneous spin currents was further theoretically studied independently by Heinonen \textit{et al}.~\cite{heinonen2016generation73kk,lin2016edge} and Liu \textit{et al}.~\cite{liu2016topological} confirming its efficient mechanism and device compatibility suitable for future applications. In the case of corner constrictions, the nucleation is found to be largely assisted by the Oersted field. A weak magnetic field ( $\mu_0 H_z \approx 8 \mathrm{mT}$) might be sufficient to nucleate and stabilise skyrmion not only on the corner but also inside the track. Consequently, after nucleation, they propagate away from the corner under the action of current-induced torques. Note that current-induced spin torques may also play a role in the nucleation, by destabilising the perpendicular magnetization [see FIG.~\ref{fig:nucdel}(g)]. \\
    Skyrmion can also be nucleated by locally heating the sample~\cite{koshibae2017theory103c}, high-frequency bipolar excitations in combination with pinning sites~\cite{woo2016observation46k}.\\
    The reports mentioned above indicate that topological charge is created during the formation of a skyrmion. Since topological charge $Q$ cannot change smoothly, a singularity in the magnetization must be introduced during the nucleation process. However, this is not necessary for spins on a lattice, where the topological charge can change without a singularity. In a lattice spin system~\cite{haldane19883}, energy is needed to overcome the barrier between the current magnetic configuration and the skyrmion texture. Different mechanisms can provide the energy injection required to overcome this barrier, as seen in the aforementioned studies. Even in the absence of thermal fluctuations or external disturbances, the system can still escape a metastable state through quantum tunneling. The process of quantum nucleation of a magnetic skyrmion can therefore be described as a \textit{quantum tunneling process}.\\
    Destruction, on the other hand, being generally simpler than creation, can, as mentioned above, be managed once skyrmion is created by in our out-of-plane magnetic fields, similar to the magnetic vortices in magnetic disks~\cite{wachowiak2002direct,van2006magnetic}. For example, B\"{u}ttner \textit{et al}. showed that the gigahertz gyrotropic eigenmode dynamics of a single skyrmion bubble can be excited by an external magnetic field pulse and corresponding magnetic field gradient~\cite{buttner2015dynamics}. Magnetic skyrmions in nanostructures with confined geometries can also be switched by microwave magnetic fields~\cite{zhang2015microwave} or magnetic field pulses~\cite{li2014tailoring,heo2016switching}. Another simple method is pushing, using magnetic fields or electric current, skyrmions against a boundary~\cite{romming2013writing48}. Spin-polarized current can also be used to delete skyrmions. In 2017, De Lucia theoretically studied the annihilation of skyrmions induced by spin current pulses ($j \sim 5\times 10^{12} \mathrm{Am^{-2}}$) and suggested that skyrmions can be reliably deleted by designing the pulse shape and later, was experimentally demonstrated by S. Woo \textit{et al}. in 2018~\cite{woo2018deterministic}. Other research areas include-the imprinting of skyrmions via the interlayer exchange of fields from a patterned layer to the skyrmion medium~\cite{fraerman2015skyrmion}, electric field-driven switching~\cite{hsu2017electric}
    and the use of ultrafast laser pulses~\cite{finazzi2013laser}.
    Over the following years, several alternative writing mechanisms have been realized
    or predicted. Besides the reported generation due to spin torques~\cite{sampaio2013nucleation88c,romming2013writing48, bessarab2018lifetime}, locally applied magnetic fields~\cite{zhang2018direct0021, flovik2017generation} and electric fields~\cite{hsu2017electric}, laser
    pulses~\cite{finazzi2013laser}, electron pulses~\cite{schaffer2017ultrafast} or by defects and at boundaries~\cite{jiang2015blowing, lin2016edge, iwasaki2013current85c} assisted nucleation has also observed. An overview
    of generating mechanisms is given in several review articles on magnetic skyrmions~\cite{jiang2017skyrmions, everschor2012rotating91c,nagaosa2013topological36}.\\
    The spin-polarized current can also result in the creation of skyrmions or skyrmion bubbles by other different mechanisms. In 2016, Yuan and Wang theoretically demonstrated that a skyrmion can be created in a ferromagnetic nanodisk by applying a nano-second current pulse ($j \sim 2.0\times 10^{12} \mathrm{ A m^{-2}}$)~\cite{yuan2016skyrmion}. Yin \textit{et al}. suggested in a theoretical study that it is possible to create a single skyrmion in helimagnetic thin films using the dynamical excitations induced by the Oersted field and the STT given by a vertically injected spinpolarized current ($j \sim 1.7\times 10^{11} \mathrm{Am^{-2}}$)~\cite{yin2016topological}. 
    In 2017, as shown in FIG.~\ref{fig:nucdel}(h), Legrand \textit{et al}. experimentally realized the creation of magnetic skyrmions by applying a uniform spin current directly into nanotracks ($j \sim 2.38\times 10^{11} \mathrm{Am^{-2}}$)~\cite{legrand2017room}. Later in 2017, Woo \textit{et al.} conducted an experiment demonstrating the creation of skyrmions at room temperature and zero external magnetic field~\cite{woo2017spin0019}. They achieved this by applying bipolar spin current pulses ($j \sim 1.6\times 10^{11} \mathrm{Am^{-2}}$) directly into a Pt/CoFeB/MgO multilayer, resulting in thermally-induced skyrmion generation, as later studied by Lemesh \textit{et al}~\cite{lemesh2018current}. Hrabec \textit{et al.} also reported the experimental creation of skyrmions by applying electric current ($j \sim 2.8\times 10^{11} \mathrm{Am^{-2}}$) through an electric contact on a symmetric magnetic bilayer system~\cite{hrabec2017current}. The generation of skyrmions using spin-polarized current can be deterministic and systematic when utilizing pinning sites or patterned notches with reduced PMA and are used as a source of generation. This was reported by Buttner \textit{et al.} in 2017~\cite{buttner2017field} and Woo \textit{et al.} in 2018~\cite{woo2018deterministic} [see FIG.~\ref{fig:nucdel}(f)]. More recently, Finizio \textit{et al}. found that localized strong thermal fluctuations could also lead to systematic skyrmion generation at a designed location~\cite{finizio2019deterministic}.
\subsection{\label{sec:level33} Detection}
\begin{figure}[t]
        \centering
        \includegraphics[width =\linewidth]{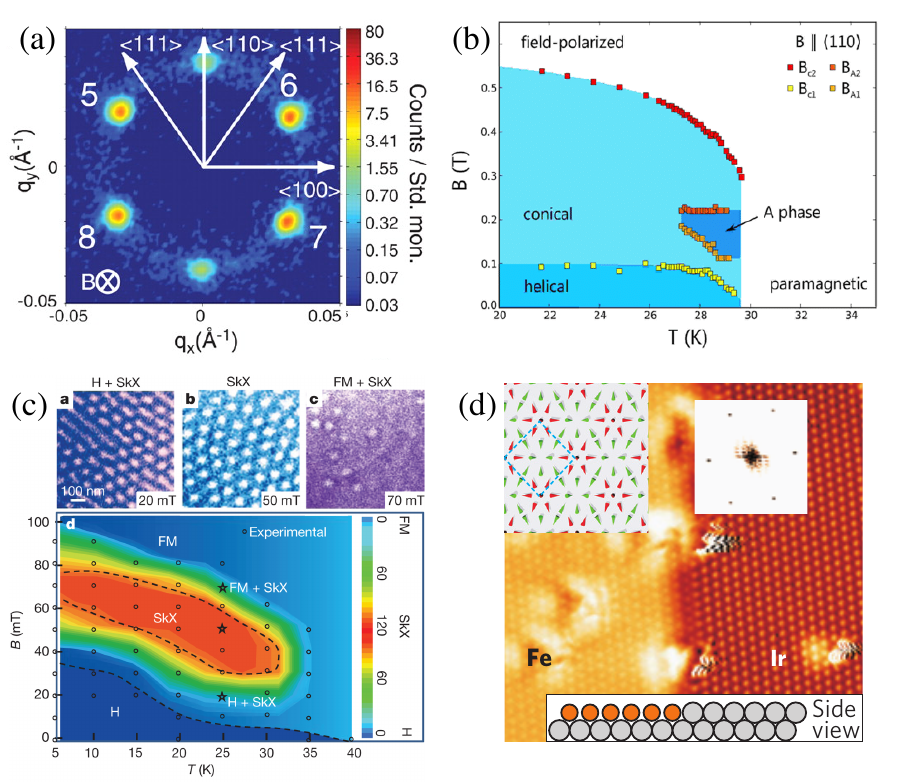}
        \caption{\textbf{First observation of magnetic skyrmions.} (a) SANS Image of MnSi skyrmion crystal. (b) Measured phase diagram of MnSi, with permission from Ref.~\cite{muhlbauer2009skyrmion38c}, in 2009.
        (c) \textit{Bloch Skyrmion.} in Thin film of $\mathrm{Fe_{1-x}Co_xSi}$ and phase diagram, with permission from Ref~\cite{yu2010real41c}, in 2010.
        (d) SP-STEM image of 2D square lattice of \textit{N\'{e}el Skyrmion.}, with permission from Ref.~\cite{heinze2011spontaneous47}, in 2011.}
        \label{fig:firstsky}
    \end{figure}

    In 1989, the existence of magnetic skyrmions was initially predicted, setting the stage for a remarkable scientific journey. It was in 2009 that a groundbreaking milestone was achieved when M\"{u}hlbauer \textit{et al}. conducted neutron scattering experiments on a MnSi sample, making them the pioneers in observing magnetic skyrmions experimentally~\cite{muhlbauer2009skyrmion38c}. Their observations revealed the presence of a helical phase known as the `A-phase' [FIG.~\ref{fig:firstsky}(b)]. At high temperatures, the system exhibited three energetically degenerate spin spirals, each offset by 120 degrees, resulting in six intensity maxima in reciprocal space giving rise to a periodic lattice of skyrmions [FIG.~\ref{fig:firstsky}(a)].
    A year after the initial findings, further confirmation of the presence of skyrmions in a sample was obtained through Lorentz transmission electron microscopy (TEM) images~\cite{yu2010real41c} ( FIG.~\ref{fig:firstsky}(c)). The images showed that the skyrmions were of Bloch type, with a helicity of $\gamma = \pi/2$, and were stabilized by the bulk DMI. Following this, more skyrmion hosts in the B20 materials family, which are non-centrosymmetric, were identified. One year later, in 2011, a different type of skyrmion, called N\'{e}el skyrmion, was observed [FIG.~\ref{fig:firstsky}(d)] in Fe and Ir(111) interface~\cite{heinze2011spontaneous47}. The presence of heavy Ir atoms at the interface caused an interfacial DMI. Unlike Bloch skyrmions, all magnetic moments of N\'{e}el skyrmions are rotated around the perpendicular direction, giving them a helicity of $\gamma = 0$.
    With the idea of enhancement of DMI by creating two interfaces with opposite signs of the DMI constants in a ``sandwich'' structure, such as $\mathrm{Ir/Co/Pt}$ allowed them to be detected even at room temperature in these multilayer systems. This has been demonstrated in several studies~\cite{moreau2016additive, woo2016observation46k, boulle2016room, soumyanarayanan2017tunable49k}.
    Skyrmions have been observed not only in ferromagnetic materials but also in ferrimagnets~\cite{woo2018deterministic}, multiferroics~\cite{seki2012magnetoelectric44c}, and ferroelectric materials~\cite{nahas2015discovery}. Although skyrmion tubes with varying helicity have been observed~\cite{legrand2018hybrid}, those with a fixed helicity remain unobserved. In-plane magnetized materials yet lack observation of skyrmions. \\
    \begin{figure*}[htbp]
        \centering
        \includegraphics[width = \linewidth]{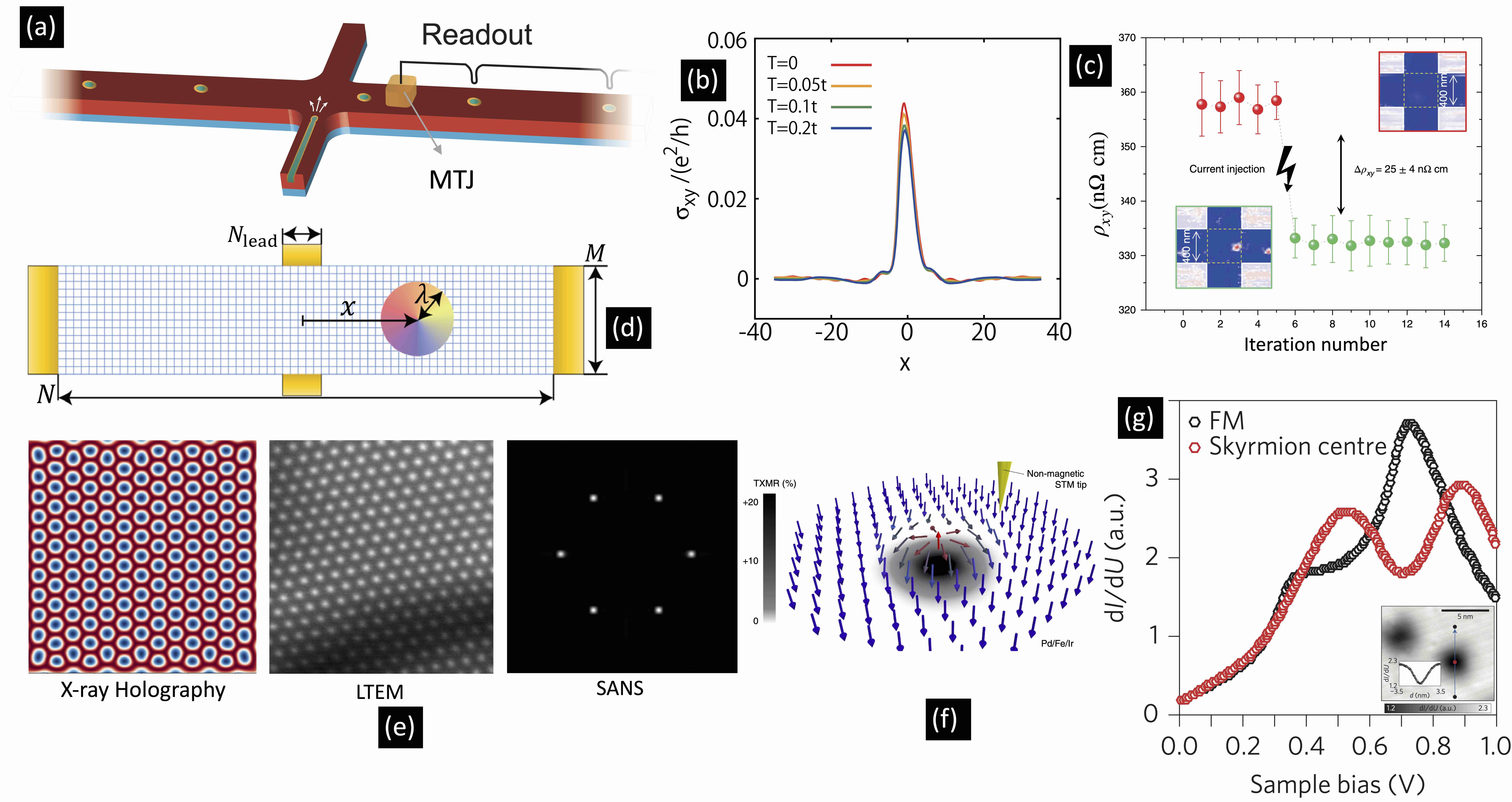}
        \caption{\textbf{Detection of Skyrmion}. 
        {(a)} Schematic depiction of a skyrmion racetrack with a magnetic tunnel junction for skyrmion readout, with permission from Ref.~\cite{tomasello2014strategy108c};
        {(b)} Calculated THE conductance as a function of the distance $x$ between a skyrmion and the Hall contacts [Depicted in {(d)}], with permission from Ref.~\cite{hamamoto2016purely};
        {(c)} Variation in Hall due to a single skyrmion. The RT Hall resistivity electrical detection when one single skyrmion is nucleated in a track of width 400 nm. The red and green dots correspond to the Hall resistivity of the uniform ferromagnetic and single skyrmion states, respectively, shown in the MFM images, with permission from Ref.~\cite{maccariello2018electrical};
        {(d)} Schematic of detection of a skyrmion position by topological Hall effect, with permission from Ref.~\cite{hamamoto2016purely};
        {(e)} Snapshots of computational Imaging Methods of experiments generated for Skyrmion Lattice.
        \textbf{(f)} Perpendicular reading of single magnetic skyrmions, with permission from Ref.~\cite{crum2015perpendicular74kk};
        {(g)} Non-collinear magnetoresistance. The $\mathrm{dI/dU}$ tunnel spectra in the centre of a skyrmion (red) and outside the skyrmion in the FM background (black) (inset shows the signal change caused by the non-collinear magnetoresistance $\mathrm{dI/dU}$ map of two skyrmions in PdFe/Ir(111)), with permission from Ref.~\cite{hanneken2015electrical75kk}}
        \label{fig:det}
    \end{figure*}
    In the context of future spintronic applications centered around magnetic skyrmions, one of the key hurdles is the ability to electrically detect and read individual skyrmions. Researchers have explored two notable approaches to tackle this hurdle-by harnessing the
    electrical property either through the magnetoresistance effect~\cite{hanneken2015electrical75kk, du2015electrical, tomasello2014strategy108c, kubetzka2017impact} or the topological Hall effect (THE)~\cite{kanazawa2017topological,zhang2015topological,nagaosa2013topological36, buhl2017topological} associated with magnetic texture.
        \begin{figure}[htbp]
        \centering
        \includegraphics[width = \linewidth]{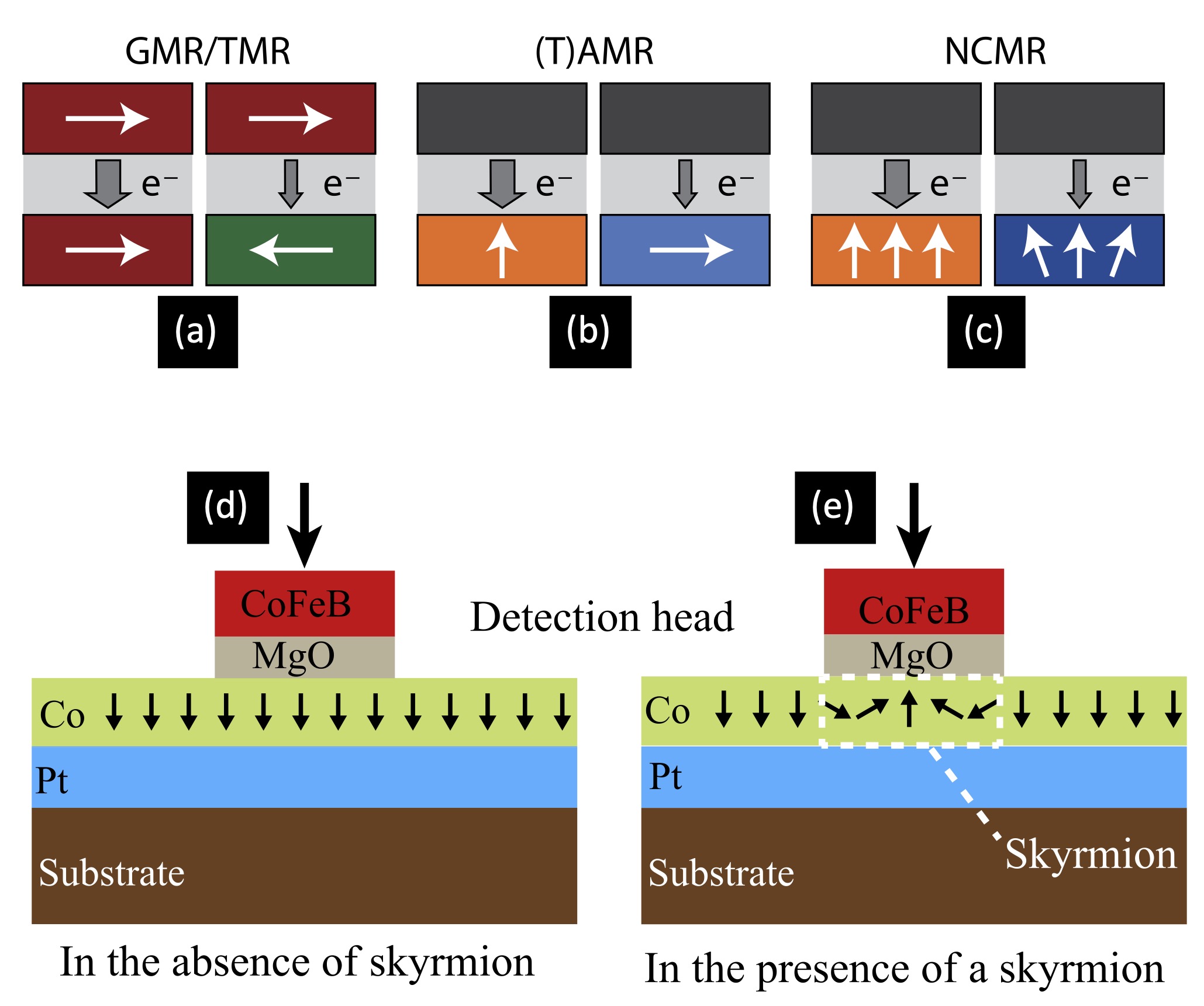}
        \caption{\textbf{Magnetoresistance effects and Skyrmion detection.} {(a)} Diagram of the GMR and TMR effects; {(b)} The (T)AMR effect; {(c)} The NCMR effect and {(d)} skyrmion detection with a perpendicular current and the magnetization change of the system in the (d) absence of or (e) presence of a skyrmion, with permission from Ref.~\cite{hanneken2015electrical75kk, kang2016skyrmion}}
        \label{fig:GMRdete}
    \end{figure}
    By employing magnetoresistance measurements, one approach involves integrating magnetic tunnel junction (MTJ) sensors into skyrmion-based devices and circuits. In 2015, Du \textit{et al}.~\cite{du2015electrical} successfully identified individual skyrmions by analyzing the magnetoresistance in MnSi nanowires. The presence of skyrmions was revealed by observing quantized jumps in the magnetoresistance curves. 
    Hanneken \textit{et al}.~\cite{hanneken2015electrical75kk} proposed a purely electrical method for detecting nanoscale skyrmions by measuring the tunnelling non-collinear magnetoresistance (NCMR effect) [FIG.~\ref{fig:det}(g)]. This technique exploits the sensitivity of the electric conductance signal to the local magnetic environment, allowing a direct differentiation between collinear and noncollinear magnetic states without the need for a magnetic electrode~\cite{hanneken2015electrical75kk, kubetzka2017impact}. A closer view of the dI/dU map (Inset of FIG.~\ref{fig:det}(g)), where I is the tunneling current and U is the bias voltage, shows that a skyrmion with a diameter of approximately 3 nm can be detectable.  Additionally, Crum \textit{et al}. investigated the electric reading of isolated single skyrmions with variation of more than 20\% in magnetoresistance in a current-perpendicular-to-plane geometry using first-principles calculations~\cite{crum2015perpendicular74kk}. Their method utilizes the variation in spin mixing of electronic states at different sites caused by the presence of a skyrmion. This variation is influenced by the noncollinear spin textures compared to the ferromagnetic background. As a result, the tunneling conductance or resistance is affected, leading to different magnetoresistance effects that enable the detection of a skyrmion [see FIG.~\ref{fig:det}(f) and \ref{fig:GMRdete}(d-e)]. Another magnetoresistance effect is the (tunneling) anisotropic magnetoresistance (T)AMR which offers only a minor variation, making it less suitable for skyrmion detection. A newly proposed method for detecting skyrmions solely using electrical means relies on the tunneling non-collinear magnetoresistance (NCMR) effect. This effect leads to a significant change in the differential tunneling conductance for magnetic skyrmions due to their noncollinearity, which causes mixing of spin channels and locally alters the electronic structure. As a result, the skyrmion becomes electronically distinct from the FM background. This effect allows individual skyrmions as small as approximately 3 nm in diameter to be observed experimentally using scanning tunneling microscopy (STM) [See Fig.~\ref{fig:GMRdete}(a-c)].
    In 2017, an innovative protocol for the electrical detection of magnetic skyrmions was proposed by Tomasello \textit{et al}~\cite{tomasello2014strategy108c}. Their theoretical proposal involved measuring the change in TMR signal through a point-contact MTJ in a three-terminal device. This method, which has been successfully employed for other magnetic solitons such as DWs and magnetic bubbles in racetrack-type devices~\cite{loreto2018creation}, offers a promising avenue for skyrmion detection. By adopting this approach, it becomes feasible to design a multi-level memory system where the number of skyrmions represents distinct states. Each individual magnetic tunnel junction element can be read by assessing the tunnel resistance, which correlates with the number of skyrmions present [see FIG.~\ref{fig:det}(a)]. Notably in 2019, Penthorn \textit{et al}. demonstrated that a single skyrmion with a diameter smaller than 100 nm can induce a substantial 10\% change in MTJ resistance~\cite{penthorn2019experimental} which is substantial for accurate skyrmion detection. Moreover, in the same year, Kasai \textit{et al}. also experimentally realized the electrical detection of skyrmions in MTJ~\cite{kasai2019voltage}, where the skyrmion diameter is about 200 nm.\\
    The other promising method we mentioned to read skyrmions is through the THE~\cite{hamamoto2016purely, neubauer2009topological,lee2009unusual,kanazawa2011large,huang2012extended226}, which is directly linked to the topological properties of skyrmions, detailed in Section (\ref{THE}). An emergent field leads to a measurable Hall voltage, enabling a purely electrical detection of skyrmions~\cite{nagaosa2013topological36} [see FIG.~\ref{fig:det}(b,d)]. In 2009, Neubauer \textit{et al}.~\cite{neubauer2009topological} experimentally investigated the THE of the skyrmion crystal phase in MnSi. Their pioneering work revealed a distinct and anomalous contribution to the Hall effect specifically in the presence of skyrmion crystal phases. Building upon this progress, in 2012, researchers successfully identified magnetic skyrmions in epitaxial B20 FeGe(111) thin films~\cite{huang2012extended226}.\\
    While electrical measurements related to the THE (THE) have predominantly focused on crystalline materials, where the presence of a skyrmion lattice leads to a substantial collective electrical signal, recent studies by Maccariello \textit{et al}.~\cite{maccariello2018electrical} [see FIG.~\ref{fig:det}(c)] and Zeissler \textit{et al}.~\cite{zeissler2018discrete} have observed electrical Hall measurements of individual room temperature skyrmions in sputter-grown films and noncrystalline nanostructures. 
    Furthermore, in 2016, Hamamoto \textit{et al}. put forth a theoretical approach for precisely detecting the position of a skyrmion using a purely electrical method by measuring the Hall conductance in a constricted geometry~\cite{hamamoto2015quantized} where Hall conductance exhibits a distinctive peak when a skyrmion is situated at the lead position. In contrast, the Hall conductance diminishes when the skyrmion moves away from the lead location. These advancements in electrical detection methods offer promising avenues for precisely characterizing and manipulating individual skyrmions in various device architectures.\\
    
    \noindent \textbf{Computational Detection/Imaging of Experimental Techniques:}
     \begin{table*}[htbp]
    \caption{\textbf{Imaging techniques for magnetic structures} based on Ref.~\cite{yu2021magnetic}}
        \label{tab:Imaging} 
        \centering
        \resizebox{\linewidth}{!}{%
    \begin{tabular}{ccccc}
    \hline \hline
    \rowcolor[HTML]{FAFCCE}
    \textbf{} & \textbf{MFM} & \textbf{LTEM} & \textbf{Electron Holography} & \textbf{X-ray Holography}\\
    \hline
    \rowcolor[HTML]{F6CDCD}
     Contrast Origin & $\nabla \mathbf{B}$ & $\mathbf{B}$ & Phase Shift & $\mathbf{M}$ \\
     \rowcolor[HTML]{FAFCCE}
    Approximate resolution & 40 nm & 5 nm & 5 nm & 15 nm \\
    \rowcolor[HTML]{F6CDCD}
    Acquisition time & 5-30 min & 0.01-10 sec & 0.01-10 sec & 0.01 sec \\
    \rowcolor[HTML]{FAFCCE}
    Sample thickness & N/A & $< 150$ nm & $<150$ nm & $< 200$ nm \\
    \hline \hline
    \end{tabular}%
    }
    \end{table*}
    Due to the progress in computation technologies, a variety of techniques have been developed to detect and characterize small magnetic configurations such as skyrmions with their unique properties. Neutron scattering(SANS)~\cite{muhlbauer2009skyrmion38c} and X-ray scattering~\cite{zhang2017direct54k, zhang2017direct55k} are commonly used to detect skyrmion lattice phases in reciprocal space. Real-space observation can be achieved through imaging techniques such as Lorentz transmission electron microscopy(LTEM), which is effective for bulk~\cite{yu2010real41c} and interfacial systems~\cite{soumyanarayanan2017tunable49k, montoya2017tailoring56k, mcvitie2018transmission57k}. Scanning probe techniques like spin-polarized tunnelling microscopy~\cite{okubo2012multiple26k, wiesendanger2016nanoscale} and Magnetic Force Microscopy (MFM)~\cite{milde2013unwinding59k} are also useful for studying magnetic textures and skyrmions. Measurement of skyrmion profiles through spin-polarized tunnelling microscopy~\cite{romming2015field60k} has validated models of interactions stabilizing magnetic skyrmions~\cite{rohart2013skyrmion}.\\
    Table~\ref{tab:Imaging} summarises a handful of techniques highlighting the important features. While there are a plethora of techniques capable of measuring magnetism and magnetic properties, focus has been placed on the techniques listed in Table~\ref{tab:Imaging} as they are in common use, with scope for further additions at a later stage.
    The number of experimental techniques and their capability to measure and image magnetic behaviour is constantly increasing along with the ability to computationally simulate different magnetic structures. This chapter details how a simulation software package has been developed in Python to be readily available for simulations of techniques such as LTEM, MFM, X-ray holography and SANS. An outlook of a few methods using Ubermag-assisted computational images is shown in FIG.~\ref{fig:det}(e).
\section{\label{SkyrmionDynamics}Skyrmion Dynamics}
\subsection{LLG Equation}
    Landau and Lifshitz~\cite{landau1980lifshitz11}, and later modified by T. Gilbert, also known as LLG, equation in the field of micromagnetic reads, 
    \begin{equation}
        \frac{\partial \boldsymbol{m}}{\partial t} = -\gamma' \boldsymbol{m} \times \mathbf{H}_{\mathrm{eff}} -\frac{\alpha'}{M_s }\boldsymbol{m}\times (\boldsymbol{m} \times \mathbf{H}_{\mathrm{eff}})
        \label{Eqn. Llg}
    \end{equation}
    where $\gamma'$ and $\alpha'$ are defined as,
    \begin{align*}
        \gamma' & = \frac{\gamma}{1+\alpha^2} \\
        \alpha' & = \frac{\alpha \gamma}{1+\alpha^2}
    \end{align*}
    We can rearrange Eqn.~(\ref{Eqn. Llg}) as
    \begin{equation}
        \frac{\partial \boldsymbol{m}}{\partial t} = -\frac{\gamma}{1+\alpha^2} \boldsymbol{m} \times \mathbf{H}_{\mathrm{eff}} -\frac{\alpha \gamma}{(1+\alpha^2) M_s }\boldsymbol{m}\times (\boldsymbol{m} \times \mathbf{H}_{\mathrm{eff}})
    \end{equation}
    where the first term is Larmor precessional motio ad the second term is a damping term due to dissipative processes such as spin diffusion, spin-orbit coupling etc~\cite{aharoni2000introduction32}.
\subsection{Spin Transfer Torque (STT)}
    The field of spintronics seeks to manipulate magnetic configurations through electric effects. The use of electric currents allows for short switching times and localized application using specific circuitry, making it highly advantageous in the development of future information technology devices compared to magnetic field control. With the discovery of the Giant Magnetoresistance (GMR) effect in 1988 by Gr\"{u}nberg~\cite{binasch1989enhanced} and Fert~\cite{baibich1988giant}, the alteration of magnetic structures in thin-film stacks consisting of alternating ferromagnetic and non-magnetic conductive layers is an example of the successful transfer of knowledge from the solid-state community to technology companies. The effect leads to a significant change in electrical resistance depending on the parallel or antiparallel alignment of adjacent ferromagnetic layers. GMR is mainly used in magnetic field sensors for reading data in hard disk drives.\\
    Moving charges in a ferromagnet can induce a magnetic field (known as the Oersted field) that can affect the state of the magnetization. This traditional interaction between electrical currents and magnetization is incorporated into the LLG equation as an effective Zeeman field. However, Slonczewski and Berger~\cite{berger1996emission,slonczewski1996current} have predicted that a spin-polarized electric current can directly impact magnetization. As the current flows through the ferromagnet, the magnetization exerts a torque on the spins of the conduction electrons, aligning them with the magnetization direction. Due to the conservation of spin, the spins exert a reaction torque on the magnetization~\cite{brataas2012current,abert2019micromagnetics}, which is called spin-transfer torque (STT) and can displace the magnetic structure. Usually, the electron dynamics are faster than the magnetization, so the conduction electron spins align with the local magnetization, leading to adiabatic spin torque. 
    Here, we will consider two types of torque: current perpendicular to plane (CPP) and current in plane
    (CIP).
        \begin{figure*}[t]
        \centering
        \includegraphics[width=\linewidth]{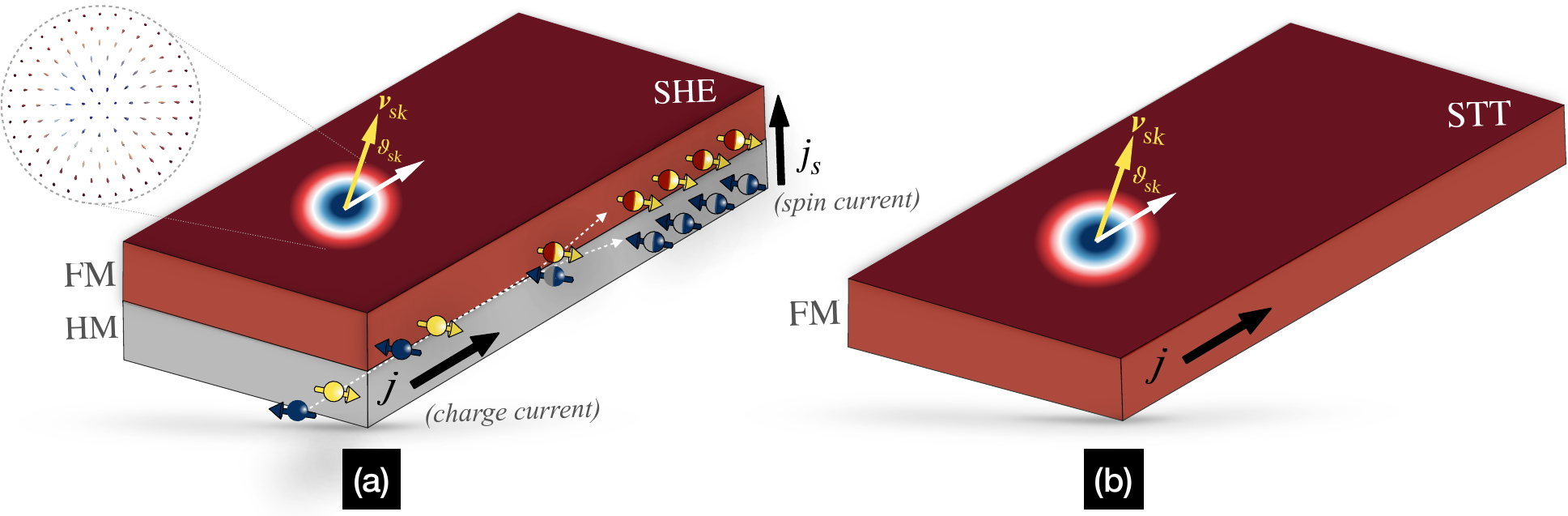}
        \caption{\textbf{Schematic Racetrack geometries with STT and SOT.} \textbf{(a)} Spin-orbit-torque scenario of perpendicular spin injection $\boldsymbol{j_s}$ \textbf{(b)} Spin-transfer-torque scenario where a current $\boldsymbol{j}$ is applied directly along the FM layer. Both setups allow to move magnetic skyrmion but at an angle $\vartheta_{\mathrm{sk}}$ with respect to $\boldsymbol{j}$. This figure is partially adapted with permission from Ref.~\cite{tomasello2014strategy108c}.
        \label{fig:racetrack}}
    \end{figure*}
    \subsubsection{Slonczewski STT:}
    J. Slonczewski introduced CPP torque into LLG in 1996, specifically for magnetic multilayers that were approximately 1 nm or thinner~\cite{slonczewski1996current}. In this setup, a ``hard'' or reference magnetic layer polarized a current passing through it, normal to the layer's plane. This polarized current acted upon a ``soft'' or free magnetic layer, imparting spin and affecting the magnetic moments. This was experimentally confirmed by M. Tsoi \textit{et al}. in 1998~\cite{tsoi1998excitation}.\\
    The corresponding term in the LLG equation is derived for the most simple
    case reads
    \begin{equation}
    \begin{aligned}
        \frac{\partial \boldsymbol{m}}{\partial t} = -{\gamma} \boldsymbol{m} \times \mathbf{H}_{\mathrm{eff}}
        & + \frac{\alpha }{ M_s }\boldsymbol{m}\times \frac{\partial \boldsymbol{m}}{\partial t}+\tau_{\mathrm{FL}} \boldsymbol{m} \times \boldsymbol{m}_p \\
        & + \tau_{\mathrm{IP}} \boldsymbol{m} \times (\boldsymbol{m}_p \times \boldsymbol{m})
        \end{aligned}
    \end{equation}
    The first torque term is called `field-like'(FL) or `out-of-plane' torque as shown in the figure. It is commonly small compared to the second torque term, called `in-plane'(IP) torque, and has no
    qualitative influence on the motion of non-collinear spin textures, as will be shown in
    the next section. Although not applicable to some classes of multilayer systems \cite{berkov2008spin}, the Slonczewski model is still valid for many types of multilayer and pillar-constructed materials.
    \begin{equation}
        \begin{aligned}
            \vec{\boldsymbol{\tau}}_{\mathrm{SL}} & =\beta \frac{\epsilon-\alpha \epsilon^{\prime}}{1+\alpha^{2}}\left(\boldsymbol{m} \times\left(\boldsymbol{m}_{p} \times \boldsymbol{m}\right)\right) -\beta \frac{\epsilon^{\prime}-\alpha \epsilon}{1+\alpha^{2}} \boldsymbol{m} \times \boldsymbol{m}_{p} \\ 
            \beta & =\frac{j_{e} \hbar}{M_{s} e t} \\ 
            \epsilon & =\frac{P(\overrightarrow{\boldsymbol{r}}, t) \Lambda^{2}}{\left(\Lambda^{2}+1\right)+\left(\Lambda^{2}-1\right)\left(\boldsymbol{m} \cdot \boldsymbol{m}_{p}\right)}
        \end{aligned}
    \end{equation}
    where $j_e$ is the current density, $e$ is the elementary charge, $\hbar$ is the reduced Planck constant, $t$ the free layer thickness, $\boldsymbol{m}_p$ the fixed layer magnetisation, $P$ the spin polarisation, $\Lambda$ the Slonczewski parameter with $\epsilon$ the primary and $\epsilon_0$
    the secondary spin torque parameters.
\subsubsection{Zhang-Li STT:}
    In 2004, S. Zhang and Z. Li considered the CIP case~\cite{zhang2004roles}. Zhang-Li is similar to Slonczewski's multilayer model but without the interface or layer effects, see FIG.~\ref{fig:racetrack}(b). Since the torque only depends on the first derivatives of the magnetisation and the diffusion of the spin polarisation is neglected~\cite{abert2019micromagnetics}
    \begin{equation}
    \begin{aligned} \vec{\boldsymbol{\tau}}_{\mathrm{ZL}} =\frac{1}{1+\alpha^{2}}((1&+\xi \alpha) \boldsymbol{m} \times(\boldsymbol{m}  \times(\boldsymbol{u} \cdot \nabla)) \boldsymbol{m} \\
    &+(\xi-\alpha) \boldsymbol{m} \times(\boldsymbol{u} \cdot \nabla) \boldsymbol{m})
    \end{aligned}
    \end{equation}
    \begin{align*}\boldsymbol{u} & =\frac{\mu_{B} \mu_{0}}{2 e \gamma_{0} \mathbf{B}_{\mathrm{sat}}\left(1+\xi^{2}\right)} \boldsymbol{j}
    \end{align*}
    where $j$ is the current density, $\gamma_0$ is the gyromagnetic ratio, $\mu_0$ is the vacuum permeability constant, $\xi$ is the degree of non-adiabaticity from~\cite{zhang2004roles}, $\mathbf{B}_{\mathrm{sat}}$ the saturation magnetisation expressed in Tesla and $\mu_B$ the Bohr magnet.\\
     Experimental results demonstrate that skyrmions can be displaced using relatively low spin-polarized current densities compared to domain walls~\cite{yu2012skyrmion}. This property has contributed to the suggestion of using skyrmions as a basic unit of information in nonvolatile memory devices such as race-track memory~\cite{fert2013skyrmions,parkin2008magnetic}.
\subsection{Spin-Orbit Torque (SOT)}
    The spin-transfer torque refers to the transfer of angular momentum between electron spins and the magnetization in a ferromagnet. It can occur within the ferromagnet itself or when spin currents permeate into the ferromagnet. The Slonczewski spin-transfer torque~\cite{slonczewski1996current} specifically deals with spin currents that enter a ferromagnet, which can arise from processes like spin injection from magnetic tunnel junctions~\cite{barraud2010unravelling,wolf2001spintronics} or spin diffusion through the Spin Hall Effect (SHE)~\cite{liu2012spin1}. The Slonczewski torque has been studied for various applications, including magnetic tunnel junction~\cite{fukami2016spin} switching and the propagation of chiral spin textures at high velocities~\cite{mihai2010current}.\\
    Spin-orbit Torque (SOT), on the other hand, refers to the torque exerted on the magnetization of a magnetic material by the transfer of angular momentum between the spin and orbital degrees of freedom of electrons. SOT can be generated through various mechanisms, including the Spin Hall effect, Rashba effect, Dresselhaus effect, and effects based on topological insulators. In one approach, SOT can be achieved by interfacing the ferromagnetic material(FM) with a heavy metal(HM) layer as shown in FIG.~\ref{fig:racetrack}(a). In this configuration, a charge current $\boldsymbol{j}$ passes through the heavy metal layer, and the spin Hall effect generates a pure spin current in the perpendicular direction~\cite{manchon2015new}. These spins denoted as $\boldsymbol{s}$, are injected into the ferromagnetic layer, exerting a torque on its magnetization. This torque, known as spin-orbit torque (SOT), arises from the spin-orbit interaction.\\
     This specific type of SOT is called Spin Hall Torque (SHT). SOT is a more general term that encompasses a broader range of mechanisms that can exert torque on the magnetization via spin-orbit coupling. The magnitude of the in-plane torque can be determined by analyzing the conduction electrons in the ferromagnetic layer. This allows us to write the version of the LLG equation that is considered for the rest of this paper
    \begin{equation}
    \frac{d \boldsymbol{m}}{d t}=-\gamma \boldsymbol{m} \times \mathbf{H}_{\mathrm{eff}}+\frac{\alpha}{M_s}\left(\boldsymbol{m} \times \frac{d \boldsymbol{m}}{d t}\right)- \frac{\epsilon \beta}{M_s} \boldsymbol{m} \times (\boldsymbol{m} \times \hat{\boldsymbol{m}_p})
    \label{Eq: 71}
    \end{equation}
    with spin-orbit torque constant
    \begin{equation}
        \epsilon \beta = \frac{\gamma \hbar \boldsymbol{j} \Theta_{\mathrm{SH}}}{2e\mu_0 t_{\mathrm{FM}}}
    \end{equation}
    So, for a thin film with a current flowing in the in-plane direction, the torque is given by
    \begin{equation}
        \tau_{\mathrm{SHE}} = \frac{\mu_B \hbar \theta_{\mathrm{SH}} \boldsymbol{j} }{2e M_s t_{\mathrm{FM}}} \boldsymbol{m} \times (\boldsymbol{m} \times \vec{\boldsymbol{m}}_p )
    \end{equation}
    where $\theta_{\mathrm{SH}}$ is the spin Hall angle which quantizes the efficiency of spin current conversion from charge current, $t_{\mathrm{FM}}$ is the thickness of the ferromagnetic layer, $\boldsymbol{j}$ is the current density, $M_s$ is the saturation magnetization and $\vec{\boldsymbol{m}}_p$ is the polarization of the spin current.
\section{Motion of Skyrmion}
\begin{figure*}[ht]
        \centering
        \includegraphics[width = \linewidth]{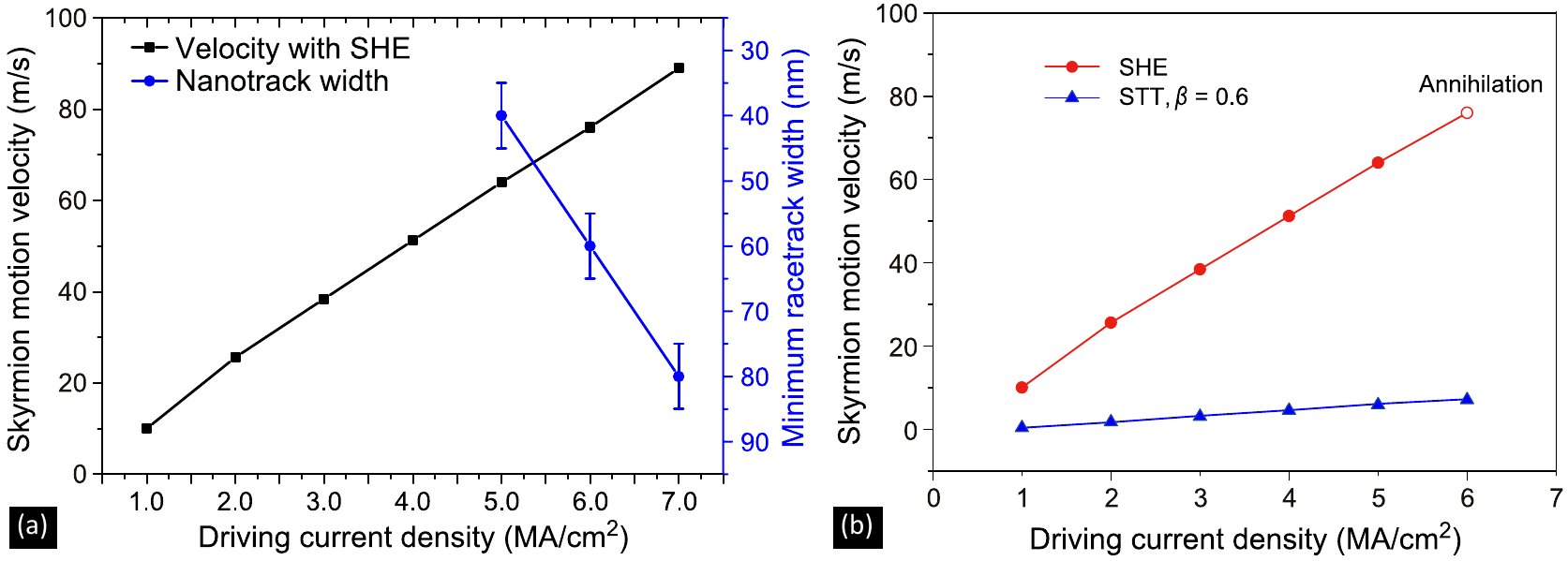}
        \caption{\textbf{Skyrmion's velocity with Torque and track-width width.} \textbf{(a)} Depiction of minimum nano track width and skyrmion's velocity with respect to the driving current density; \textbf{(b)} Dependence of the longitudinal velocity with respect to the driving current density, with permission from Ref.~\cite{kang2016voltage} }
        \label{fig:velotrack}
    \end{figure*}

    Skyrmions (Sks) and Skyrmion Lattices (SkLs) are dynamic structures with diverse behavior over time. They, as quasiparticles, can move in space without altering their internal texture due to their magnetic properties. Although they are often conceptualized as rigid objects with shifting center coordinates, it's crucial to understand that the dynamics of magnetic skyrmions involve the collective evolution of magnetic spins in the system. To study the time evolution of a skyrmion, the field equations for the magnetization field $\boldsymbol{m}$ need to be solved. Despite undergoing deformations caused by variations in the strength of the DMI, skyrmions are topologically protected and stable, meaning they are not destroyed by these deformations. Their axial symmetry is generally maintained during low-energy distortions and long-term dynamics. Two main mechanisms we explained above, spin-transfer torque (STT)~\cite{brataas2012current} and spin-orbit torque (SOT)~\cite{gambardella2011current, brataas2014spin}, have been extensively studied for moving nanoscale magnetic textures. However, SOT has shown~\cite{tomasello2014strategy108c} to be more efficient than STT for current-induced motion in an in-plane track~\cite{sampaio2013nucleation88c} [see FIG.~\ref{fig:velotrack}(b)]. This efficiency is defined by the velocity of the moving skyrmion relative to the required current density. SOT is particularly important for a controlled displacement of skyrmions, which is crucial for memory shift and computational operations utilizing skyrmion-based information storage.
\subsection{\label{EffectiveDynamics}Effective Skyrmion Dynamics: Thiele's Equation}
    In 1973, A. A. Thiele proposed a collective coordinate approach to describe the dynamics of magnetic textures, which is still widely used today~\cite{thiele1973steady}. He introduced the Thiele equation by integrating quadratic functions of the special derivatives of the magnetization from the Landau-Lifshitz-Gilbert equation, resulting in a simplified dynamics equation. Thiele assumed a stationary limit where the magnetization texture moves with a constant velocity and the rigidity of the texture. The presence of spin-orbit torques in a skyrmion-enabled film leads to the emergence of a force that depends on the skyrmion's configuration and size~\cite{thiele1973steady}. The model initially predicts poor scalability for skyrmionic devices, but it is shown that by choosing appropriate skyrmion types, amplification in the driving force can be achieved to compensate for the poor scaling of skyrmions.\\
    Consider a rigid skyrmion of center of mass $\boldsymbol{R} (x,y)$ so it can be located as
    \begin{equation}
        \boldsymbol{m} = \boldsymbol{m}(\boldsymbol{r}, t) = \boldsymbol{m}(\boldsymbol{r}-\boldsymbol{R}(t))
    \end{equation}
    and with the assumption of steady motion and rigid texture
    \begin{equation}
        \frac{\partial \boldsymbol{m}}{\partial t} = \frac{d \boldsymbol{m}}{d (\boldsymbol{r} - \boldsymbol{R})} \frac{\partial \boldsymbol{R} }{\partial t} = -\left(\frac{\partial \boldsymbol{R} }{\partial t} \cdot \nabla \right)\boldsymbol{m} = -(\boldsymbol{v} \cdot \nabla)\boldsymbol{m}
    \end{equation}
    By integrating the effect of the torque term $\boldsymbol{m} \times(\boldsymbol{m} \times \hat{\boldsymbol{m}_p})$ and of other terms in Eqn.~(\ref{Eq: 71}) over the whole magnetic texture of the skyrmion $\boldsymbol{m}(x,y)$, some effective forces acting on the skyrmion can be derived~\cite{sampaio2013nucleation88c}. The components of these effective forces describe the steady-regime velocity of the skyrmion (which is reached immediately under the hypothesis of a strictly rigid skyrmion, and below the ns timescale in practice).
    \subsubsection{Thiele Equation: STT-driven}
    \begin{enumerate}
        \item \textbf{Zhang-Li } \\
    We anticipate only the results pertaining to our discussion. The velocity of skyrmion moving along a wire or racetrack for current vector $\boldsymbol{j} = (\boldsymbol{j_x},0)$ is, 
    \begin{equation}
    \begin{aligned}
               \boldsymbol{v}_{\parallel, x} &= \frac{\alpha \beta \frac{{D}^{2}}{{G}^{2}}+1}{\alpha^2  \frac{{D}^{2}}{{G}^{2}}+1 } \boldsymbol{j}_x = \left(\frac{\beta}{\alpha}+ \frac{(\alpha-\beta)}{\alpha^3({D/G})^2+\alpha}\right)\boldsymbol{v}_s \\
                       \boldsymbol{v}_{\perp, y} &= \frac{\frac{{D}}{{G}} (\alpha - \beta)}{\alpha^2  \frac{{D}^{2}}{{G}^{2}}+1 } \boldsymbol{j}_x = \frac{(\alpha-\beta)({D/G})}{\alpha^2({D/G})^2+1}(\hat{z} \times \boldsymbol{v}_s)
    \end{aligned}
    \label{76}
    \end{equation}
    where $\boldsymbol{v}_{\parallel,x}$ and $\boldsymbol{v}_{\perp,y}$ are the components of $\boldsymbol{v}$ parallel and perpendicular to $\boldsymbol{v}_s$, the velocity of electrons, respectively.
    \textit{When $\alpha = \beta$, the textures move in the direction of the current. The extent of deviation from the current direction is influenced by the discrepancy between $alpha$ and $beta$ as well as the $\frac{{D}}{{G}}$ ratio.} Here, we will examine two distinct forms of torque: current perpendicular to the plane (CPP) and current in the plane (CIP). \\
    \item \textbf{Slonczewski's } \\
    Similarly, the results from Slonczewski's equation of motion for the current $\boldsymbol{j}=(\boldsymbol{j}_x, 0) $, 
    \begin{equation}
    \begin{aligned} \boldsymbol{v}_{x} & =\frac{u}{a j} \frac{\alpha {D} \mathscr{I}_{x y}}{(\alpha {D})^{2}+{G}^{2}} \boldsymbol{j}_{x} \\ \boldsymbol{v}_{y} & =\frac{u}{a j} \frac{{G} \mathscr{I}_{x y}}{(\alpha {D})^{2}+{G}^{2}} \boldsymbol{j}_{x}\end{aligned}
    \label{77}
    \end{equation}
    It should be noted that the transverse velocity component $\boldsymbol{v}_y$ is only non-zero for ${g} \neq 0$, meaning that magnetic textures with \textit{zero topological degree ($Q = 0$), such as skyrmionium, will not experience any transverse motion} in the direction of the current. The presence of the gyrocoupling vector, $G$, as shown in the Eqns.~(\ref{76}) and (\ref{77}), accounts for an effective Magnus force that pushes the skyrmion perpendicular to motion, resulting in the skyrmion Hall effect (SHE)~\cite{neubauer2009topological}. This effect is directly proportional to the topological degree $Q$.
    \end{enumerate}
\subsubsection{\label{SOT-driven}Thiele Equation: SOT-driven}
    SOT generates magnetization dynamics in FM layers using a vertical pure spin-current $\boldsymbol{j}_s$ injected from other layers, typically HM layers through various mechanism~\cite{amin2016spin1,amin2016spin,amin2018interface} illustrated in FIG.~\ref{fig:racetrack}(a). The sign and magnitude of the produced spin-current $\boldsymbol{j}_s$ is determined by the spin Hall angle $\theta_{\mathrm{SHE}}$ of the material, which quantifies the charge-to-spin conversion efficiency
    \begin{equation}
        \boldsymbol{j}_s = \theta_{\mathrm{SHE}} \left(\frac{\hbar}{2e}\right) \boldsymbol{j} \times \hat{\boldsymbol{s}}
    \end{equation}
    where $\boldsymbol{j}$ is the in-plane charge current vector and $\hat{\boldsymbol{s}}$ indicates the polarization direction of spin current. This spin-current interacts with the local magnetization, resulting in a transfer of angular momentum and the creation of a torque on the magnetization. The magnitude and sign of the injected spin-current depend on the combination of materials used and are quantified by the effective spin Hall angle $\theta_{\mathrm{eff}}$, which differs from the inherent spin Hall angle $\theta_{\mathrm{SHE}}$ of the heavy metal layer due to partial reflection or dissipation at the interface~\cite{rojas2014spin, zhang2015skyrmion250,pai2015dependence}. The effective spin Hall angle can change sign depending on the relative positions of the heavy metal and ferromagnetic layers. The values of the $|\theta_{\mathrm{eff}}|$ are widely distributed and depend on the specific materials used, with magnitudes up to 0.1-0.4~\cite{liu2012current,garello2013symmetry} for certain combinations such as Pt (with positive spin Hall angle) or Ta~\cite{liu2012spin1}, W ~\cite{pai2012spin1} and Hf~\cite{ramaswamy2016hf} (with negative spin Hall angle). The effective spin Hall angle is also influenced by the crystalline structure of the materials~\cite{pai2012spin1}.
    The way the injected spin current then acts on the magnetization $\boldsymbol{m}$ can be described by modifying the Landau-Lifshitz-Gilbert equation for
    \begin{equation}
    \begin{aligned}
    \frac{d \boldsymbol{m}}{d t}=-\gamma \mu_{0} \boldsymbol{m} \times \mathbf{H}_{\mathrm{eff}}& +\alpha\left(\boldsymbol{m} \times \frac{d \boldsymbol{m}}{d t}\right)\\
    &-\gamma \frac{\hbar}{2 e} \frac{{j}}{\mu_{0} M_{\mathrm{s}} t_{\mathrm{FM}}} \theta_{\mathrm{eff}} \boldsymbol{m} \times(\boldsymbol{m} \times \hat{\mathbf{s}})
    \end{aligned}
    \end{equation}
    with $\gamma$ the electron gyromagnetic ratio, $\mathbf{H}_{\mathrm{eff}}$ the sum of all effective fields acting on the magnetization, and $\alpha$ the Gilbert damping parameter, which quantifies the dissipation of magnetic energy in the system.\\
    \begin{figure*}[htbp]
        \centering
        \includegraphics[width = \linewidth]{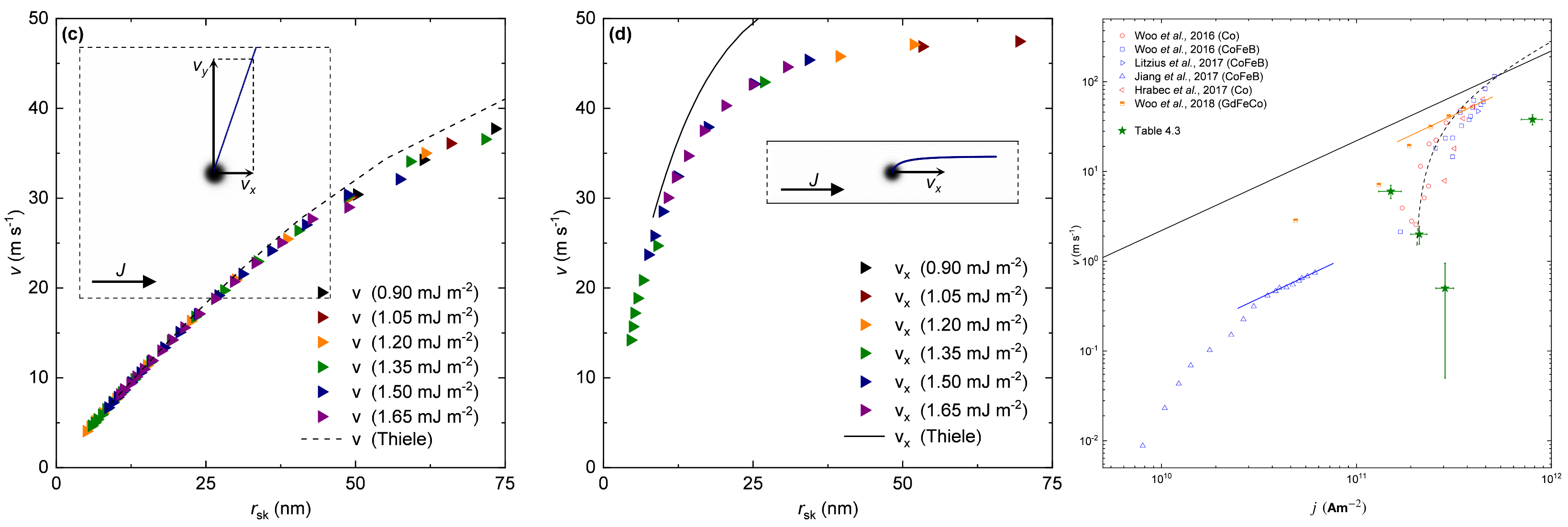}
        \caption{\textbf{Velocity of a skyrmion in a multilayer system.} {(a)} $\boldsymbol{v}$ in an infinite plane and {(b)} ${v}_x$ in track; {(c)} Comparison of motion results with other systems, with permission from Ref.~\cite{legrand2020room}}
        \label{fig:22}
    \end{figure*}
    The Thiele equation for the velocity vector $\boldsymbol{v} = (\boldsymbol{v}_x, \boldsymbol{v}_y)$ can be written as
    \begin{equation}
        \mathscr{G} \times \boldsymbol{v}-\alpha [\mathscr{D}]\boldsymbol{v}+\mathscr{F}+\mathscr{F}^{\mathrm{ext}} = 0
        \label{Eqn: 81}
    \end{equation}
    with
    \begin{equation}
    \begin{aligned} G & =-\frac{M_{{s}} t_{\mathrm{FM}}}{\gamma} \iint\left\{\left(\frac{\partial \boldsymbol{m}}{\partial x} \times \frac{\partial \boldsymbol{m}}{\partial y}\right) \cdot \boldsymbol{m}\right\} d x d y \\
    [{\mathscr{D} }] & = \begin{bmatrix} {D}_{{xx}} & {D}_{{xy}} \\ {D}_{{yx}} & {D}_{{yy}} \end{bmatrix}, {D}_{i j}=\frac{M_{{s}} t_{\mathrm{FM}}}{\gamma} \iint\left\{\frac{\partial \boldsymbol{m}}{\partial i} \cdot \frac{\partial \boldsymbol{m}}{\partial j}\right\} d x d y \\ F_{{x}, {y}} & =\frac{\mu_{0} M_{{s}} t_{\mathrm{FM}}}{\gamma} \iint\left\{(\boldsymbol{m} \times \boldsymbol{\tau}) \cdot \frac{\partial \boldsymbol{m}}{\partial x, y}\right\} d x d y \\
    \mathscr{F}^{\mathrm{ext}} & = -\int \frac{\delta \mathscr{U}}{\delta \boldsymbol{m}} \cdot \frac{\partial \boldsymbol{m}}{\partial_{x,y}} d x dy = -\frac{\partial \mathscr{U}}{\partial_{x,y}} = -\nabla \mathscr{U}(r)
    \end{aligned}
    \end{equation}
    where $\mathscr{G} = G \hat{z}$ is the gyrovector, $\mathscr{D}$ the dissipation matrix, $\mathscr{F} = (F_x, F_y)$ the force exerted on the skyrmion magnetization by the SOT torque $\boldsymbol{\tau}$, itself defined as the last term on the right-hand side of Eqn.(~\ref{Eqn: 81}). In additionally to the gyroscopic, dissipative and torque terms, an extrinsic
    force $\mathscr{F}^{\mathrm{ext}} = -\nabla \mathscr{U}(r)$ is considered. The interaction potential $\mathscr{U}(r)$ accounts for the interaction of a magnetic quasiparticle with other non-collinearities or the sample's edge which is effective when we consider some interactions in the system. Exploiting the cylindrical symmetry of the skyrmion, with $\boldsymbol{m} = [{m}_x(r),{m}_{y}(r),{m}_{z}(r)]$, and considering, for example, a skyrmion of core magnetization pointing down [${m}_z(0) = -\hat{z}$ and ${m}_z(\infty) = +\hat{z}$] with an electrical current directed along $+\hat{x}$, it results that off-diagonal terms in $\mathscr{D}$ are zero and
    \begin{equation}
        \begin{aligned}
            G & = 4 \pi \frac{M_s t_{\mathrm{FM}}}{\gamma}\\
            D & = D_{xx} = D_{yy} = \pi \frac{M_s t_{\mathrm{FM}} }{\gamma}\delta_{xy}a \\
            F_{x,y} &= \mp \theta_{\mathrm{eff}} \boldsymbol{j} \frac{\pi \hbar}{2e} b_{x,y}
        \end{aligned}
        \label{Eqn:constsot}
    \end{equation}
    where $a$ is a dimensionless coefficient related to the magnetic texture, while $b_x$ and $b_y$ are homogeneous to different characteristic sizes of the skyrmion related to its geometry and magnetic texture. It is important to note that, as appears in Eqn.~(\ref{Eqn:constsot}), a N\'{e}el skyrmion ($b_x  \neq 0; b_y = 0$) is driven along the current, while a Bloch skyrmion ($b_x = 0; b_y \neq 0$) is driven orthogonal to the current. However, due to the non-zero gyrovector, the motion is deflected with respect to the direction of the driving force, which is known as the skyrmion Hall effect ~\cite{nagaosa2013topological36}.\\
    For enhancing the limitation, skyrmion's velocity for two possible geometries is studied in Ref.~\cite{legrand2018hybrid}. In the case of an infinite plane [see FIG.~\ref{fig:22}], which is usually when the skyrmion is located far from any edge of the film, the angle of deflection of motion with respect to the driving force, known as skyrmion Hall angle $\Theta_{\mathrm{sk}}$, is 
    \begin{equation}
        \Theta_{\mathrm{sk}} = \tan^{-1}\frac{1}{\eta}
    \end{equation}
   where $\eta = \alpha D/G$ is a dimensionless ration quantifying the relative strength of dissipation effects vs deflection effects while the skyrmion velocity in this direction is
    \begin{equation}
        \boldsymbol{v}_{\mathrm{sk}} = \sqrt{\boldsymbol{v}^2_x+\boldsymbol{v}^2_y} = \frac{F}{\sqrt{1+\eta^2}G}
        \label{Eqn:85}
    \end{equation}
    Second, in the case of a nanostructure patterned into a track geometry~\cite{tomasello2014strategy108c}, after some time the motion of the skyrmion is repelled from the edge due to an additional confining potential by interaction with the track edge~\cite{sampaio2013nucleation88c, iwasaki2014colossal,yoo2017current} and forced along the $x-$ direction, resulting in $\boldsymbol{v}_y = 0$ [see FIG.~\ref{fig:22}]. The resulting velocity
    \begin{equation}
        \boldsymbol{v}_x = \frac{1}{\eta} \frac{F_x}{G}
    \end{equation}
    When considering the case of infinity, it's important to note that the direction of current, denoted by $\hat{x}$, determines the direction of motion. A N\'{e}el skyrmion moves along the $\pm \hat{y}$ direction when dissipation is negligible ($\eta \rightarrow 0$), but moves along the $\pm \hat{x}$ direction when dissipation dominates ($\eta \rightarrow \infty$). Conversely, a Bloch skyrmion moves along the $\mp \hat{x}$ direction when dissipation is negligible ($\eta \rightarrow 0$), but moves along the $\pm \hat{y}$ direction when dissipation dominates ($\eta \rightarrow \infty$). As a result, dissipation acts as a counteracting force against the Skyrmion Hall effect.\\
    In practical experiments and projected skyrmion devices, there are limitations imposed by one-dimensional tracks with limited width, studied in Ref.~\cite{kang2016voltage} [see FIG.~\ref{fig:velotrack}(a)]. N\'{e}el skyrmions in tracks can be accelerated along the edges by a factor of $(1 + \eta^2/\eta)$, as compared to Bloch skyrmions. This acceleration factor can be very large for small dissipation factors ($\eta$), but utilizing edge-enhanced velocity may not be advisable currently, as experiments have shown that skyrmions can be significantly slowed down by edge defects~\cite{jiang2017skyrmions}, and may even be destroyed by large driving currents that overcome the repulsive potential.\\
    Applicability of skyrmion motion is restricted by skyrmion hall effect as shown by Eqn.~(\ref{Eqn:85}). A reduction in the size of skyrmions moving along the edge causes their deacceleration [FIG.~\ref{fig:22}(a-b)], and it may even lead to their temperature-activated annihilation. In such cases, multilayer skyrmions have certain advantages over other types of skyrmions extensively studied in Ref.~\cite{legrand2020room}. Interaction between multiple layers in the multilayer stack helps to stabilize the skyrmions, making them less susceptible to thermal fluctuations and external perturbations which strengthens the stability compared to single-layer skyrmions. They can be created and manipulated at smaller sizes compared to single-layer skyrmions, offer better scalability in the design and fabrication of magnetic memory technologies. In the current-induced motion experiment depicted in Fig.~\ref{fig:22}(c), the author studied a group of multilayers with the aim of attaining higher velocities and reduced sizes.
\subsection{Current-driven Motion of Magnetic Skyrmions}
    Two methods are typically used for the propulsion of skyrmions. The first method involves applying a spin-polarized electric current directly to the skyrmion or skyrmion crystal. This current interacts with the magnetic moments in the skyrmion due to the spin-transfer torque, causing the skyrmion to move along a magnetic racetrack. This method was first observed in bulk MnSi \cite{jonietz2010spin80c, schulz2012emergent} and has a lower critical current density ($\sim 10^6 \mathrm{A/cm^2}$) for initiating motion compared to domain walls, making it useful for technological applications. The second method involves using a system with multiple layers, where a charge current generates a spin current through the spin Hall effect, which is then injected into the ferromagnetic layer. This generates larger spin torques because the skyrmion's magnetic moments can have large angles with the injected spins, resulting in faster skyrmion motion~\cite{jiang2017skyrmions}, but also causes the skyrmion to be pushed towards the edge of the racetrack due to the skyrmion Hall effect, which arises from the topological charge of the skyrmion.
    \subsubsection{Simulation-~Current-driven motion of a magnetic skyrmion:}
    Researchers have derived the Thiele equation, we described in previous section, a system of algebraic equations that describes the velocity of a magnetic quasiparticle, from the analytically unsolvable LLG equation of magnetic moments. They have investigated~\cite{tomasello2014strategy108c, iwasaki2013current85c, bode2007chiral, neubauer2009topological} the behavior of different types of skyrmions (N\'{e}el and Bloch) driven by current-induced torques in various scenarios for both STT and SOT torques based on DMI and SOC. The study found that different types of skyrmions exhibit different motion directions depending on the nature of the torque. Among the investigated scenarios, the most promising strategy for designing a skyrmion racetrack memory involves N\'{e}el skyrmions and SHE~\cite{tomasello2014strategy108c}. To demonstrate the theory, the current-driven motion of a magnetic N\'{e}el skyrmion with interfacial DMI in a Co/Pt bilayer system is presented as an example~\cite{sampaio2013nucleation88c}. The interfacial DMI arises from the inversion-symmetry breaking at the interface and the presence of strong spin-orbit coupling~\cite{sampaio2013nucleation88c, rohart2013skyrmion,ferriani2008atomic}.\\
    \begin{table}[b]
    \caption{\textbf{Micromagnetic parameters of a Co/Pt interface}, with permission based on Ref.~\cite{sampaio2013nucleation88c,simon2018magnetism,tao2018self}.}
        \label{tab:parameter}
        \centering
        \resizebox{\linewidth}{!}{%
    \begin{tabular}{lr}
    \hline \hline
    \rowcolor[HTML]{FAFCCE}
    \textbf{Parameters} & \textbf{Value}\\
    \hline
    \rowcolor[HTML]{F6CDCD}
    Saturation magnetization & $\mathrm{0.58~ MA/m}$ \\
    \rowcolor[HTML]{FAFCCE}
    Exchange stiffness & $\mathrm{15~ pJ/m}$ \\
    \rowcolor[HTML]{F6CDCD}
    DMI constant & $\mathrm{3~ mJ/m^2}$ \\
    \rowcolor[HTML]{FAFCCE}
    Uniaxial anisotropy $(z)$ & $\mathrm{0.8~ MJ/m^3}$ \\
    \rowcolor[HTML]{F6CDCD}
    Gilbert damping & $\mathrm{0.3}$ \\
    \rowcolor[HTML]{FAFCCE}
    Spin Hall angle & $\mathrm{0.4} $ \\
    \hline \hline
    \end{tabular}%
    }
    \end{table}
    The parameters for the Co/Pt system, as shown in Table \ref{tab:parameter}, are derived from Ref.~\cite{sampaio2013nucleation88c}. We examine the behaviour with both the theoretical and simulated approaches, also in~\cite{fert2013skyrmions, sampaio2013nucleation88c, iwasaki2013current85c}. As the system involves a boundary between magnetic atoms and heavy metal atoms, an interfacial Dzyaloshinskii-Moriya interaction (DMI) emerges, which serves to stabilize N\'{e}el skyrmions. 
    \begin{figure*}[t]
        \centering
        \includegraphics[width = \linewidth]{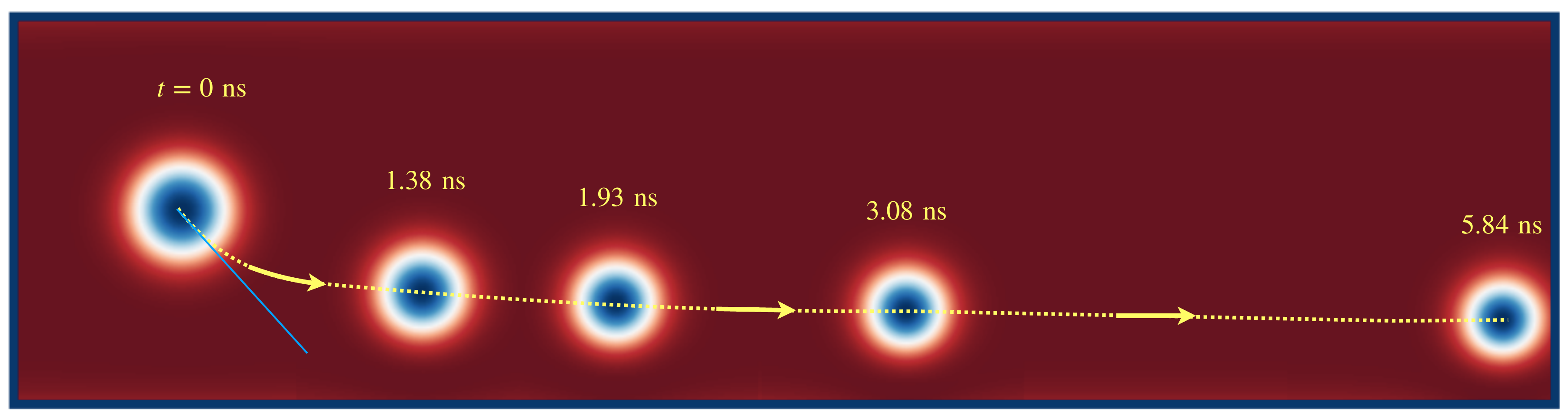}
        \caption{\textbf{Current-driven motion of a racetrack skyrmion.} The stabilized N\'{e}el
        skyrmion is driven by the spin-orbit torque in
        a $256 \text{nm} \times 64 \text{nm} \times 0.5 \text{nm}$ $\mathrm{Co/Pt}$ racetrack with periodic boundary conditions along the track. The trajectory is based on the Thiele equation without considering the skyrmion-edge interaction.}
        \label{fig:sim-sky}
    \end{figure*}
    For a N\'{e}el skyrmion characterized by $N_{\mathrm{sk}} = -1$ and $\gamma = 0$ the tensors $\mathscr{D}$ and $\mathscr{I}$ have the following shapes.
    \begin{equation}
    \begin{aligned}
         D_{xx} & = D_{yy}, \qquad D_{ij} = 0 \quad \text{else} \\
         I_{xy} & = -I_{yx}, \qquad I_{ij} = 0 \quad \text{else}
    \end{aligned}  
    \end{equation}
    Even in a confined sample, where the rotational symmetry of the skyrmion is broken, the above equations are good approximations. For a N\'{e}el skyrmion in a racetrack [periodic boundary conditions along $x$, therefore $\mathscr{U}(r) = \mathscr{U}(y)$], the Thiele equation Eqn.~(\ref{Eqn: 81}) for injected spins along $-y$ ($\boldsymbol{s}_x = 0$ and $\boldsymbol{s}_y = -1$) and for skyrmions far from the edge [$\partial_y \mathscr{U}(y) = 0$] simplifies to
    \begin{equation}
        -4\pi N_{\mathrm{sk}}c \begin{pmatrix}
            -\boldsymbol{v}_y \\
            -\boldsymbol{v}_x \\
            0 \\
        \end{pmatrix} = c D_{xx}\alpha \begin{pmatrix}
            \boldsymbol{v}_x \\
            \boldsymbol{v}_y \\
            0
        \end{pmatrix}-C I_{xy}\begin{pmatrix}
            j \\
            0 \\
            0
        \end{pmatrix}j
    \end{equation}
    where $c = \frac{M_s t_{\text{FM}}}{\gamma}$ and $C = \frac{\hbar}{2e} \Theta_{\text{SH}}$. Skyrmion Hall angle can be calculated from the $y$ component of this equation as,
    \begin{equation}
        \tan \Theta_{\mathrm{sk}} = \frac{\boldsymbol{v}_y}{\boldsymbol{v}_x} = -\frac{4\pi N_{\mathrm{sk}}}{D_{xx}\alpha}
    \end{equation}
    Similarly, the $x$ component yields the skyrmion velocity characterizing the motion along the track
    \begin{equation}
        \boldsymbol{v}_x = \frac{C I_{xy}}{c D_{xx} \alpha}j-\tan \Theta_{\mathrm{sk}}
        \label{Eq:91}
    \end{equation}
    The skyrmion exhibits motion at the skyrmion Hall angle as it moves towards the confined region of the racetrack. As the skyrmion approaches the edge, both the potential energy $\mathscr{U}$ and the corresponding force acting on the skyrmion $-\partial_y \mathscr{U}$ increase. This increase in force continues until either the transverse motion of the skyrmion is balanced, resulting in stable motion [as shown in FIG.~\ref{fig:sim-sky}], or until the maximum repulsion force is exceeded, leading to edge annihilation [as depicted in FIG.~\ref{fig:sim-sky}(b)]. For $j<j_c$, the motion in $y$ direction is suppressed if
    \begin{equation}
        4 \pi c \boldsymbol{v}_x = -\partial_y \mathscr{U}(y)|_{y = y_{\mathrm{comp}}}
    \end{equation}
    The skyrmion will then travel along the confinement region at a constant speed, as described by the first term in Eqn.~(\ref{Eq:91}). However, if the force from the confinement potential reaches its maximum value, denoted as $f_c = -\partial_x \mathscr{U}(x)|_{\mathrm{max}}$, the skyrmion will be annihilated at the edge of the racetrack.\\
    The highest possible velocity is therefore
    \begin{equation}
        \boldsymbol{v}_c = \pm |f_c|\frac{\gamma}{4\pi M_s t_{\mathrm{FM}}}
    \end{equation}
    at a critical current density of
    To determine the critical current density and the highest possible velocity, $\mathscr{U}(y)$ has
    to be determined. The numerically computed value of the critical force is around $|f_c| = 8$ meV/nm. This results is
    \begin{equation}
        \boldsymbol{v}_c = 42 \mathrm{\frac{m}{s}} \qquad \boldsymbol{j}_c = 7.3 \mathrm{\frac{MA}{cm^2}}
    \end{equation}
    based on this consideration, when the current density $\boldsymbol{j}$ is set to $7.0 \mathrm{\frac{MA}{cm^2}}$, which is below the critical value $\boldsymbol{j}_c$, the skyrmion exhibits continuous motion along the confinement region, maintaining a velocity of $\boldsymbol{v}_x = 39.7$ m/s, as depicted in FIG.~\ref{fig:sot-plot}. However, when the current density is further increased to $7.8 \mathrm{\frac{MA}{cm^2}}$, the skyrmion undergoes annihilation at the edge during its trajectory within a few nanoseconds.\\
    In summary, the maximum velocity at which a skyrmion can be propelled by SOT is constrained by the parasitic SHE which poses challenges as skyrmions are unable to move along defect-decorated edges, resulting in annihilation and loss of information in the device. Thus, it is essential to find ways to suppress the skyrmion Hall effect.
        \begin{equation}
        \boldsymbol{j}_c = \pm |f_c| \frac{1}{4\pi} \frac{D_{xx} \alpha}{I_{xx}} \frac{2e}{\hbar \Theta_{\mathrm{SH}}}
    \end{equation}
    The behaviour depicted in FIG.~\ref{fig:sim-sky} agrees well with the result from the micromagnetic simulations. The stabilized N\'{e}el skyrmion is described by $N_{sk} = -1$, $D_{xx} = 15$ and $I_{xy} = 60$ nm. This gives a skyrmion Hall anngle of $\Theta_{\mathrm{sk}} = 70.3^{\circ}$.
    \begin{figure}[t]
        \centering
        \includegraphics[width = \linewidth]{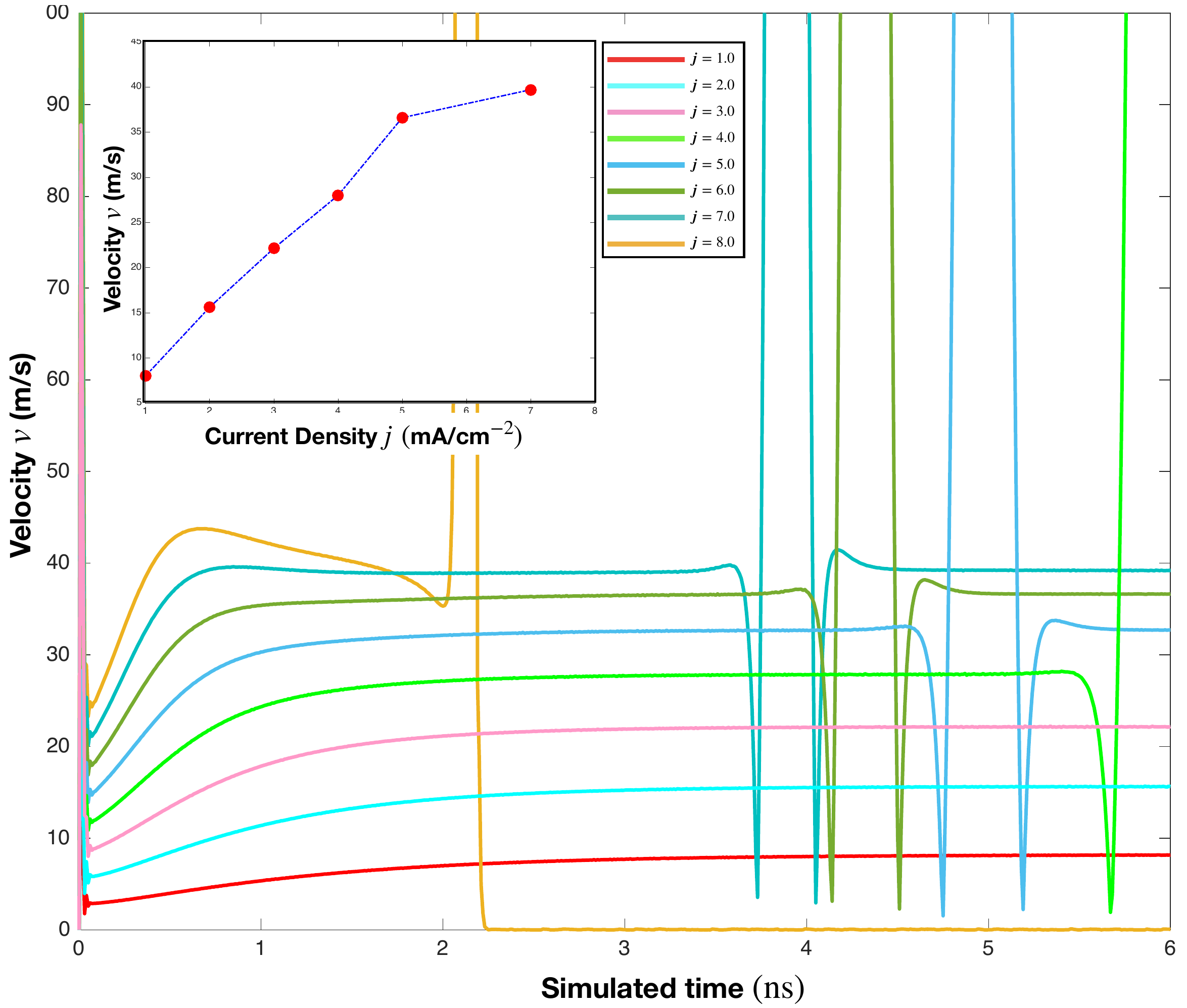}
        \caption{\textbf{Skyrmion's velocity $v$ \& Current density $j$.} Current Density dependence of skyrmion's velocity on a nanotrack.}
        \label{fig:sot-plot}
    \end{figure}
\section{Electronic Properties}
    The preceding section laid the foundation for simulating the dynamics of non-collinear spin textures propelled by current. Using the example of a N\'{e}el skyrmion, it was shown that spin textures with complex real-space topology do not travel in the same direction as the applied current, but instead experience a transverse force, to which we owe a detailed explanation now. \\
    We now focus on analyzing the behavior of the electrons that make up the current. In condensed matter physics, the topological properties of the band structure are identified through the Berry phases of Bloch electrons. For instance, the topological winding number associated with the integer quantum Hall state is related to the Berry phases of Bloch electrons in the magnetic Brillouin Zone. When these itinerant electrons moving adiabatically encounter a non-collinear spin texture, they get ferromagnetically coupled to the skyrmion texture that their spins partially reorient along the spatially-dependent magnetic texture $\boldsymbol{m(r)}$. In case of chiral magnets, the position-space winding number translates to Berry phases picked up by conduction electrons as they traverse the magnetization texture resulting in the topological properties being reflected in the phase factor of the wave function of the current electrons [see FIG.~\ref{THE}(a)]. Mathematically, the Berry phase in position space is equivalent to a spin-dependent Aharonov-Bohm phase. This phase manifests itself in an emergent (spin-dependent) Lorentz force acting on the electrons by generating an ``emergent'' magnetic field  associated with the skyrmion. This leads to the efficient coupling of spin currents and skyrmions which raises expectations for applications in novel spintronic devices. The forced-led rotation of the electron spin introduces a \textit{Berry curvature that appears as an emergent magnetic field} and quantifies the resulting topological Hall conductivity.\\
    For a time-dependent Hamiltonian $H(\boldsymbol{R})$ that depends on a set of parameters $\boldsymbol{R}(t)$ varying adiabatically along a path in parameter space. If $| n(\boldsymbol{R}(t))\rangle$ is the $n^{\mathrm{th}}$ eigenstate with eigenenergy $E_n$ then by solving the time-dependent Schr\"{o}dinger equation for a wave function $|\psi(t, \boldsymbol{R}(t))\rangle$, apart from the dynamical phase, we find that the wave function accumulates a geometric phase along the contour on which $\boldsymbol{R}$ is varied, known as the \emph{Berry phase}.
    Along a closed path in parameter space, it can be expressed as
    \begin{equation}
        \gamma_{n} = \oint i \langle n|\nabla_R|n \rangle \cdot d\boldsymbol{R}
    \end{equation}
    The Berry phase is a closed loop on parameter space that can be observed through interference effects in various physical systems.
    To investigate the underlying geometric structure and the associated gauge fields that govern the wavefunctions a related mathematical parameter \textit{Berry connection} is defined by integrating the above equation, 
    \begin{equation}
        \boldsymbol{A}_n(\boldsymbol{R}) = i \langle n|\nabla_R|n \rangle
    \end{equation}
    Although Berry connection cannot be a physically measurable quantity since it depends on the choice of phases for the instantaneous eigenstates $|n(\boldsymbol{R}(t))\rangle$. The geometric vector potential $\boldsymbol{A}_n(\boldsymbol{R})$ is not gauge invariant which takes us to Berry curvature calculated according to Stoke's theorem along $C$
    \begin{equation}
        \gamma_{n|_{C}} = \int_{{\delta}S} d\boldsymbol{R}  \cdot  \boldsymbol{A}_n(\boldsymbol{R}) = \frac{1}{\hbar} \int_S  \boldsymbol{\Omega}_n
    \end{equation}
    where the two-form, \\
    \begin{equation}
            \boldsymbol{\Omega}_n = \frac{1}{2} \sum_{i,j = 1,2}^{N}  \boldsymbol{\Omega}_{n, ij}(\boldsymbol{R})d\boldsymbol{R}_i \wedge d\boldsymbol{R}_j
    \end{equation}
    where $$\boldsymbol{\Omega}_{n, ij}(\boldsymbol{R}) = \frac{\partial \boldsymbol{A}_{n,j}}{\partial \boldsymbol{R}_i}- \frac{\partial \boldsymbol{A}_{n,i}}{\partial \boldsymbol{R}_j}$$
    So we can write the corresponding `field', the \emph{Berry curvature} as,
    \begin{equation}
        \boldsymbol{\Omega}_n(\boldsymbol{R}) = i \langle \nabla_R n| \times |\nabla_R n \rangle 
    \end{equation}
        \begin{figure}[t]
        \centering
        \includegraphics[width = \linewidth]{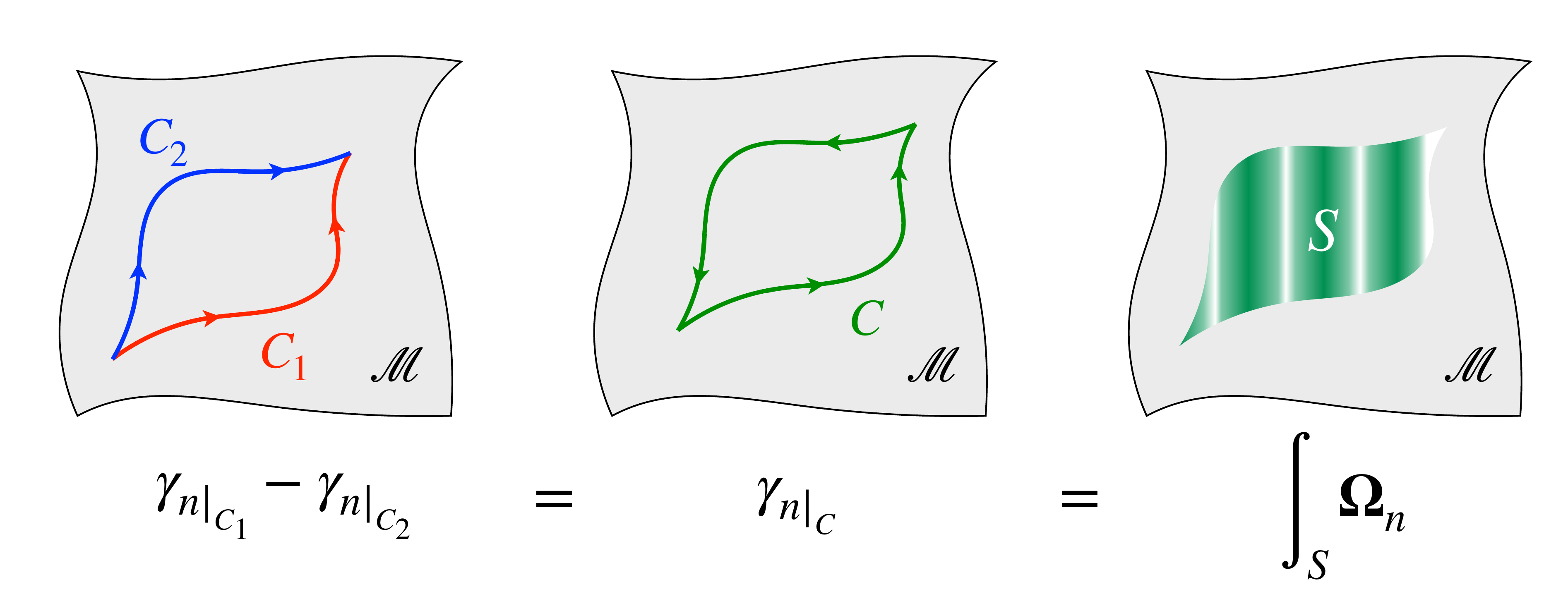}
        \caption{\textbf{Berry Phase and Curvature.} The gauge invariant difference in Berry phases $\gamma_{n|{C_1}}-\gamma{n|{C_2}}$ along two part is equivalent to the Berry phase $\gamma{n|{C}}$  along a closed loop $C$, formed by attaching the reverse of $C_2$ to the end of $C_1$. When loop is contractible, $\gamma{n|_{C}}$ can be evaluated from Berry curvature  $\Omega_n$ over any surface $S$ where $\mathcal{\delta}S = C$.}
        \label{fig:berryphase-curv}
    \end{figure}
    The Berry curvature is a geometric vector potential and is gauge-invariant. The importance of the Berry curvature in describing the topological properties of an electron system is due to the fact that the integral over the Brillouin zone, which is known as the Chern number, 
    \begin{equation}
        \mathcal{C}_n = \frac{1}{2\pi}\int_{\mathrm{BZ} }{\boldsymbol{\Omega}^z_n(\boldsymbol{k})} d^2k
    \end{equation}
    that can only have integer values (here written for a two-dimensional system).
    The Berry curvature is antisymmeteric
    under time-reversal symmetry and symmetric under inversion symmetry which results in the non-vanishing of Berry curvature. On the other hand, physical observables (like the Hall conductivity) are obtained by integrating over an energy range or contour. When time-reversal symmetry is conserved, even in the case of broken inversion symmetry, $E_n(\boldsymbol{k}) =-E_n(\boldsymbol{k})$ leads to the cancellation of the Berry curvature contributions from $\boldsymbol{k}_0$ and $-\boldsymbol{k}_0$, so there is no actual effect.
    However, when time-reversal symmetry is broken, for example, by applying a magnetic field by accounting for a magnetic texture, the Berry curvature-induced effects become significant. Therefore, it is necessary to take into account the effects of the Berry curvature when studying the electronic properties of magnetic systems with non-collinear spin textures, such as skyrmion crystals. 
    \subsection{Hall Conductivity}
    The concept of Hall conductivity emerges from the reciprocal-space Berry curvature. When an electric field $\mathbf{E}$ is applied to a metal, electrons move with an electric current density $\boldsymbol{j}$. In the most simple case, the electric field and the current density are parallel but this is not strictly enforced by Ohm's law
    \begin{equation}
        \boldsymbol{j} = \sigma \mathbf{E}, \qquad\mathbf{E} = \rho  \boldsymbol{j}
    \end{equation}
    Here, $\sigma$ and $\rho$ are inverse tensors: the conductivity and the resistivity, respectively. In
    the case of broken time-reversal symmetry, i.e. when a magnetic field is applied or a magnetic texture is present, a transverse transport-tensor element can appear. \\
    In 1934 Bloch, Peierls, Jones and Zener derived the semiclassical equations of motion
    of a Bloch electron wave packet in electromagnetic fields~\cite{bloch1929quantenmechanik,peierls1929theorie,jones1934general}. In 1999 a crucial missing term, the so-called `anomalous velocity', was found, modifying the equations of motion to~\cite{sundaram1999wave}
    \begin{equation}
    \begin{aligned}
        \dot{r} &=  v_n(\boldsymbol{k}) - \dot{\boldsymbol{k}} \times \boldsymbol{\Omega}_n(\boldsymbol{k}) \\
        \hbar \dot{\boldsymbol{k}} &=q\mathbf{E} + q \dot{r} \times \mathbf{B}
        \end{aligned}
        \label{131}
    \end{equation}
    Here, $v_n = \frac{1}{\hbar} \frac{\partial E_n (\boldsymbol{k})}{\partial \boldsymbol{k}}$
    is the group velocity and, by analogy, $-\dot{\boldsymbol{k}} \times \boldsymbol{\Omega}_n(\boldsymbol{k})$ is the anomalous velocity. The second equation accounts for the Lorentz force. Due to the similarity of the two equations, the Berry curvature $\boldsymbol{\Omega}_n(\boldsymbol{k})$ can be considered a `reciprocal
    magnetic field'. This implies a transverse deflection of electrons even in the absence of a
    magnetic field. In this case,  $\dot{\boldsymbol{k}} \parallel \mathbf{E}$ point along the Hall bar. Thus the Berry curvature, which is always perpendicular to the two-dimensional plane, induces an anomalous velocity in the transverse direction. Consequently, the anomalous Hall effect and topological Hall effect can be observed without applying a magnetic field.\\
    The relationship between conductivity and Berry curvature can be derived using the Boltzmann equation~\cite{suh2023semiclassical}. The derived result for a two-dimensional system is,
    \begin{equation}
    \begin{aligned}\sigma_{x x} & =-\left.e^{2} \frac{1}{2 \pi} \sum_{n} \int_{\mathrm{BZ}} v_{n, x}(\boldsymbol{k})^{2} \tau_{n}(\boldsymbol{k}) \frac{\partial f}{\partial E}\right|_{E=E_{n}(\boldsymbol{k})} \mathrm{d}^{2} k \\ 
    \sigma_{x y} & =-\frac{e^{2}}{h} \frac{1}{2 \pi} \sum_{n} \int_{\mathrm{BZ}} \Omega_{n}^{z}(\boldsymbol{k}) f\left(E_{n}(\boldsymbol{k})-E_{F}\right) \mathrm{d}^{2} k
    \end{aligned}
    \end{equation}
    Here, $f$ is the equilibrium Fermi-distribution function, and $\tau_n\boldsymbol{k}$ is the relaxation time, which arises due to extrinsic effects. For this reason, only the transverse element is purely intrinsic. In terms of eigenstates and energies [see details in Ref.~\cite{hatsugai1993chern}]
\begin{equation}
\begin{aligned}
    \sigma_{xy}(E_F) = -\frac{e^2}{h} \frac{1}{2\pi} 2 \operatorname{Im} \sum_{m \neq n} \int \frac{\langle n|\partial_{k_x} H|m\rangle\langle m|\partial_{k_y} H|n\rangle}{(E_n-E_m)^2} \\
    \times f(E_n(\boldsymbol{k})-E_F) \, d^2k
\end{aligned}
\end{equation}

    For $T = 0$ the transverse conductivity is only affected by states below the Fermi energy
    $E_F$
    \begin{equation}
        \sigma_{x y}\left(E_{F}\right)= -\frac{e^{2}}{h}\frac{1}{2 \pi} \sum_n \int_{E_n(\boldsymbol{k}) \leq E_F} \Omega_{n}^{z}(\boldsymbol{k}) d^2 k 
    \end{equation}
    If the Fermi energy is situated in a band gap, the Hall conductivity is given by the sum
    over the Chern numbers $\mathcal{C}_n$ of all occupied (occ) bands.
    \begin{equation}
             \sigma_{x y}=-\frac{e^{2}}{h} \sum_{n}^{\mathrm{occ}} \mathcal{C}_n
    \end{equation}
    Since $\mathcal{C}_n$ can only have integer values, non-trivial Chern numbers lead to quantized
    transport. As shown by Hatsugai~\cite{hatsugai1993edge,hatsugai1993chern}, the origin is the occurrence of topologically protected edge states, which make the system conducting at the edges (and generate a non-zero $\sigma_{x y}$), while the bulk remains insulating. The quantitative difference of left-${w}^l_n$ and right-propagating ${w}^r_n$ edge states in the gap above band $n$ is given by the winding number $w_n$
    \begin{equation}
        {w}^l_n -{w}^r_n \equiv \sum_{m \leq n} \mathcal{C}_n
    \end{equation}
    This relation allows to deduce of boundary properties purely from bulk information and is therefore called `bulk-boundary correspondence'. A material that exhibits these edge states is labelled `Chern insulator'. This establishes a connection between the winding number and the Topological Hall conductivity and its effects.
    \subsection{\label{electrodynamics}Emergent Electrodynamics}
    We discussed the idea that if an electron is moving adiabatically in the vicinity of a skyrmion, although its Hamiltonian is continuously changing, the wave function describing the conduction electron gains a phase factor, called berry phase. In the case of non-trivial spin textures like the Skyrmion, such a phase factor causes incredible (and measurable) physical effects, due to the emergence of effective electromagnetic fields directly related to it [refer to \ref{AP-1}]. In the following, we will elaborate on these effective fields and explore some of their characteristics.\\
         \begin{figure*}[htbp]
        \centering
        \includegraphics[width = \linewidth]{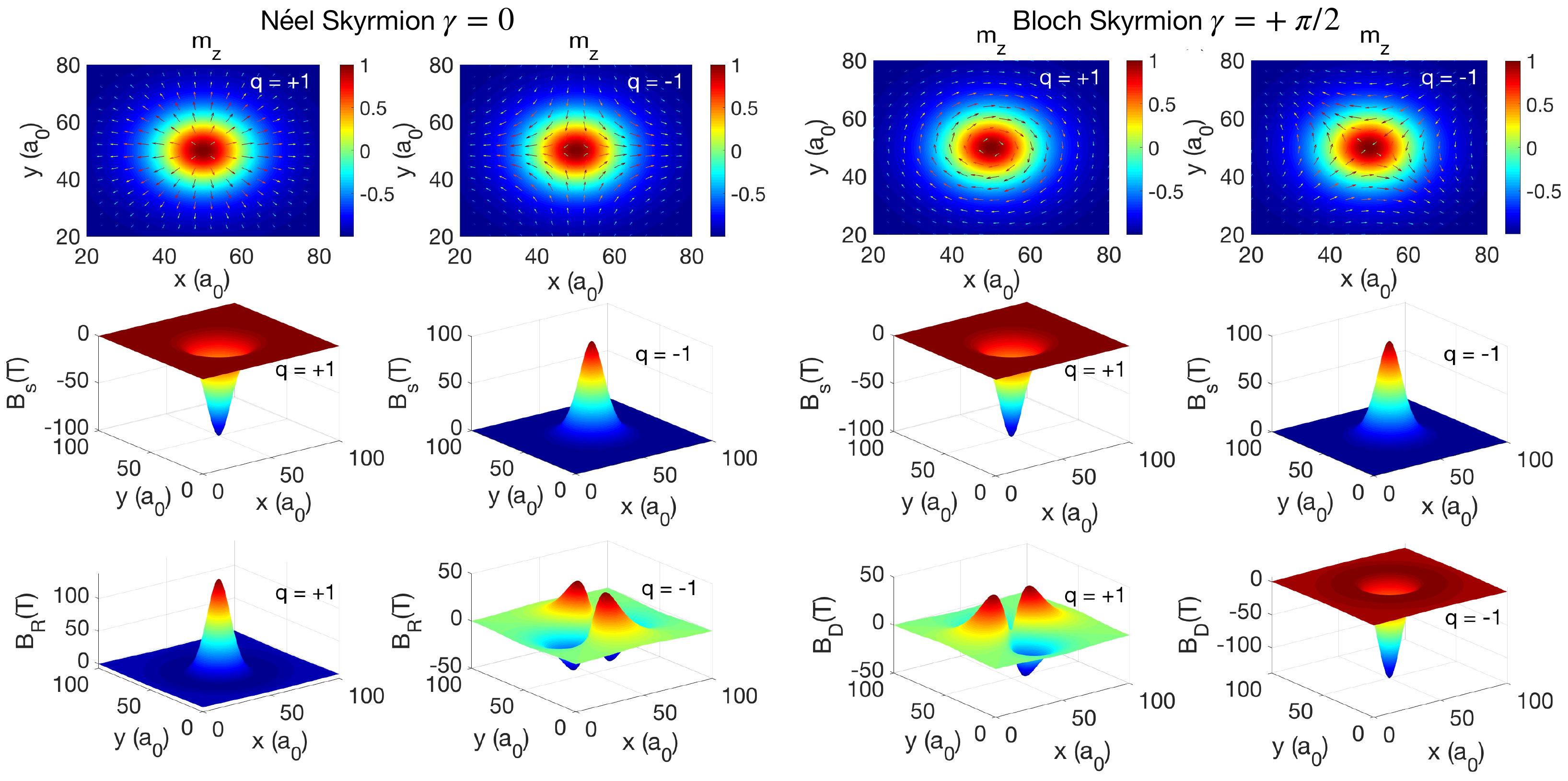}
        \caption{\textbf{Schematic diagram of skyrmion(Anti) and their emergent magnetic field.} N\'{e}el Skyrmion and their Anitskyrmion (left), Bloch Skyrmion Anitskyrmion (right), with permission from Ref.~\cite{akosa2019tuning}}
        \label{fig:emerg-field}
    \end{figure*}
    Using the ansatz defined for $\boldsymbol{m}$ from Eqn.~(\ref{eqn:9}), the emergent gauge field can be evaluated by solving the Schr\"{o}dinger equation of an electron in an EM field associated with texture $\boldsymbol{m}$
    \begin{equation}
        A_{\mu} = \zeta \langle m |\partial_{\mu}|m\rangle = \frac{\hbar}{2e}(1-\cos \theta)\partial_{\mu}\phi
        \label{Eqn:107}
    \end{equation}
    where $\zeta$ is a coefficient corresponding to the 4-vector. Eqn.~(\ref{Eqn:107}) shows that in the presence of a spatial/temporal nonuniform spin texture, the conduction electrons feel a fictitious electromagnetic field ($A_{\mu}$) generated by a non-coplanar spin texture. For typical Skyrmion radius $ r_{\mathrm{sk}} = 100$ \AA{}, this leads to a large topological magnetic field ($\sim 10^2$ Tesla), which thus provides a unique platform to study the high magnetic field response of electrons. Akosa \textit{et al}. theoretically studied (in particular N\'{e}el-type skyrmion and Bloch-type antiskyrmion) that SkHE which arises from the Magnus force or the above fictitious force acting on magnetic texture can be suppressed by tuning the spin-orbit interaction (Rashba (RSOC) and Dresselhaus (DSOC) spin-orbit couplings) strength~\cite{akosa2019tuning}. The emergent magnetic field of these textures is shown in FIG.~\ref{fig:emerg-field}. In the adiabatic limit, itinerant electrons are weakly coupled to the spin gauge fields with two contributions: (i) due to the magnetic texture $\boldsymbol{A}_s^{z}$ and (ii) as a result of an interplay between SOC and the magnetic texture $\boldsymbol{A}_{so}^{\sigma} = \boldsymbol{A}_R^{z}+\boldsymbol{A}_D^{z}$ where $\boldsymbol{A}_{R(S)}^{z}$ is RSOC(DSOC)-induced spin gauge fields 
    \begin{equation}
    \begin{aligned} \boldsymbol{A}_{s}^{z} & =\mp \frac{\hbar}{2 e}(1-\cos \theta) \boldsymbol{\nabla} \Phi \\ \boldsymbol{A}_{R}^{z} & =\mp \frac{\hbar}{2 e} \frac{\left(m_{y} \boldsymbol{e}_{x}-m_{x} \boldsymbol{e}_{y}\right)}{\lambda_{R}} \\ \boldsymbol{A}_{D}^{z} & =\mp \frac{\hbar}{2 e} \frac{\left(m_{x} \boldsymbol{e}_{x}-m_{y} \boldsymbol{e}_{y}\right)}{\lambda_{D}}\end{aligned}
    \label{Eqn108}
    \end{equation}
    and $(\boldsymbol{B}_{\eta} = \nabla \times \boldsymbol{A}_{\eta}^{z})|_{\eta = s,R,D}$, the corresponding emergent fields, [See for details \cite{akosa2019tuning}], opens avenues for experimental findings.
    \subsection{\label{THE}Topological Hall Effect (THE)}
    The topological hall effect of electrons is considered the signature feature of skyrmion phase~\cite{bruno2004topological,neubauer2009topological,lee2009unusual,hamamoto2016purely,maccariello2018electrical,schulz2012emergent,li2013robust,hamamoto2015quantized,gobel2017unconventional} which explains the emergent electromagnetic fields that arise from the interaction between conduction electrons and the skyrmion spin texture. Hall resistivity was observed when an electric current flows through a conductor in the presence of a perpendicular magnetic field, also known as the Normal Hall Effect (NHE). In non-magnetic materials and for small magnetic fields, the Hall resistivity typically increases linearly with the strength of the applied magnetic field $\mathbf{B}_z$, and the Hall conductivity reads. 
    \begin{equation}
        \sigma_{xy} = -\frac{\rho_{xy}}{\rho^2_{xy}+\rho^2_{xx}}
    \end{equation}
    In bulk MnSi (exhibiting skyrmion crystals) sample, Hall resistivity $\rho_{xy}$ was measured at various temperatures close to $T_c$ and its dependency on magnetic field reveals that in chiral magnets, the total Hall effect is influenced by several mechanisms~\cite{neubauer2009topological}. So, for skyrmion crystal, Hall resistivity $\rho_{xy}$ can be decomposed into three conventional Hall effects
    \begin{equation}
    \begin{aligned}
            \rho_{xy} & = \rho^{\mathrm{NHE}}_{xy}+\rho^{\mathrm{AHE}}_{xy}+\rho^{\mathrm{THE}}_{xy} \\
            & = R_0^{\mathrm{NHE}} \mathbf{B}_z + S_A \rho^2_{xx} \mathbf{M}_z+PR_0^{\mathrm{THE}}\mathbf{B}_{rz}^{\mathrm{em}}
    \end{aligned}
    \end{equation}
    where $\rho^{\mathrm{NHE}}_{xy}$, $\rho^{\mathrm{AHE}}_{xy}$ and $\rho^{\mathrm{THE}}_{xy}$ correspond to the normal Hall effect, the anomalous Hall effect and the topological Hall effect respectively.
     They are all oriented out of the plane ($z$), along the skyrmion tube or the external field $\mathbf{B}_z$, the net magnetization $\mathbf{M}_z$ and the emergent field $\mathbf{B}^{\mathrm{em}}$. Here, $R_0$ and $S_A$ are coupling coefficients and $P (0 < P < 1)$ is the spin polarisation ratio of conduction electrons.
     The first contribution, the normal Hall term~\cite{hall1879new} $\rho^{\mathrm{NHE}}_{xy}$, is directly proportional to the strength of the applied magnetic field $\mathbf{B}_z$, denoted as $\rho^{\mathrm{NHE}}_{xy} = R_0 \mathbf{B}_z$, where $R_0$ is the coupling constant that depends on the band structure and scattering rates in multi-band systems like MnSi ($R_0 \approx 1.7 \times 10^{10} \mathrm{\Omega m T^{-1}}$).
    The second contribution, the anomalous Hall term~\cite{nagaosa2010anomalous} (AHE), is denoted as $\rho^{\mathrm{AHE}}_{xy}$. The phenomenon arises from the interaction of spin-orbit coupling (SOC) and local electric fields at the atomic scale, causing the spin orientation to rely on momentum. This gives rise to Berry phases, which can be described as an effective magnetic field in momentum space, leading to an additional contribution to the Hall measurement. The anomalous Hall effect (ANE) is generally proportional to the total magnetization of the sample ($\mathbf{M}_z$) as $\rho^{\mathrm{AHE}}_{xy} \propto \mathbf{M}_z$, and it remains unaffected by impurity scattering. \\
    \begin{figure}[t]
        \centering
        \includegraphics[width = \linewidth]{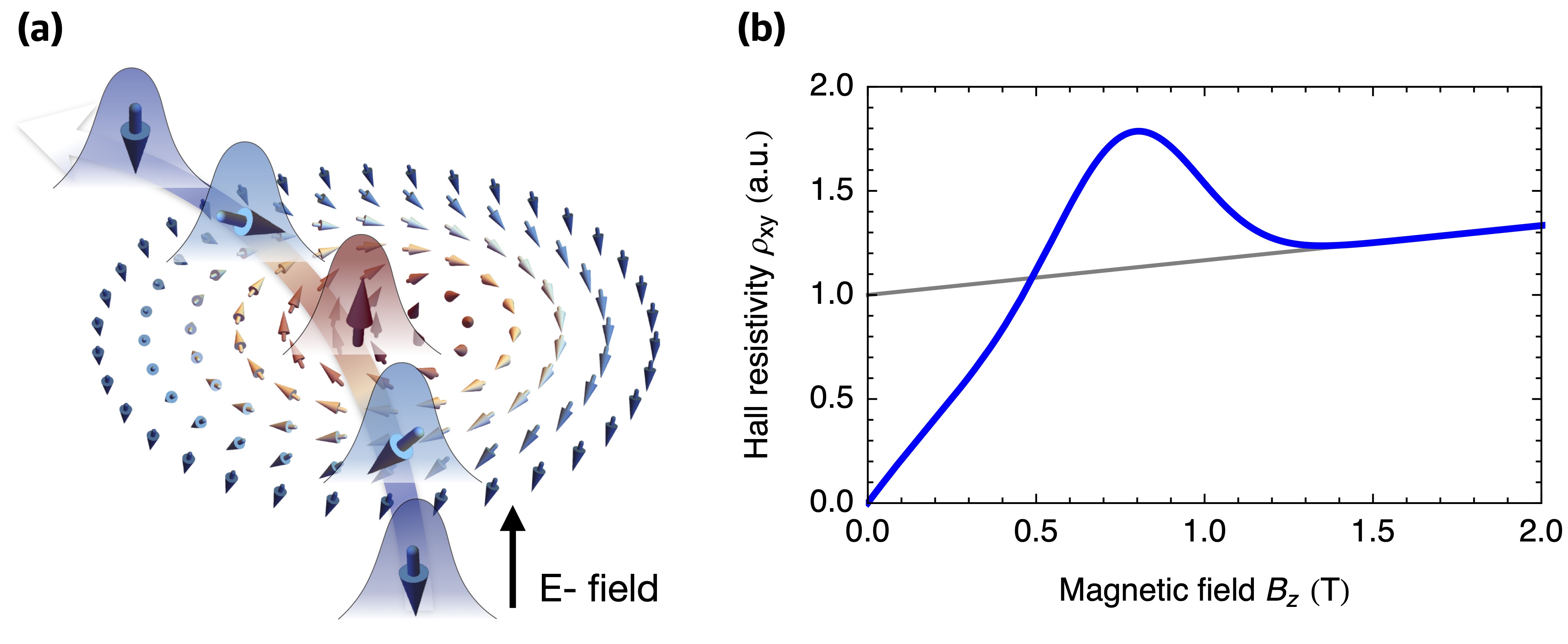}
        \caption{\textbf{Topological Hall Effect of electrons.}~{(a)} Electron traversing through the skyrmion texture accumulates the Berry Phase (shown as wavepacket); {(b)} Schematic curve of a typical experimental measurement of the transverse resistivity versus the out-of-plane magnetic field. In the range between $0.5$ T and $1$ T a skyrmion phase is assumed to form. There, an additional contribution is present: the THE, with permission from Ref.~\cite{gobel2017unconventional}}
        \label{fig:THE}
    \end{figure}
    The third contribution arises from the topological Hall term $\rho^{\mathrm{THE}}_{xy}$, which complements the AHE. In this case, Berry phases collected by the electrons when they hop between two sites and reorient their spins, cause a deflection of the charge carriers in real space, as opposed to momentum space in AHE. This phase is similar to the Peierls phase, which characterizes the magnetic field in the ordinary Hall effect. Spin textures with changing magnetization orientation on length scales larger than the Fermi wavelength of the electron result in Berry phases in real space, which is related to an effective magnetic field, called emergent field $\mathbf{B}^{\mathrm{em}} = \mathbf{B}^{\mathrm{em}}_{rz}  \propto \langle n_{\mathrm{sk}}\rangle $. A rough estimate for the size of the THE signal is given by $PR_0^{\mathrm{THE}}\mathbf{B}_{rz}^{\mathrm{em}} $ and can be inferred from the following reasoning~\cite{nagaoka1966ferromagnetism}. As per~\cite{nagaoka1966ferromagnetism}, for MnSi skyrmion lattice, considering the size of the unit cell and the ordered magnetic moment, the strength of the effective magnetic field is,
    \begin{equation}
        \boldsymbol{B}_{rz}^{\mathrm{eff}} = -\frac{h}{e}\left(\frac{\sqrt{3}}{2 \lambda_S^2}\right) \approx -13.75 T
    \end{equation}
    The negative sign indicates that the emergent magnetic field aligns antiparallel to the externally applied magnetic field. Using these values, one can approximate the absolute size of the THE contribution, $\rho^{\mathrm{THE}}_{xy} \approx -20 \mathrm{n\Omega cm}$. \\ In materials containing skyrmion crystal phases, this effect results from the interaction between electrons and the emergent electromagnetic field generated by the moving skyrmion. 
    Confirming experimentally whether the THE is solely attributable to the emergent field of chiral spin textures is challenging. However, in several samples, the measured Hall signal correlates with the number of detected skyrmions~\cite{maccariello2018electrical}. For example, the decrease in average magnetization caused by the presence of skyrmions is also proportional to number of skyrmions. Therefore, in theoretical models, depending solely on the emergent field approach might not be adequate, and more advanced techniques like the tight-bonding model or ab initio methods might become necessary. When dealing with topological spin textures, the accumulated berry phase in conduction electrons can be connected to an effective vector potential, resulting in a transverse deflection as depicted in FIG.~\ref{fig:THE}. In an adiabatic approximation, the corresponding field $\mathbf{B}_{\mathrm{em}}$ is referred as the \textit{emergent field}, as described by Nagaosa \textit{et al.} [see section~\ref{electrodynamics}]
    \begin{equation}
        \mathbf{B}^{\mathrm{em}}_{\alpha}(r) = \frac{1}{2} \epsilon_{\alpha \beta \gamma} \boldsymbol{m}(r) \cdot \left[\partial_{\beta}\boldsymbol{m}(r) \times \partial_{\gamma}\boldsymbol{m}(r) \right]
    \end{equation}
    In two dimensions it is proportional to the skyrmion density or topological charge density
    \begin{equation}
        \mathbf{B}^{\mathrm{em}}(r) =  \Phi_0 \Phi^z
    \end{equation}
    with a single emergent magnetic flux quantum $\Phi_0 = h/e$ and skyrmion density $\Phi^z$ defined as
    \begin{equation}
        \Phi^{\alpha} = \frac{1}{8\pi} \epsilon_{\alpha \beta \gamma} \boldsymbol{m} \cdot (\partial_{\beta} \boldsymbol{m} \times \partial_{\gamma} \boldsymbol{m})
    \end{equation}
    Here, $\epsilon_{\alpha \beta \gamma}$ represents the completely anti-symmetric tensor. Consequently, each skyrmion generates precisely one unit of emergent magnetic flux and the resultant topological Hall resistivity $\rho^{\mathrm{THE}}_{xy}$ is proportional to skyrmion density $\Phi^z$.\\
    Or in terms of topological charge density
    \begin{equation}
        \mathbf{B}_{\mathrm{em}}(r) =  4\pi n_{\mathrm{sk}}(r) e_z
        \label{Eqn:143}
    \end{equation}
            \begin{figure*}[htbp]
        \centering
        \includegraphics[width = \linewidth]{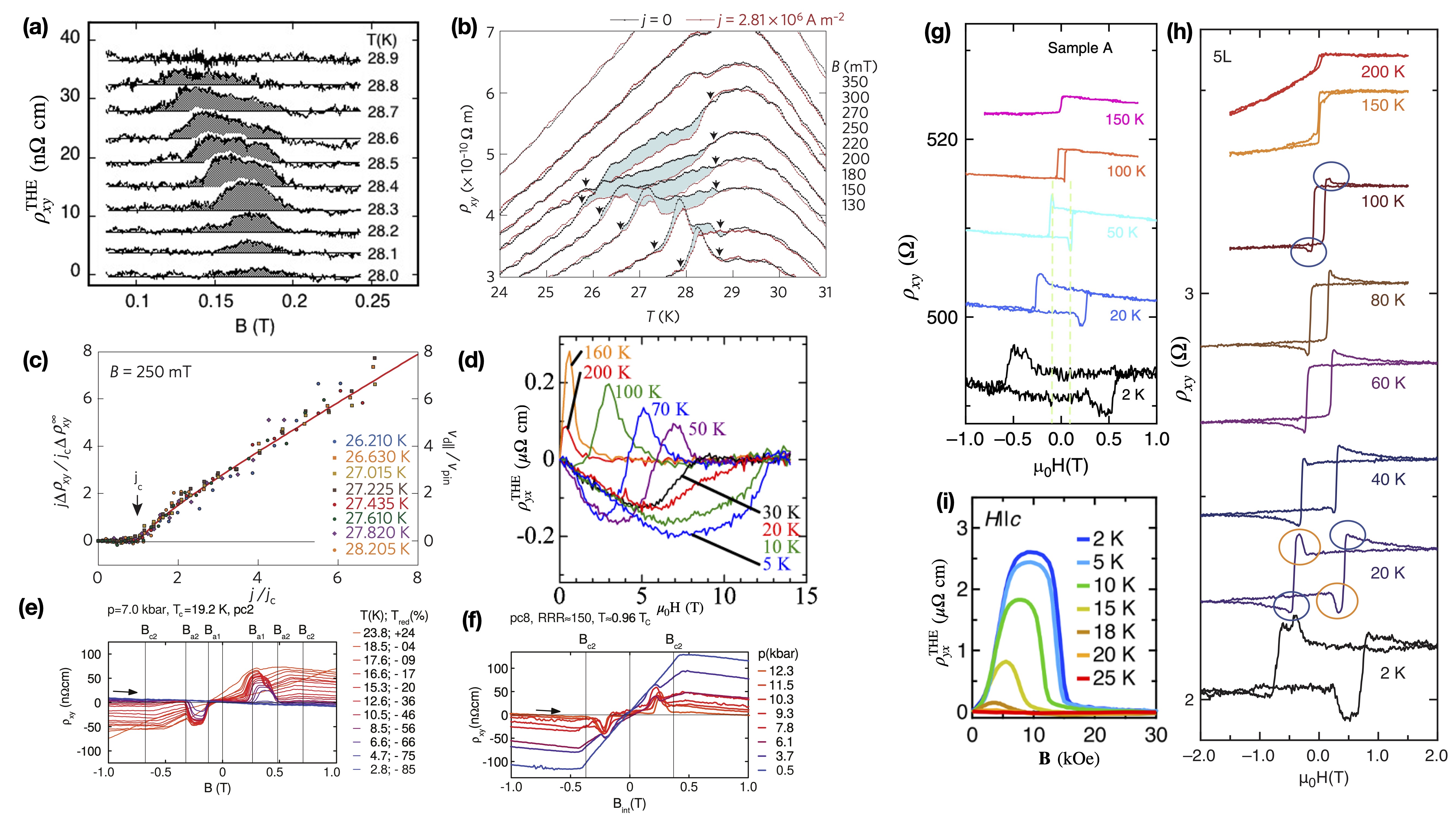}
        \caption{\textbf{Topological Hall Effect.} {(a)} Topological Hall resistivity variation with temperature due to skyrmion phase (SkL) in MnSi, with permission from Ref. \cite{neubauer2009topological};
        {(b)} Temperature dependence of the Hall resistivity of MnSi under a d.c. electric current, with permission from Ref.~\cite{schulz2012emergent};
        (c) Transverse Electric field induced by the moving skyrmion lattice, with permission from Ref.~\cite{schulz2012emergent}; 
        {(d)} Magnetic field dependence of the topological Hall resistivity $\rho^{\mathrm{THE}}_{xy}$, with permission from Ref.~\cite{kanazawa2011large};
        {(e)} THE in SkL phase at different $T$ and constant pressure $p = 7.0$ kbar within boundaries $B_{a1}$ and $B_{a2}$. Arrows indicate sweep direction from helical to FM state through transition state $B_{c2}$, with permission from Ref.~\cite{ritz2013formation};
        {(f)} At low pressures and transition $T$, enhanced anomalous Hall effect, which saturates at $B_{c2}$, with permission from Ref. \cite{ritz2013formation};
        {(g)} THE Signal in $\mathrm{WTe_2/Fe_3GeTe_2}$ heterostructures Hall hysteresis loops, with permission from Ref.~\cite{wu2020neel};
        {(h)} $\mathrm{Cr_2Te_2Ge_6/ Fe_3TeGe_2} $ van der Waals heterostructures where peaks and dips  appear below 60 K signifying the existence of a THE, with permission from Ref.~\cite{wu2022van};
        {(i)} TH resistivity variation with temperature due to skyrmion phase (SkL) in $\mathrm{Gd_2PdS_3}$, with permission from Ref.~\cite{kurumaji2019skyrmion}}
        \label{fig:Topohall}
    \end{figure*}
    FIG.~\ref{fig:THE}(b) demonstrates the contribution of $\rho_{xy}^{\mathrm{THE}}$ for MnSi, obtained by subtracting the $\mathbf{B}$-linear term from $\rho_{xy}$. The results align with the characteristics of the `A-phase' (skyrmion crystal state) and helical magnetic phase, where the skyrmion number $\int dx dy \Phi^z$ is negative and zero, respectively. The finite $\rho_{xy}^{\mathrm{THE}}$ with opposite sign to $\rho_{xy}^{\mathrm{NHE}}$ is observed solely in the former spin state. When the electron spin and the texture are strongly coupled, the THE is directly proportional to the number of skyrmions in the sample, i.e., $\rho_{xy}^{\mathrm{THE}} \propto \langle \mathbf{B}^{\mathrm{em}}_{\mathrm{z}}\rangle \propto N_{\mathrm{sk}}$.\\
    The THE bears a strong resemblance to the ordinary Hall effect. Consequently, for small skyrmions with sizes around 1 nm for atomic Fe layer on Ir(111) surface,  $\sim$3 nm for MnGe, $\sim$18 nm MnSi, and roughly 70 nm for FeGe, the resulting emergent magnetic field can be as large as thousands of Tesla, such as 4000 T, 1100 T, 28 T and 1 T, in respective cases. Theoretical predictions also suggest the existence of a quantized version of the THE~\cite{nagaosa2010anomalous,hall1879new}, akin to the quantum Hall effect, with an opposing orientation for spin-up and spin-down electrons, different from a true magnetic field. A non-zero topological Hall resistivity can arise in materials with a crystallographic lattice featuring non-trivial geometry characterized by multiple inequivalent loops in unit cell~\cite{taguchi2001spin} or due to special topology of the spin texture associated with a non-zero skyrmion number~\cite{onoda2004anomalous}. \\
    The topological hall effect dominates over the anomalous hall effect when the spin modulation period $\lambda_m$ significantly exceeds the crystallographic lattice constant $a$. This situation is observed in B20 compounds, where the lattice constant $a$ is around 0.5 nm and the skyrmion spin texture has a modulation period of 30 nm to 200 nm. Eqn.~(\ref{Eqn:143}) predicts that a higher skyrmion density, or equivalently, a smaller skyrmion size, leads to larger values of $\mathbf{B}^{\mathrm{em}}$ and $\rho_{xy}^{\mathrm{THE}}$. For instance, MnGe with a spin modulation period $\lambda_m \sim 3$ nm has been reported to exhibit a much higher $\rho^{\mathrm{THE}}_{xy}$ of approximately 0.16 $\mu \Omega$ compared to $\rho^{\mathrm{THE}}_{xy} \sim 0.004 \mu \Omega$ observed in MnSi with $\lambda_m \sim 17$ nm [see FIG.~\ref{fig:Topohall}(d) and (a) respectively].\\
    Interestingly, the THE resulting from scalar spin chirality has been investigated under hydrostatic pressure. The study reveals that above a critical pressure $p_c$, approximately $14.6$ kbar, the helical magnetic phase is suppressed, leading to intriguing transport properties such as non-Fermi liquid behavior ($\rho_{xx} \propto T^{1.5}$) and the prominence of the THE. The corresponding development of $\rho_{xy}$ at various pressures is shown in Fig.~\ref{fig:Topohall}(f), with further details available in Ref.~\cite{ritz2013formation}.
    The continuous evolution of the topological Hall resistivity from the A-phase to the non-Fermi liquid state indicates that the topological characteristics of a skyrmion lattice persist in the latter exotic, non-Fermi liquid state even without long-range magnetic order. Importantly, the temperature dependence of the THE decreases with increasing temperature due to the diminishing difference in density between minority and majority electrons. Additionally, spin-flip scattering may enhance the temperature dependence, as it typically increases with rising temperature, causing the THE to decrease even faster as the temperature increases.\\
    In 2009, Neubauer \textit{et al.} conducted the first experimental study on the THE in chiral magnets, explaining it in terms of real space Berry phases~\cite{neubauer2009topological} [see FIG.~\ref{fig:Topohall}(a)]. Another research group later improved this study by considering accuracy and additional pressure effects. To observe THE in the skyrmion lattice phase, magnetic field sweeps were performed at different pressures and temperatures. The Hall resistivity ($\rho_{xy}$) was recorded as a function of the applied magnetic field up to 1 Tesla for various temperatures at a pressure of 7 kbar and later at various pressures, depicted in FIG.~\ref{fig:Topohall}(e, f).  With such pronounced results by Ritz \textit{et al.}, it was found that at high temperatures, the transverse magnetoresistance ($\rho_{xx}$) decreases with increasing magnetic field, while the Hall resistivity ($\rho_{xy}$), as shown in FIG.~\ref{fig:Topohall}(e),  shows a gradual field dependence with a pronounced top-hat-shaped enhancement in a small field and temperature range ($B_{a1}$ and $B_{a2}$) larger than the skyrmion lattice phase at ambient pressure~\cite{ritz2013formation}. The magnitude of this top-hat-shaped signal contribution is substantially larger than a similar signal contribution in the skyrmion lattice phase at ambient pressure. \\
   To distinguish between the anomalous and topological Hall effects in the signal contribution, the researchers measured the temperature dependence of the magnetization. Surprisingly, no variations in the magnetization were observed that could account for the additional top-hat-shaped signal, typically associated with the intrinsic anomalous Hall effects. To further understand the origin of the large magnitude of the top-hat-shaped signal contribution, the Hall conductivity ($\sigma_{xy} = -\rho_{xy}/(\rho^2_{xx} + \rho^2_{xy}) \approx -\rho_{xy}/\rho^2_{xx}$) was examined in FIG.~\ref{fig:Topohall}(f). It was found that the top-hat-shaped contribution in $\sigma_{xy}$ becomes stronger at lower temperatures (and hence lower $\rho_{xx}$), compared to the signal in $\rho_{xy}$. As discussed earlier, the intrinsic anomalous Hall effect would result in a universal Hall signal in $\rho_{xy}$ that is independent of the scattering time ($\tau$), whereas for the THE, $\rho_{xy}$ is independent of $\tau$ and $\sigma_{xy}$ increases proportionally to $1/\rho^2_{xx}$. Based on these observations, it is suggested that the top-hat-shaped signal can be attributed to the THE, which is switched on and off as the system enters and leaves the skyrmion phase, respectively. In conclusion, this study experimentally observed the emergent magnetic field arising from the topologically non-trivial magnetization configuration in the skyrmion lattice phase as experienced by the conduction electrons.\\
    In 2019, a giant THE was found in the skyrmion
    state of the frustrated magnet $\mathrm{Gd_2PdSi_3}$ which has a much shorter modulated length [FIG.~\ref{fig:Topohall}(i)]~\cite{kurumaji2019skyrmion}. In 2020, N\'{e}el-type skyrmions were observed by LTEM in $\mathrm{WTe_2/Fe_3GeTe_2}$ heterostructures which display THE signals below 100 K [FIG.~\ref{fig:Topohall}(g)]~\cite{wu2020neel}. In the year 2022 \cite{wu2022van}, the distinction between signals related to temperature played a crucial role in distinguishing between two categories of skyrmions situated at the boundary of two van der Waals magnets, depicted in FIG.~\ref{fig:Topohall}(h).
    The THE of a skyrmion lattice is more readily observable than that of a single skyrmion. The contribution of emergent skyrmions to the Hall resistivity was investigated in sputtered magnetic multilayers capable of stabilizing individual skyrmions~\cite{maccariello2018electrical, zeissler2018discrete}. Although the relationship between the Hall signal change in a Pt/Co/Ir multilayer due to a single skyrmion and the THE signal is limited, it enabled the electric detection of skyrmions~\cite{maccariello2018electrical}.\\ 
    Furthermore, the anomalous Hall effect has been observed in several coplanar Kagome magnets that do not exhibit net magnetization~\cite{chen2014anomalous,kubler2014non,busch2020microscopic,nakatsuji2015large, nayak2016large}, and the crystal Hall effect has been predicted and measured in magnetic systems where the arrangement of non-magnetic atoms in the crystal breaks a set of time reversal and spatial symmetries~\cite{vsmejkal2020crystal,feng2020observation}. Although the influence of these effects on conventional skyrmions is not yet fully understood, it is challenging to make predictions on how these contributions may affect alternative magnetic quasiparticles. However, experiments on skyrmions can be adequately explained without considering these contributions, although they could potentially have a significant quantitative impact.
    \subsection{Manifesting Skyrmion Hall Effect (SkHE)}
\begin{figure}[t]
        \centering
        \includegraphics[width = \linewidth]{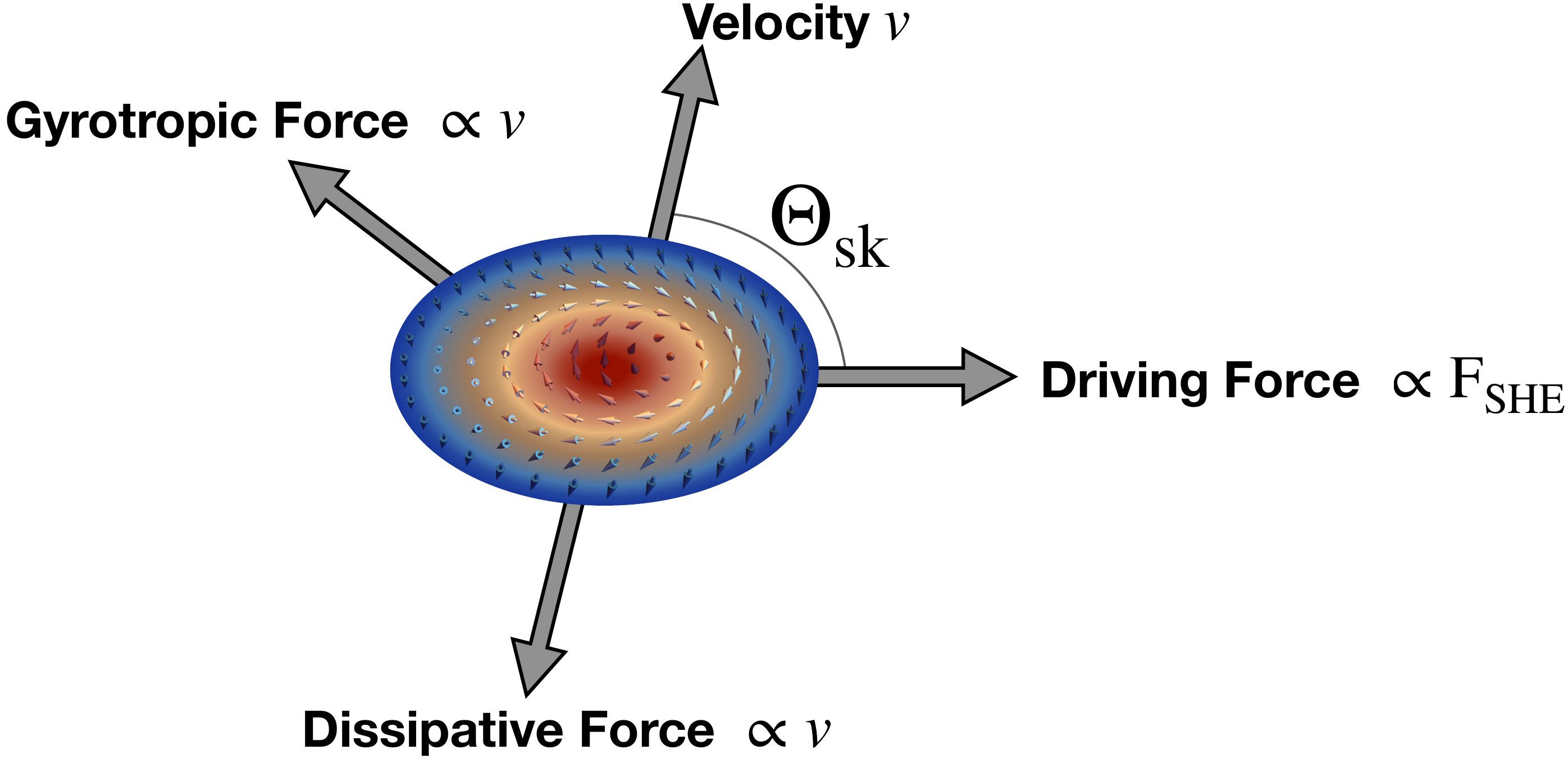}
        \caption{\textbf{Effective forces.} Forces acting on a skyrmion when a spin Hall force is
        applied. Not drawn to scale.}
        \label{fig:effective-forces}
    \end{figure}

    Skyrmions usually exhibit a distinct motion that is parasitic in nature from the direction of applied force. This effect arises from the interplay between different forces acting on it. Although there are just three effective forces, their directions can lead to intricate and complex dynamics.
    The first term in Eqn. (\ref{Eqn: 81}, studied in details) and below
\begin{equation}
\mathscr{\vec{F}_G}+\mathscr{\vec{F}_D}+\mathscr{\vec{F}}_{\text{ext}}=0
\end{equation}
    represents the gyrotropic force, which acts perpendicular to the velocity of the skyrmion. The second term corresponds to the dissipative force, which acts in the opposite direction of the velocity and can be thought of as a form of drag caused by factors such as the Gilbert damping parameter. The last term represents an equivalent conservative external driving force acting on the skyrmion by excitations such as energy gradients, electrical currents and spin currents. At equilibrium, all these forces must balance each other out. FIG.~\ref{fig:effective-forces} illustrates the different forces acting on the skyrmion. Given that $\mathscr{\vec{F}_G}$ and $\mathscr{\vec{F}_D}$ are orthogonal, it necessitates $\mathscr{\vec{F}}_{\text{ext}}$ from being colinear with $v$ and leads to the deviation in the motion of the skyrmion from its driving force, also referred to as the skyrmion Hall effect.
    One can quickly ascertain the direction of skyrmion velocity using the skyrmion hall angle $\Theta_{\mathrm{sk}}$, which represents the angle between the force vector and the velocity vector. The value of $\Theta_{\mathrm{sk}}$ is determined by material properties and the skyrmion number, and it is obtained by calculating the ratio of $\boldsymbol{v}_y$ to $\boldsymbol{v}_x$ [see section~\ref{SOT-driven}] which simplifies to
    \begin{equation}
       \Theta_{\mathrm{sk}} = \tan^{-1} \left(\frac{G}{\alpha D}\right)
    \end{equation}
    Given the skyrmion configurations, a force due to SHE ($\vec{F}_{\mathrm{SHE}}$) is always along one of the cartesian axes, depending on the phase of skyrmion with wave vector $q$ and polarity $p$. It is then possible to quickly derive the skyrmion velocity.\\
    It can be inferred from Eqn.~(\ref{Eqn108}) that the Rashba field ($\mathcal{H}_R$) is another effective field exerted by spin-orbit torque that can be expressed as,
    \begin{equation}
        \mathcal{H}_{\mathrm{rashba}} = \frac{\alpha_R j P}{\mu_B M_s} \sigma
    \end{equation}
    where $\alpha_R$ is the Rashba parameter that is averaged over the magnetic film thickness, $\mu_B$ is the Bohr magneton and $P$ is the polarization of the carriers in the ferromagnetic layer. Unlike the field generated by the Spin Hall Effect (SHE), which depends on the magnetization direction, the Rashba field is constant throughout the ferromagnetic film. As a result, the Rashba field does not directly drive the motion of the skyrmion, but rather causes a change in the magnetization configuration of the skyrmion [see FIG.~\ref{fig:emerg-field} and for theoretical details Ref.~\cite{akosa2019tuning}]. While this change may affect the calculated $\mathrm{\vec{F}}_{\text{SHE}}$ (field-like spin Hall effect), the impact is negligible compared to the other effective fields~\cite{wong2019enhanced}. \\
    Nonetheless, the skyrmion Hall effect poses a threat to the stability of magnetic skyrmion in confined structures like nanotracks, where the skyrmions can be driven to annihilation at the structure edges. The confining potential induced by DMI at the edges provides limited protection to the magnetic skyrmions and can be overcome at sufficient velocities.
    Although an interesting insight can be gleaned from experimental result FIG.~\ref{fig:racetrack_sky}(h,i) and simulated result FIG.~\ref{fig:sot-plot} that the skyrmion Hall effect predicted by the Thiele's equation was absent at low driving currents. Hall angle $\Theta_{\text{sk}}$ increases and saturates with the current density and skyrmion velocity observed in experiments. Litzius \textit{et al}. reported a linearly increasing $\Theta_{\text{sk}}$ with skyrmion velocity and proposed that this increase was caused by skyrmion deformation at high velocities~\cite{litzius2017skyrmion47k}. Moreover, they found that in an even higher drive regime, both the field-like and damping-like components of SOT induce significant deformations in the skyrmions, leading to an increase in $\Theta_{\text{sk}}$. This behavior of current density dependent $\Theta_{\text{sk}}$ has been observed in later works on different materials, such as Pt/Co/MgO and [Pt/GdFeCo/MgO].
        \begin{figure}[htbp]
        \centering
        \includegraphics[width = \linewidth]{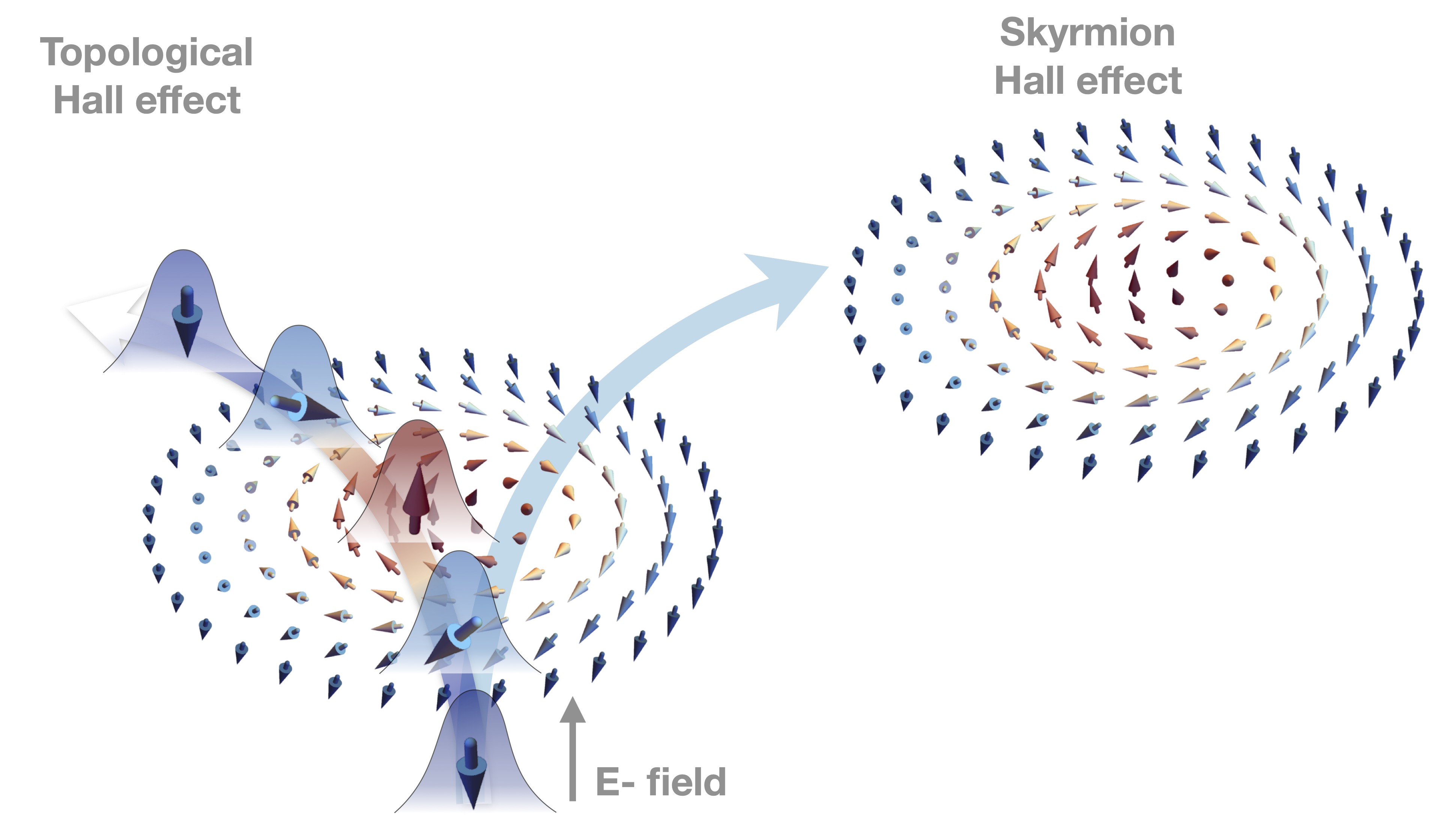}
        \caption{\textbf{Emergent electrodynamic effects in skyrmion hosts.} Semiclassical wave packet follows the texture influenced by real-space Berry curvature from skyrmions (THE). Skyrmion is propelled due to spin torques from partial alignment of current electrons' spins with the texture (skyrmion Hall effect).}
        \label{fig:SHE-53117188}
    \end{figure}
\section{Skyrmion Technologies}
Spintronic devices utilizing magnetic skyrmions offer several advantages, such as increased density and energy-efficient data storage, owing to their nanoscale dimensions and topological protection~\cite{luo2021skyrmion}. These magnetic skyrmions can be driven with remarkably low depinning current densities~\cite{zang2011dynamics83c}, and their dimensions can be scaled down to 1 nm~\cite{wang2018theory30}. These properties make them highly promising for data storage and computing applications~\cite{kang2016skyrmion}, particularly in unconventional computing techniques like neuromorphic computing~\cite{song2020skyrmion} and reversible computing~\cite{chauwin2019skyrmion}. Neuromorphic computing draws inspiration from the brain's performance and energy efficiency~\cite{8429425}, employing neuromimetic devices that mimic neurons for various computing tasks. Among these devices, magnetic tunnel junctions (MTJs) have garnered significant interest~\cite{srinivasan2016magnetic,deng2020voltage,lone2021voltage}, and recently, there has been a surge in proposals for neuromorphic computing systems that integrate MTJs based on skyrmions, such as skyrmion neurons~\cite{luo2021skyrmion,he2017developing,lone2022skyrmionneuron} and skyrmion synapses~\cite{huang2017magnetic, lone2022skyrmion, song2020skyrmion}.\\
Additionally, controlling spintronic devices using electric fields has emerged as a highly attractive approach for memory and logic applications, enabling improved data-storage density~\cite{tan2019high,wang2013voltage} and reduced switching energy consumption~\cite{ong2016electric}. However, the successful implementation of skyrmions in storage and computing, whether conventional or unconventional, presents crucial challenges, notably concerning the controlled motion and readability of skyrmions~\cite{back20202020}.

\subsection{Skyrmion Racetrack}
   \begin{figure*}[htbp]
    \centering
    \includegraphics[width  = \linewidth]{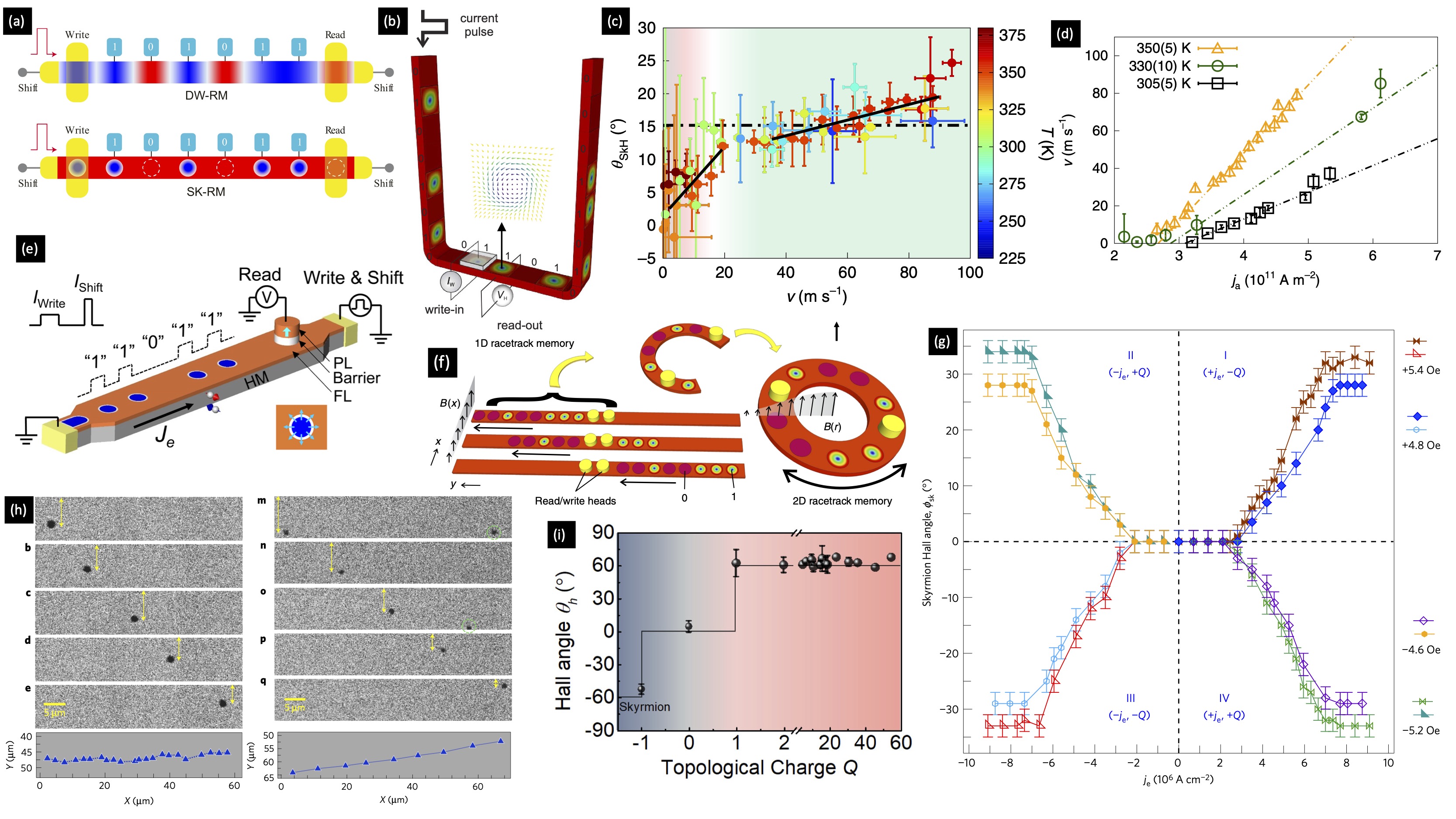}
    \caption{\textbf{Skyrmion-based Racetrack Memory.}~
    (a)~Schematic of a DW-racetrack memory and Sk-racetrack memory, with permission from Ref.~\cite{krause2016skyrmionics}
    {(b)}~STT-driven Skyrmion-based racetrack memory, with permission from Ref.~\cite{zhang2015topological}; 
    (c)~Skyrmion Hall angle (SHA) $\Theta_{\text{sk}}$ as a function of skyrmion velocity at various temperatures in the same structure as (d), with permission from Ref.~\cite{litzius2017skyrmion47k};
    (d)~skyrmion velocity as a function of current density at several sample temperatures stack, with permission from Ref.~\cite{litzius2017skyrmion47k}; 
    (e)~1D Sks memory device, writing is realized by applying a low amplitude current pulse ($I_{\mathrm{Write}}$) and readout is accomplished using an MTJ, with permission from Ref.~\cite{yu2017room};
    (f)~Schematic of Circular skyrmion racetrack memory concept, with permission from Ref.~\cite{zhang2018manipulation};
    (g)~MOKE images of the pulse-driven motion of skyrmion showing zero and enhanced transverse motion for low (left) and high current density (right), with permission from Ref.~\cite{jiang2017direct};
    (h)~Skyrmion Hall angle in $\mathrm{Ta/CoFeB/TaO_x}$ across all four combinations between current directions and skyrmion polarities, with permission from Ref.~\cite{jiang2017direct};
    (i)~Plot of SHA as a function of the topological charge $Q$ of skyrmion bundles, with permission from Ref.~\cite{tang2021magnetic}}
    \label{fig:racetrack_sky}
\end{figure*}
Stuart S.P Parkin and his IBM team initially proposed a highly promising magnetic memory architecture known as the racetrack in 2008, wherein data bits are encoded utilising a sequence of a magnetic domain of spin-up and spin-down within magnetic regions separated by domain walls~\cite{parkin2008magnetic}, and has been widely studied ever since. This encoded information is then transmitted through a nanowire, employing an electric current (STT mechanism), in either horizontal or vertical  configurations as illustrated in FIG.~\ref{fig:racetrack_sky}(a) (top). However, the motion of DWs requires current densities exceeding $1 \times 10^{11}~\si{\ampere\per\meter^2}$ and is susceptible to domain wall pinning at the edges and defects.
In recent years, magnetic skyrmions have been envisioned as potential carriers of information in racetrack storage devices. In a proposal by Fert \textit{et al}.~\cite{fert2013skyrmions,fert2017magnetic}, these magnetic textures due to their remarkable properties such as their size, quantized topological charge, statics and dynamics originating from their nontrivial topology can replace domain walls in ferromagnetic systems as shown in FIG.~\ref{fig:racetrack_sky}{(a) (bottom)}. Due to the topological protection, scalability and magnetic particle-like character, skyrmions can be driven by STT and SOT current densities around $1\times 10^6\mathrm{A/m^2}$ which is orders smaller than the current requirements of the DW motion [refer to section~\ref{EffectiveDynamics} for details]. Hence, skyrmions have demonstrated substantial promise in racetrack memories. Within a skyrmion-based racetrack memory setup, the storage of binary data follows this pattern - a skyrmion signifies a ``1'', while its absence signifies a ``0'' as illustrated in FIGs.~\ref{fig:racetrack_sky}(a,b,e,f)].
The skyrmions are nucleated at one edge of the device and driven by current towards the reading MTJ. In section (\ref{sec:level33}), we thoroughly examined the techniques for detecting or reading skyrmions, which form the basis for the subsequent detailed discussions. The MTJ detects the presence or absence of the skyrmion and generates corresponding binary output at its reading terminal [see FIG.~\ref{fig:racetrack_sky}(e)],  which is further resolved by the read circuitry of the MTJ. One option for confining skyrmions at specific points along the track is the use of artificial pinning structures~\cite{sampaio2013nucleation88c}. An alternative approach, proposed by Zheng \textit{et al}., involves employing a continuous magnetic skyrmion and bobbers chain. In this scheme, a bobber takes the place of a skyrmion's absence to maintain their relative positions along the track~\cite{zheng2018experimental}. Another variation of this idea was demonstrated by Jena \textit{et al}., who used antiskyrmions instead of chiral bobbers~\cite{jena2020evolution}. In racetrack memory, data needs to be shifted to the desired position for reading and writing, which requires extra space to accommodate these shifts. To address this issue, Zhang \textit{et al}. proposed a circular racetrack design that eliminates the need for excessive space and can be driven using magnetic field gradients [see FIG.~\ref{fig:racetrack_sky}(f)]~\cite{zhang2018manipulation}. The authors further introduced a concept of a perpendicular track architecture to enhance information density, as depicted in FIG.~\ref{fig:racetrack_sky}{(b)} and experimentally demonstrated the feasibility of this architecture using a thin BDMI material, $\mathrm{Cu_2OSeO_3}$~\cite{zhang2020robust}.\\
Apart from the MTJ reading~\cite{li2022experimental,zhang2018skyrmions,kang2016voltage}, the alternate readout mechanism is based on the topological Hall resistivity as discussed in section~(\ref{THE}). Although the skyrmion racetrack has shown substantial potential for both storage class memory and CPU memory. It still faces certain challenges such as the Skyrmion Hall Effect (SkHE) which deviates skyrmion from the desired path and can cause data loss to edges.
As evidenced by experimental findings in a $\mathrm{[Pt/CoFeB/MgO]_{15}}$ multilayer stack racetrack configuration~\cite{litzius2017skyrmion47k}, along with the longitudinal motion of skyrmion, a transverse velocity component is also present, as illustrated in FIG.~\ref{fig:racetrack_sky}(c). Furthermore, FIG.~\ref{fig:racetrack_sky}(d) depicts that skyrmions exhibit linear response against the hall angle and can achieve significant velocities ($\geq$ 100 m/s). Another similar experimental study directly observed the SkHE in $\mathrm{Ta/CoFeB/TaO_x}$ multilayers using MOKE, demonstrated in FIG.~\ref{fig:racetrack_sky}(g). Authors showed that at a low driving current density ($j \sim 1.3 \times 10^6 \mathrm{A/m^2}$), the transverse motion of the skyrmion is negligible and the motion of the skyrmion exhibits stochastic behavior in tandem with the current. However, upon increasing the current density ($j \sim 2.8 \times 10^6 \mathrm{A/m^2}$), the transverse motion of the skyrmion becomes pronounced. The result of linear skyrmion Hall angle dependence on density until reaching saturation is illustrated in FIG.~\ref{fig:racetrack_sky}(h), and found similar to the FIG.~\ref{fig:racetrack_sky}(d). Moreover, at the edge of the tracks, skyrmions exhibit oscillatory motion as a result of the interaction between the driving force and the repulsive force originating from the edge.
To overcome this parasitic SkHE, numerous theoretical and empirical models have been proposed. A noticeable approach involves the combination of two skyrmions with opposite topological charges, for example, results in antiferromagnetic skyrmions~\cite{zhang2016antiferromagnetic, buhl2017topological,gobel2017antiferromagnetic} [discussed in the subsequent section~(\ref{AFMRacetrack})], bilayer skyrmions~\cite{barker2016static, zhang2016magnetic}, or $2\pi$ skyrmions~\cite{finazzi2013laser,zhang2018skyrmions}. Their lack of topological charge results in an absence of the SkHE, but either their experimental observation remains elusive or their instability during motion affects their detection~\cite{zhang2016control}. FIG.~\ref{fig:racetrack_sky}(g) shows a graph of the hall angle for different topological charge ($Q$) of skyrmion. Another idea is to modify the racetrack setup by interfacing the actual racetrack with a second ferromagnetic. The magnetization direction can be deliberately selected to enable the skyrmion's motion along the racetrack. G\"{o}ebel \textit{et al.} presented the idea where the skyrmion Hall angle can be engineered to zero utilising the method in Ref.~\cite{gobel2019overcoming}.
\subsection{\label{AFMRacetrack} Antiferromagnetic Skyrmion THE (AFSkHE)}
     Skyrmion-based racetrack memory is promising for future magnetic memories due to its topological protection, small size~\cite{buttner2015dynamics}, weak sensitivity to defects~\cite{lin2013particle243}, ultra-low driving current density~\cite{iwasaki2013current85c,iwasaki2013universal86c,iwasaki2014colossal,sampaio2013nucleation88c,everschor2011current,litzius2017skyrmion47k}, and high mobility, as discussed in various studies~\cite{fert2013skyrmions, wiesendanger2016nanoscale, romming2013writing48, hsu2017electric, zhang2015magnetic,zhang2015magnetic2,jiang2015blowing, boulle2016room, seki2012magnetoelectric44c, woo2016observation46k}. In spite of these remarkable properties of skyrmion, skyrmion crystals (SkXs) exhibit THE~\cite{neubauer2009topological, schulz2012emergent, kanazawa2011large, lee2009unusual,li2013robust,bruno2004topological,ndiaye2017topological,hamamoto2015quantized} which in turn gives rise to SkHE (Skyrmion Hall Effect, also observed in single skyrmion)~\cite{nagaosa2013topological36, zang2011dynamics83c,jiang2017skyrmions,litzius2017skyrmion47k}. However, from the perspective of data storage applications, the SkHE is considered undesirable. Various approaches have been suggested to mitigate this parasitic effect, and among them is the utilization of skyrmions in an antiferromagnetic (AFM) background instead of a ferromagnetic (FM) as we concluded in the previous section. Due to the inherent texture of AFM, dipolar fields are merely present unlike in ferromagnets where they can hinder the ultrasmall size nucleation of skyrmions and their stability without external magnetic fields~\cite{woo2016observation46k,moreau2016additive, boulle2016room}. As a result, both THE and SkHE~\cite{nagaosa2013topological36, zang2011dynamics83c,jiang2017skyrmions,litzius2017skyrmion47k} vanish ~\cite{barker2016static,jin2016dynamics,velkov2016phenomenology,gobel2017antiferromagnetic} for such distinct textures. Despite the continuous progress of various topological textures in pertaining research, antiferromagnetic (AFM) textures currently hold the most promising potential as the preferred data carrier in racetrack storage devices.\\
     \begin{figure*}[htbp]
        \centering
        \includegraphics[width = \linewidth]{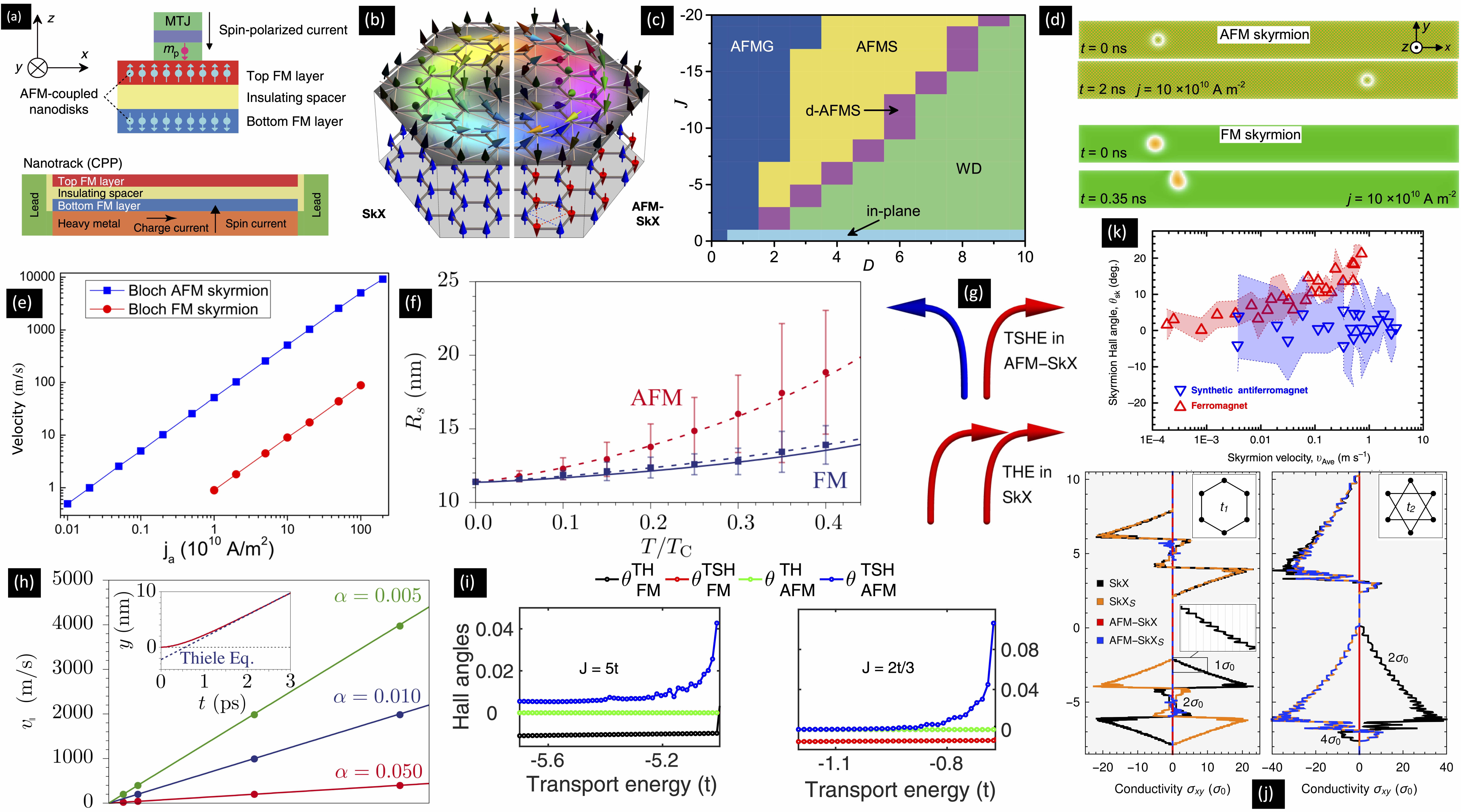}
        \caption{\textbf{Antiferromagnetic Skyrmion THE.}~(a)~Schematics of AFM coupled bilayer-skyrmion system, with permission from Ref.~\cite{zhang2016magnetic};
        (b)~FM (left) and AFM (right) skyrmion on a honeycomb lattice with dotted triangular sublattices, with permission from Ref.~\cite{gobel2017antiferromagnetic}; 
        (c)~Phase diagram of the AFM skyrmion, with permission from Ref.~\cite{zhang2016antiferromagnetic};
        (d)~Illustration of Zero(non-zero) SkHE of AFM(FM) skyrmion moving on an AFM(FM) track, with permission from Ref.~\cite{zhang2016antiferromagnetic};
        (e)~Velocity of Bloch AFM skyrmion and Bloch FM skyrmion, with permission from Ref.~\cite{jin2016dynamics};
        (f)~Radius of AFM and FM Skyrmion $R_s$ dependenc on temperature $T/T_C$, with permission from Ref. ~\cite{barker2016static};
        (g)~Signature of substantial TSHE (top) and THE (bottom) in AFM Sks and SkX, with permission from Ref.\cite{gobel2017antiferromagnetic};
        (h)~Current induced AFM Skyrmion longitudinal velocity for different combinations of $\alpha$ and $\beta$, with permission from Ref.~\cite{barker2016static};
        (i)~Schematic of the significant TSHE and zero THE in an AFM-Sk, with permission from Ref.~\cite{akosa2018theory};
        (j)~Topological Hall conductivity in SkX (black: charge conductivity, orange: spin conductivity (SkX$_\text{S}$)) and AFM-SkX (red: charge conductivity, blue: spin conductivity ($\text{AFM-SkX}_\text{S}$)), skyrmion coupling is $J = 5t$ and conductivities are quantized in units of $\sigma_0 = e^2/h$ (charge) and $\sigma_0 = e^2/{4\pi}$ (spin), with permission from Ref.\cite{gobel2017antiferromagnetic}
        (k)~Dependence of skyrmion Hall angle on velocity for AFSk and FMSk, with permission from Ref.~\cite{dohi2019formation};}
        \label{fig:TSHE_sky}
    \end{figure*}
     In subsequent developments~\cite{nunez2006theory,haney2008current,gomonay2010spin, gomonay2012symmetry}, AFM Skyrmions have also been predicted in AFM materials. As no such antiferromagnetic skyrmion crystal has yet been discovered, researchers have studied surrogate systems comprising two skyrmion layers~\cite{zhang2016magnetic,barker2016static} and two sublattice~\cite{zhang2016antiferromagnetic, buhl2017topological, gobel2017antiferromagnetic} systems. In the former case, authors have proposed generating a bilayer-skyrmion [see FIG.~\ref{fig:TSHE_sky}(a)] from antiferromagnetically coupled two perpendicularly magnetized FM layers with a bottom layer atop a heavy metal layer where a vertical current injection nucleates skyrmion in one sublayer that subsequently nucleates a skyrmion in another sublayer when AFM coupling is strong. On the other hand, in the latter case, authors have demonstrated generating an AFM-SkX crystal [see FIG.~\ref{fig:TSHE_sky}(b)] with two intrinsic sub-lattices (A and B triangular sublattices on a honeycomb lattice) coupled by a weak inter-sublattice $J_1^{\text{AB}}$ while within the sublattices, there exists nearest $J_1$ and third nearest coupling $J_3$, that stabilize skyrmions in an external magnetic field; all energies measured in terms of $J$. Within the limit $J_1^{\text{AB}}<<J$, both sublattice skyrmions are more or less equivalent. To assist with physical parameters, the SkX system is now coupled with an AFM layer. That generates an AFM-SkX using Monte Carlo Simulations (for more details refer to Ref.~\cite{gobel2017antiferromagnetic}).
    Authors have shown generating generated both SkX and AFM-SkX, depicted in FIG.~\ref{fig:TSHE_sky}{b}(left, right) respectively. For AFM-SkX, the local input parameters like magnetization etc. are reversed for one sublattice. Authors further \cite{akosa2018theory, gobel2017antiferromagnetic} employed the tight-bonding model~\cite{hamamoto2015quantized} to understand the band structure and origins of proposed effects
     \begin{equation}
    \mathcal{H}= \sum_{\langle i j\rangle} t_{ij}\hat{c}_{i}^{\dagger} \hat{c}_{j}+J \sum_{i} \hat{c}_{i}^{\dagger} \boldsymbol{m}_{i} \cdot \hat{\boldsymbol{\sigma}} \hat{c}_{i}
    \end{equation}
     where $t_{ij}=t$ is the hopping strength and $J$ (measured in terms of hopping parameters $t$). In the SkX case, both sublattices can have either a parallel or antiparallel spin alignment, which in coupled case (two sublattices coupled by nearest-hopping $t_1 = t$ and second nearest-hopping $t_2 = 0$) and uncoupled cases ($t_1 = 0$ and $t_2 = t$) project different properties [see inset of FIG.~\ref{fig:TSHE_sky}(j)] with a strong ($J=5t$), intermediate ($J=2t/3$) and weak ($J=0$) electron coupling represented by $J$. In the former coupled case [see FIG.~\ref{fig:TSHE_sky}(j) (left)] and strong coupling ($J = 5t$), having equivalent sublattice results in asymmetric conductivity while in the latter case [FIG.~\ref{fig:TSHE_sky}(j) (right)], since sublattices are decoupled and degenerate, resulting in a transverse spin-polarized current. 
     On the other hand, concerning AFM-SkX, for both strongly ($J = 5t$) and intermediately ($J=2t/3$) coupled, the former case shows zero transverse current due to zero effective field and the latter case shows zero THE with zero effective emergent magnetic field [green lines in FIG.~\ref{fig:TSHE_sky}(i)]. However, there have been found signatures of the topological spin hall effect (TSHE) in strong ($J = 5t$) and intermediate ($J = 2t/3$) coupled system [see blue lines FIG.~\ref{fig:TSHE_sky}(i)]. When an electron traverses through the AFM crystal in the latter decoupled case, it localises itself with either of the parallel or antiparallel sublattices leading to a spin-up from positive magnetization and spin-down current from negative magnetization and in total a pure spin current means no charge transport or THE, unlike the THE in SkX case [see FIG.~\ref{fig:TSHE_sky}(g)].
    In a manner akin to the THE for the detection of SkX, a spin-based counterpart known as the Topological Spin Hall Effect (TSHE) could act as a detection method for AFM-SkX. Moreover, an intriguing observation of made by the authors when the AFM-SkX is generated of asymmetric nature modelled on inequivalent sublattice, revealing the presence of a non-zero THE which can be employed in Spintronics devices.\\
    Similar to FM, the Hamiltonian for antiferromagnetic (AFM) system is
    \begin{equation}
        \mathcal{H}=\frac{1}{2} \sum_{\langle i j\rangle}\bigl\{-\mathcal{J}_{i j} {\boldsymbol{m}}_{i}\cdot {\boldsymbol{m}}_{j}+ \mathcal{D}_{i j}\left({\boldsymbol{m}}_{i}\times {\boldsymbol{m}}_{j}\right)- \mathcal{K}_{z}{\boldsymbol{m}}_{i}^{2}-\mathbf{H} \cdot \boldsymbol{m}_{i}\}
        \end{equation}
    where the initial term denotes the AFM exchange interaction with positive AFM exchange stiffness $\mathcal{J}_{ij}$, the second term represents the DMI with the DMI vector $\mathcal{D}_{ij}$, the third term denotes the PMA with the anisotropic constant $\mathcal{K}_z$ and the final term $\mathbf{H}$ signifies the external magnetic field.\\
    Zhang \textit{et al.} \cite{zhang2016antiferromagnetic}, demonstrated the nucleation of AFM skyrmion from AFM DW pairs, similar to FM skyrmion from FM DW pairs~\cite{zhou2014reversible}, through a phase diagram [see FIG.~\ref{fig:TSHE_sky}(c)] in a disk shape region by applying a spin-polarised current perpendicularly to the disk which flips the spin in the region. In a nanostrip, a stable AFM skyrmion is nucleated in the DMI range $0.7~\mathrm{mJ^2/m^2}<D<1.2~\mathrm{mJ^2/m^2}$, also been observed in Ref.~\cite{jin2016dynamics}. Furthermore, the dynamics of AFM skyrmions are also examined by~\cite{cheng2014dynamics,zhang2016antiferromagnetic, zhang2016magnetic} where, unlike FM skyrmions that easily get destroyed at the edges due to the SkHE, AFM skyrmion have shown remarkable uni-directional velocity promising for future information bits~\cite{yu2017room,zhang2015topological}, depicted in FIG.~\ref{fig:TSHE_sky}(d). Similarly, through a skyrmion bag simulation \cite{tang2021magnetic}, it is found that the skyrmion bundle with zero topological charges (similar to AFM-Sk $Q=0$) exhibits a colinear motion with zero skyrmion hall effect.\\
    The corresponding terminal velocity for AFM-Sk from the Theile equation is evaluated in Ref.~\cite{barker2016static}
    \begin{equation}
        \boldsymbol{v}_{\parallel} = \frac{\beta}{\alpha} \boldsymbol{j}
    \end{equation}
    which is only along the direction of the current.
    In a comparison of velocities, the maximum velocity of a skyrmion in a ferromagnetic (FM) nano track is constrained by a confining force approximately proportional to $\sim (D^2/J)$. Typically, this velocity is much less than $10^2$ m/s or $\boldsymbol{v} \approx 200$ m/s for a current $\boldsymbol{j} = 200$ m/s~\cite{barker2016static} and small values of $\alpha$ and $\beta$ (damping parameters). On the other hand, the antiferromagnetic (AFM) skyrmion has been observed to move directly along an AFM nano track with a velocity around $\approx 10^2-10^3$ m/s, as reported by Barker \text{et al.} Additionally, AFM skyrmion can exhibit high-speed velocities up to km/s while maintaining their stability for low $\alpha$ and high $\beta$, shown in FIG.~\ref{fig:TSHE_sky}(h). A similar comparable outcomes were obtained by \cite{jin2016dynamics} for Bloch type AFM and FM skyrmions, see FIG.~\ref{fig:TSHE_sky}(e) and Ref.~\cite{zhang2016antiferromagnetic}.\\
    Theoretical studies of AFM dynamics have been made by many scholars~\cite{tveten2014antiferromagnetic,cheng2014spin, akosa2018theory}. One challenging threat to AFM skyrmion studied by Barker \textit{et al.} is the thermal perturbation that may cause excessive randomness in the motion for AFM skyrmions~\cite{barker2016static} which was observed with a derived expression for temperature dependence radius [Eqn.~(\ref{thermalAFM})] with possible known parameters ${A}, {K}$ and $D$ dependencies. AFM skyrmions demonstrated a more pronounced temperature dependence and exhibited larger fluctuations compared to FM skyrmions.
    \begin{equation}
    R_{s}(T)=\sqrt{\frac{2 A \lambda}{4 \sqrt{A K} m(T)-\pi D m^{-3 / 2}(T)}}
    \label{thermalAFM}
    \end{equation}
    FIG.~\ref{fig:TSHE_sky}(f) reports the plots for FM and AFM skyrmions temperature dependence mean radius. \\
    Recently~\cite{legrand2020room}, synthetic antiferromagnet (SAF) is seen as a substitute for antiferromagnets comprised of antiferromagnetically coupled ferromagnetic layers of nanometer thickness through non-magnetic space layers. The authors examined the device structure (Pt/Co/Pt) and corresponding phase development, including spin spirals and antiferromagnetic skyrmions (AFMSk) in their study. One key advantage of bilayer SAFs (BL-SAFs) is that they can stabilise skyrmions without a magnetic field, and they exhibit significant compensation of dipolar fields within their structure. Additionally, BL-SAFs exhibit stability up to high temperatures which earlier we saw a threat proposed by~\cite{barker2016static}, allowing for the possibility of inducing skyrmion motion using large current pulses. Employing the method, in an experimental work in Ref.~\cite{dohi2019formation}, the authors have successfully achieved a suppressed Skyrmion Hall Effect (SkHE) in a synthetic antiferromagnetic [see FIG.~\ref{fig:TSHE_sky}(k)].\\
\subsection{Skyrmion-based Neuromorphic Devices}
\begin{figure}[b]
    \centering
    \includegraphics[width = \linewidth]{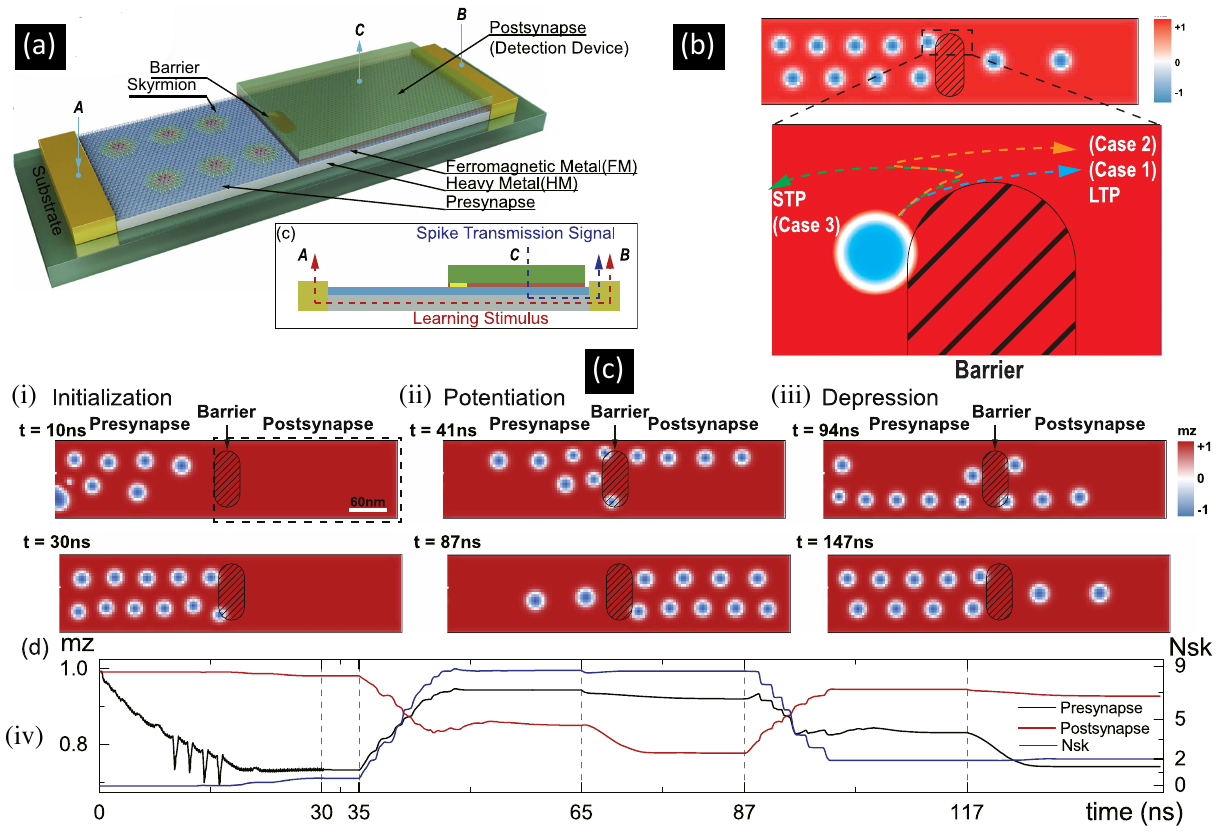}
    \caption{\textbf{Schematic of the skyrmionic synaptic device.}
    (a)~Device Structure;
    (b)~Depiction of the STP and LTP of the device;
    (c)~Operation modes (i, ii, iii), normalized $\boldsymbol{m}_z$ and the skyrmion number $N_{\text{sk}}$ of the postsynapse (iv), with permission from Ref.~\cite{huang2017magnetic}}
    \label{fig:synap-2}
\end{figure}
\begin{figure*}[hbtp]
    \centering
    \includegraphics[width = \linewidth]{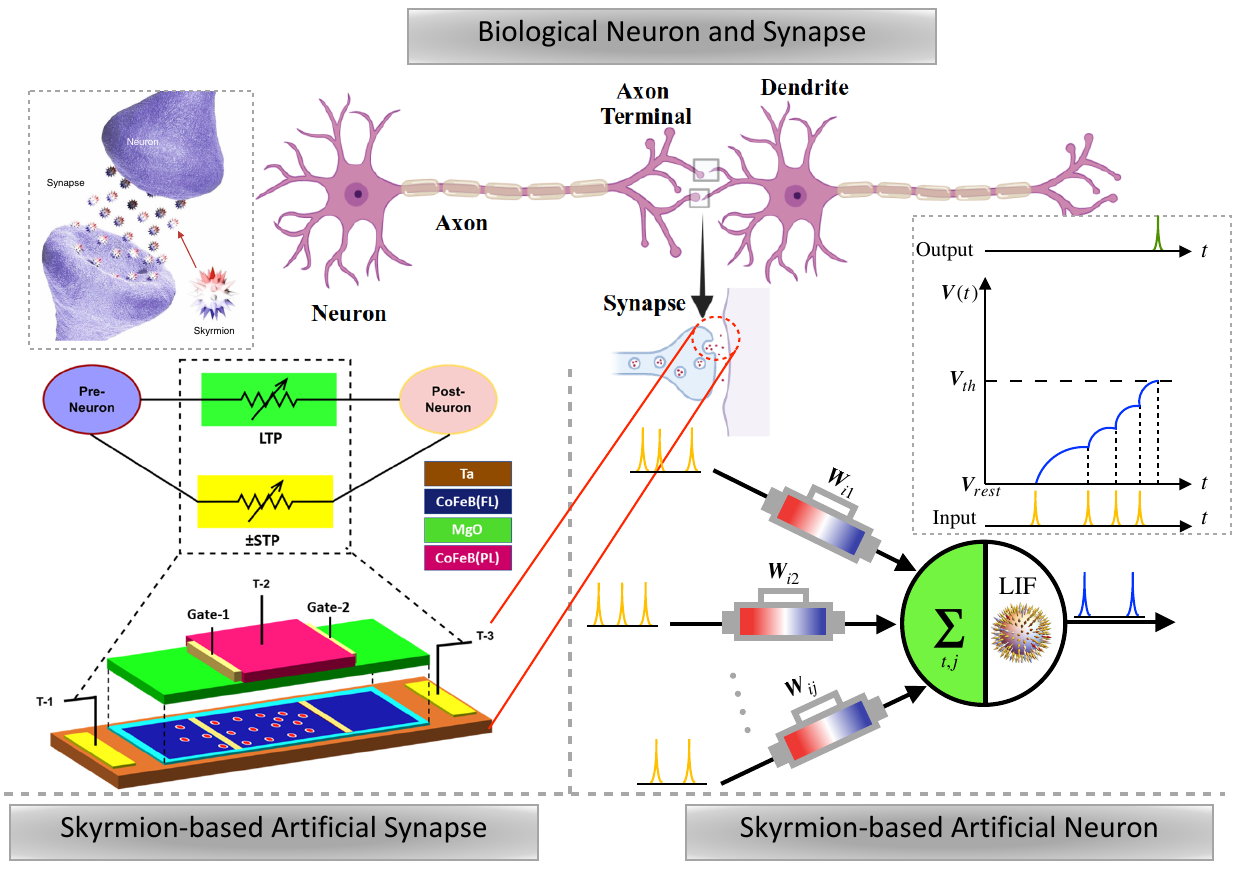}
    \caption{\textbf{Schematic of Biological and Skyrmion-based Neuromorphic Devices.} Biological neurons connected by two synapses (top), Skyrmion-based Synapse (bottom-left) and Neuronb(bottom-right), with permission from Ref.~\cite{lone2022skyrmion, das2023bilayer}}
    \label{fig:s-1}
\end{figure*}
Neuromorphic computing, inspired by the biological nervous system, has attracted considerable attention. Intensive research has been conducted in this field for developing artificial synapses and neurons, attempting to mimic the behaviours of biological synapses and neurons, which are two basic elements of the human brain. So two essential and fundamental devices are artificial neurons and synapses to implement a neuromorphic computing system. In recent decades, a promising technical advance has been made with the introduction of emerging nano-electronic devices, such as phase change devices, resistive memory devices, and spintronic devices, to implement artificial synapses. However, the implementation of an artificial neuron still mostly relies on silicon-based circuits through the integration of many transistors, greatly sacrificing power efficiency and integration density compared with the advancement of artificial synapses. Therefore a single device-based artificial neuron is much preferable and is currently under intensive research for designing an advanced neuromorphic computing system.\\
The computing architecture of the human brain [see FIG.~\ref{fig:s-1}(top)] consists of billions of neurons that are highly connected by junctions turned into synapses. Neurons generate stimuli (spikes) when they are activated by integrating stimuli from other connected neurons. The spikes are modulated during propagating through the synapses, also called \textit{synaptic transmission}. The synapses function by changing their connection strength as a result of neural activities, known as \textit{synaptic plasticity}.
\subsubsection{Skyrmion-based Synapses}
\begin{figure*}[htbp]
    \centering
    \includegraphics[width = \linewidth]{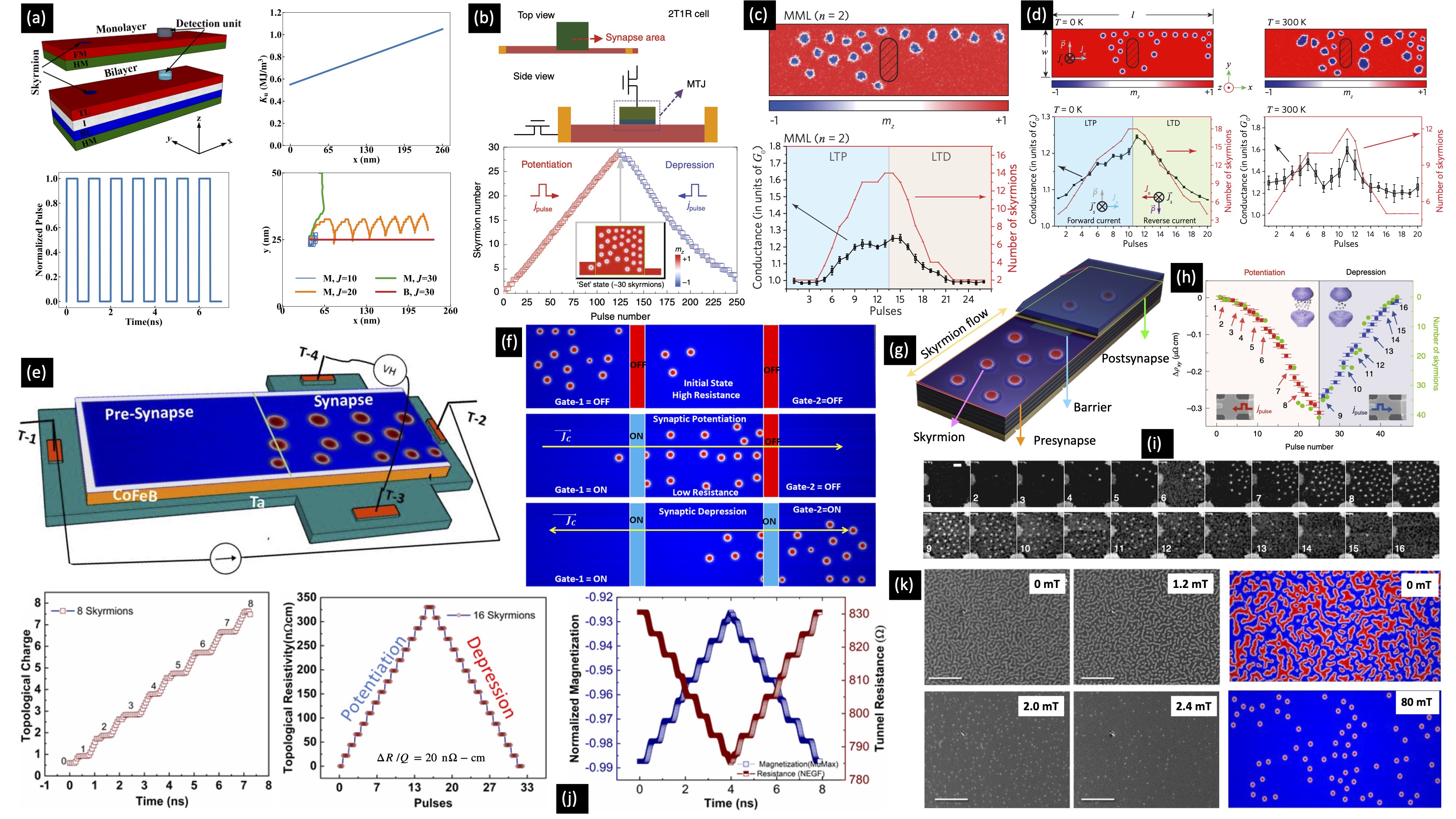}
    \caption{\textbf{Magnetic skyrmion-based artificial synapses.}
    (a)~Schematic of the monolayer and bilayer device, PMA constant variation with position, normalised square input pulse train and locus of the skyrmion center, with permission from Ref.~\cite{das2023bilayer};
    (b)~Schematic of the synapse device in a 2T1R configuration (top), simulated potentiation and depression (bottom), with permission from Ref. ~\cite{song2020skyrmion};
    (c)~Skyrmionic synapses (g) composed of the MML ($n=2$)structure;
    (d)~The characteristic conductance modulation curves of the device (g) within the postsynapse region during the whole LTP-LTD process at $T = 0 $K (left) and $T = 300$ K (right), with permission from Ref.~\cite{chen2020nanoscale};
    (e)~Topological resistivity-based skyrmion synapse, with permission from Ref.~\cite{lone2023controlling};
    (f)~Operating modes of the VCMA controlled device [FIG.~\ref{fig:s-1}(left-bottom)] for realizing synaptic potentiation and synaptic depression, with permission from Ref.~\cite{lone2022skyrmion};
    (g) Illustration of nanoscale multilayer skyrmionic synaptic device, with permission from Ref.~\cite{chen2020nanoscale};
    (h)~Hall resistivity change and skyrmion number as a function of injected pulse number;
    (i)~Sequential STXM images of skyrmion populations with pulse along the track, with permission from Ref.~\cite{song2020skyrmion}
    (j)~Topological charge evolution in the presence of continuous current ($8 \times 10^{11} \mathrm{A/m^2}$) (left), Discrete topological resistivity of the device performing potentiation and depression (middle), TMR in the MTJ configuration (right), with permission from Ref.~\cite{lone2023controlling};
    (k)~MOKE and corresponding simulation images showing MF dependence of skyrmion density of device (e), with permission from Ref.~\cite{lone2023controlling}}
    \label{fig:synapse_sky}
\end{figure*}
Owing to their small size, robustness against
pinning defects (topological stability), and low depinning current density, skyrmions have considerable potential for use as information carriers in future ultra-dense, high-speed and low-power spintronic devices.  One of the intrinsic features of skyrmions is their particle-like behaviour, due to which, multiple skyrmions can aggregate, exhibiting potential for multi-valued storage devices with skyrmions as information carriers. The states of such a device can be modulated by the magnetic field, an electric current or more efficiently with the voltage (VCMA). The predominant mode of operation for most neuromorphic devices based on skyrmions involves the following process. By applying an electric current, either through Spin Transfer Torque (STT) or Spin-orbit Torque (SOT), the skyrmions are propelled through the magnetic thin film. The thin film is divided into different regions such as pre-synapse, active region and post-synapse region. These skyrmions when driven into the active region, are detected by majorly two reading mechanisms: (1) Tunnel Magnetoresistance (TMR) and (2) Topological/Anomalous Hall measurement (THE/AHE). With proper design engineering, the skyrmions can be fine-controlled and depending upon the number of current pulses we can control the resistance of the active synapse. Such characteristics are analogous to those of a biological synapse, in which the weight can be dynamically adapted to a changing environment, i.e., the synaptic plasticity (connection strength or conductance). In a biological synapse, synaptic plasticity is achieved based on signal propagation by the release of neurotransmitters. For artificial synapses, the capability responsible for Short-Term Plasticity (STP), Long-Term Potentiation (LTP), and temporal dynamics is the foundation of learning in neuromorphic applications.
In Ref. \cite{huang2017magnetic} as shown in FIG.~\ref{fig:synap-2}(a), the authors have proposed a skyrmion-based artificial synaptic device and based on the micromagnetic simulations and analytical TMR calculations, they showed the STP and LTP synaptic behaviour of the device presented in FIG.~\ref{fig:synap-2}(b). The conductance of the device is regulated based on the number of skyrmions present in the post-synaptic region. Increasing the skyrmion density in the post-synapse leads to an increase in conductance (long-term potentiation, LTP), as depicted in FIG.~\ref{fig:synap-2}(d). Conversely, reducing the skyrmion density in the post-synapse results in reduced conductance (long-term depression, LTD). The experimental study conducted by Song et al. \cite{song2020skyrmion} illustrated the synaptic characteristics of a skyrmion device [see FIG.~\ref{fig:synapse_sky}(b,h-i)].
Skyrmions are induced within the multilayer Hallbar $\mathrm{[Pt (3nm)/GdFeCo(9nm)/MgO(1nm)]\times20}$. By manipulating the external magnetic field, the authors effectively demonstrated the creation and annihilation of skyrmions within the system. The authors propose the reading of skyrmion resistance via topological Hall resistivity measurements. Regarding this outcome, the assessment of Hall resistivity's influence by a solitary skyrmion has been demonstrated by a different research group \cite{zeissler2018discrete}. Additionally, the researchers utilize both experimental and simulated approaches to exhibit the linear synaptic response within these devices. These synaptic weights have subsequently been employed for incorporating basic artificial neural networks (ANNs).
\begin{figure}[t]
    \centering
    \includegraphics[width = \linewidth]{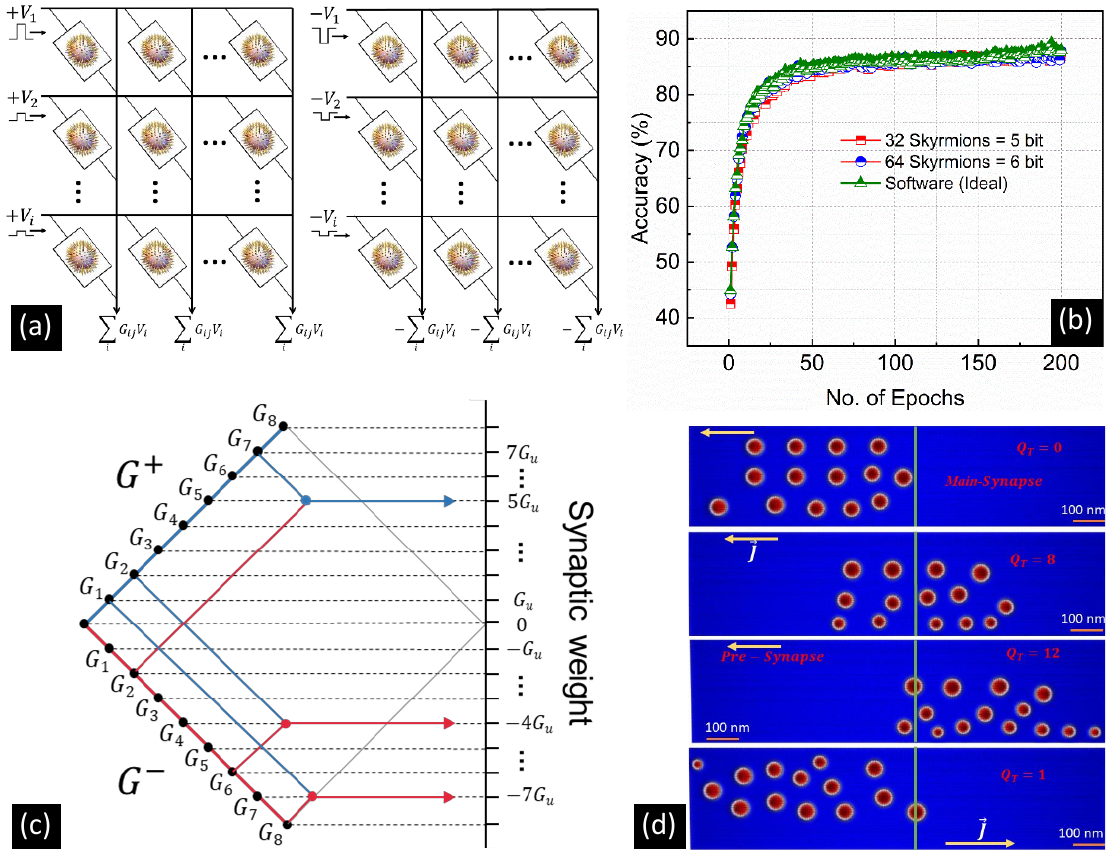}
    \caption{\textbf{Schematic of circuit diagram \& performance.}~(a)~Schematic of the circuit diagram comprising skyrmion-based synapses;
    (b)~System-level performance in terms of recognition accuracy;
    (c)~G diamond plot mapping two skyrmion devices to the synaptic weight;
    (d)~Skyrmion motion from the pre-synapse region into the main-synapse region and vice-versa for the device presented in FIG.~\ref{fig:synapse_sky}(e), from Ref.~\cite{lone2023controlling}}
    \label{s-2}
\end{figure}
However, we used both SOT and voltage-controlled magnetic anisotropy (VCMA) switching techniques [as depicted in FIG.~\ref{fig:s-1}(bottom-left)] to propose a skyrmion-based synapse design capable of exhibiting long-term plasticity (LTP) driven by SOT and short-term plasticity (STP) controlled by the VCMA effect, illustrated in FIG.~\ref{fig:synapse_sky}(f). The LTP conductance of the linear skyrmion device is demonstrated to be advantageous for static pattern recognition, and the STP component aids in dynamic pattern recognition within the device's functionality. In our collaborative study involving both experimental and simulation approaches~\cite{lone2023controlling}, we showed the quantized conductance of the skyrmion device, as depicted in FIG.~\ref{fig:synapse_sky}(j). The skyrmions are driven by SOT and magnetic fields while we adopt the topological resistivity for reading the skyrmions, as shown in FIG.~\ref{s-2}(d). The device structure is further elucidated in FIG.~\ref{fig:synapse_sky}(e). We integrate this device in a Quantized Convolutional Neural Network (QCNN) as shown in FIG.~\ref{s-2}(a-c), showing the pattern classification accuracy around $\sim 89\%$, presented in FIG.~\ref{s-2}(b). Furthermore, in another skyrmion-based synaptic devices-related work~\cite{das2023bilayer}, the authors used antiferromagnetic coupling between two ferromagnetic layers to reduce the scattering due to the skyrmion Hall effect. The SOT pulses drive the skyrmions in the ferromagnetic layers, due to the reduced skyrmion Hall effect the conductance linearity is improved as shown in FIG.~\ref{fig:synapse_sky}(a). The improved linearity adds to improvement in the pattern recognition. 
In a different study focused on skyrmion-based synapses~\cite{chen2020nanoscale} employing micromagnetic simulations, the authors demonstrate the realization of a skyrmion synapse utilizing a magnetic multi-layer stack (MML) [depicted in FIG.~\ref{fig:synapse_sky}(g)]. The stack consists of four iterations of a trilayer structure [HM1|FM|HM2], where HM denotes the heavy metal and FM represents the ferromagnetic thin film. The synapse has shown six discrete conductance states as shown in FIG.~\ref{fig:synapse_sky}(c,d) with energy consumption per update equal to approximately 300fJ. Authors have done room temperature micromagnetic simulations for this task claiming to be more realistic.
\begin{figure}[b]
    \centering
    \includegraphics[width = \linewidth]{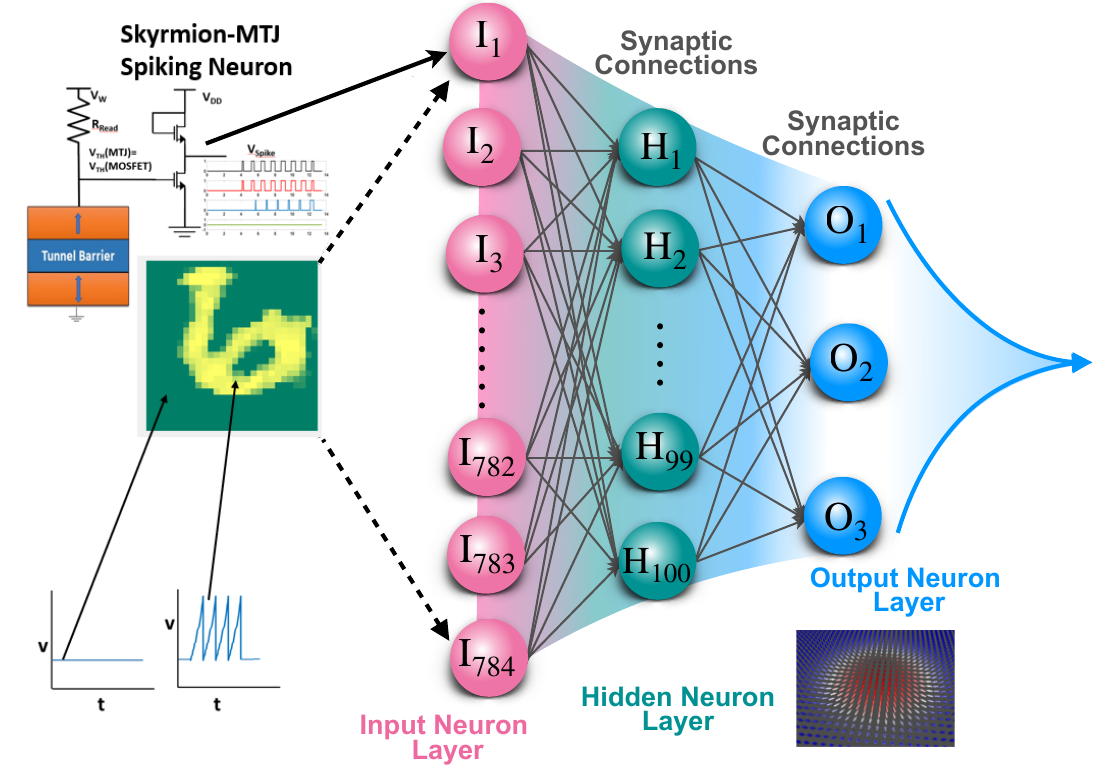}
    \caption{The training and testing of MNIST handwritten digit-(6), from Ref.~\cite{lone2022skyrmion}}
    \label{fig:training}
\end{figure}
\begin{figure*}[htbp]
    \centering
    \includegraphics[width  = \linewidth]{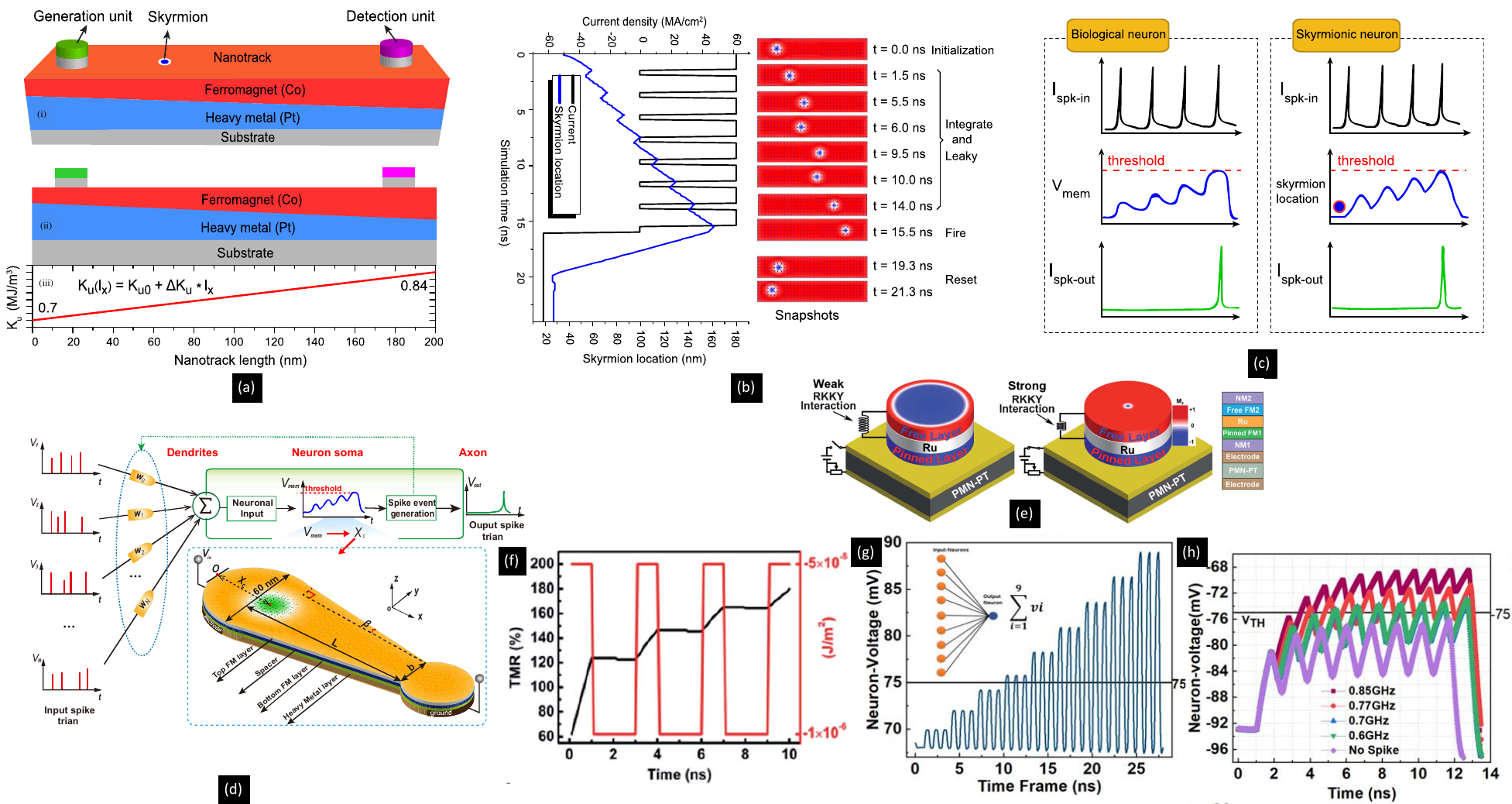}
    \caption{\textbf{Magnetic skyrmion-based artificial neuron.}~(a)~Schematic of the skyrmion-based artificial neuron device; (ii) the front cross-section view; (iii) the linear increase of the PMA value along the nano track;
    (b)~Micromagnetic simulations of the skyrmionic neuron device (a) exhibiting `leaky-integrate' behaviors;
    (c)~LIF behaviors of a biological neuron(left) and a skyrmion-based artificial neuron(right), with permission form Ref.~\cite{li2017magnetic};
    (d) A novel compact neuron device exploiting the current-driven skyrmion dynamics in a wedge-shaped nanowire, with permission from Ref.~\cite{chen2018compact}; 
    (e)~Manipulating Skyrmion Size with Interlayer RKKY Exchange Control under an external voltage;
    (f)~Voltage-pulse induced LIF model based on size variations of the skyrmion in a SAF, with permission from Ref.~\cite{yu2020voltage};
    (g)~Skyrmion output voltage mimics the LIF neuron behaviour for different input spike frequencies;
    (h)~9 input neurons feeding one output neuron in different time frames, with permission from Ref.~\cite{lone2022skyrmionneuron}}
    \label{fig:neuron_sky}
\end{figure*}
Although skyrmion-based synaptic devices have shown the potential for neuromorphic computing. However, these devices still suffer from a low on/off conductance ratio. For a reliable weight operation higher the on/off or maximum achievable conductance modulation, the better it reflects in the circuit implementations. Clearly, for a big memory window, we can achieve more synaptic states. So far, in most of these proposals/demonstrations, the low memory window can be clearly observed. This challenge needs a lot of attention and can be overcome by novel high-spin polarized materials, advances in device design engineering and the development of novel interface circuitry. The other most important challenge faced by these devices is the reading of skyrmions. The two schemes used for the reading are TMR reading via MTJ and topological Hall resistivity. Although TMR reading is a viable option but in present MTJ devices the TMR ratio is around 200\% which is still small compared to other beyond CMOS devices such as RRAM and PCRAM. Moreover, the skyrmions don't have a uniform magnetization thus effective TMR is even smaller for these devices. Considering the skyrmion reading via topological Hall resistivity, again a novel method but it is almost impossible to extract exact topological conductivity contribution by an individual skyrmion due to myriad effects such as anisotropic magneto-resistance, domain wall contributions and so on. Thus, research is still needed to realize better ways of extracting the topological conductivity contribution of the skyrmions.
\subsubsection{Skyrmion-based Neurons}
In biological neural systems, the neuron serves as the basic computational unit. It receives input signals from other pre-neurons, and these inputs are weighted by the synaptic strength. The weighted inputs are then summed up in the neuron's body. As a result, the neuron's membrane potential gradually increases. However, the neuron's behavior also includes a leaky component, causing a gradual decrease in the membrane potential. When the accumulated input spikes exceed a certain threshold potential, the neuron may or may not fire. The neuron generates spikes which propagate to the next layers as shown in FIG.~\ref{fig:s-1} (bottom-left). This behavior is mathematically expressed by what we call the leaky integrate and fire LIF neuron model.
\begin{equation}
    \tau_{\mathrm{mem}}\frac{dV}{dt} = -(V-V_{\mathrm{rest}})+\sum_j\delta(t-t_j)w_j
    \label{Eqn:130}
\end{equation}
In recent literature, many groups have shown the potential use of skyrmion devices as leaky integrate and fire neurons in spiking neural networks. Most of these proposals have considered the current (SOT) driven motion of skyrmion and its interplay with the device edges or in general device geometry for the realization of the skyrmion LIF behavior. As shown in FIG.~\ref{fig:neuron_sky}(d) \cite{chen2018compact} authors designed a skyrmion LIF device by considering the SOT-driven motion of skyrmion in a laterally tapered HM/FM geometry. Skyrmion starts from the wider edge region and as the skyrmion translates across the FM region with decreasing width. Skyrmion begins to feel an edge repulsive force. In the presence of a current pulse, it can move but in the absence of a pulse, the skyrmion is pushed back due to this edge force. This brings the leaky component into the device's operating behavior. Here, the authors have considered the skyrmion position as the membrane potential or computing variable. Finally, with the application of more input spikes/pulses the skyrmion reaches a threshold region and generates the output spike as shown in FIG.~\ref{fig:neuron_sky}(c). In another skyrmion LIF work the authors considered an almost similar approach as \cite{li2017magnetic} but the device geometry is tapered vertically as shown in FIG.~\ref{fig:neuron_sky}(a).
The varying thickness of the FM results in a magnetic anisotropy gradient across the film. Because the skyrmion's static and dynamic properties depend upon the anisotropy. So, an anisotropy gradient realizes the extra repulsive force necessary for mimicking the leaky behavior of the device, see FIG.~\ref{fig:neuron_sky}(b). In other related works using micromagnetic simulations, we showed LIF neuron behaviour in a confined skyrmion controlled by voltage pulses as shown in FIG.~\ref{fig:training}. The SNN is constructed with the LIF neuron unit model -based on our skyrmion device. The network was able to show recognition up to 89\%
\cite{lone2022skyrmionneuron}. Instead of considering skyrmion position as a neuron variable we took into account the actual voltage output of the voltage-controlled MTJ, see FIG.~\ref{fig:neuron_sky}(g-h). 
This makes the skyrmion device more realistic for actual circuit applications. Also, the skyrmion motion is avoided so the skyrmion Hall effect is not a concern. Most importantly with better scalability, the confined approach promises high integration capability. In another voltage-controlled skyrmion neuromorphic device~\cite{yu2020voltage} the authors used synthetic anti-ferromagnetic geometry SAF consisting of Co/Pt or ultra-thin CoFeB layers with PMA. The skyrmion MTJ fabricated over PMN piezoelectric controlled by voltage acts as LIF. By applying the voltage pulse the $J_{\text{ex}}$ alternates between two values. Due to the different relaxation constant for expanding and shrinking, the skyrmion MTJ behaves as a LIF neuron as shown in FIG.~\ref{fig:neuron_sky}(f).\\
\begin{figure*}[htbp]
    \centering
    \includegraphics[width  = \linewidth]{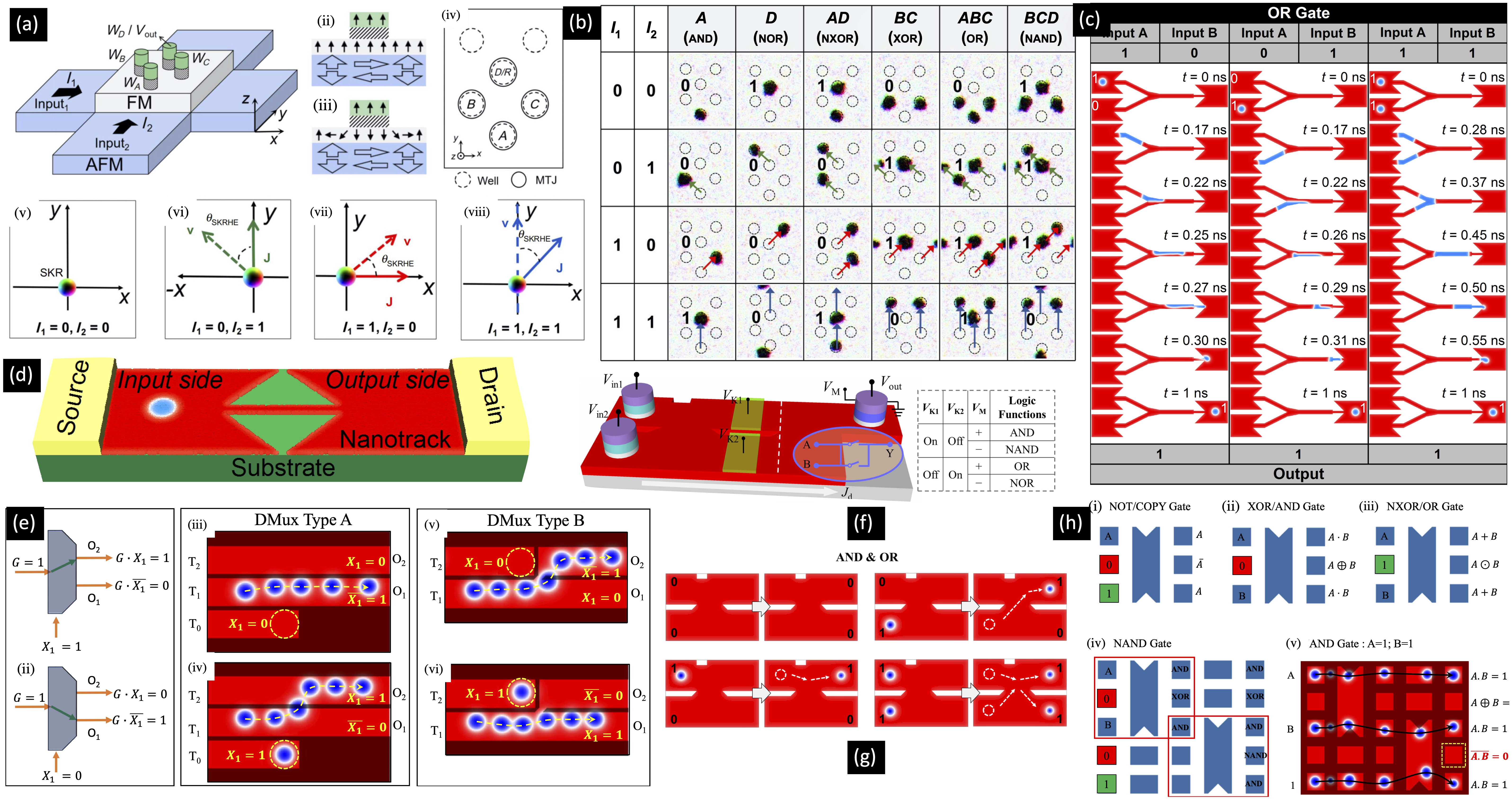}
    \caption{\textbf{Skyrmion logic gates.}~(a)~Device setup and skyrmion motion: (i) Configuration and current input. (ii)-(iii) Spin structures in MTJ. (iv) Pinning sites and MTJ layout. (v)-(viii) Skyrmion motion under different currents;
    (b)~Simulated results of different logic functions, with permission from Ref. \cite{yan2021skyrmion};
    (c)~Skyrmion logical OR operation of the device in (d);
    (d)~Sketch of a Nanowire Simulation, with permission from Ref.~\cite{zhang2015magnetic};
    (e)~(i)-(ii): 1-to-2 DMux gate schematic and operation, (iii)-(iv): Skyrmion-based DMux gate micromagnetic simulation,(e) and (f): Alternative skyrmion-based DMux design with an additional channel in the barrier, with permission from Ref.~\cite{sisodia2022programmable};
    (f)~ Structure of a reconfigurable skyrmion logic (RSL) gate and the resulting logic functions in an RSL;
    (g)~Evolution of the position of skyrmion(s) in FM films for four different cases of inputs and outputs, with permission from Ref. \cite{luo2018reconfigurable}; 
    (h)~Logic gate designs derived from FA: NOT/COPY, XOR/AND, NXOR/OR, universal NAND gate, NAND operation with both inputs as 1, with permission from Ref.~\cite{sisodia2022robust}}
    \label{fig:logic-gates}
\end{figure*}
Although the skyrmion devices possess the realization of highly integrative single-device neurons. Research is still needed when it comes to industry-scale skyrmion-based neuron chips. The stabilization of small nanometer room temperature skyrmions, the reading of these skyrmions and better writing mechanisms are important. Furthermore, these devices must show good integration with the CMOS technology nodes.
\subsection{Skyrmion Conventional and Unconventional Computing Paradigms}
\begin{figure*}[htbp]
    \centering
    \includegraphics[width = \linewidth]{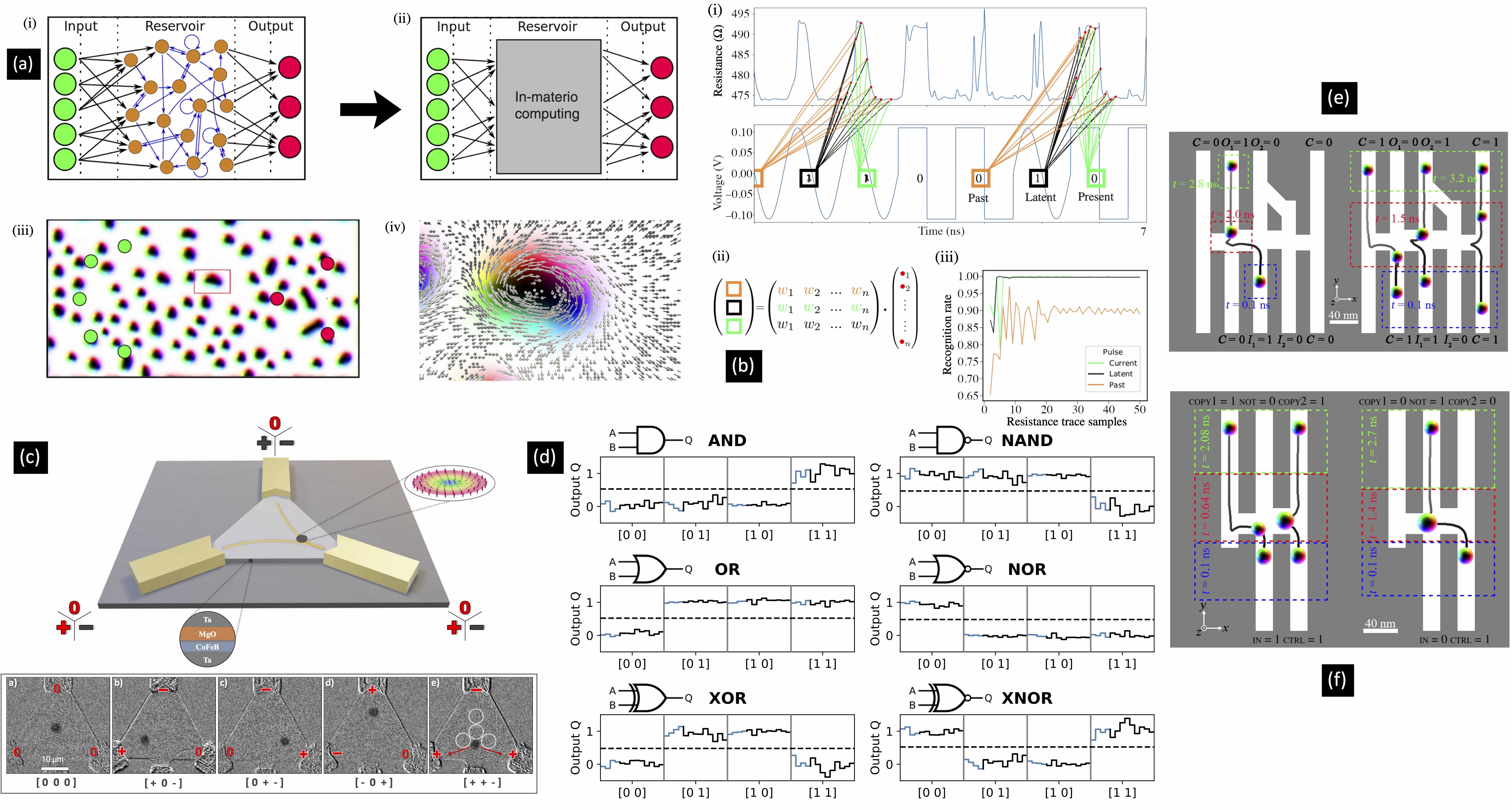}
    \caption{\textbf{Reservoir Computing.}~(a)~Set up for a-(i) recurrent neural network, (ii) material-based reservoir, (iii) skyrmion fabric reservoir with the locations of input (output) contacts identified by green (red) circles, (iv) Bloch skyrmion identified by the red frame in (iii);
    (b)~Temporal tracing and pulse recognition with resistance sampling, with permission from Ref.~\cite{pinna2020reservoir}
    (c)~3D Device Schematic: Stack Structure with Applied Potentials and Boolean Logic Demonstration;
    (d)~Optimizing logic operations: skyrmion probabilities and read-out threshold, with permission from Ref.~\cite{raab2022brownian};
    (e) Micromagnetic simulation results for the Fredkin gate for two different input combinations, with permission from Ref.~\cite{chauwin2019skyrmion};
    (f) Micromagnetic simulation results of INV/COPY gate for two different input combinations,  with permission from Ref.~\cite{chauwin2019skyrmion}; }
    \label{fig:uc1}
\end{figure*}
Apart from racetrack memory and neuromorphic computing, the skyrmions have held promise for other conventional and unconventional logic applications. In Ref.~\cite{yan2021skyrmion} authors realized the skyrmion-based logic gates by considering the device structure as shown in FIG.~\ref{fig:logic-gates}(a). The device consists of a bottom AFM layer, FM layer and 4 MTJs. Two equal amplitude orthogonal currents $I_1$ and $I_2$ which act as the input signals, flow in the AFM Hall bar arms. MTJ (A, B and C) are used for writing skyrmions by STT while MTJ (D) reads the skyrmions. By properly programming the different skyrmion configurations at the pinning sites, the authors have demonstrated the achievement of distinct logic gates, as illustrated in FIG.~\ref{fig:logic-gates}(b). The authors in Ref.~\cite{zhang2015magnetic} have proposed a device structure based on the conversion, duplication and merging of skyrmions for the realization of the logic (AND, OR) gates as shown in FIG.~\ref{fig:logic-gates}(c-d). In Ref.~\cite{luo2018reconfigurable} using micromagnetic simulations, the authors utilize combinations of SOT, skyrmion Hall effect, edge repulsive effects and skyrmion-skyrmion interactions for the realization of reconfigurable logic gates, see FIG.~\ref{fig:logic-gates}(f). The skyrmion trajectories are controlled by external techniques, especially VCMA. Illustrated in FIG.~\ref{fig:logic-gates}(g), the visualization of the AND gate reveals that, owing to the interplay of edge force and the skyrmion Hall effect, the skyrmion fails to reach the output (detector) for input combinations $(0,0)$, $(0,1)$, and $(1,0)$. However, when both skyrmions are propelled, their repulsion causes one skyrmion to reach the output, leading to an output of $Y=1$ for the input combination $(1,1)$. A similar logic gate design was proposed in Ref.~\cite{sisodia2022robust} including the universal NAND and NOR gates, depicted in FIG.~\ref{fig:logic-gates}(h). In the work by Sisodia \textit{et al.}~\cite{sisodia2022programmable}, micromagnetic simulations were employed to demonstrate that skillfully manipulating the barrier height and width of the magnetic nano track enables the specific tunneling of skyrmions between parallel nano tracks, prompted by skyrmion-skyrmion interactions, as illustrated in FIG.~\ref{fig:logic-gates}(e). This can be leveraged to design a skyrmion demultiplexer logic gate that works solely using skyrmions as logic inputs.\\
Apart from conventional binary logic, skyrmion-based unconventional computing devices have also been in discussion recently. In Ref.~\cite{pinna2020reservoir} the skyrmion-based reversible logic devices and circuits have been depicted. The skyrmions remain conserved during device operation, and input/output reversal is achievable, as depicted in FIG.~\ref{fig:uc1}(a). To achieve this reversible logic utilizing the skyrmion-Hall effect, spin-Hall effect, and skyrmion-skyrmion interactions, the authors have introduced reversible AND, OR, and INV/COPY gates, as illustrated in FIG.~\ref{fig:uc1}(f). Furthermore, the Fredkin reversible gate and Toffoli Hadmard universal quantum gates are realized [see FIG.~\ref{fig:uc1}(e)]. Another highly significant application of skyrmions involves reservoir computing. Reservoir computing harnesses the nonlinear dynamics of a system (known as a reservoir) to transform intricate input signals into a higher-dimensional space, ultimately leading to a linear 1D output. This 1D readout can be trained to discern the input's distinctive features. In Ref.~\cite{pinna2020reservoir} the authors used the rectangular film containing skyrmions as the reservoir. The grain inhomogeneities are considered in the simulations which ensure the skyrmion fabric pinning. The pinning of skyrmion fabric is important for maintaining the echo state of the reservoir system. The skyrmion fabric when excited by input patterns generates a particular current distribution which is the signature of the particular magnetization texture. Here the difference in the relaxation time scales magnetization ($\sim 10^{-9}$) and electron ($\sim 10^{-9}$) guarantees this behaviour. As shown in FIG.~\ref{fig:uc1}(b) the output of the system is read in terms of the AMR signals, by properly analyzing the response of the system to different combinations of the input sequences. It is clearly observed that the output response depends both upon the current input as well as the previous input. This shows that the skyrmion fabric has a memory. By providing sine and square pulse combinations the short-term memory correlation in the skyrmion system is revealed. The authors in Ref.~\cite{raab2022brownian} advanced the skyrmion-based reservoir computing and have experimentally demonstrated the realization of the confined skyrmion-based non-linearly separable functions. For this authors use the voltage gating and thermal skyrmion motion (Brownian motion), SOT and skyrmion-edge repulsion. Depending upon the voltage applied at the contacts the skyrmion position (state) is altered and read via MOKE imaging as shown in FIG.~\ref{fig:uc1}(c). Using this concept the authors have realized different logic gates such as AND, OR, NAND, NOR, XOR and XNOR as shown in FIG.~\ref{fig:uc1}(d). The probability of occurrence of the skyrmion in the 4 circular regions is processed via readout as the linear functions. The output is the weighted sum of the probabilities plus and offset. The realization of the XOR function by this single device proves its completeness for the non-linearly separable tasks. The use of single skyrmion and thermal effects driven operation promise high memory integration and low power operation. Like other skyrmion-based technologies although these unconventional computing devices promise a better future for computing and data storage. But the reading of the skyrmions needs a robust solution. Like here in Ref.~\cite{raab2022brownian} although the proposal is scalable and interesting, but the reading is done via MOKE images. To integrate these devices with present CMOS technology, it is inevitable to realize a proper current/voltage-based writing and reading mechanism for these devices.
\subsection{Skyrmion Qubit}
By harnessing the fundamental principles of Quantum mechanics, Quantum computer holds the potential to revolutionize computation capacities~\cite{feynman2018simulating, divincenzo1995quantum, nielsen2002quantum}. 
Recently, Qubits based on magnetic skyrmions~\cite{psaroudaki2021skyrmion,xia2023universal} and later meron~\cite{xia2022qubits} have been proposed pacing the advancement in the field. These magnetic textures as we discussed earlier exist in 2D and 3D magnetic materials with certain stability conditions and are highly promising in both classical and quantum-related applications.\\
\begin{figure}[t]
    \centering
    \includegraphics[width = \linewidth]{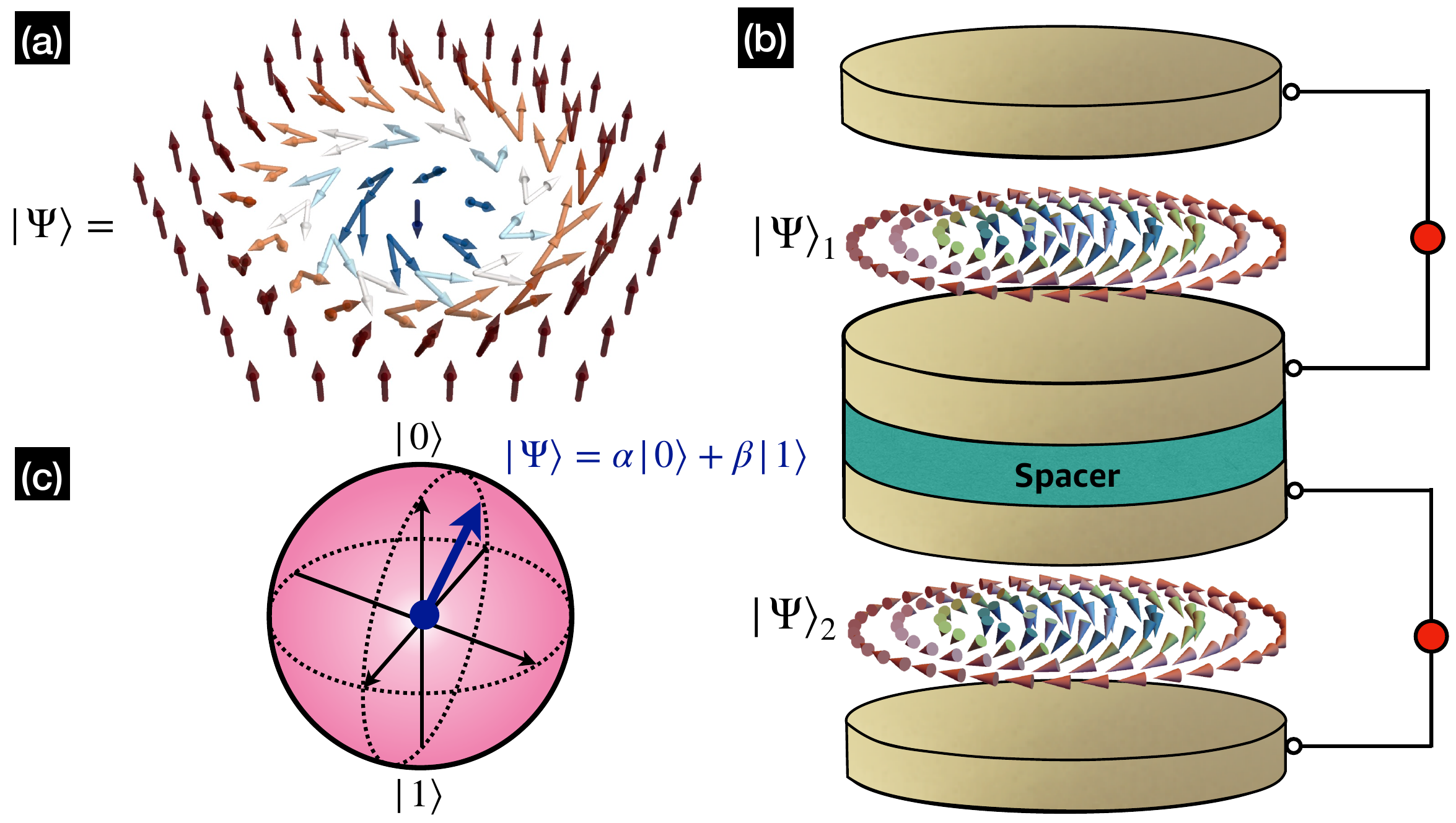}
    \caption{\textbf{Qubit Representation.}~(a)~Skyrmionic quantum state $|\Psi\rangle$ as an arbitrary superposition of skyrmion  with distinct helicities $\gamma$;
    (b)~Bloch sphere representation of $|\Psi\rangle= \alpha|0\rangle+\beta|1\rangle$
    (c)~Skyrmion qubit coupling in a bilayer of magnetic materials, tuned by a nonmagnetic spacer (blue) and adjusted by electric fields (yellow plates), adapted with permission from Ref.~\cite{psaroudaki2021skyrmion}}
    \label{fig:bilayer-qubit}
\end{figure}
Single spin-based quantum qubits are the most basic structure where the quantum state $|0\rangle$ is associated with an up spin ($\uparrow$) and the state $|1\rangle$ is associated with a down spin ($\downarrow$) [see FIG.~\ref{fig:bilayer-qubit}(c)]. Manipulating a single qubit involves applying a magnetic field that causes the spin to undergo Rabi precession, resulting in a change in its direction. On the other hand, the interaction described by the Heisenberg model allows for two-qubit gate operations. 
Now in the case of quantum skyrmion, instead of a single spin, we focus on the helicity degree of freedom of these magnetic textures which encapsulates the idea of the direction of swirling arrows in such magnetic textures created in frustrated magnets. Skyrmion~\cite{xia2023universal} and  meron~\cite{xia2022qubits} both have four-fold degeneracy in the absence of magnetic dipole-dipole interaction (DDI)~\cite{leonov2015multiply,lin2016ginzburg,batista2016frustration,diep2019phase} which can be lifted on the application of external means. However, in the presence of DDI, helicity is reduced to a two-fold degenerate state for a Bloch skyrmion~\cite{kurumaji2019skyrmion} which the authors~\cite{xia2023universal} have utilised for realising skyrmion-based qubits. Furthermore, a linear combination of the two-fold helicity signifies a quantum-mechanical state $|\Psi\rangle$ [depicted in the FIG.~\ref{fig:bilayer-qubit}(a)] that enables qubit operations (construction of gates) originally by controllable Magnetic Field Gradients (MFGs) and microwave fields studied in Ref.~\cite{psaroudaki2021skyrmion} and later on, by manipulating the electric field and the spin current without a magnetic field in Ref.~\cite{xia2023universal}. Since the controllable creation of manipulation of an electric field and spin current is easy compared to a magnetic field, we will focus our attention on this approach.\\
Authors~\cite{xia2023universal, xia2022qubits} considered the ansatz [Eqn.~(\ref{eqn:9})] of skyrmion with angular function $\phi = \varphi+\eta+\pi/2$, a difference by angle $\pi/2$ from the conventional ansatz. Based on this, they defined an expression for the magnetic DDI in the frustrated magnet as
\begin{equation}
    H_V = -V\cos 2\eta
\end{equation}
where $V$ denotes the DDI coupling constant, typically in the order of $\sim 10^{-21}$J at 100K. The helicity of skyrmion $\eta = 0$ or $\pi$ represent the classical bit ($|0\rangle, |1\rangle$) as shown in FIG.~\ref{fig:QC_sky}(a).\\
Nevertheless, Schr\"{o}dinger equation of the helicity dynamics can be written as
\begin{equation}
    i\hbar \frac{d}{dt} |\Psi\rangle = H |\Psi \rangle
\end{equation}
where Hamiltonian of a single qubit is represented by
\begin{equation}
    H = H_{\mathrm{J}_{\text{current}}}+H_{\mathrm{E}_z} = -\alpha_{J} \mathrm{J}_{\text {current }} \sigma_{x}+\alpha_E \mathrm{E}_z\sigma_z
\end{equation}
where $H_{\mathrm{J}_{\text{current}}}$ performs the helicity of rotation when spin current is applied and $H_{\mathrm{E}_z}$ is the coupling of electric dipole moment induced by skyrmion with perpendicular electric field $\mathrm{E}_z$.
Solving the Hamiltonian [see for details~\cite{xia2023universal}], the authors presented the construction of the basic skyrmion-based single qubit gates as follows\\
\textit{$z$ rotation gate}
\begin{equation}
 \mathrm{U}_{Z}(\theta) = \exp \left[-\frac{i \theta}{2} \sigma_{z}\right]
 \label{137}
\end{equation}
\textit{$x$ rotation gate}
\begin{equation} \mathrm{U}_{X}(\theta) \equiv \exp \left[-\frac{i \theta}{2} \sigma_{x}\right]
\end{equation}
\textit{$\pi/4$ phase-shift gate}
\begin{equation} \mathrm{U}_{T}(\theta)  = e^{i\pi/8} U_Z \left(\frac{-\pi}{8}\right)
\end{equation}
and the \textit{Hadamard gate} which is realised by a sequential application of the $z$ rotation gate and $x$ rotation gate
\begin{equation}
    \mathrm{U}_H  = i\mathrm{U}_Z\left(\frac{\pi}{2} \right) \mathrm{U}_X\left(\frac{\pi}{2} \right)\mathrm{U}_Z\left(\frac{\pi}{2} \right)
\end{equation}
Furthermore, the construction of \textit{multiqubit gates} was proposed with an N-layered system with individual skyrmion in each layer [see Fig.~\ref{fig:bilayer-qubit}(b) and FIG.~\ref{fig:QC_sky} in the case of $N = 2$]. As depicted, when two skyrmions are horizontally separated, there is no interaction between them. However, as they approach each other in neighboring layers, an exchange interaction ($\propto -J_{\text{int}}(d_m)$) starts to operate between them. This interaction between two neighboring skyrmions ($m, m+1$) can be represented as an Ising coupling gate, given as
\begin{equation}
    \mathrm{U}^m_{ZZ} (\theta) \equiv \left[-\frac{i\theta}{2}\sigma^m_z \sigma^{m+1}_z\right]
\end{equation}
which acts on a 2-qubit in adjacent layers. \\
Similarly, the \textit{controlled-Z (CZ) gate} is constructed as~\cite{makhlin2002nonlocal}
\begin{equation}
    \mathrm{U}_{CZ} = e^{i\pi/4} \mathrm{U}^m_Z\left(\frac{\pi}{2}\right)\mathrm{U}^{m+1}_Z\left(\frac{\pi}{2}\right)\mathrm{U}^{m}_{ZZ}\left(-\frac{\pi}{2}\right)
\end{equation}
whose quantum circuitry is depicted in FIG.~\ref{fig:QC_sky}(c). The CNOT is formed through the successive application of the CZ gate followed by the Hadamard gate as
\begin{equation}
    \mathrm{U}^{m\rightarrow m+1}_{\mathrm{CNOT}} = \mathrm{U}^{m+1}_H \mathrm{U}_{CZ}U^{m+1}_H
\end{equation}
where the control qubit is represented by the skyrmion in the $m-$layer, and the target qubit is represented by the neighboring skyrmion in the $(m+1)-$layer. So, the SWAP gate is defined as the swapping of two adjacent qubits
\begin{equation}
U^{m \longleftrightarrow m+1}_{\mathrm{SWAP}} = |s_m s_{m+1}\rangle = |s_{m+1} s_m\rangle
\end{equation}
This concludes the construction of Qubit gates with circuitry illustrated in FIG.~\ref{fig:QC_sky}(c) but in order to execute quantum computations, the initialization and readout processes are a must. To begin the computation, an electric field is applied to lift the degeneracy between the two Bloch-type skyrmions in each layer, leading to the lower-energy state $|00 \dots 0\rangle$. Cooling the sample further brings each qubit to its ground state $|0\rangle$. During the readout process, the skyrmion helicity is observed, fixing it at either $\eta = 0$ or $\pi$ due to their two-fold degenerate ground states, resulting in a standard representation $|s_1 s_2 \dots s_N \rangle$ of the quantum computation's readout. The presented proposal by the authors requires precise control of the skyrmion's position during the implementation of the Ising coupling gate. A recent theoretical report suggests that this control can be achieved by digitizing the position of the skyrmion, which was initially designed for nanoscale skyrmions in ferromagnets with DMI. This approach could also be applicable to nanoscale skyrmions in frustrated magnets, potentially enhancing the accuracy of gate operations.\\
Similar formalism was idealised in Ref.~\cite{xia2022qubits} for \textit{meron} characterized by the direction of spin-circulation, chirality ($c$), and core spin polarity ($p$). In the absence of DMI, four degenerate meron states with $cp=\pm1$ are considered right-handed and left-handed respectively. DMI's presence lifts this degeneracy which energetically favours the left-handed merons~\cite{im2012symmetry}. These merons are used as qubits, with core spin up ($p=1$) as $|0\rangle$, and core spin down ($p=-1$) as $|1\rangle$, allowing quantum superposition for qubits.
Much like skyrmions, the concept of meron core spin control is also put forth for the design of quantum gates. Although the advantage of meron over skyrmion is its vortex-like in-plane spin texture except at its core that ensures its stability for a smaller size ($\sim$ 3 nm). Further, skyrmion number conservation ensures core spin stability for an infinitely large sample, resulting in infinite relaxation time (expected below 3 K).

\subsection{Topological Superconductivity(TSC) \& Q-Computing(TQC)}
\begin{figure*}[htbp]
    \centering
    \includegraphics[width  = \linewidth]{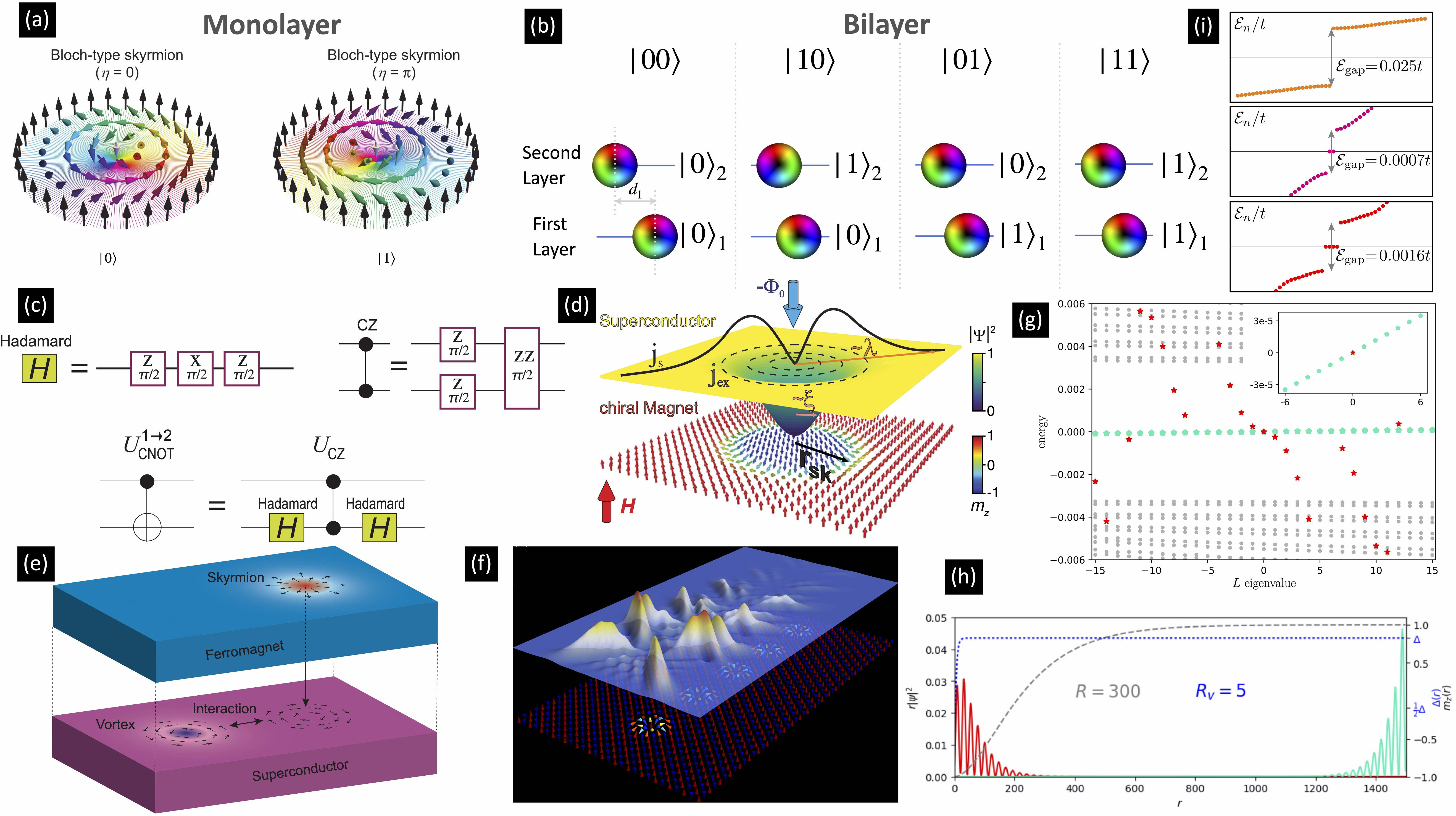}
    \caption{(a) Blochtype skyrmion representing qubit states $|0\rangle$ and $|1\rangle$ with helicity degeneracy $\eta = 0$ and $\pi$, with permission from Ref.~\cite{xia2023universal};
    (b) Bloch-type skyrmions in two layers representing qubit states like $|01\rangle = |0\rangle_1 |1\rangle_2$, with exchange interaction $J_{\mathrm{int}}(d_1)$ between them, with permission from Ref.~\cite{xia2023universal};
    (c) Quantum circuit representation of the Hadamard, CZ, and CNOT gates, with permission from Ref.~\cite{xia2022qubits};
    (d) Schematic of a N\'{e}el skyrmion creating an antivortex with flux $-\Phi_0 \equiv -h/2e$ antiparallel to the external magnetic field $\mathbf{H}$, with permission from Ref.~\cite{petrovic2021skyrmion}; 
    (e) Exchange field coupled SC in the proximity of skyrmion generates a supercurrent in the SC due to the magnetoelectric effect, leading to the binding of SVPs, with permission from Ref.~\cite{hals2016composite};
    (f) Probability density of MBSs (top) localized at one end of a chain of AFM-Sk embedded in a collinear AFM (bottom), with permission from Ref.~\cite{diaz2021majorana};
    (g) Energy levels in a single skyrmion-vortex pair with $n = 1$ and $b = 1$ with $R = 300$ and $R_v = 5$, with permission from Ref.~\cite{rex2019majorana};
    (h) Radial probability density of inner (red solid line) and outer (turquoise solid line) Majorana modes, radial shape of the skyrmion texture in terms of $\boldsymbol{m}_z$ (grey dashed line), and the profile of the vortex (blue dotted line), with permission from Ref.~\cite{rex2019majorana};
    (i) Energy spectrum of selected topological phases induced by an AFM skyrmion chain, with permission from Ref.~\cite{diaz2021majorana}}
    \label{fig:QC_sky}
\end{figure*}
Topological superconductivity has emerged as a promising platform for quantum computation. It combines the advantages of superconductivity, which enables the efficient manipulation and storage of quantum information, with the unique properties of topological states of matter i.e. MZMs, which provide inherent protection against errors. However, the practical implementation of quantum computers is complicated by errors that originate from the computing system and its surroundings.
To overcome this challenge, researchers are actively exploring the concept of topological quantum computing as a potential avenue for constructing fault-tolerant quantum computers. In topological quantum computing, quantum operations are performed on anyons which are quasiparticles whose collective states remain largely unaffected by surroundings. However, generating these quasiparticles within materials has posed significant challenges.\\
In particular, anyons exist in two dimensions and possess unique characteristics by exhibiting non-abelian statistics distinct from conventional particles like fermions and bosons~\cite{ivanov2001non,nayak2008non,alicea2011non}. When two non-abelian anyons are exchanged, their quantum state is not simply multiplied by a phase factor as in the case of abelian anyons, but can instead undergo a more complex transformation known as \textit{braiding}. The non-abelian nature of these anyons means that the result of braiding depends on the order in which the anyons are exchanged. As a result, the braiding process yields a quantum information encoding and manipulation that is highly resilient against decoherence and perturbations which is why non-abelian anyons present a compelling substitute for qubits. This property to manipulate quantum information through braiding holds promise for developing robust and fault-tolerant quantum computing architectures and is of great interest for the field of topological quantum computing~\cite{stern2013topological, sarma2015majorana}, as it provides a means for performing quantum operations and implementing quantum algorithms.\\
This history of anyons is related to quantum phenomena such as the fraction quantum Hall effect observed in 2D electron gas under a strong magnetic field which exhibited fractional statistics or non-abelian statistics. However, these magnetic materials are not suitable candidates for quantum computation due to their practical limitations. A proposed alternative system to host anyons is the two-dimensional p-wave superconductors (SC)~\cite{kitaev2001unpaired}. Unlike conventional s-wave superconductors, these superconductors have unique spin alignments in the Cooper pairs carrying the electrical current.
Physicists anticipate that the building blocks of anyons, known as Majorana Zero Modes (MZMs), could emerge within a p-wave superconductor at the core of vortices formed in the material. However, the practical realization of these p-wave superconductors has further proven to be challenging. But in principle, it is possible to transform an s-wave SC into a p-wave SC, simply by coupling it to a semiconducting heteronanowire~\cite{kitaev2001unpaired, lutchyn2010majorana, oreg2010helical} characterized by a robust spin-orbit interaction and a Zeeman field. So, to date, a hetero-nanowire of semiconductor-superconductor structure with strong SOC and Zeeman is the prominent platform for the realization of p-wave SC and topological phases.  Compelling experimental evidence of zero-energy modes consistent with Majorana Bound States (MBSs) has been observed ~\cite{mourik2012signatures,das2012zero,deng2016majorana,suominen2017zero,vaitiekenas2021zero}, although real-space braiding in this system was found theoretically challenging.
\begin{figure}[t]
    \centering
    \includegraphics[width = \linewidth]{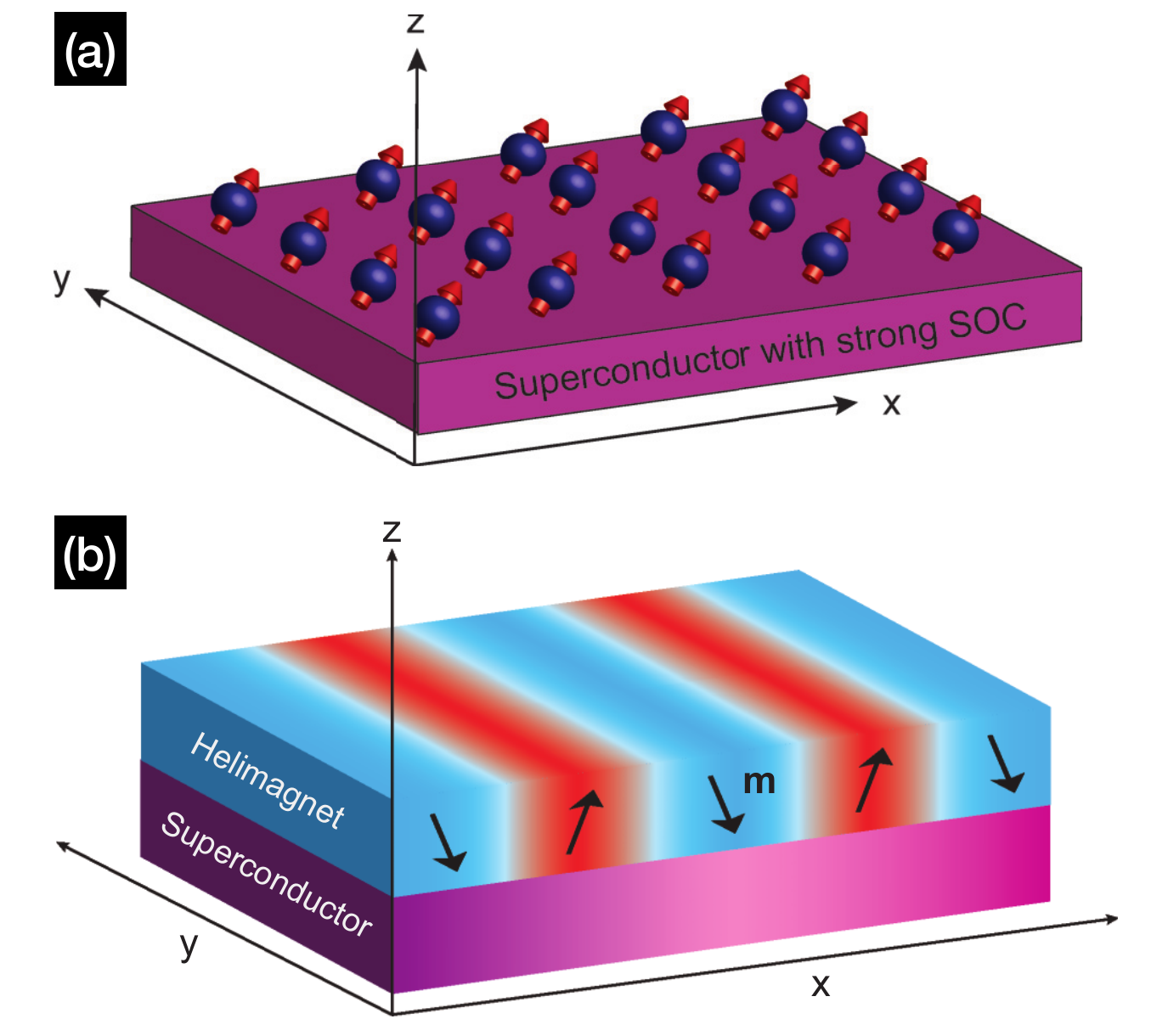}
    \caption{\textbf{Schematic Diagram of}~(a) Ferromagnetically ordered atom on a SC with strong SOC, with permission from Ref.~\cite{hals2016supercurrent}; (b) A superconductor-helimagnet (conical spin texture) heterostructure, with permission from Ref.~\cite{hals2017magnetoelectric} }
    \label{fig:supercond}
\end{figure}
Later on, atomic chains (Shiba chains) on SC have been considered which are expected to exhibit one-dimensional topological superconductivity~\cite{nadj2014observation,ruby2015end,pawlak2016probing,kim2018toward,lutchyn2010majorana,oreg2010helical,klinovaja2013topological,vazifeh2013self,braunecker2013interplay,pientka2013topological}. In such chains, the external Zeeman field are replaced by using a magnetized material (for instance, nanowires here) with potential emergent magnetic field coupled to s-wave SC, which together as a system can produce a p-wave SC and host Majorana Zero Modes (MZMs) [see FIG.~\ref{fig:supercond}(a)]. One-dimensional topological superconductors have been found to host localized zero energy states, also known as majorana bound states~\cite{kitaev2001unpaired,alicea2012new,pawlak2019majorana,prada2020andreev}. These states are localized to specific regions or structures, such as vortices or edges in two-dimensional materials. Subsequently, researchers started utilizing rotating or helical magnetic fields instead of ferromagnetic fields, which further eliminated the requirement of SO (Spin-Orbit) interactions [see FIG.~\ref{fig:supercond}(b)]. Building upon the idea of a swirling magnetic field, an intriguing proposal for establishing a framework capable of accommodating MZMs involves connecting a vortex in a conventional superconductor with noncollinear magnetic textures~\cite{klinovaja2013topological,vazifeh2013self,braunecker2013interplay,rex2019majorana,desjardins2019synthetic,matos2017tunable}, such as skyrmion~\cite{poyhonen2016skyrmion, gungordu2018stabilization,rex2019majorana,garnier2019topological,kubetzka2020towards}, described by the `winding number' that inherently incorporates robust SO coupling was made [see FIG.~\ref{fig:QC_sky}(e)]. The emergent field [see section~(\ref{electrodynamics})] can be effectively represented as a combination of Zeeman and SO coupling, leading to the emergence of a topological superconductor that harbours majorana fermions at its boundary and vortex core.\\
Theoretical studies have explored the potential for creating a coupling between skyrmions and vortices in heterostructures comprising magnetic and superconducting layers. Both single skyrmion (Sk)~\cite{yang2016majorana} and skyrmion lattice (SkL)~\cite{nakosai2013two,mascot2021topological,mohanta2021skyrmion} have been found promising hosting MZMs. When the magnetic layer under specific conditions transitions into a skyrmion lattice phase, the boundaries between regions with different skyrmion winding numbers act as one-dimensional channels for the flow of supercurrent. Within these channels, majorana zero modes can form. These MZMs~\cite{alicea2012new, beenakker2015random} are localized at the ends of the channels and possess nontrivial topological protection, making them robust against decoherence and local perturbations. The MZMs are then used to perform quantum operations, while the skyrmions provide a means for controlling their positions and interactions.
However, there is still a lack of strong experimental investigations in this area due to the inability of skyrmion's controllable dynamics and experimental precision. To achieve the desired topological phase (MZMs), both components of the heterostructure need to be maintained within specific temperature and magnetic field ranges. To equip with such fining control, several theoretical models have been proposed involving magnetic skyrmion crystal and topological insulator surface and josephson junction~\cite{yokoyama2015josephson} creating and controlling the majorana bound states.\\
Broadly speaking, two primary mechanisms govern the interaction between a vortex and a skyrmion. The first mechanism is referred to as the Rasbha-Edelstein or Exchange-field method~\cite{edelstein1995magnetoelectric}, which occurs when the superconductor and magnet come into contact or are in close proximity, see FIG~\ref{fig:QC_sky}(e). The strength and characteristics of this interaction depend on the magnitude and sign of SOC and exchange field. Several studies have investigated this effect \cite{hals2016composite,baumard2019generation,hals2017magnetoelectric,hals2016supercurrent}. The second mechanism is stray field coupling, which operates over distances greater than the exchange length~\cite{dahir2019interaction}. This type of coupling is modulated by the magnetic field generated by the skyrmions and vortices themselves. The interaction between skyrmions and vortices is influenced by the current profile induced in the superconductor, which is dependent on factors such as the magnetic layer thickness, skyrmion chirality, and skyrmion-vortex separation~\cite{dahir2020meissner}.\\
Recently, reported in Ref.~\cite{petrovic2021skyrmion} on the stray field approach by Petrovi\'{c} and his colleagues that they have experimentally developed a heterostructure that hosts stable skyrmion-vortex coexistence at low fields and temperatures below the superconducting transition $T_c$ with significant control over coupling. The fabricated and simulated design [see details~\cite{petrovic2021skyrmion}] is a chiral MIS (Magnet-Insulator-Superconductor) structure. Once the skyrmion is nucleated in the magnetic layer with the core of opposite polarity to the direction of the applied magnetic field $\mathbf{H}$ which repels the vortices present in SC, its size and interaction can be controlled by modelling skyrmion/vortex length scales. Nevertheless, when the skyrmion's size in the magnetic layer is increased and reaches a certain threshold, the authors found that the stray field induces the formation of an antivortex. Further, the SC (Nb) layer thickness was tuned to optimize skyrmion-vortex coupling such that with a fine-tuning in the range $\xi<\xi_{\text{eff}}<r_{\text{sk}}<\lambda$ (usual notation), skyrmion of radius $r_{\text{sk}} \sim 50$nm was observed forming (anti)vortices and vortices (bound pair solution) separated by the screening currents in the superconductor [illustrated in FIG.~\ref{fig:Fig37}].\\
\begin{figure}[!ht]
    \centering
    \includegraphics[width = \linewidth]{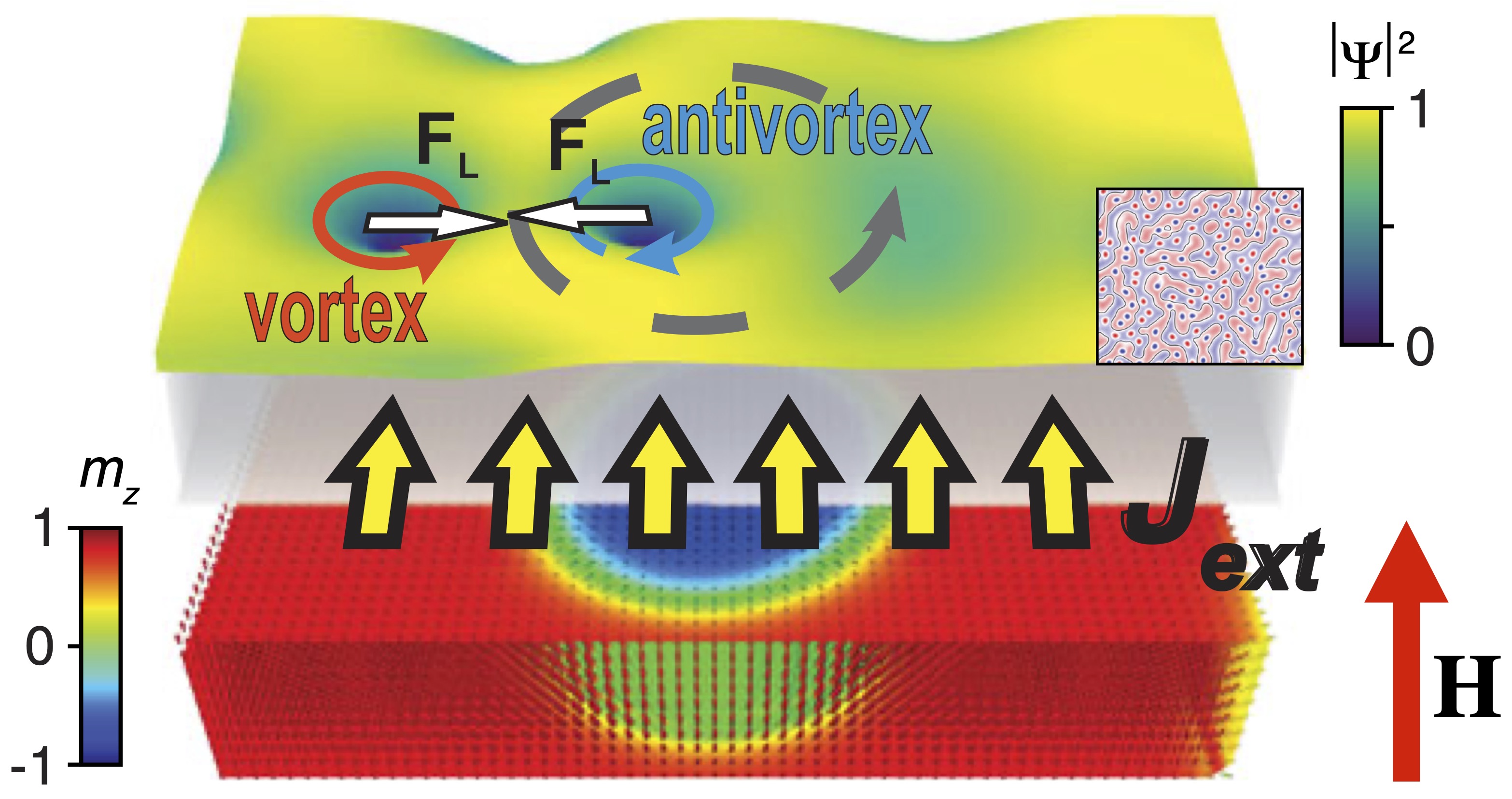}
    \caption{\textbf{Antivortex-facilitated vortex with N\'{e}el Skyrmion}. Blue arrows indicate antivortex currents induced above skyrmion cores; red arrows depict vortex currents outside skyrmion domains. Vortex-antivortex annihilation is prevented by supercurrents screening the skyrmion stray field (grey arrows), maximal at the skyrmion domain wall, inset shows the MFM images of Vortices/Antivortices(red/blue dots resp.), with permission from Ref. \cite{petrovic2021skyrmion}}
    \label{fig:Fig37}
\end{figure}
On the other hand, the exchange field mechanism was studied by Yang \textit{et al.}~\cite{yang2016majorana} as a means to create majorana bound states in a superconductor when placed in the proximity of a small skyrmion of even (double) winding number (DWS), i.e. $R < \xi_{\text{sc}}$ similar to Yu-Shiba-Rusinov (YSR) states~\cite{luh1965bound,shiba1968classical,rusinov1969superconductivity,balatsky2006impurity}. The skyrmion-bound state found is a long-range state with a power law decay YSR. Therefore, in the presence of multiple skyrmions, the SC could mediate an effective long-range interaction between the skyrmions \cite{yao2014enhanced} when the bound state wavefunctions overlap. However, due to the challenges in stabilizing higher-winding skyrmions in chiral magnets, Rex \textit{at al.}~\cite{rex2019majorana} proposed an alternative approach by forming a skyrmion-vortex pair instead of a standalone skyrmion, which creates a platform to obtain MBS from realistic skyrmions with a winding number of one.\\
Nonetheless, grasping the theoretical groundwork is essential for the progression of MZM realization. In light of this, we will emphasize several significant findings. Consider a heterostructure comprising of a SC and FM with a skyrmion described as $\boldsymbol{m}_{\text{sk}}(r,\varphi)=(\cos f(\varphi) \sin g(r), \sin f(\varphi)\sin g(r), \cos g(r))$ as schematically depicted in FIG.~\ref{fig:QC_sky}(d-e). The superconductor is characterized by the $4\times 4$ Bogoliubov-de Gennes (BdG) Hamiltonian
\begin{equation}
    \mathcal{H} = \biggl[\left(\frac{\boldsymbol{p}^2}{2m}-\mu\right) \tau_z- \lambda \boldsymbol{m}_{\text{sk}}(\boldsymbol{r}) \cdot \boldsymbol{\sigma}\\
    +\mathrm{Re}(\Delta) \tau_x-\mathrm{Im}(\Delta) \tau_y \biggr]
\end{equation}
Here, $\boldsymbol{p}=-i(\nabla_x, \nabla_y)$ describes the kinetic energy, $m$ mass of electron at a chemical potential $\mu$ exchange-coupled to the magnetic skyrmion texture $\boldsymbol{m}_{\text{sk}}(\boldsymbol{r})$, in the presence of a superconducting s-wave gap $\Delta(\boldsymbol{r})$; the term $\boldsymbol{m}_{\text{sk}}(\boldsymbol{r}) \cdot \boldsymbol{\sigma}$ describes the proximity coupling between the FM film and SC. The Pauli matrices $\boldsymbol{\tau}$ and $\boldsymbol{\sigma}$ act, respectively, in the particle-hole and spin subspaces
of the four-component spinor $\Psi = (\psi_{\uparrow},\psi_{\downarrow}, \psi^{\dagger}_{\downarrow},-\psi^{\dagger}_{\uparrow})^{\mathrm{T}}$. Hence proximity-induced superconductivity is described by the Hamiltonian
\begin{equation}
    H = \int d^2r \Psi^{\dagger} \mathcal{H} \Psi
\end{equation}
The superconducting order parameter is spatially dependent to account for the vortex, $\Delta(\mathbf{r}) = e^{ib\phi}\Delta(r)$ with integer $b$ and $\Delta(r) = \Delta(1 - e^{-r/R_v} )$, where the vortex and the skyrmion cores are both located at the origin making it a energetically stable structure~\cite{hals2016composite,baumard2019generation,dahir2019interaction}.\\
The next step is to look for zero-energy solutions to the BdG equation satisfying the Majorana condition
\begin{equation}
    LC \Psi^l = -lC\Psi^l
\end{equation}
where the angular momentum operator $L = -i\partial_\varphi+\frac{n}{2}\sigma_z-\frac{b}{2}\tau_z$ with eigenvalues $l$ and the particle-hole operator $C = \sigma_y \tau_y K$ (with complex conjugation $K$) have the commutation relation $[L, C] = 2LC$. The angular momentum commutes with the Hamiltonian such that the eigenstates $\Psi^l$ of the system can be separated into radial and angular parts as
\begin{equation}
    \Psi^l(r, \varphi) =\Psi^l_r \Psi^l_\varphi 
\end{equation}
with angular part $\Psi^l_\varphi(\varphi) =e^{i\varphi}(l-\frac{n}{2}\sigma_z+\frac{b}{2}\tau_z)$. Hence, whenever $\Psi^l$ represents an eigenstate of $C$ that is a Majorana mode, it must be the case that $l = 0$. As a result, the occurrence of MBSs is limited to skyrmion-vortex pairs where the sum $n+b$ is even, such as in the realistic scenario where $n = b = 1$~\cite{rex2019majorana}. Therefore, the presence of Double-Winding Skyrmions (DWS) is no longer necessary for the appearance of MBS in such a pair, as previously proposed by Ref.~\cite{yang2016majorana}.\\
The eigenproblem for $\mathcal{H}$ can be reduced to one dimension for the radial component, $H^l(r)\Psi_r^l(r) = \epsilon^l \Psi_r^l(r)$, with
\begin{equation}
\begin{aligned} \mathcal{H}^{l}(r)= & -\frac{1}{2 m}\left[\partial_{r}^{2}+\frac{1}{r} \partial_{r}+\frac{1}{r^{2}}\left(l-\frac{n}{2} \sigma_{z}+\frac{b}{2} \tau_{z}\right)^{2}\right] \tau_{z} \\ & -\mu \tau_{z}+\lambda \sigma_{z} \cos g(r)+\lambda \sigma_{x} \sin g(r)+\Delta(r) \tau_{x}\end{aligned}
\end{equation}
Because the radial probability density is $r|\Psi^l_r (r)|^2$, it is convenient to consider $\Phi^l_r(r) = \sqrt{r}\Psi^l_r(r)$. So the new Hamiltonian reads
\begin{equation}
    \tilde{\mathcal{H}}^l = \frac{1}{\sqrt{r}}\left(\mathcal{H}^l(r)+\frac{1}{2mr}\partial_r \tau_z-\frac{1}{8mr^2}\tau_z\right)
    \label{Eqn:206}
\end{equation}
Despite the absence of SOC term ($\mathcal{H}_{\mathrm{soc}}$) and the electromagnetic vector potential responsible for the orbital effects and screening supercurrent, these factors have not been observed to impact majorana criterian~\cite{rex2019majorana}. $\mathcal{H}_{\mathrm{soc}}$ plays a crucial role in minimizing the overlap between the two Majorana modes located at the skyrmion core and the system's periphery.\\
To present the Skyrmion-Vortex Pair (SVP) solution, authors~\cite{rex2019majorana} diagonalize the Hamiltonian Eqn.~(\ref{Eqn:206}) including the $\mathcal{H}_{\mathrm{soc}}$ term. In FIG.~\ref{fig:QC_sky}(g), authors observed the case with realistic winding numbers $n = 1$ and $b = 1$, along with an exponential radial shape of both the skyrmion and the vortex in the presence of background SOC. Notably, a pair of zero-energy states appears exclusively at $l = 0$ ( a closer examination of the spectrum [inset in FIG.~\ref{fig:QC_sky}(g)] reveals that this band displays a weak linear dispersion, and none of the states at $l \neq 0$ truly possess zero energy), where among the zero-energy states, one is localized at the core of the skyrmion, while the other is situated at the rim of the system. Radial probability density of Majorana modes, the radial shape of the skyrmion texture and the vortex profile (represented by the blue dotted line with a blue right $y-$scale) are depicted in FIG.~\ref{fig:QC_sky}(h).\\
Researchers worldwide are vigorously investigating the challenges associated with moving skyrmions, which requires incorporating them into a two-dimensional collinear ferromagnetic background. However, this embedding process disrupts the proximity-induced superconducting gap and causes the delocalization of zero-energy states. To address this concern, the authors in~\cite{diaz2021majorana} conducted an investigation using a collinear antiferromagnetic (AFM) background in which a chain of AFM skyrmions~\cite{barker2016static, zhang2016antiferromagnetic} is incorporated, as shown in FIG.~\ref{fig:QC_sky}(f). The significant advantage of employing AFM skyrmions is their ability to remain stable without requiring external magnetic fields, which could otherwise disrupt superconductivity. Importantly, the presence of the collinear antiferromagnetic region does not suppress superconductivity or induce topological superconductivity, thereby enabling the localization of MBSs at the chain ends, as depicted in FIG.~\ref{fig:QC_sky}(i). This offers promising opportunities for the definitive observation of MBSs.\\
\appendix
\section{\label{AP-1}:Electron-Skyrmion Interaction}
When an electron travels across a magnetic texture, an interaction between the electron's spin and magnetic spin texture is unavoidable. The action of the fully interacted system is
\begin{equation}
    S = S_{\mathrm{el}}+S_{\mathrm{spin}}+S_{\mathrm{Hund}}
\end{equation}
where the respective terms are; the free electron's action
\begin{equation}
    S_{\mathrm{el}}= \int dt \mathcal{L}_{\mathrm{el}} = \int dt d^3x \left[i\hbar\Psi^{\dagger}\frac{\partial}{\partial t} {\Psi}-\Psi^{\dagger}\frac{p^2}{2m_e}{\Psi}\right]
\end{equation}
where $p = -i\hbar\nabla$, the spin action is
\begin{equation}
    S_{\text{spin}} = \int dt \mathcal{L}_{\text{spin}} = \int dt d^3x(\mathcal{L}_B - \mathcal{H})
\end{equation}
where $\mathcal{L}_B$ is the berry phase of the form $\mathcal{L}_B = q_e \mathbf{A}\cdot \dot{\boldsymbol{m}}$ with berry potential $\mathbf{A} \sim \langle \boldsymbol{m}| \dot{\boldsymbol{m}}\rangle$ and $\mathcal{H}$ is the Hamiltonian of $\boldsymbol{m}$. With our parametrization Eqn.~(\ref{eqn:9}), the berry phase $\mathcal{L}_B$ can be expressed as
\begin{equation}
    \mathcal{L}_B = -\frac{\hbar S}{a^3}(1-\cos \theta)\frac{d\psi}{dt}
\end{equation}
where $S$ is the magnitude of momentum, $a$ is the lattice constant, and the last term is where the electron interacts with the local magnetic moment by Hund Coupling
\begin{equation}
    S_{\mathrm{Hund}} = -\int dt d^3x \mathcal{J}_{H} \boldsymbol{m}_a \cdot j_e = -\int dt d^3x J_H \boldsymbol{m}^a \cdot \Psi^{\dagger}_a \sigma_a \Psi_a
\end{equation}
where $\Psi_a(r,t) = (\Psi_{\uparrow},\Psi_{\downarrow})$ is the electron operator, $\mathcal{J}_{H}$ is the coupling strength and $\boldsymbol{m}^a(r, t)$ is the skyrmionic field with $a = 1, 2, 3$.\\
In large coupling limit $\mathcal{J}_H >> 0$ (or adiabatic approximation), the skyrmion local magnetic moment and electron's spin align with each other so the dynamics of the electron-skyrmion system can be obtained by solving the Lagrangian
\begin{equation}
    \mathcal{L}_{\text{electron}} = \hbar\Psi^{\dagger}\frac{i\partial}{\partial t} \Psi -  \Psi^{\dagger}\left[\frac{p^2}{2m_e}-J_H \sigma \cdot \boldsymbol{m}\right]\Psi
\end{equation}
where $\sigma \cdot \boldsymbol{m} = \sigma^a \cdot \boldsymbol{m}_a$.
\section*{\label{SnA}List of Symbols and Abbreviations}
\addcontentsline{toc}{section}{List of Symbols and Abbreviations}
\begin{table}[htbp]
    \centering
    \begin{adjustbox}{width= 1.05\linewidth}
    \begin{tabular}{ll}
     \textbf{1D, 2D, 3D} &  One, two and three-dimensional \\
    \textbf{STT/SOT} & Spin-transfer/Spin-Orbit Torque\\
    \textbf{DMI} & Dzyaloinshii-Moriya Interaction \\
    \textbf{NHE/AHE/THE} & Normal/Anomalous/Topological Hall Effect \\
    \textbf{LLG} & Landau-Lifshitz and Gilbert \\
    \multirow{2}{*}{\textbf{Sk/SkX/AFM-Sk}} & \multirow{1}{*}{Skyrmion/Skyrmion Crystal/}\\
    & Anti-ferromagnetic Skyrmion \\
    \textbf{Sk/SkX/AFM-Sk} & Skyrmion/Skyrmion Crystal/Anti-ferromagnetic Skyrmion \\
    \textbf{SHE/SkHE} & Spin/Skyrmion Hall Effect\\
    \textbf{TMR/GMR/NCMR} & Tunnel/Giant/Non-Collinear Magnetoresistance \\
    \textbf{LTEM} & Lorentz Transmission Electron Microscopy \\
    \textbf{MFM} & Magnetic force Transmission \\
    \textbf{MTXM} & Magnetic transmission soft X-ray microscopy \\
    \textbf{RT} & Room Temperature \\
    \textbf{RXS} & Resonant X-ray scattering \\
    \textbf{SANS} & Small angle neutron scattering \\
    \textbf{SPLEEM} & Spin-polarized low-energy electron microscopy \\
    \textbf{SP-STM} &  Spin-polarized scanning electron microscopy \\
    \textbf{STXM} &  Scanning transmission X-ray microscopy \\
    \multirow{2}{*}{\textbf{XMCD(XMLD)-PEEM}} & \multirow{1}{*}{X-ray Magnetic Circular (Linear) Dichroism-assisted} \\
    & Photoemission Electron Microscopy \\
    \textbf{TM} & Transport Mechanics \\
    \textbf{CS/NCS} & Centrosymmetic/Non-centrosymmetric Single
    \end{tabular}
    \end{adjustbox}
\end{table}

\section*{Acknowledgements}
\addcontentsline{toc}{section}{Acknowledgements}

This work acknowledges the support from the University of Delhi (DU), IN and King Abdullah University of Science and Technology (KAUST), SA. 

\subsection*{Author Contribution}
K. Mishra initiated the project with AL and SS. The micromagnetic simulations and post-processing were carried out by KM. KM wrote the manuscript with contributions from AL. The project was supervised by S. Srinivasan, H. Fariborzi and G. Setti.

\section*{References}
\addcontentsline{toc}{section}{References}
\bibliographystyle{unsrt}
\bibliography{References}

\providecommand{\noopsort}[1]{}\providecommand{\singleletter}[1]{#1}%
\begin{thebibliography}{100}

\bibitem{skyrme1962unified}
Tony Hilton~Royle Skyrme.
\newblock A unified field theory of mesons and baryons.
\newblock {\em Nuclear Physics}, 31:556--569, 1962.

\bibitem{skyrme1961non}
Tony Hilton~Royle Skyrme.
\newblock A non-linear field theory.
\newblock {\em Proceedings of the Royal Society of London. Series A.
  Mathematical and Physical Sciences}, 260(1300):127--138, 1961.

\bibitem{bogdanov1989thermodynamically28c}
Alexei~N Bogdanov and DA~Yablonskii.
\newblock Thermodynamically stable ''vortices'' in magnetically ordered
  crystals. the mixed state of magnets.
\newblock {\em Zh. Eksp. Teor. Fiz}, 95(1):178, 1989.

\bibitem{stefanovich1986two29c}
VA~Stefanovich et~al.
\newblock Two-dimensional small-radius solitons in magnets.
\newblock {\em Soviet Journal of Experimental and Theoretical Physics},
  64(2):376, 1986.

\bibitem{ivanov1990magnetic30c}
BA~Ivanov, VA~Stephanovich, and AA~Zhmudskii.
\newblock Magnetic vortices the microscopic analogs of magnetic bubbles.
\newblock {\em Journal of Magnetism and Magnetic Materials}, 88(1-2):116--120,
  1990.

\bibitem{81c}
Ulrich~K Roessler, AN~Bogdanov, and C~Pfleiderer.
\newblock Spontaneous skyrmion ground states in magnetic metals.
\newblock {\em Nature}, 442(7104):797--801, 2006.

\bibitem{derrick}
GH~Derrick.
\newblock Comments on nonlinear wave equations as models for elementary
  particles.
\newblock {\em Journal of Mathematical Physics}, 5(9):1252--1254, 1964.

\bibitem{29}
Igor Dzyaloshinsky.
\newblock A thermodynamic theory of ''weak ferromagnetism of
  antiferromagnetics.
\newblock {\em Journal of physics and chemistry of solids}, 4(4):241--255,
  1958.

\bibitem{30}
T{\^o}ru Moriya.
\newblock Anisotropic superexchange interaction and weak ferromagnetism.
\newblock {\em Physical review}, 120(1):91, 1960.

\bibitem{58c}
Shivaji~Lal Sondhi, A~Karlhede, SA~Kivelson, and EH~Rezayi.
\newblock Skyrmions and the crossover from the integer to fractional quantum
  hall effect at small zeeman energies.
\newblock {\em Physical Review B}, 47(24):16419, 1993.

\bibitem{59c}
L~Brey, HA~Fertig, R~C{\^o}t{\'e}, and AH~MacDonald.
\newblock Skyrme crystal in a two-dimensional electron gas.
\newblock {\em Physical review letters}, 75(13):2562, 1995.

\bibitem{60c}
Tarun Grover and T~Senthil.
\newblock Topological spin hall states, charged skyrmions, and
  superconductivity in two dimensions.
\newblock {\em Physical review letters}, 100(15):156804, 2008.

\bibitem{61c}
A~Knigavko, B~Rosenstein, and YF~Chen.
\newblock Magnetic skyrmions and their lattices in triplet superconductors.
\newblock {\em Physical Review B}, 60(1):550, 1999.

\bibitem{62c}
A~Knigavko and Baruch Rosenstein.
\newblock Magnetic skyrmion lattices in heavy fermion superconductor upt 3.
\newblock {\em Physical review letters}, 82(6):1261, 1999.

\bibitem{63c}
Julien Garaud and Egor Babaev.
\newblock Skyrmionic state and stable half-quantum vortices in chiral p-wave
  superconductors.
\newblock {\em Physical Review B}, 86(6):060514, 2012.

\bibitem{64c}
Daniel~F Agterberg, Egor Babaev, and Julien Garaud.
\newblock Microscopic prediction of skyrmion lattice state in clean interface
  superconductors.
\newblock {\em Physical Review B}, 90(6):064509, 2014.

\bibitem{65c}
Eun-Gook Moon.
\newblock Skyrmions with quadratic band touching fermions: A way to achieve
  charge 4 e superconductivity.
\newblock {\em Physical Review B}, 85(24):245123, 2012.

\bibitem{yokoyama2015josephson}
Takehito Yokoyama and Jacob Linder.
\newblock Josephson effect through magnetic skyrmions.
\newblock {\em Physical Review B}, 92(6):060503, 2015.

\bibitem{68}
Qi~Li, John Toner, and D~Belitz.
\newblock Elasticity and melting of skyrmion flux lattices in p-wave
  superconductors.
\newblock {\em Physical review letters}, 98(18):187002, 2007.

\bibitem{ho1998spinor69c}
Tin-Lun Ho.
\newblock Spinor bose condensates in optical traps.
\newblock {\em Physical review letters}, 81(4):742, 1998.

\bibitem{al2001skyrmions70c}
Usama Al~Khawaja and Henk Stoof.
\newblock Skyrmions in a ferromagnetic bose--einstein condensate.
\newblock {\em Nature}, 411(6840):918--920, 2001.

\bibitem{marzlin2000creation71c}
Karl-Peter Marzlin, Weiping Zhang, and Barry~C Sanders.
\newblock Creation of skyrmions in a spinor bose-einstein condensate.
\newblock {\em Physical Review A}, 62(1):013602, 2000.

\bibitem{battye2002stable72c}
Richard~A Battye, NR~Cooper, and Paul~M Sutcliffe.
\newblock Stable skyrmions in two-component bose-einstein condensates.
\newblock {\em Physical review letters}, 88(8):080401, 2002.

\bibitem{savage2003energetically73c}
CM~Savage and J~Ruostekoski.
\newblock Energetically stable particlelike skyrmions in a trapped
  bose-einstein condensate.
\newblock {\em Physical review letters}, 91(1):010403, 2003.

\bibitem{leslie2009creation74c}
LS~Leslie, A~Hansen, KC~Wright, BM~Deutsch, and NP~Bigelow.
\newblock Creation and detection of skyrmions in a bose-einstein condensate.
\newblock {\em Physical review letters}, 103(25):250401, 2009.

\bibitem{choi2012observation75c}
Jae-yoon Choi, Woo~Jin Kwon, and Yong-il Shin.
\newblock Observation of topologically stable 2d skyrmions in an
  antiferromagnetic spinor bose-einstein condensate.
\newblock {\em Physical review letters}, 108(3):035301, 2012.

\bibitem{vilenkin1994cosmic}
Alexander Vilenkin and E~Paul~S Shellard.
\newblock {\em Cosmic strings and other topological defects}.
\newblock Cambridge University Press, 1994.

\bibitem{fert2013skyrmions}
Albert Fert, Vincent Cros, and Joao Sampaio.
\newblock Skyrmions on the track.
\newblock {\em Nature nanotechnology}, 8(3):152--156, 2013.

\bibitem{kiselev2011chiral68kk}
Nikolai~S Kiselev, AN~Bogdanov, R~Sch{\"a}fer, and UK~R{\"o}{\ss}ler.
\newblock Chiral skyrmions in thin magnetic films: new objects for magnetic
  storage technologies?
\newblock {\em Journal of Physics D: Applied Physics}, 44(39):392001, 2011.

\bibitem{jonietz2010spin80c}
Florian Jonietz, Sebastain m{\"u}hlbauer, Christian Pfleiderer, Andreas
  Neubauer, Wolfgang M{\"u}nzer, Andreas Bauer, T~Adams, Robert Georgii, Peter
  B{\"o}ni, Rembert~A Duine, et~al.
\newblock Spin transfer torques in mnsi at ultralow current densities.
\newblock {\em Science}, 330(6011):1648--1651, 2010.

\bibitem{yu2012skyrmion}
XZ~Yu, Naoya Kanazawa, WZ~Zhang, T~Nagai, Toru Hara, Koji Kimoto, Yoshio
  Matsui, Yoshinori Onose, and Yoshinori Tokura.
\newblock Skyrmion flow near room temperature in an ultralow current density.
\newblock {\em Nature communications}, 3(1):988, 2012.

\bibitem{sampaio2013nucleation88c}
Jo{\~a}o Sampaio, Vincent Cros, Stanislas Rohart, Andr{\'e} Thiaville, and
  Albert Fert.
\newblock Nucleation, stability and current-induced motion of isolated magnetic
  skyrmions in nanostructures.
\newblock {\em Nature nanotechnology}, 8(11):839--844, 2013.

\bibitem{buttner2015dynamics}
Felix B{\"u}ttner, C~Moutafis, M~Schneider, B~Kr{\"u}ger, CM~G{\"u}nther,
  J~Geilhufe, C~v~Korff Schmising, J~Mohanty, B~Pfau, S~Schaffert, et~al.
\newblock Dynamics and inertia of skyrmionic spin structures.
\newblock {\em Nature Physics}, 11(3):225--228, 2015.

\bibitem{luo2021skyrmion}
Shijiang Luo and Long You.
\newblock Skyrmion devices for memory and logic applications.
\newblock {\em APL Materials}, 9(5), 2021.

\bibitem{jiang2015blowing}
Wanjun Jiang, Pramey Upadhyaya, Wei Zhang, Guoqiang Yu, M~Benjamin Jungfleisch,
  Frank~Y Fradin, John~E Pearson, Yaroslav Tserkovnyak, Kang~L Wang, Olle
  Heinonen, et~al.
\newblock Blowing magnetic skyrmion bubbles.
\newblock {\em Science}, 349(6245):283--286, 2015.

\bibitem{iwasaki2013current85c}
Junichi Iwasaki, Masahito Mochizuki, and Naoto Nagaosa.
\newblock Current-induced skyrmion dynamics in constricted geometries.
\newblock {\em Nature nanotechnology}, 8(10):742--747, 2013.

\bibitem{heinonen2016generation73kk}
Olle Heinonen, Wanjun Jiang, Hamoud Somaily, Suzanne~GE Te~Velthuis, and Axel
  Hoffmann.
\newblock Generation of magnetic skyrmion bubbles by inhomogeneous spin hall
  currents.
\newblock {\em Physical Review B}, 93(9):094407, 2016.

\bibitem{crum2015perpendicular74kk}
Dax~M Crum, Mohammed Bouhassoune, Juba Bouaziz, Benedikt Schweflinghaus, Stefan
  Bl{\"u}gel, and Samir Lounis.
\newblock Perpendicular reading of single confined magnetic skyrmions.
\newblock {\em Nature communications}, 6(1):8541, 2015.

\bibitem{hanneken2015electrical75kk}
Christian Hanneken, Fabian Otte, Andr{\'e} Kubetzka, Bertrand Dup{\'e}, Niklas
  Romming, Kirsten Von~Bergmann, Roland Wiesendanger, and Stefan Heinze.
\newblock Electrical detection of magnetic skyrmions by tunnelling
  non-collinear magnetoresistance.
\newblock {\em Nature nanotechnology}, 10(12):1039--1042, 2015.

\bibitem{parkin2008magnetic}
Stuart~SP Parkin, Masamitsu Hayashi, and Luc Thomas.
\newblock Magnetic domain-wall racetrack memory.
\newblock {\em Science}, 320(5873):190--194, 2008.

\bibitem{yu2014biskyrmion47c}
XZ~Yu, Y~Tokunaga, Y~Kaneko, WZ~Zhang, K~Kimoto, Y~Matsui, Y~Taguchi, and
  Y~Tokura.
\newblock Biskyrmion states and their current-driven motion in a layered
  manganite.
\newblock {\em Nature Communications}, 5(1):3198, 2014.

\bibitem{nagaosa2013topological36}
Naoto Nagaosa and Yoshinori Tokura.
\newblock Topological properties and dynamics of magnetic skyrmions.
\newblock {\em Nature nanotechnology}, 8(12):899--911, 2013.

\bibitem{yan2021skyrmion}
ZR~Yan, YZ~Liu, Y~Guang, K~Yue, JF~Feng, RK~Lake, GQ~Yu, and XF~Han.
\newblock Skyrmion-based programmable logic device with complete boolean logic
  functions.
\newblock {\em Physical Review Applied}, 15(6):064004, 2021.

\bibitem{zhang2015magnetic}
Xichao Zhang, Motohiko Ezawa, and Yan Zhou.
\newblock Magnetic skyrmion logic gates: conversion, duplication and merging of
  skyrmions.
\newblock {\em Scientific reports}, 5(1):1--8, 2015.

\bibitem{sisodia2022programmable}
Naveen Sisodia, Johan Pelloux-Prayer, Liliana~D Buda-Prejbeanu, Lorena Anghel,
  Gilles Gaudin, and Olivier Boulle.
\newblock Programmable skyrmion logic gates based on skyrmion tunneling.
\newblock {\em Physical Review Applied}, 17(6):064035, 2022.

\bibitem{luo2018reconfigurable}
Shijiang Luo, Min Song, Xin Li, Yue Zhang, Jeongmin Hong, Xiaofei Yang,
  Xuecheng Zou, Nuo Xu, and Long You.
\newblock Reconfigurable skyrmion logic gates.
\newblock {\em Nano letters}, 18(2):1180--1184, 2018.

\bibitem{zhang2015topological}
Shilei Zhang, Alexander~A Baker, Stavros Komineas, and Thorsten Hesjedal.
\newblock Topological computation based on direct magnetic logic communication.
\newblock {\em Scientific reports}, 5(1):15773, 2015.

\bibitem{zhang2018manipulation}
SL~Zhang, WW~Wang, DM~Burn, H~Peng, H~Berger, A~Bauer, C~Pfleiderer, G~Van
  Der~Laan, and T~Hesjedal.
\newblock Manipulation of skyrmion motion by magnetic field gradients.
\newblock {\em Nature communications}, 9(1):2115, 2018.

\bibitem{litzius2017skyrmion47k}
Kai Litzius, Ivan Lemesh, Benjamin Kr{\"u}ger, Pedram Bassirian, Lucas Caretta,
  Kornel Richter, Felix B{\"u}ttner, Koji Sato, Oleg~A Tretiakov, Johannes
  F{\"o}rster, et~al.
\newblock Skyrmion hall effect revealed by direct time-resolved x-ray
  microscopy.
\newblock {\em Nature Physics}, 13(2):170--175, 2017.

\bibitem{yu2017room}
Guoqiang Yu, Pramey Upadhyaya, Qiming Shao, Hao Wu, Gen Yin, Xiang Li, Congli
  He, Wanjun Jiang, Xiufeng Han, Pedram~Khalili Amiri, et~al.
\newblock Room-temperature skyrmion shift device for memory application.
\newblock {\em Nano letters}, 17(1):261--268, 2017.

\bibitem{woo2016observation46k}
Seonghoon Woo, Kai Litzius, Benjamin Kr{\"u}ger, Mi-Young Im, Lucas Caretta,
  Kornel Richter, Maxwell Mann, Andrea Krone, Robert~M Reeve, Markus Weigand,
  et~al.
\newblock Observation of room-temperature magnetic skyrmions and their
  current-driven dynamics in ultrathin metallic ferromagnets.
\newblock {\em Nature materials}, 15(5):501--506, 2016.

\bibitem{neubauer2009topological}
A~Neubauer, C~Pfleiderer, B~Binz, A~Rosch, R~Ritz, PG~Niklowitz, and
  P~B{\"o}ni.
\newblock Topological hall effect in the a phase of mnsi.
\newblock {\em Physical review letters}, 102(18):186602, 2009.

\bibitem{kanazawa2011large}
N~Kanazawa, Y~Onose, T~Arima, D~Okuyama, K~Ohoyama, S~Wakimoto, K~Kakurai,
  S~Ishiwata, and Y~Tokura.
\newblock Large topological hall effect in a short-period helimagnet mnge.
\newblock {\em Physical review letters}, 106(15):156603, 2011.

\bibitem{kanazawa2015discretized87k}
N~Kanazawa, M~Kubota, A~Tsukazaki, Y~Kozuka, KS~Takahashi, M~Kawasaki,
  M~Ichikawa, F~Kagawa, and Y~Tokura.
\newblock Discretized topological hall effect emerging from skyrmions in
  constricted geometry.
\newblock {\em Physical Review B}, 91(4):041122, 2015.

\bibitem{87}
ea~Y Ishikawa, K~Tajima, D~Bloch, and M~Roth.
\newblock Helical spin structure in manganese silicide mnsi.
\newblock {\em Solid State Communications}, 19(6):525--528, 1976.

\bibitem{aharoni2000introduction32}
Amikam Aharoni et~al.
\newblock {\em Introduction to the Theory of Ferromagnetism}, volume 109.
\newblock Clarendon Press, 2000.

\bibitem{muhlbauer2009skyrmion38c}
Sebastian m{\"u}hlbauer, Benedikt Binz, F~Jonietz, Christian Pfleiderer, Achim
  Rosch, Anja Neubauer, Robert Georgii, and Peter B\"{o}ni.
\newblock Skyrmion lattice in a chiral magnet.
\newblock {\em Science}, 323(5916):915--919, 2009.

\bibitem{yu2011near42c}
XZ~Yu, Naoya Kanazawa, Yoshinori Onose, K~Kimoto, WZ~Zhang, Shintaro Ishiwata,
  Yoshio Matsui, and Yoshinori Tokura.
\newblock Near room-temperature formation of a skyrmion crystal in thin-films
  of the helimagnet fege.
\newblock {\em Nature materials}, 10(2):106--109, 2011.

\bibitem{seki2012magnetoelectric44c}
S~Seki, S~Ishiwata, and Y~Tokura.
\newblock Magnetoelectric nature of skyrmions in a chiral magnetic insulator cu
  2 oseo 3.
\newblock {\em Physical Review B}, 86(6):060403, 2012.

\bibitem{yu2010real41c}
XZ~Yu, Yoshinori Onose, Naoya Kanazawa, Joung~Hwan Park, JH~Han, Yoshio Matsui,
  Naoto Nagaosa, and Yoshinori Tokura.
\newblock Real-space observation of a two-dimensional skyrmion crystal.
\newblock {\em Nature}, 465(7300):901--904, 2010.

\bibitem{tokunaga2015new}
Y~Tokunaga, XZ~Yu, JS~White, Henrik~M R{\o}nnow, D~Morikawa, Y~Taguchi, and
  Y~Tokura.
\newblock A new class of chiral materials hosting magnetic skyrmions beyond
  room temperature.
\newblock {\em Nature communications}, 6(1):7638, 2015.

\bibitem{kurumaji2019skyrmion}
Takashi Kurumaji, Taro Nakajima, Max Hirschberger, Akiko Kikkawa, Yuichi
  Yamasaki, Hajime Sagayama, Hironori Nakao, Yasujiro Taguchi, Taka-hisa Arima,
  and Yoshinori Tokura.
\newblock Skyrmion lattice with a giant topological hall effect in a frustrated
  triangular-lattice magnet.
\newblock {\em Science}, 365(6456):914--918, 2019.

\bibitem{hirschberger2019skyrmion}
Max Hirschberger, Taro Nakajima, Shang Gao, Licong Peng, Akiko Kikkawa, Takashi
  Kurumaji, Markus Kriener, Yuichi Yamasaki, Hajime Sagayama, Hironori Nakao,
  et~al.
\newblock Skyrmion phase and competing magnetic orders on a breathing
  kagom{\'e} lattice.
\newblock {\em Nature communications}, 10(1):5831, 2019.

\bibitem{khanh2020nanometric}
Nguyen~Duy Khanh, Taro Nakajima, Xiuzhen Yu, Shang Gao, Kiyou Shibata, Max
  Hirschberger, Yuichi Yamasaki, Hajime Sagayama, Hironori Nakao, Licong Peng,
  et~al.
\newblock Nanometric square skyrmion lattice in a centrosymmetric tetragonal
  magnet.
\newblock {\em Nature Nanotechnology}, 15(6):444--449, 2020.

\bibitem{takagi2022square}
Rina Takagi, Naofumi Matsuyama, Victor Ukleev, Le~Yu, Jonathan~S White, Sonia
  Francoual, Jos{\'e}~RL Mardegan, Satoru Hayami, Hiraku Saito, Koji Kaneko,
  et~al.
\newblock Square and rhombic lattices of magnetic skyrmions in a
  centrosymmetric binary compound.
\newblock {\em Nature communications}, 13(1):1472, 2022.

\bibitem{kezsmarki2015neel}
Istv{\'a}n K{\'e}zsm{\'a}rki, S{\'a}ndor Bord{\'a}cs, Peter Milde, E~Neuber,
  LM~Eng, JS~White, Henrik~M R{\o}nnow, CD~Dewhurst, M~Mochizuki, K~Yanai,
  et~al.
\newblock N{\'e}el-type skyrmion lattice with confined orientation in the polar
  magnetic semiconductor gav4s8.
\newblock {\em Nature materials}, 14(11):1116--1122, 2015.

\bibitem{zhang2022room}
Hongrui Zhang, David Raftrey, Ying-Ting Chan, Yu-Tsun Shao, Rui Chen, Xiang
  Chen, Xiaoxi Huang, Jonathan~T Reichanadter, Kaichen Dong, Sandhya Susarla,
  et~al.
\newblock Room-temperature skyrmion lattice in a layered magnet (fe0. 5co0. 5)
  5gete2.
\newblock {\em Science advances}, 8(12):eabm7103, 2022.

\bibitem{heinze2011spontaneous47}
Stefan Heinze, Kirsten Von~Bergmann, Matthias Menzel, Jens Brede, Andr{\'e}
  Kubetzka, Roland Wiesendanger, Gustav Bihlmayer, and Stefan Bl{\"u}gel.
\newblock Spontaneous atomic-scale magnetic skyrmion lattice in two dimensions.
\newblock {\em nature physics}, 7(9):713--718, 2011.

\bibitem{romming2013writing48}
Niklas Romming, Christian Hanneken, Matthias Menzel, Jessica~E Bickel, Boris
  Wolter, Kirsten von Bergmann, Andr{\'e} Kubetzka, and Roland Wiesendanger.
\newblock Writing and deleting single magnetic skyrmions.
\newblock {\em Science}, 341(6146):636--639, 2013.

\bibitem{romming2015field60k}
Niklas Romming, Andr{\'e} Kubetzka, Christian Hanneken, Kirsten von Bergmann,
  and Roland Wiesendanger.
\newblock Field-dependent size and shape of single magnetic skyrmions.
\newblock {\em Physical review letters}, 114(17):177203, 2015.

\bibitem{boulle2016room}
Olivier Boulle, Jan Vogel, Hongxin Yang, Stefania Pizzini, Dayane
  de~Souza~Chaves, Andrea Locatelli, Tevfik~Onur Mente{\c{s}}, Alessandro Sala,
  Liliana~D Buda-Prejbeanu, Olivier Klein, et~al.
\newblock Room-temperature chiral magnetic skyrmions in ultrathin magnetic
  nanostructures.
\newblock {\em Nature nanotechnology}, 11(5):449--454, 2016.

\bibitem{moreau2016additive}
Constance Moreau-Luchaire, Christoforos Moutafis, Nicolas Reyren, Jo{\~a}o
  Sampaio, CAF Vaz, N~Van~Horne, Karim Bouzehouane, K~Garcia, C~Deranlot,
  P~Warnicke, et~al.
\newblock Additive interfacial chiral interaction in multilayers for
  stabilization of small individual skyrmions at room temperature.
\newblock {\em Nature nanotechnology}, 11(5):444--448, 2016.

\bibitem{soumyanarayanan2016emergent}
Anjan Soumyanarayanan, Nicolas Reyren, Albert Fert, and Christos Panagopoulos.
\newblock Emergent phenomena induced by spin--orbit coupling at surfaces and
  interfaces.
\newblock {\em Nature}, 539(7630):509--517, 2016.

\bibitem{chen2015room}
Gong Chen, Arantzazu Mascaraque, Alpha~T N'Diaye, and Andreas~K Schmid.
\newblock Room temperature skyrmion ground state stabilized through interlayer
  exchange coupling.
\newblock {\em Applied Physics Letters}, 106(24):242404, 2015.

\bibitem{wu2020neel}
Yingying Wu, Senfu Zhang, Junwei Zhang, Wei Wang, Yang~Lin Zhu, Jin Hu, Gen
  Yin, Kin Wong, Chi Fang, Caihua Wan, et~al.
\newblock N{\'e}el-type skyrmion in wte2/fe3gete2 van der waals
  heterostructure.
\newblock {\em Nature communications}, 11(1):3860, 2020.

\bibitem{wu2022van}
Yingying Wu, Brian Francisco, Zhijie Chen, Wei Wang, Yu~Zhang, Caihua Wan,
  Xiufeng Han, Hang Chi, Yasen Hou, Alessandro Lodesani, et~al.
\newblock A van der waals interface hosting two groups of magnetic skyrmions.
\newblock {\em Advanced Materials}, 34(16):2110583, 2022.

\bibitem{yang2020creation}
M~Yang, Q~Li, RV~Chopdekar, R~Dhall, J~Turner, JD~Carlstr{\"o}m, C~Ophus,
  C~Klewe, P~Shafer, AT~N’Diaye, et~al.
\newblock Creation of skyrmions in van der waals ferromagnet fe3gete2 on
  (co/pd) n superlattice.
\newblock {\em Science advances}, 6(36):eabb5157, 2020.

\bibitem{nayak2017magnetic}
Ajaya~K Nayak, Vivek Kumar, Tianping Ma, Peter Werner, Eckhard Pippel, Roshnee
  Sahoo, Franoise Damay, Ulrich~K R{\"o}{\ss}ler, Claudia Felser, and Stuart~SP
  Parkin.
\newblock Magnetic antiskyrmions above room temperature in tetragonal heusler
  materials.
\newblock {\em Nature}, 548(7669):561--566, 2017.

\bibitem{peng2020controlled}
Licong Peng, Rina Takagi, Wataru Koshibae, Kiyou Shibata, Kiyomi Nakajima,
  Taka-hisa Arima, Naoto Nagaosa, Shinichiro Seki, Xiuzhen Yu, and Yoshinori
  Tokura.
\newblock Controlled transformation of skyrmions and antiskyrmions in a
  non-centrosymmetric magnet.
\newblock {\em Nature nanotechnology}, 15(3):181--186, 2020.

\bibitem{karube2021room}
Kosuke Karube, Licong Peng, Jan Masell, Xiuzhen Yu, Fumitaka Kagawa, Yoshinori
  Tokura, and Yasujiro Taguchi.
\newblock Room-temperature antiskyrmions and sawtooth surface textures in a
  non-centrosymmetric magnet with s 4 symmetry.
\newblock {\em Nature Materials}, 20(3):335--340, 2021.

\bibitem{caretta2018fast}
Lucas Caretta, Maxwell Mann, Felix B{\"u}ttner, Kohei Ueda, Bastian Pfau,
  Christian~M G{\"u}nther, Piet Hessing, Alexandra Churikova, Christopher
  Klose, Michael Schneider, et~al.
\newblock Fast current-driven domain walls and small skyrmions in a compensated
  ferrimagnet.
\newblock {\em Nature nanotechnology}, 13(12):1154--1160, 2018.

\bibitem{woo2018current}
Seonghoon Woo, Kyung~Mee Song, Xichao Zhang, Yan Zhou, Motohiko Ezawa, Xiaoxi
  Liu, S~Finizio, J~Raabe, Nyun~Jong Lee, Sang-Il Kim, et~al.
\newblock Current-driven dynamics and inhibition of the skyrmion hall effect of
  ferrimagnetic skyrmions in gdfeco films.
\newblock {\em Nature communications}, 9(1):959, 2018.

\bibitem{dohi2019formation}
Takaaki Dohi, Samik DuttaGupta, Shunsuke Fukami, and Hideo Ohno.
\newblock Formation and current-induced motion of synthetic antiferromagnetic
  skyrmion bubbles.
\newblock {\em Nature communications}, 10(1):5153, 2019.

\bibitem{legrand2018hybrid}
William Legrand, Jean-Yves Chauleau, Davide Maccariello, Nicolas Reyren, Sophie
  Collin, Karim Bouzehouane, Nicolas Jaouen, Vincent Cros, and Albert Fert.
\newblock Hybrid chiral domain walls and skyrmions in magnetic multilayers.
\newblock {\em Science advances}, 4(7):eaat0415, 2018.

\bibitem{chen2020realization}
Ruyi Chen, Yang Gao, Xichao Zhang, Ruiqi Zhang, Siqi Yin, Xianzhe Chen,
  Xiaofeng Zhou, Yongjian Zhou, Jing Xia, Yan Zhou, et~al.
\newblock Realization of isolated and high-density skyrmions at room
  temperature in uncompensated synthetic antiferromagnets.
\newblock {\em Nano letters}, 20(5):3299--3305, 2020.

\bibitem{jani2021antiferromagnetic}
Hariom Jani, Jheng-Cyuan Lin, Jiahao Chen, Jack Harrison, Francesco
  Maccherozzi, Jonathon Schad, Saurav Prakash, Chang-Beom Eom, Ariando Ariando,
  Thirumalai Venkatesan, et~al.
\newblock Antiferromagnetic half-skyrmions and bimerons at room temperature.
\newblock {\em Nature}, 590(7844):74--79, 2021.

\bibitem{li2023magnetic}
Sheng Li, Xuewen Wang, and Theo Rasing.
\newblock Magnetic skyrmions: Basic properties and potential applications.
\newblock {\em Interdisciplinary Materials}, 2(2):260--289, 2023.

\bibitem{beg2015ground16}
Marijan Beg, Rebecca Carey, Weiwei Wang, David Cort{\'e}s-Ortu{\~n}o, Mark
  Vousden, Marc-Antonio Bisotti, Maximilian Albert, Dmitri Chernyshenko, Ondrej
  Hovorka, Robert~L Stamps, et~al.
\newblock Ground state search, hysteretic behaviour and reversal mechanism of
  skyrmionic textures in confined helimagnetic nanostructures.
\newblock {\em Scientific Reports}, 5(1):17137, 2015.

\bibitem{89}
AN~Bogdanov and UK~R{\"o}{\ss}ler.
\newblock Chiral symmetry breaking in magnetic thin films and multilayers.
\newblock {\em Physical review letters}, 87(3):037203, 2001.

\bibitem{bogdanov2002magnetic79c}
AN~Bogdanov, Ulrich~K Roessler, M~Wolf, and K-H M{\"u}ller.
\newblock Magnetic structures and reorientation transitions in
  noncentrosymmetric uniaxial antiferromagnets.
\newblock {\em Physical Review B}, 66(21):214410, 2002.

\bibitem{braun2012topological}
Hans-Benjamin Braun.
\newblock Topological effects in nanomagnetism: from superparamagnetism to
  chiral quantum solitons.
\newblock {\em Advances in Physics}, 61(1):1--116, 2012.

\bibitem{hagemeister2015stability}
J~Hagemeister, N~Romming, K~Von~Bergmann, EY~Vedmedenko, and R~Wiesendanger.
\newblock Stability of single skyrmionic bits.
\newblock {\em Nature communications}, 6(1):8455, 2015.

\bibitem{mascot2021topological}
Eric Mascot, Jasmin Bedow, Martin Graham, Stephan Rachel, and Dirk~K Morr.
\newblock Topological superconductivity in skyrmion lattices.
\newblock {\em npj Quantum Materials}, 6(1):6, 2021.

\bibitem{fukuda2011quasi}
Jun-ichi Fukuda and Slobodan {\v{Z}}umer.
\newblock Quasi-two-dimensional skyrmion lattices in a chiral nematic liquid
  crystal.
\newblock {\em Nature communications}, 2(1):246, 2011.

\bibitem{ge2021observation}
Hao Ge, Xiang-Yuan Xu, Le~Liu, Rui Xu, Zhi-Kang Lin, Si-Yuan Yu, Ming Bao,
  Jian-Hua Jiang, Ming-Hui Lu, and Yan-Feng Chen.
\newblock Observation of acoustic skyrmions.
\newblock {\em Physical Review Letters}, 127(14):144502, 2021.

\bibitem{du2019deep}
Luping Du, Aiping Yang, Anatoly~V Zayats, and Xiaocong Yuan.
\newblock Deep-subwavelength features of photonic skyrmions in a confined
  electromagnetic field with orbital angular momentum.
\newblock {\em Nature Physics}, 15(7):650--654, 2019.

\bibitem{yu2018transformation}
XZ~Yu, W~Koshibae, Y~Tokunaga, K~Shibata, Y~Taguchi, N~Nagaosa, and Y~Tokura.
\newblock Transformation between meron and skyrmion topological spin textures
  in a chiral magnet.
\newblock {\em Nature}, 564(7734):95--98, 2018.

\bibitem{hayami2021meron}
Satoru Hayami and Ryota Yambe.
\newblock Meron-antimeron crystals in noncentrosymmetric itinerant magnets on a
  triangular lattice.
\newblock {\em Physical Review B}, 104(9):094425, 2021.

\bibitem{lin2015skyrmion}
Shi-Zeng Lin, Avadh Saxena, and Cristian~D Batista.
\newblock Skyrmion fractionalization and merons in chiral magnets with
  easy-plane anisotropy.
\newblock {\em Physical Review B}, 91(22):224407, 2015.

\bibitem{capic2019stabilty}
Daniel Capic, Dmitry~A Garanin, and Eugene~M Chudnovsky.
\newblock Stabilty of biskyrmions in centrosymmetric magnetic films.
\newblock {\em Physical Review B}, 100(1):014432, 2019.

\bibitem{kanazawa2017topological}
N~Kanazawa, JS~White, HM~R{\o}nnow, CD~Dewhurst, D~Morikawa, K~Shibata,
  T~Arima, F~Kagawa, A~Tsukazaki, Y~Kozuka, et~al.
\newblock Topological spin-hedgehog crystals of a chiral magnet as engineered
  with magnetic anisotropy.
\newblock {\em Physical Review B}, 96(22):220414, 2017.

\bibitem{finazzi2013laser}
Marco Finazzi, Matteo Savoini, AR~Khorsand, A~Tsukamoto, A~Itoh, Lamberto Duo,
  Andrei Kirilyuk, Th~Rasing, and M~Ezawa.
\newblock Laser-induced magnetic nanostructures with tunable topological
  properties.
\newblock {\em Physical review letters}, 110(17):177205, 2013.

\bibitem{zhang2017skyrmion}
Xichao Zhang, Jing Xia, Yan Zhou, Xiaoxi Liu, Han Zhang, and Motohiko Ezawa.
\newblock Skyrmion dynamics in a frustrated ferromagnetic film and
  current-induced helicity locking-unlocking transition.
\newblock {\em Nature communications}, 8(1):1717, 2017.

\bibitem{derras2022dynamics}
A~Derras-Chouk, EM~Chudnovsky, and DA~Garanin.
\newblock Dynamics of the collapse of a ferromagnetic skyrmion in a
  centrosymmetric lattice.
\newblock {\em Physical Review B}, 105(13):134432, 2022.

\bibitem{hoffmann2017antiskyrmions}
Markus Hoffmann, Bernd Zimmermann, Gideon~P M{\"u}ller, Daniel Sch{\"u}rhoff,
  Nikolai~S Kiselev, Christof Melcher, and Stefan Bl{\"u}gel.
\newblock Antiskyrmions stabilized at interfaces by anisotropic
  dzyaloshinskii-moriya interactions.
\newblock {\em Nature communications}, 8(1):308, 2017.

\bibitem{jena2019observation}
Jagannath Jena, Rolf Stinshoff, Rana Saha, Abhay~K Srivastava, Tianping Ma,
  Hakan Deniz, Peter Werner, Claudia Felser, and Stuart~SP Parkin.
\newblock Observation of magnetic antiskyrmions in the low magnetization
  ferrimagnet mn2rh0. 95ir0. 05sn.
\newblock {\em Nano letters}, 20(1):59--65, 2019.

\bibitem{dzyaloshinskii1964theory}
IE~Dzyaloshinskii.
\newblock The theory of helicoidal structures in antiferromagnets. ii.
\newblock {\em Metals. J. Exptl. Theoret. Phys.(USSR)}, 47:336, 1964.

\bibitem{bogdanov1994thermodynamically16}
A~Bogdanov and A~Hubert.
\newblock Thermodynamically stable magnetic vortex states in magnetic crystals.
\newblock {\em Journal of magnetism and magnetic materials}, 138(3):255--269,
  1994.

\bibitem{kataoka1981helical}
Mitsuo Kataoka and Osamu Nakanishi.
\newblock Helical spin density wave due to antisymmetric exchange interaction.
\newblock {\em Journal of the Physical Society of Japan}, 50(12):3888--3896,
  1981.

\bibitem{brown1963micromagnefics}
WF~Brown.
\newblock Micromagnefics, john wiley and sons.
\newblock {\em New York}, 1963.

\bibitem{ex3}
Oleg~N Mryasov, Ulrich Nowak, K~Yu Guslienko, and Roy~W Chantrell.
\newblock Temperature-dependent magnetic properties of fept: Effective spin
  hamiltonian model.
\newblock {\em Europhysics Letters}, 69(5):805, 2005.

\bibitem{100}
P~Gr{\"u}nberg, CM~Mayr, W~Vach, and M~Grimsditch.
\newblock Determination of magnetic parameters by means of brillouinscattering.
  examples: Fe, ni, ni0. 8fe0. 2.
\newblock {\em Journal of Magnetism and Magnetic Materials}, 28(3):319--325,
  1982.

\bibitem{101}
X~Liu, MM~Steiner, R~Sooryakumar, GA~Prinz, RFC Farrow, and G~Harp.
\newblock Exchange stiffness, magnetization, and spin waves in cubic and
  hexagonal phases of cobalt.
\newblock {\em Physical Review B}, 53(18):12166, 1996.

\bibitem{ie}
SSP Parkin and D~Mauri.
\newblock Spin engineering: Direct determination of the
  ruderman-kittel-kasuya-yosida far-field range function in ruthenium.
\newblock {\em Physical Review B}, 44(13):7131, 1991.

\bibitem{12}
Melvin~A Ruderman and Charles Kittel.
\newblock Indirect exchange coupling of nuclear magnetic moments by conduction
  electrons.
\newblock {\em Physical Review}, 96(1):99, 1954.

\bibitem{13}
Tadao Kasuya.
\newblock A theory of metallic ferro-and antiferromagnetism on zener's model.
\newblock {\em Progress of theoretical physics}, 16(1):45--57, 1956.

\bibitem{14}
Kei Yosida.
\newblock Magnetic properties of cu-mn alloys.
\newblock {\em Physical Review}, 106(5):893, 1957.

\bibitem{10}
Stephen Blundell.
\newblock Magnetism in condensed matter, 2003.

\bibitem{88}
Y~Ishikawa, G~Shirane, JA~Tarvin, and M~Kohgi.
\newblock Magnetic excitations in the weak itinerant ferromagnet mnsi.
\newblock {\em Physical Review B}, 16(11):4956, 1977.

\bibitem{32}
Adeline Cr{\'e}pieux and Claudine Lacroix.
\newblock Dzyaloshinsky--moriya interactions induced by symmetry breaking at a
  surface.
\newblock {\em Journal of magnetism and magnetic materials}, 182(3):341--349,
  1998.

\bibitem{34}
MJ~Benitez, A~Hrabec, AP~Mihai, TA~Moore, G~Burnell, Damien McGrouther,
  CH~Marrows, and Stephen McVitie.
\newblock Magnetic microscopy and topological stability of homochiral n{\'e}el
  domain walls in a pt/co/alo x trilayer.
\newblock {\em Nature Communications}, 6(1):8957, 2015.

\bibitem{wang2018theory30}
XS~Wang, HY~Yuan, and XR~Wang.
\newblock A theory on skyrmion size.
\newblock {\em Communications Physics}, 1(1):31, 2018.

\bibitem{35}
Soong-Geun Je, Duck-Ho Kim, Sang-Cheol Yoo, Byoung-Chul Min, Kyung-Jin Lee, and
  Sug-Bong Choe.
\newblock Asymmetric magnetic domain-wall motion by the dzyaloshinskii-moriya
  interaction.
\newblock {\em Physical Review B}, 88(21):214401, 2013.

\bibitem{36}
A~Hrabec, NA~Porter, A~Wells, MJ~Benitez, G~Burnell, S~McVitie, D~McGrouther,
  TA~Moore, and CH~Marrows.
\newblock Measuring and tailoring the dzyaloshinskii-moriya interaction in
  perpendicularly magnetized thin films.
\newblock {\em Physical Review B}, 90(2):020402, 2014.

\bibitem{37}
Kai Di, Vanessa~Li Zhang, Hock~Siah Lim, Ser~Choon Ng, Meng~Hau Kuok, Xuepeng
  Qiu, and Hyunsoo Yang.
\newblock Asymmetric spin-wave dispersion due to dzyaloshinskii-moriya
  interaction in an ultrathin pt/cofeb film.
\newblock {\em Applied Physics Letters}, 106(5):052403, 2015.

\bibitem{rohart2013skyrmion}
S~Rohart and A~Thiaville.
\newblock Skyrmion confinement in ultrathin film nanostructures in the presence
  of dzyaloshinskii-moriya interaction.
\newblock {\em Physical Review B}, 88(18):184422, 2013.

\bibitem{brown1962magnetostatic45}
William~Fuller Brown.
\newblock {\em Magnetostatic principles in ferromagnetism}, volume~1.
\newblock North-Holland Publishing Company, 1962.

\bibitem{coey2010magnetism43}
John~MD Coey.
\newblock {\em Magnetism and magnetic materials}.
\newblock Cambridge university press, 2010.

\bibitem{johnson1996magnetic54}
MT~Johnson, PJH Bloemen, FJA Den~Broeder, and JJ~De~Vries.
\newblock Magnetic anisotropy in metallic multilayers.
\newblock {\em Reports on Progress in Physics}, 59(11):1409, 1996.

\bibitem{parlak2015thickness53}
U~Parlak, M~Ak{\"o}z, S~Tokdemir~{\"O}zt{\"u}rk, and M~Erkovan.
\newblock Thickness dependent magnetic properties of polycrystalline nickel
  thin films.
\newblock {\em Acta Physica Polonica A}, 127(4):995--997, 2015.

\bibitem{den1991magnetic5}
FJA Den~Broeder, W~Hoving, and PJH Bloemen.
\newblock Magnetic anisotropy of multilayers.
\newblock {\em Journal of magnetism and magnetic materials}, 93:562--570, 1991.

\bibitem{yoshimura2016soliton}
Yoko Yoshimura, Kab-Jin Kim, Takuya Taniguchi, Takayuki Tono, Kohei Ueda, Ryo
  Hiramatsu, Takahiro Moriyama, Keisuke Yamada, Yoshinobu Nakatani, and Teruo
  Ono.
\newblock Soliton-like magnetic domain wall motion induced by the interfacial
  dzyaloshinskii--moriya interaction.
\newblock {\em Nature Physics}, 12(2):157--161, 2016.

\bibitem{mishra2020cross}
Aijaz~H Lone, Shivangi Shringi, Kishan Mishra, and Srikant Srinivasan.
\newblock Cross-sectional area dependence of tunnel magnetoresistance, thermal
  stability, and critical current density in mtj.
\newblock {\em IEEE Transactions on Magnetics}, 57(2):1--10, 2020.

\bibitem{bloch1932theorie93}
Felix Bloch and Felix Bloch.
\newblock {\em Zur theorie des austauschproblems und der remanenzerscheinung
  der ferromagnetika}.
\newblock Springer, 1932.

\bibitem{103}
Louis N{\'e}el.
\newblock L'anisotropie superficielle des substances ferromagn{\'e}tiques.
\newblock {\em Comptes Rendus Hebdomadaires Des Seances De L Academie Des
  Sciences}, 237(23):1468--1470, 1953.

\bibitem{kravchuk2014influence}
Volodymyr~P Kravchuk.
\newblock Influence of dzialoshinskii--moriya interaction on static and dynamic
  properties of a transverse domain wall.
\newblock {\em Journal of magnetism and magnetic materials}, 367:9--14, 2014.

\bibitem{landeros2009equilibrium}
P~Landeros, OJ~Suarez, A~Cuchillo, and P~Vargas.
\newblock Equilibrium states and vortex domain wall nucleation in ferromagnetic
  nanotubes.
\newblock {\em Physical Review B}, 79(2):024404, 2009.

\bibitem{ryu2013chiral}
Kwang-Su Ryu, Luc Thomas, See-Hun Yang, and Stuart Parkin.
\newblock Chiral spin torque at magnetic domain walls.
\newblock {\em Nature nanotechnology}, 8(7):527--533, 2013.

\bibitem{craik1995magnetism}
D~Craik.
\newblock Magnetism: Principles and applications, john wiley and sons.
\newblock {\em Inc., New York}, 1995.

\bibitem{heide2008dzyaloshinskii95}
M~Heide, G~Bihlmayer, and Stefan Bl{\"u}gel.
\newblock Dzyaloshinskii-moriya interaction accounting for the orientation of
  magnetic domains in ultrathin films: Fe/w (110).
\newblock {\em Physical Review B}, 78(14):140403, 2008.

\bibitem{thiaville2012dynamics96}
Andr{\'e} Thiaville, Stanislas Rohart, {\'E}milie Ju{\'e}, Vincent Cros, and
  Albert Fert.
\newblock Dynamics of dzyaloshinskii domain walls in ultrathin magnetic films.
\newblock {\em Europhysics Letters}, 100(5):57002, 2012.

\bibitem{bogdanov1999stability}
A~Bogdanov and A~Hubert.
\newblock The stability of vortex-like structures in uniaxial ferromagnets.
\newblock {\em Journal of magnetism and magnetic materials}, 195(1):182--192,
  1999.

\bibitem{wilson2014chiral217}
Murray~Neff Wilson, AB~Butenko, AN~Bogdanov, and TL~Monchesky.
\newblock Chiral skyrmions in cubic helimagnet films: The role of uniaxial
  anisotropy.
\newblock {\em Physical Review B}, 89(9):094411, 2014.

\bibitem{keesman2015degeneracies219}
Rick Keesman, AO~Leonov, P~van Dieten, Stefan Buhrandt, GT~Barkema, Lars Fritz,
  and RA~Duine.
\newblock Degeneracies and fluctuations of n{\'e}el skyrmions in confined
  geometries.
\newblock {\em Physical Review B}, 92(13):134405, 2015.

\bibitem{rybakov2013three235}
FN~Rybakov, AB~Borisov, and AN~Bogdanov.
\newblock Three-dimensional skyrmion states in thin films of cubic helimagnets.
\newblock {\em Physical Review B}, 87(9):094424, 2013.

\bibitem{rybakov2015new236}
Filipp~N Rybakov, Aleksandr~B Borisov, Stefan Bl{\"u}gel, and Nikolai~S
  Kiselev.
\newblock New type of stable particlelike states in chiral magnets.
\newblock {\em Physical review letters}, 115(11):117201, 2015.

\bibitem{bub1}
Stavros Komineas and Nikos Papanicolaou.
\newblock Skyrmion dynamics in chiral ferromagnets.
\newblock {\em Physical Review B}, 92(6):064412, 2015.

\bibitem{butenko2010stabilization242}
AB~Butenko, AA~Leonov, UK~R{\"o}{\ss}ler, and AN~Bogdanov.
\newblock Stabilization of skyrmion textures by uniaxial distortions in
  noncentrosymmetric cubic helimagnets.
\newblock {\em Physical Review B}, 82(5):052403, 2010.

\bibitem{lin2013particle243}
Shi-Zeng Lin, Charles Reichhardt, Cristian~D Batista, and Avadh Saxena.
\newblock Particle model for skyrmions in metallic chiral magnets: Dynamics,
  pinning, and creep.
\newblock {\em Physical Review B}, 87(21):214419, 2013.

\bibitem{kim2014breathing245}
Joo-Von Kim, Felipe Garcia-Sanchez, Joao Sampaio, Constance Moreau-Luchaire,
  Vincent Cros, and Albert Fert.
\newblock Breathing modes of confined skyrmions in ultrathin magnetic dots.
\newblock {\em Physical Review B}, 90(6):064410, 2014.

\bibitem{zhang2015skyrmion250}
Xichao Zhang, GP~Zhao, Hans Fangohr, J~Ping Liu, WX~Xia, J~Xia, and FJ~Morvan.
\newblock Skyrmion-skyrmion and skyrmion-edge repulsions in skyrmion-based
  racetrack memory.
\newblock {\em Scientific reports}, 5(1):7643, 2015.

\bibitem{hubert2008magnetic}
Alex Hubert and Rudolf Sch{\"a}fer.
\newblock {\em Magnetic domains: the analysis of magnetic microstructures}.
\newblock Springer Science \& Business Media, 2008.

\bibitem{leonov2016properties}
AO~Leonov, TL~Monchesky, N~Romming, A~Kubetzka, AN~Bogdanov, and
  R~Wiesendanger.
\newblock The properties of isolated chiral skyrmions in thin magnetic films.
\newblock {\em New Journal of Physics}, 18(6):065003, 2016.

\bibitem{braun1994fluctuations}
Hans-Benjamin Braun.
\newblock Fluctuations and instabilities of ferromagnetic domain-wall pairs in
  an external magnetic field.
\newblock {\em Physical Review B}, 50(22):16485, 1994.

\bibitem{romming2015field232}
Niklas Romming, Andr{\'e} Kubetzka, Christian Hanneken, Kirsten von Bergmann,
  and Roland Wiesendanger.
\newblock Field-dependent size and shape of single magnetic skyrmions.
\newblock {\em Physical review letters}, 114(17):177203, 2015.

\bibitem{castro2016skyrmion}
MA~Castro and S~Allende.
\newblock Skyrmion core size dependence as a function of the perpendicular
  anisotropy and radius in magnetic nanodots.
\newblock {\em Journal of Magnetism and Magnetic Materials}, 417:344--348,
  2016.

\bibitem{vidal2017stability}
Nicol{\'a}s Vidal-Silva, Alejandro Riveros, and Juan Escrig.
\newblock Stability of neel skyrmions in ultra-thin nanodots considering
  dzyaloshinskii-moriya and dipolar interactions.
\newblock {\em Journal of Magnetism and Magnetic Materials}, 443:116--123,
  2017.

\bibitem{wiesendanger2016nanoscale}
Roland Wiesendanger.
\newblock Nanoscale magnetic skyrmions in metallic films and multilayers: a new
  twist for spintronics.
\newblock {\em Nature Reviews Materials}, 1(7):1--11, 2016.

\bibitem{jiang2017skyrmions}
Wanjun Jiang, Gong Chen, Kai Liu, Jiadong Zang, Suzanne~GE Te~Velthuis, and
  Axel Hoffmann.
\newblock Skyrmions in magnetic multilayers.
\newblock {\em Physics Reports}, 704:1--49, 2017.

\bibitem{kang2016skyrmion}
Wang Kang, Yangqi Huang, Xichao Zhang, Yan Zhou, and Weisheng Zhao.
\newblock Skyrmion-electronics: An overview and outlook.
\newblock {\em Proceedings of the IEEE}, 104(10):2040--2061, 2016.

\bibitem{fert2017magnetic}
Albert Fert, Nicolas Reyren, and Vincent Cros.
\newblock Magnetic skyrmions: advances in physics and potential applications.
\newblock {\em Nature Reviews Materials}, 2(7):1--15, 2017.

\bibitem{hsu2017electric}
Pin-Jui Hsu, Andr{\'e} Kubetzka, Aurore Finco, Niklas Romming, Kirsten
  Von~Bergmann, and Roland Wiesendanger.
\newblock Electric-field-driven switching of individual magnetic skyrmions.
\newblock {\em Nature nanotechnology}, 12(2):123--126, 2017.

\bibitem{zhou2014reversible}
Yan Zhou and Motohiko Ezawa.
\newblock A reversible conversion between a skyrmion and a domain-wall pair in
  a junction geometry.
\newblock {\em Nature communications}, 5(1):4652, 2014.

\bibitem{woo2018deterministic}
Seonghoon Woo, Kyung~Mee Song, Xichao Zhang, Motohiko Ezawa, Yan Zhou, Xiaoxi
  Liu, Markus Weigand, Simone Finizio, J{\"o}rg Raabe, Min-Chul Park, et~al.
\newblock Deterministic creation and deletion of a single magnetic skyrmion
  observed by direct time-resolved x-ray microscopy.
\newblock {\em Nature Electronics}, 1(5):288--296, 2018.

\bibitem{milde2013unwinding59k}
Peter Milde, Denny K{\"o}hler, Joachim Seidel, LM~Eng, Andreas Bauer, Alfonso
  Chacon, Jonas Kindervater, Sebastian m{\"u}hlbauer, Christian Pfleiderer,
  Stefan Buhrandt, et~al.
\newblock Unwinding of a skyrmion lattice by magnetic monopoles.
\newblock {\em Science}, 340(6136):1076--1080, 2013.

\bibitem{huang2012extended226}
SX~Huang and CL~Chien.
\newblock Extended skyrmion phase in epitaxial fege (111) thin films.
\newblock {\em Physical review letters}, 108(26):267201, 2012.

\bibitem{mochizuki2015writing94c}
Masahito Mochizuki and Yoshio Watanabe.
\newblock Writing a skyrmion on multiferroic materials.
\newblock {\em Applied Physics Letters}, 107(8):082409, 2015.

\bibitem{tomasello2014strategy108c}
Riccardo Tomasello, E~Martinez, Roberto Zivieri, Luis Torres, Mario
  Carpentieri, and Giovanni Finocchio.
\newblock A strategy for the design of skyrmion racetrack memories.
\newblock {\em Scientific reports}, 4(1):1--7, 2014.

\bibitem{muller2016edge0018}
Jan M{\"u}ller, Achim Rosch, and Markus Garst.
\newblock Edge instabilities and skyrmion creation in magnetic layers.
\newblock {\em New Journal of Physics}, 18(6):065006, 2016.

\bibitem{garanin2018writing0020}
Dmitry~A Garanin, Daniel Capic, Senfu Zhang, Xixiang Zhang, and Eugene~M
  Chudnovsky.
\newblock Writing skyrmions with a magnetic dipole.
\newblock {\em Journal of Applied Physics}, 124(11):113901, 2018.

\bibitem{zhang2018direct0021}
Senfu Zhang, Junwei Zhang, Qiang Zhang, Craig Barton, Volker Neu, Yuelei Zhao,
  Zhipeng Hou, Yan Wen, Chen Gong, Olga Kazakova, et~al.
\newblock Direct writing of room temperature and zero field skyrmion lattices
  by a scanning local magnetic field.
\newblock {\em Applied Physics Letters}, 112(13):132405, 2018.

\bibitem{woo2017spin0019}
Seonghoon Woo, Kyung~Mee Song, Hee-Sung Han, Min-Seung Jung, Mi-Young Im,
  Ki-Suk Lee, Kun~Soo Song, Peter Fischer, Jung-Il Hong, Jun~Woo Choi, et~al.
\newblock Spin-orbit torque-driven skyrmion dynamics revealed by time-resolved
  x-ray microscopy.
\newblock {\em Nature communications}, 8(1):15573, 2017.

\bibitem{ralph2008spin}
Daniel~C Ralph and Mark~D Stiles.
\newblock Spin transfer torques.
\newblock {\em Journal of Magnetism and Magnetic Materials}, 320(7):1190--1216,
  2008.

\bibitem{sinova2015spin}
Jairo Sinova, Sergio~O Valenzuela, J{\"o}rg Wunderlich, CH~Back, and
  T~Jungwirth.
\newblock Spin hall effects.
\newblock {\em Reviews of modern physics}, 87(4):1213, 2015.

\bibitem{everschor2011current}
Karin Everschor, Markus Garst, RA~Duine, and Achim Rosch.
\newblock Current-induced rotational torques in the skyrmion lattice phase of
  chiral magnets.
\newblock {\em Physical Review B}, 84(6):064401, 2011.

\bibitem{tchoe2012skyrmion}
Youngbin Tchoe and Jung~Hoon Han.
\newblock Skyrmion generation by current.
\newblock {\em Physical Review B}, 85(17):174416, 2012.

\bibitem{koshibae2014creation}
Wataru Koshibae and Naoto Nagaosa.
\newblock Creation of skyrmions and antiskyrmions by local heating.
\newblock {\em Nature communications}, 5(1):5148, 2014.

\bibitem{heo2016switching}
Changhoon Heo, Nikolai~S Kiselev, Ashis~Kumar Nandy, Stefan Bl{\"u}gel, and
  Theo Rasing.
\newblock Switching of chiral magnetic skyrmions by picosecond magnetic field
  pulses via transient topological states.
\newblock {\em Scientific reports}, 6(1):1--11, 2016.

\bibitem{wang2022electrical}
Weiwei Wang, Dongsheng Song, Wensen Wei, Pengfei Nan, Shilei Zhang, Binghui Ge,
  Mingliang Tian, Jiadong Zang, and Haifeng Du.
\newblock Electrical manipulation of skyrmions in a chiral magnet.
\newblock {\em Nature Communications}, 13(1):1593, 2022.

\bibitem{lin2016edge}
Shi-Zeng Lin.
\newblock Edge instability in a chiral stripe domain under an electric current
  and skyrmion generation.
\newblock {\em Physical Review B}, 94(2):020402, 2016.

\bibitem{liu2016topological}
Yan Liu, Huan Yan, Min Jia, HaiFeng Du, and An~Du.
\newblock Topological analysis of spin-torque driven magnetic skyrmion
  formation.
\newblock {\em Applied Physics Letters}, 109(10):102402, 2016.

\bibitem{koshibae2017theory103c}
Wataru Koshibae and Naoto Nagaosa.
\newblock Theory of skyrmions in bilayer systems.
\newblock {\em Scientific reports}, 7(1):42645, 2017.

\bibitem{haldane19883}
F~Duncan~M Haldane.
\newblock O (3) nonlinear $\sigma$ model and the topological distinction
  between integer-and half-integer-spin antiferromagnets in two dimensions.
\newblock {\em Physical review letters}, 61(8):1029, 1988.

\bibitem{wachowiak2002direct}
A~Wachowiak, J~Wiebe, M~Bode, O~Pietzsch, M~Morgenstern, and R~Wiesendanger.
\newblock Direct observation of internal spin structure of magnetic vortex
  cores.
\newblock {\em science}, 298(5593):577--580, 2002.

\bibitem{van2006magnetic}
Bartel Van~Waeyenberge, A~Puzic, H~Stoll, KW~Chou, T~Tyliszczak, R~Hertel,
  M~F{\"a}hnle, H~Br{\"u}ckl, Karsten Rott, G{\"u}nter Reiss, et~al.
\newblock Magnetic vortex core reversal by excitation with short bursts of an
  alternating field.
\newblock {\em Nature}, 444(7118):461--464, 2006.

\bibitem{zhang2015microwave}
Bin Zhang, Weiwei Wang, Marijan Beg, Hans Fangohr, and Wolfgang Kuch.
\newblock Microwave-induced dynamic switching of magnetic skyrmion cores in
  nanodots.
\newblock {\em Applied Physics Letters}, 106(10):102401, 2015.

\bibitem{li2014tailoring}
Jia Li, A~Tan, KW~Moon, A~Doran, MA~Marcus, AT~Young, E~Arenholz, S~Ma,
  RF~Yang, C~Hwang, et~al.
\newblock Tailoring the topology of an artificial magnetic skyrmion.
\newblock {\em Nature communications}, 5(1):4704, 2014.

\bibitem{fraerman2015skyrmion}
AA~Fraerman, OL~Ermolaeva, EV~Skorohodov, NS~Gusev, VL~Mironov, SN~Vdovichev,
  and ES~Demidov.
\newblock Skyrmion states in multilayer exchange coupled ferromagnetic
  nanostructures with distinct anisotropy directions.
\newblock {\em Journal of Magnetism and Magnetic Materials}, 393:452--456,
  2015.

\bibitem{bessarab2018lifetime}
Pavel~F Bessarab, Gideon~P M{\"u}ller, Igor~S Lobanov, Filipp~N Rybakov,
  Nikolai~S Kiselev, Hannes J{\'o}nsson, Valery~M Uzdin, Stefan Bl{\"u}gel,
  Lars Bergqvist, and Anna Delin.
\newblock Lifetime of racetrack skyrmions.
\newblock {\em Scientific reports}, 8(1):3433, 2018.

\bibitem{flovik2017generation}
Vegard Flovik, Alireza Qaiumzadeh, Ashis~K Nandy, Changhoon Heo, and Theo
  Rasing.
\newblock Generation of single skyrmions by picosecond magnetic field pulses.
\newblock {\em Physical Review B}, 96(14):140411, 2017.

\bibitem{schaffer2017ultrafast}
Alexander~F Sch{\"a}ffer, Hermann~A D{\"u}rr, and Jamal Berakdar.
\newblock Ultrafast imprinting of topologically protected magnetic textures via
  pulsed electrons.
\newblock {\em Applied Physics Letters}, 111(3):032403, 2017.

\bibitem{everschor2012rotating91c}
Karin Everschor, Markus Garst, Benedikt Binz, Florian Jonietz, Sebastian
  m{\"u}hlbauer, Christian Pfleiderer, and Achim Rosch.
\newblock Rotating skyrmion lattices by spin torques and field or temperature
  gradients.
\newblock {\em Physical Review B}, 86(5):054432, 2012.

\bibitem{yuan2016skyrmion}
HY~Yuan and XR~Wang.
\newblock Skyrmion creation and manipulation by nano-second current pulses.
\newblock {\em Scientific reports}, 6(1):1--8, 2016.

\bibitem{yin2016topological}
Gen Yin, Yufan Li, Lingyao Kong, Roger~K Lake, Chia-Ling Chien, and Jiadong
  Zang.
\newblock Topological charge analysis of ultrafast single skyrmion creation.
\newblock {\em Physical Review B}, 93(17):174403, 2016.

\bibitem{legrand2017room}
William Legrand, Davide Maccariello, Nicolas Reyren, Karin Garcia, Christoforos
  Moutafis, Constance Moreau-Luchaire, Sophie Collin, Karim Bouzehouane,
  Vincent Cros, and Albert Fert.
\newblock Room-temperature current-induced generation and motion of sub-100 nm
  skyrmions.
\newblock {\em Nano letters}, 17(4):2703--2712, 2017.

\bibitem{lemesh2018current}
Ivan Lemesh, Kai Litzius, Marie B{\"o}ttcher, Pedram Bassirian, Nico Kerber,
  Daniel Heinze, Jakub Z{\'a}zvorka, Felix B{\"u}ttner, Lucas Caretta, Maxwell
  Mann, et~al.
\newblock Current-induced skyrmion generation through morphological thermal
  transitions in chiral ferromagnetic heterostructures.
\newblock {\em Advanced materials}, 30(49):1805461, 2018.

\bibitem{hrabec2017current}
Ales Hrabec, Joao Sampaio, Mohamed Belmeguenai, Isabell Gross, Raphael Weil,
  Salim~Mourad Ch{\'e}rif, A~Stashkevich, Vincent Jacques, Andre Thiaville, and
  Stanislas Rohart.
\newblock Current-induced skyrmion generation and dynamics in symmetric
  bilayers.
\newblock {\em Nature communications}, 8(1):15765, 2017.

\bibitem{buttner2017field}
Felix B{\"u}ttner, Ivan Lemesh, Michael Schneider, Bastian Pfau, Christian~M
  G{\"u}nther, Piet Hessing, Jan Geilhufe, Lucas Caretta, Dieter Engel,
  Benjamin Kr{\"u}ger, et~al.
\newblock Field-free deterministic ultrafast creation of magnetic skyrmions by
  spin--orbit torques.
\newblock {\em Nature Nanotechnology}, 12(11):1040--1044, 2017.

\bibitem{finizio2019deterministic}
Simone Finizio, Katharina Zeissler, Sebastian Wintz, Sina Mayr, Teresa
  We{\ss}els, Alexandra~J Huxtable, Gavin Burnell, Christopher~H Marrows, and
  Jorg Raabe.
\newblock Deterministic field-free skyrmion nucleation at a nanoengineered
  injector device.
\newblock {\em Nano letters}, 19(10):7246--7255, 2019.

\bibitem{soumyanarayanan2017tunable49k}
Anjan Soumyanarayanan, M~Raju, AL~Gonzalez~Oyarce, Anthony~KC Tan, Mi-Young Im,
  Alexander~Paul Petrovi{\'c}, Pin Ho, KH~Khoo, M~Tran, CK~Gan, et~al.
\newblock Tunable room-temperature magnetic skyrmions in ir/fe/co/pt
  multilayers.
\newblock {\em Nature materials}, 16(9):898--904, 2017.

\bibitem{nahas2015discovery}
Y~Nahas, S~Prokhorenko, L~Louis, Z~Gui, Igor Kornev, and Laurent Bellaiche.
\newblock Discovery of stable skyrmionic state in ferroelectric nanocomposites.
\newblock {\em Nature communications}, 6(1):8542, 2015.

\bibitem{hamamoto2016purely}
Keita Hamamoto, Motohiko Ezawa, and Naoto Nagaosa.
\newblock Purely electrical detection of a skyrmion in constricted geometry.
\newblock {\em Applied Physics Letters}, 108(11):112401, 2016.

\bibitem{maccariello2018electrical}
Davide Maccariello, William Legrand, Nicolas Reyren, Karin Garcia, Karim
  Bouzehouane, Sophie Collin, Vincent Cros, and Albert Fert.
\newblock Electrical detection of single magnetic skyrmions in metallic
  multilayers at room temperature.
\newblock {\em Nature nanotechnology}, 13(3):233--237, 2018.

\bibitem{du2015electrical}
Haifeng Du, Dong Liang, Chiming Jin, Lingyao Kong, Matthew~J Stolt, Wei Ning,
  Jiyong Yang, Ying Xing, Jian Wang, Renchao Che, et~al.
\newblock Electrical probing of field-driven cascading quantized transitions of
  skyrmion cluster states in mnsi nanowires.
\newblock {\em Nature communications}, 6(1):7637, 2015.

\bibitem{kubetzka2017impact}
Andr{\'e} Kubetzka, Christian Hanneken, Roland Wiesendanger, and Kirsten
  Von~Bergmann.
\newblock Impact of the skyrmion spin texture on magnetoresistance.
\newblock {\em Physical Review B}, 95(10):104433, 2017.

\bibitem{buhl2017topological}
Patrick~M Buhl, Frank Freimuth, Stefan Bl{\"u}gel, and Yuriy Mokrousov.
\newblock Topological spin hall effect in antiferromagnetic skyrmions.
\newblock {\em physica status solidi (RRL)--Rapid Research Letters},
  11(4):1700007, 2017.

\bibitem{loreto2018creation}
Renan~Pires Loreto, WA~Moura-Melo, AR~Pereira, X~Zhang, Y~Zhou, M~Ezawa, and
  CIL de~Araujo.
\newblock Creation, transport and detection of imprinted magnetic solitons
  stabilized by spin-polarized current.
\newblock {\em Journal of Magnetism and Magnetic Materials}, 455:25--31, 2018.

\bibitem{penthorn2019experimental}
NE~Penthorn, X~Hao, Z~Wang, Y~Huai, and HW~Jiang.
\newblock Experimental observation of single skyrmion signatures in a magnetic
  tunnel junction.
\newblock {\em Physical Review Letters}, 122(25):257201, 2019.

\bibitem{kasai2019voltage}
Shinya Kasai, Satoshi Sugimoto, Yoshinobu Nakatani, Ryo Ishikawa, and Yukiko~K
  Takahashi.
\newblock Voltage-controlled magnetic skyrmions in magnetic tunnel junctions.
\newblock {\em Applied Physics Express}, 12(8):083001, 2019.

\bibitem{lee2009unusual}
Minhyea Lee, W~Kang, Y~Onose, Y~Tokura, and Nai~Phuan Ong.
\newblock Unusual hall effect anomaly in mnsi under pressure.
\newblock {\em Physical review letters}, 102(18):186601, 2009.

\bibitem{zeissler2018discrete}
Katharina Zeissler, Simone Finizio, Kowsar Shahbazi, Jamie Massey, Fatma~Al
  Ma’Mari, David~M Bracher, Armin Kleibert, Mark~C Rosamond, Edmund~H
  Linfield, Thomas~A Moore, et~al.
\newblock Discrete hall resistivity contribution from n{\'e}el skyrmions in
  multilayer nanodiscs.
\newblock {\em Nature Nanotechnology}, 13(12):1161--1166, 2018.

\bibitem{hamamoto2015quantized}
Keita Hamamoto, Motohiko Ezawa, and Naoto Nagaosa.
\newblock Quantized topological hall effect in skyrmion crystal.
\newblock {\em Physical Review B}, 92(11):115417, 2015.

\bibitem{yu2021magnetic}
Xiuzhen Yu.
\newblock Magnetic imaging of various topological spin textures and their
  dynamics.
\newblock {\em Journal of Magnetism and Magnetic Materials}, 539:168332, 2021.

\bibitem{zhang2017direct54k}
SL~Zhang, G~Van Der~Laan, and T~Hesjedal.
\newblock Direct experimental determination of the topological winding number
  of skyrmions in cu2oseo3.
\newblock {\em Nature Communications}, 8(1):14619, 2017.

\bibitem{zhang2017direct55k}
SL~Zhang, G~van~der Laan, and T~Hesjedal.
\newblock Direct experimental determination of spiral spin structures via the
  dichroism extinction effect in resonant elastic soft x-ray scattering.
\newblock {\em Physical Review B}, 96(9):094401, 2017.

\bibitem{montoya2017tailoring56k}
SA~Montoya, S~Couture, JJ~Chess, JCT Lee, N~Kent, D~Henze, SK~Sinha, M-Y Im,
  SD~Kevan, P~Fischer, et~al.
\newblock Tailoring magnetic energies to form dipole skyrmions and skyrmion
  lattices.
\newblock {\em Physical Review B}, 95(2):024415, 2017.

\bibitem{mcvitie2018transmission57k}
Stephen McVitie, Sean Hughes, Kayla Fallon, Sam McFadzean, Damien McGrouther,
  Matus Krajnak, W~Legrand, D~Maccariello, S~Collin, K~Garcia, et~al.
\newblock A transmission electron microscope study of n{\'e}el skyrmion
  magnetic textures in multilayer thin film systems with large interfacial
  chiral interaction.
\newblock {\em Scientific Reports}, 8(1):5703, 2018.

\bibitem{okubo2012multiple26k}
Tsuyoshi Okubo, Sungki Chung, and Hikaru Kawamura.
\newblock Multiple-q states and the skyrmion lattice of the triangular-lattice
  heisenberg antiferromagnet under magnetic fields.
\newblock {\em Physical review letters}, 108(1):017206, 2012.

\bibitem{landau1980lifshitz11}
LD~Landau.
\newblock Em lifshitz statistical physics-course of theoretical physics.
\newblock {\em Pergamon Press,}, 19:80, 1980.

\bibitem{binasch1989enhanced}
Gr{\"u}nberg Binasch, Peter Gr{\"u}nberg, F~Saurenbach, and W~Zinn.
\newblock Enhanced magnetoresistance in layered magnetic structures with
  antiferromagnetic interlayer exchange.
\newblock {\em Physical review B}, 39(7):4828, 1989.

\bibitem{baibich1988giant}
Mario~Norberto Baibich, Jean~Marc Broto, Albert Fert, F~Nguyen Van~Dau,
  Fr{\'e}d{\'e}ric Petroff, P~Etienne, G~Creuzet, A~Friederich, and J~Chazelas.
\newblock Giant magnetoresistance of (001) fe/(001) cr magnetic superlattices.
\newblock {\em Physical review letters}, 61(21):2472, 1988.

\bibitem{berger1996emission}
Luc Berger.
\newblock Emission of spin waves by a magnetic multilayer traversed by a
  current.
\newblock {\em Physical Review B}, 54(13):9353, 1996.

\bibitem{slonczewski1996current}
John~C Slonczewski.
\newblock Current-driven excitation of magnetic multilayers.
\newblock {\em Journal of Magnetism and Magnetic Materials}, 159(1-2):L1--L7,
  1996.

\bibitem{brataas2012current}
Arne Brataas, Andrew~D Kent, and Hideo Ohno.
\newblock Current-induced torques in magnetic materials.
\newblock {\em Nature materials}, 11(5):372--381, 2012.

\bibitem{abert2019micromagnetics}
Claas Abert.
\newblock Micromagnetics and spintronics: models and numerical methods.
\newblock {\em The European Physical Journal B}, 92:1--45, 2019.

\bibitem{tsoi1998excitation}
M~Tsoi, AGM Jansen, J~Bass, W-C Chiang, M~Seck, V~Tsoi, and P~Wyder.
\newblock Excitation of a magnetic multilayer by an electric current.
\newblock {\em Physical Review Letters}, 80(19):4281, 1998.

\bibitem{berkov2008spin}
Dmitry~V Berkov and Jacques Miltat.
\newblock Spin-torque driven magnetization dynamics: Micromagnetic modeling.
\newblock {\em Journal of Magnetism and Magnetic Materials}, 320(7):1238--1259,
  2008.

\bibitem{zhang2004roles}
Sh~Zhang and Z~Li.
\newblock Roles of nonequilibrium conduction electrons on the magnetization
  dynamics of ferromagnets.
\newblock {\em Physical review letters}, 93(12):127204, 2004.

\bibitem{barraud2010unravelling}
Cl{\'e}ment Barraud, Pierre Seneor, Richard Mattana, St{\'e}phane Fusil, Karim
  Bouzehouane, Cyrile Deranlot, Patrizio Graziosi, Luis Hueso, Ilaria Bergenti,
  Valentin Dediu, et~al.
\newblock Unravelling the role of the interface for spin injection into organic
  semiconductors.
\newblock {\em Nature Physics}, 6(8):615--620, 2010.

\bibitem{wolf2001spintronics}
SA~Wolf, DD~Awschalom, RA~Buhrman, JM~Daughton, von~S von Moln{\'a}r,
  ML~Roukes, A~Yu Chtchelkanova, and DM~Treger.
\newblock Spintronics: a spin-based electronics vision for the future.
\newblock {\em science}, 294(5546):1488--1495, 2001.

\bibitem{liu2012spin1}
Luqiao Liu, Chi-Feng Pai, Y~Li, HW~Tseng, DC~Ralph, and RA~Buhrman.
\newblock Spin-torque switching with the giant spin hall effect of tantalum.
\newblock {\em Science}, 336(6081):555--558, 2012.

\bibitem{fukami2016spin}
Shunsuke Fukami, T~Anekawa, C~Zhang, and H~Ohno.
\newblock A spin--orbit torque switching scheme with collinear magnetic easy
  axis and current configuration.
\newblock {\em nature nanotechnology}, 11(7):621--625, 2016.

\bibitem{mihai2010current}
Ioan Mihai~Miron, Gilles Gaudin, St{\'e}phane Auffret, Bernard Rodmacq, Alain
  Schuhl, Stefania Pizzini, Jan Vogel, and Pietro Gambardella.
\newblock Current-driven spin torque induced by the rashba effect in a
  ferromagnetic metal layer.
\newblock {\em Nature materials}, 9(3):230--234, 2010.

\bibitem{manchon2015new}
Aurelien Manchon, Hyun~Cheol Koo, Junsaku Nitta, Sergey~M Frolov, and Rembert~A
  Duine.
\newblock New perspectives for rashba spin--orbit coupling.
\newblock {\em Nature materials}, 14(9):871--882, 2015.

\bibitem{gambardella2011current}
Pietro Gambardella and Ioan~Mihai Miron.
\newblock Current-induced spin--orbit torques.
\newblock {\em Philosophical Transactions of the Royal Society A: Mathematical,
  Physical and Engineering Sciences}, 369(1948):3175--3197, 2011.

\bibitem{brataas2014spin}
Arne Brataas and Kjetil~MD Hals.
\newblock Spin--orbit torques in action.
\newblock {\em Nature nanotechnology}, 9(2):86--88, 2014.

\bibitem{kang2016voltage}
Wang Kang, Yangqi Huang, Chentian Zheng, Weifeng Lv, Na~Lei, Youguang Zhang,
  Xichao Zhang, Yan Zhou, and Weisheng Zhao.
\newblock Voltage controlled magnetic skyrmion motion for racetrack memory.
\newblock {\em Scientific reports}, 6(1):23164, 2016.

\bibitem{thiele1973steady}
AA~Thiele.
\newblock Steady-state motion of magnetic domains.
\newblock {\em Physical Review Letters}, 30(6):230, 1973.

\bibitem{amin2016spin1}
Vivek~P Amin and Mark~D Stiles.
\newblock Spin transport at interfaces with spin-orbit coupling: Formalism.
\newblock {\em Physical Review B}, 94(10):104419, 2016.

\bibitem{amin2016spin}
VP~Amin and MD~Stiles.
\newblock Spin transport at interfaces with spin-orbit coupling: Phenomenology.
\newblock {\em Physical Review B}, 94(10):104420, 2016.

\bibitem{amin2018interface}
Vivek~P Amin, Jan Zemen, and Mark~D Stiles.
\newblock Interface-generated spin currents.
\newblock {\em Physical review letters}, 121(13):136805, 2018.

\bibitem{rojas2014spin}
J-C Rojas-S{\'a}nchez, N~Reyren, P~Laczkowski, W~Savero, J-P Attan{\'e},
  C~Deranlot, M~Jamet, J-M George, Laurent Vila, and H~Jaffr{\`e}s.
\newblock Spin pumping and inverse spin hall effect in platinum: the essential
  role of spin-memory loss at metallic interfaces.
\newblock {\em Physical review letters}, 112(10):106602, 2014.

\bibitem{pai2015dependence}
Chi-Feng Pai, Yongxi Ou, Luis~Henrique Vilela-Le{\~a}o, DC~Ralph, and
  RA~Buhrman.
\newblock Dependence of the efficiency of spin hall torque on the transparency
  of pt/ferromagnetic layer interfaces.
\newblock {\em Physical Review B}, 92(6):064426, 2015.

\bibitem{liu2012current}
Luqiao Liu, OJ~Lee, TJ~Gudmundsen, DC~Ralph, and RA~Buhrman.
\newblock Current-induced switching of perpendicularly magnetized magnetic
  layers using spin torque from the spin hall effect.
\newblock {\em Physical review letters}, 109(9):096602, 2012.

\bibitem{garello2013symmetry}
Kevin Garello, Ioan~Mihai Miron, Can~Onur Avci, Frank Freimuth, Yuriy
  Mokrousov, Stefan Bl{\"u}gel, St{\'e}phane Auffret, Olivier Boulle, Gilles
  Gaudin, and Pietro Gambardella.
\newblock Symmetry and magnitude of spin--orbit torques in ferromagnetic
  heterostructures.
\newblock {\em Nature nanotechnology}, 8(8):587--593, 2013.

\bibitem{pai2012spin1}
Chi-Feng Pai, Luqiao Liu, Y~Li, HW~Tseng, DC~Ralph, and RA~Buhrman.
\newblock Spin transfer torque devices utilizing the giant spin hall effect of
  tungsten.
\newblock {\em Applied Physics Letters}, 101(12):122404, 2012.

\bibitem{ramaswamy2016hf}
Rajagopalan Ramaswamy, Xuepeng Qiu, Tanmay Dutta, Shawn~David Pollard, and
  Hyunsoo Yang.
\newblock Hf thickness dependence of spin-orbit torques in hf/cofeb/mgo
  heterostructures.
\newblock {\em Applied Physics Letters}, 108(20):202406, 2016.

\bibitem{legrand2020room}
William Legrand, Davide Maccariello, Fernando Ajejas, Sophie Collin, Aymeric
  Vecchiola, Karim Bouzehouane, Nicolas Reyren, Vincent Cros, and Albert Fert.
\newblock Room-temperature stabilization of antiferromagnetic skyrmions in
  synthetic antiferromagnets.
\newblock {\em Nature materials}, 19(1):34--42, 2020.

\bibitem{iwasaki2014colossal}
Junichi Iwasaki, Wataru Koshibae, and Naoto Nagaosa.
\newblock Colossal spin transfer torque effect on skyrmion along the edge.
\newblock {\em Nano letters}, 14(8):4432--4437, 2014.

\bibitem{yoo2017current}
Myoung-Woo Yoo, Vincent Cros, and Joo-Von Kim.
\newblock Current-driven skyrmion expulsion from magnetic nanostrips.
\newblock {\em Physical Review B}, 95(18):184423, 2017.

\bibitem{schulz2012emergent}
Tomek Schulz, R~Ritz, Andreas Bauer, Madhumita Halder, Martin Wagner, Chris
  Franz, Christian Pfleiderer, Karin Everschor, Markus Garst, and Achim Rosch.
\newblock Emergent electrodynamics of skyrmions in a chiral magnet.
\newblock {\em Nature Physics}, 8(4):301--304, 2012.

\bibitem{bode2007chiral}
Matthias Bode, M~Heide, K~Von~Bergmann, P~Ferriani, Stefan Heinze, G~Bihlmayer,
  A~Kubetzka, O~Pietzsch, Stefan Bl{\"u}gel, and R~Wiesendanger.
\newblock Chiral magnetic order at surfaces driven by inversion asymmetry.
\newblock {\em Nature}, 447(7141):190--193, 2007.

\bibitem{ferriani2008atomic}
P~Ferriani, K~Von~Bergmann, EY~Vedmedenko, S~Heinze, M~Bode, M~Heide,
  G~Bihlmayer, S~Bl{\"u}gel, and R~Wiesendanger.
\newblock Atomic-scale spin spiral with a unique rotational sense: Mn monolayer
  on w (001).
\newblock {\em Physical review letters}, 101(2):027201, 2008.

\bibitem{simon2018magnetism}
E~Simon, L~R{\'o}zsa, K~Palot{\'a}s, and L~Szunyogh.
\newblock Magnetism of a co monolayer on pt (111) capped by overlayers of 5 d
  elements: A spin-model study.
\newblock {\em Physical Review B}, 97(13):134405, 2018.

\bibitem{tao2018self}
Xinde Tao, Qi~Liu, Bingfeng Miao, Rui Yu, Zheng Feng, Liang Sun, Biao You, Jun
  Du, Kai Chen, Shufeng Zhang, et~al.
\newblock Self-consistent determination of spin hall angle and spin diffusion
  length in pt and pd: The role of the interface spin loss.
\newblock {\em Science advances}, 4(6):eaat1670, 2018.

\bibitem{bloch1929quantenmechanik}
Felix Bloch.
\newblock {\"U}ber die quantenmechanik der elektronen in kristallgittern.
\newblock {\em Zeitschrift f{\"u}r physik}, 52(7-8):555--600, 1929.

\bibitem{peierls1929theorie}
Rudolf Peierls.
\newblock Zur theorie der galvanomagnetischen effekte.
\newblock {\em Zeitschrift f{\"u}r Physik}, 53:255--266, 1929.

\bibitem{jones1934general}
Harry Jones and Clarence Zener.
\newblock The general proof of certain fundamental equations in the theory of
  metallic conduction.
\newblock {\em Proceedings of the Royal Society of London. Series A, Containing
  Papers of a Mathematical and Physical Character}, 144(851):101--117, 1934.

\bibitem{sundaram1999wave}
Ganesh Sundaram and Qian Niu.
\newblock Wave-packet dynamics in slowly perturbed crystals: Gradient
  corrections and berry-phase effects.
\newblock {\em Physical Review B}, 59(23):14915, 1999.

\bibitem{suh2023semiclassical}
Jeonghyeon Suh, Sanghyun Park, and Hongki Min.
\newblock Semiclassical boltzmann magnetotransport theory in anisotropic
  systems with a nonvanishing berry curvature.
\newblock {\em New Journal of Physics}, 25(3):033021, 2023.

\bibitem{hatsugai1993chern}
Yasuhiro Hatsugai.
\newblock Chern number and edge states in the integer quantum hall effect.
\newblock {\em Physical review letters}, 71(22):3697, 1993.

\bibitem{hatsugai1993edge}
Yasuhiro Hatsugai.
\newblock Edge states in the integer quantum hall effect and the riemann
  surface of the bloch function.
\newblock {\em Physical Review B}, 48(16):11851, 1993.

\bibitem{akosa2019tuning}
Collins~Ashu Akosa, Hang Li, Gen Tatara, and Oleg~A Tretiakov.
\newblock Tuning the skyrmion hall effect via engineering of spin-orbit
  interaction.
\newblock {\em Physical Review Applied}, 12(5):054032, 2019.

\bibitem{bruno2004topological}
P~Bruno, VK~Dugaev, and M~Taillefumier.
\newblock Topological hall effect and berry phase in magnetic nanostructures.
\newblock {\em Physical review letters}, 93(9):096806, 2004.

\bibitem{li2013robust}
Yufan Li, N~Kanazawa, XZ~Yu, A~Tsukazaki, M~Kawasaki, M~Ichikawa, XF~Jin,
  F~Kagawa, and Y~Tokura.
\newblock Robust formation of skyrmions and topological hall effect anomaly in
  epitaxial thin films of mnsi.
\newblock {\em Physical review letters}, 110(11):117202, 2013.

\bibitem{gobel2017unconventional}
B{\"o}rge G{\"o}bel, Alexander Mook, J{\"u}rgen Henk, and Ingrid Mertig.
\newblock Unconventional topological hall effect in skyrmion crystals caused by
  the topology of the lattice.
\newblock {\em Physical Review B}, 95(9):094413, 2017.

\bibitem{hall1879new}
Edwin~H Hall et~al.
\newblock On a new action of the magnet on electric currents.
\newblock {\em American Journal of Mathematics}, 2(3):287--292, 1879.

\bibitem{nagaosa2010anomalous}
Naoto Nagaosa, Jairo Sinova, Shigeki Onoda, Allan~H MacDonald, and Nai~Phuan
  Ong.
\newblock Anomalous hall effect.
\newblock {\em Reviews of modern physics}, 82(2):1539, 2010.

\bibitem{nagaoka1966ferromagnetism}
Yosuke Nagaoka.
\newblock Ferromagnetism in a narrow, almost half-filled s band.
\newblock {\em Physical Review}, 147(1):392, 1966.

\bibitem{ritz2013formation}
R~Ritz, M~Halder, M~Wagner, C~Franz, A~Bauer, and C~Pfleiderer.
\newblock Formation of a topological non-fermi liquid in mnsi.
\newblock {\em Nature}, 497(7448):231--234, 2013.

\bibitem{taguchi2001spin}
Y~Taguchi, Y~Oohara, H~Yoshizawa, N~Nagaosa, and Y~Tokura.
\newblock Spin chirality, berry phase, and anomalous hall effect in a
  frustrated ferromagnet.
\newblock {\em Science}, 291(5513):2573--2576, 2001.

\bibitem{onoda2004anomalous}
Masaru Onoda, Gen Tatara, and Naoto Nagaosa.
\newblock Anomalous hall effect and skyrmion number in real and momentum
  spaces.
\newblock {\em Journal of the Physical Society of Japan}, 73(10):2624--2627,
  2004.

\bibitem{chen2014anomalous}
Hua Chen, Qian Niu, and Allan~H MacDonald.
\newblock Anomalous hall effect arising from noncollinear antiferromagnetism.
\newblock {\em Physical review letters}, 112(1):017205, 2014.

\bibitem{kubler2014non}
J{\"u}rgen K{\"u}bler and Claudia Felser.
\newblock Non-collinear antiferromagnets and the anomalous hall effect.
\newblock {\em Europhysics Letters}, 108(6):67001, 2014.

\bibitem{busch2020microscopic}
Oliver Busch, B{\"o}rge G{\"o}bel, and Ingrid Mertig.
\newblock Microscopic origin of the anomalous hall effect in noncollinear
  kagome magnets.
\newblock {\em Physical Review Research}, 2(3):033112, 2020.

\bibitem{nakatsuji2015large}
Satoru Nakatsuji, Naoki Kiyohara, and Tomoya Higo.
\newblock Large anomalous hall effect in a non-collinear antiferromagnet at
  room temperature.
\newblock {\em Nature}, 527(7577):212--215, 2015.

\bibitem{nayak2016large}
Ajaya~K Nayak, Julia~Erika Fischer, Yan Sun, Binghai Yan, Julie Karel,
  Alexander~C Komarek, Chandra Shekhar, Nitesh Kumar, Walter Schnelle,
  J{\"u}rgen K{\"u}bler, et~al.
\newblock Large anomalous hall effect driven by a nonvanishing berry curvature
  in the noncolinear antiferromagnet mn3ge.
\newblock {\em Science advances}, 2(4):e1501870, 2016.

\bibitem{vsmejkal2020crystal}
Libor {\v{S}}mejkal, Rafael Gonz{\'a}lez-Hern{\'a}ndez, Tom{\'a}{\v{s}}
  Jungwirth, and Jairo Sinova.
\newblock Crystal time-reversal symmetry breaking and spontaneous hall effect
  in collinear antiferromagnets.
\newblock {\em Science advances}, 6(23):eaaz8809, 2020.

\bibitem{feng2020observation}
Zexin Feng, Xiaorong Zhou, Libor {\v{S}}mejkal, Lei Wu, Zengwei Zhu, Huixin
  Guo, Rafael Gonz{\'a}lez-Hern{\'a}ndez, Xiaoning Wang, Han Yan, Peixin Qin,
  et~al.
\newblock Observation of the anomalous hall effect in a collinear
  antiferromagnet.
\newblock {\em arXiv preprint arXiv:2002.08712}, 2020.

\bibitem{wong2019enhanced}
QY~Wong, C~Murapaka, WC~Law, WL~Gan, GJ~Lim, and WS~Lew.
\newblock Enhanced spin-orbit torques in rare-earth pt/[co/ni] 2/co/tb systems.
\newblock {\em Physical Review Applied}, 11(2):024057, 2019.

\bibitem{zang2011dynamics83c}
Jiadong Zang, Maxim Mostovoy, Jung~Hoon Han, and Naoto Nagaosa.
\newblock Dynamics of skyrmion crystals in metallic thin films.
\newblock {\em Physical review letters}, 107(13):136804, 2011.

\bibitem{song2020skyrmion}
Kyung~Mee Song, Jae-Seung Jeong, Biao Pan, Xichao Zhang, Jing Xia, Sunkyung
  Cha, Tae-Eon Park, Kwangsu Kim, Simone Finizio, J{\"o}rg Raabe, et~al.
\newblock Skyrmion-based artificial synapses for neuromorphic computing.
\newblock {\em Nature Electronics}, 3(3):148--155, 2020.

\bibitem{chauwin2019skyrmion}
Maverick Chauwin, Xuan Hu, Felipe Garcia-Sanchez, Neilesh Betrabet, Alexandru
  Paler, Christoforos Moutafis, and Joseph~S Friedman.
\newblock Skyrmion logic system for large-scale reversible computation.
\newblock {\em Physical Review Applied}, 12(6):064053, 2019.

\bibitem{8429425}
Sai Li, Wang Kang, Xing Chen, Jinyu Bai, Biao Pan, Youguang Zhang, and Weisheng
  Zhao.
\newblock Emerging neuromorphic computing paradigms exploring magnetic
  skyrmions.
\newblock In {\em 2018 IEEE Computer Society Annual Symposium on VLSI
  (ISVLSI)}, pages 539--544, 2018.

\bibitem{srinivasan2016magnetic}
Gopalakrishnan Srinivasan, Abhronil Sengupta, and Kaushik Roy.
\newblock Magnetic tunnel junction based long-term short-term stochastic
  synapse for a spiking neural network with on-chip stdp learning.
\newblock {\em Scientific reports}, 6(1):29545, 2016.

\bibitem{deng2020voltage}
Jiefang Deng, Venkata Pavan~Kumar Miriyala, Zhifeng Zhu, Xuanyao Fong, and
  Gengchiau Liang.
\newblock Voltage-controlled spintronic stochastic neuron for restricted
  boltzmann machine with weight sparsity.
\newblock {\em IEEE Electron Device Letters}, 41(7):1102--1105, 2020.

\bibitem{lone2021voltage}
Aijaz~H Lone, Selma Amara, and Hossein Fariborzi.
\newblock Voltage-controlled domain wall motion-based neuron and stochastic
  magnetic tunnel junction synapse for neuromorphic computing applications.
\newblock {\em IEEE Journal on Exploratory Solid-State Computational Devices
  and Circuits}, 8(1):1--9, 2021.

\bibitem{he2017developing}
Zhezhi He and Deliang Fan.
\newblock Developing all-skyrmion spiking neural network.
\newblock {\em arXiv preprint arXiv:1705.02995}, 2017.

\bibitem{lone2022skyrmionneuron}
Aijaz~H Lone, Selma Amara, Fernando Aguirre, Mario Lanza, and Hossein
  Fariborzi.
\newblock Skyrmion-based leaky integrate and fire neurons for neuromorphic
  applications.
\newblock {\em arXiv preprint arXiv:2205.14913}, 2022.

\bibitem{huang2017magnetic}
Yangqi Huang, Wang Kang, Xichao Zhang, Yan Zhou, and Weisheng Zhao.
\newblock Magnetic skyrmion-based synaptic devices.
\newblock {\em Nanotechnology}, 28(8):08LT02, 2017.

\bibitem{lone2022skyrmion}
Aijaz~H Lone, Arnab Ganguly, Selma Amara, Gobind Das, and Hossein Fariborzi.
\newblock Skyrmion-magnetic tunnel junction synapse with mixed synaptic
  plasticity for neuromorphic computing.
\newblock {\em arXiv preprint arXiv:2205.14915}, 2022.

\bibitem{tan2019high}
Fu~Nan Tan, Wei~Liang Gan, Calvin Ching~Ian Ang, GDH Wong, HX~Liu, F~Poh, and
  WS~Lew.
\newblock High velocity domain wall propagation using voltage controlled
  magnetic anisotropy.
\newblock {\em Scientific Reports}, 9(1):7369, 2019.

\bibitem{wang2013voltage}
WG~Wang and CL~Chien.
\newblock Voltage-induced switching in magnetic tunnel junctions with
  perpendicular magnetic anisotropy.
\newblock {\em Journal of Physics D: Applied Physics}, 46(7):074004, 2013.

\bibitem{ong2016electric}
PV~Ong, Nicholas Kioussis, P~Khalili Amiri, and KL~Wang.
\newblock Electric-field-driven magnetization switching and nonlinear
  magnetoelasticity in au/feco/mgo heterostructures.
\newblock {\em Scientific reports}, 6(1):29815, 2016.

\bibitem{back20202020}
Christian Back, Vincent Cros, Hubert Ebert, Karin Everschor-Sitte, Albert Fert,
  Markus Garst, Tianping Ma, Sergiy Mankovsky, TL~Monchesky, Maxim Mostovoy,
  et~al.
\newblock The 2020 skyrmionics roadmap.
\newblock {\em Journal of Physics D: Applied Physics}, 53(36):363001, 2020.

\bibitem{krause2016skyrmionics}
Stefan Krause and Roland Wiesendanger.
\newblock Skyrmionics gets hot.
\newblock {\em Nature materials}, 15(5):493--494, 2016.

\bibitem{jiang2017direct}
Wanjun Jiang, Xichao Zhang, Guoqiang Yu, Wei Zhang, Xiao Wang,
  M~Benjamin~Jungfleisch, John~E Pearson, Xuemei Cheng, Olle Heinonen, Kang~L
  Wang, et~al.
\newblock Direct observation of the skyrmion hall effect.
\newblock {\em Nature Physics}, 13(2):162--169, 2017.

\bibitem{tang2021magnetic}
Jin Tang, Yaodong Wu, Weiwei Wang, Lingyao Kong, Boyao Lv, Wensen Wei, Jiadong
  Zang, Mingliang Tian, and Haifeng Du.
\newblock Magnetic skyrmion bundles and their current-driven dynamics.
\newblock {\em Nature Nanotechnology}, 16(10):1086--1091, 2021.

\bibitem{zheng2018experimental}
Fengshan Zheng, Filipp~N Rybakov, Aleksandr~B Borisov, Dongsheng Song, Shasha
  Wang, Zi-An Li, Haifeng Du, Nikolai~S Kiselev, Jan Caron, Andr{\'a}s
  Kov{\'a}cs, et~al.
\newblock Experimental observation of chiral magnetic bobbers in b20-type fege.
\newblock {\em Nature nanotechnology}, 13(6):451--455, 2018.

\bibitem{jena2020evolution}
Jagannath Jena, B{\"o}rge G{\"o}bel, Vivek Kumar, Ingrid Mertig, Claudia
  Felser, and Stuart Parkin.
\newblock Evolution and competition between chiral spin textures in nanostripes
  with d 2d symmetry.
\newblock {\em Science Advances}, 6(49):eabc0723, 2020.

\bibitem{zhang2020robust}
Shilei Zhang, David~M Burn, Nicolas Jaouen, Jean-Yves Chauleau, Amir~A
  Haghighirad, Yizhou Liu, Weiwei Wang, Gerrit van~der Laan, and Thorsten
  Hesjedal.
\newblock Robust perpendicular skyrmions and their surface confinement.
\newblock {\em Nano letters}, 20(2):1428--1432, 2020.

\bibitem{li2022experimental}
Sai Li, Ao~Du, Yadong Wang, Xinran Wang, Xueying Zhang, Houyi Cheng, Wenlong
  Cai, Shiyang Lu, Kaihua Cao, Biao Pan, et~al.
\newblock Experimental demonstration of skyrmionic magnetic tunnel junction at
  room temperature.
\newblock {\em Science Bulletin}, 67(7):691--699, 2022.

\bibitem{zhang2018skyrmions}
Xueying Zhang, Wenlong Cai, Xichao Zhang, Zilu Wang, Zhi Li, Yu~Zhang, Kaihua
  Cao, Na~Lei, Wang Kang, Yue Zhang, et~al.
\newblock Skyrmions in magnetic tunnel junctions.
\newblock {\em ACS applied materials \& interfaces}, 10(19):16887--16892, 2018.

\bibitem{zhang2016antiferromagnetic}
Xichao Zhang, Yan Zhou, and Motohiko Ezawa.
\newblock Antiferromagnetic skyrmion: stability, creation and manipulation.
\newblock {\em Scientific reports}, 6(1):24795, 2016.

\bibitem{gobel2017antiferromagnetic}
B{\"o}rge G{\"o}bel, Alexander Mook, J{\"u}rgen Henk, and Ingrid Mertig.
\newblock Antiferromagnetic skyrmion crystals: Generation, topological hall,
  and topological spin hall effect.
\newblock {\em Physical Review B}, 96(6):060406, 2017.

\bibitem{barker2016static}
Joseph Barker and Oleg~A Tretiakov.
\newblock Static and dynamical properties of antiferromagnetic skyrmions in the
  presence of applied current and temperature.
\newblock {\em Physical review letters}, 116(14):147203, 2016.

\bibitem{zhang2016magnetic}
Xichao Zhang, Yan Zhou, and Motohiko Ezawa.
\newblock Magnetic bilayer-skyrmions without skyrmion hall effect.
\newblock {\em Nature communications}, 7(1):10293, 2016.

\bibitem{zhang2016control}
Xichao Zhang, Jing Xia, Yan Zhou, Daowei Wang, Xiaoxi Liu, Weisheng Zhao, and
  Motohiko Ezawa.
\newblock Control and manipulation of a magnetic skyrmionium in nanostructures.
\newblock {\em Physical Review B}, 94(9):094420, 2016.

\bibitem{gobel2019overcoming}
B{\"o}rge G{\"o}bel, Alexander Mook, J{\"u}rgen Henk, and Ingrid Mertig.
\newblock Overcoming the speed limit in skyrmion racetrack devices by
  suppressing the skyrmion hall effect.
\newblock {\em Physical Review B}, 99(2):020405, 2019.

\bibitem{iwasaki2013universal86c}
Junichi Iwasaki, Masahito Mochizuki, and Naoto Nagaosa.
\newblock Universal current-velocity relation of skyrmion motion in chiral
  magnets.
\newblock {\em Nature communications}, 4(1):1463, 2013.

\bibitem{zhang2015magnetic2}
Xichao Zhang, Yan Zhou, Motohiko Ezawa, GP~Zhao, and Weisheng Zhao.
\newblock Magnetic skyrmion transistor: skyrmion motion in a voltage-gated
  nanotrack.
\newblock {\em Scientific reports}, 5(1):11369, 2015.

\bibitem{ndiaye2017topological}
Papa~Birame Ndiaye, Collins~Ashu Akosa, and Aur{\'e}lien Manchon.
\newblock Topological hall and spin hall effects in disordered skyrmionic
  textures.
\newblock {\em Physical Review B}, 95(6):064426, 2017.

\bibitem{jin2016dynamics}
Chendong Jin, Chengkun Song, Jianbo Wang, and Qingfang Liu.
\newblock Dynamics of antiferromagnetic skyrmion driven by the spin hall
  effect.
\newblock {\em Applied Physics Letters}, 109(18):182404, 2016.

\bibitem{velkov2016phenomenology}
Hristo Velkov, Olena Gomonay, Maarten Beens, Georg Schwiete, Arne Brataas,
  Jairo Sinova, and Rembert~A Duine.
\newblock Phenomenology of current-induced skyrmion motion in antiferromagnets.
\newblock {\em New Journal of Physics}, 18(7):075016, 2016.

\bibitem{akosa2018theory}
Collins~Ashu Akosa, OA~Tretiakov, G~Tatara, and Aurelien Manchon.
\newblock Theory of the topological spin hall effect in antiferromagnetic
  skyrmions: Impact on current-induced motion.
\newblock {\em Physical review letters}, 121(9):097204, 2018.

\bibitem{nunez2006theory}
Alvaro~S N{\'u}nez, RA~Duine, Paul Haney, and AH~MacDonald.
\newblock Theory of spin torques and giant magnetoresistance in
  antiferromagnetic metals.
\newblock {\em Physical Review B}, 73(21):214426, 2006.

\bibitem{haney2008current}
Paul~M Haney and AH~MacDonald.
\newblock Current-induced torques due to compensated antiferromagnets.
\newblock {\em Physical review letters}, 100(19):196801, 2008.

\bibitem{gomonay2010spin}
Helen~V Gomonay and Vadim~M Loktev.
\newblock Spin transfer and current-induced switching in antiferromagnets.
\newblock {\em Physical Review B}, 81(14):144427, 2010.

\bibitem{gomonay2012symmetry}
Helen~V Gomonay, Roman~V Kunitsyn, and Vadim~M Loktev.
\newblock Symmetry and the macroscopic dynamics of antiferromagnetic materials
  in the presence of spin-polarized current.
\newblock {\em Physical Review B}, 85(13):134446, 2012.

\bibitem{cheng2014dynamics}
Ran Cheng and Qian Niu.
\newblock Dynamics of antiferromagnets driven by spin current.
\newblock {\em Physical Review B}, 89(8):081105, 2014.

\bibitem{tveten2014antiferromagnetic}
Erlend~G Tveten, Alireza Qaiumzadeh, and Arne Brataas.
\newblock Antiferromagnetic domain wall motion induced by spin waves.
\newblock {\em Physical review letters}, 112(14):147204, 2014.

\bibitem{cheng2014spin}
Ran Cheng, Jiang Xiao, Qian Niu, and Arne Brataas.
\newblock Spin pumping and spin-transfer torques in antiferromagnets.
\newblock {\em Physical review letters}, 113(5):057601, 2014.

\bibitem{das2023bilayer}
Debasis Das, Yunuo Cen, Jianze Wang, and Xuanyao Fong.
\newblock Bilayer-skyrmion-based design of neuron and synapse for spiking
  neural network.
\newblock {\em Physical Review Applied}, 19(2):024063, 2023.

\bibitem{chen2020nanoscale}
Runze Chen, Chen Li, Yu~Li, James~J Miles, Giacomo Indiveri, Steve Furber,
  Vasilis~F Pavlidis, and Christoforos Moutafis.
\newblock Nanoscale room-temperature multilayer skyrmionic synapse for deep
  spiking neural networks.
\newblock {\em Physical Review Applied}, 14(1):014096, 2020.

\bibitem{lone2023controlling}
Aijaz~H Lone, Arnab Ganguly, Hanrui Li, Nazek~El Atab, Gobind Das, and
  H~Fariborzi.
\newblock Controlling the skyrmion density and size for quantized convolutional
  neural networks.
\newblock {\em arXiv preprint arXiv:2302.01390}, 2023.

\bibitem{li2017magnetic}
Sai Li, Wang Kang, Yangqi Huang, Xichao Zhang, Yan Zhou, and Weisheng Zhao.
\newblock Magnetic skyrmion-based artificial neuron device.
\newblock {\em Nanotechnology}, 28(31):31LT01, 2017.

\bibitem{chen2018compact}
Xing Chen, Wang Kang, Daoqian Zhu, Xichao Zhang, Na~Lei, Youguang Zhang, Yan
  Zhou, and Weisheng Zhao.
\newblock A compact skyrmionic leaky--integrate--fire spiking neuron device.
\newblock {\em Nanoscale}, 10(13):6139--6146, 2018.

\bibitem{yu2020voltage}
Ziyang Yu, Maokang Shen, Zhongming Zeng, Shiheng Liang, Yong Liu, Ming Chen,
  Zhenhua Zhang, Zhihong Lu, Long You, Xiaofei Yang, et~al.
\newblock Voltage-controlled skyrmion-based nanodevices for neuromorphic
  computing using a synthetic antiferromagnet.
\newblock {\em Nanoscale Advances}, 2(3):1309--1317, 2020.

\bibitem{sisodia2022robust}
Naveen Sisodia, Johan Pelloux-Prayer, Liliana~D Buda-Prejbeanu, Lorena Anghel,
  Gilles Gaudin, and Olivier Boulle.
\newblock Robust and programmable logic-in-memory devices exploiting skyrmion
  confinement and channeling using local energy barriers.
\newblock {\em Physical Review Applied}, 18(1):014025, 2022.

\bibitem{pinna2020reservoir}
Daniele Pinna, George Bourianoff, and Karin Everschor-Sitte.
\newblock Reservoir computing with random skyrmion textures.
\newblock {\em Physical Review Applied}, 14(5):054020, 2020.

\bibitem{raab2022brownian}
Klaus Raab, Maarten~A Brems, Grischa Beneke, Takaaki Dohi, Jan Roth{\"o}rl,
  Fabian Kammerbauer, Johan~H Mentink, and Mathias Kl{\"a}ui.
\newblock Brownian reservoir computing realized using geometrically confined
  skyrmion dynamics.
\newblock {\em Nature Communications}, 13(1):6982, 2022.

\bibitem{feynman2018simulating}
Richard~P Feynman.
\newblock Simulating physics with computers.
\newblock In {\em Feynman and computation}, pages 133--153. CRC Press, 2018.

\bibitem{divincenzo1995quantum}
David~P DiVincenzo.
\newblock Quantum computation.
\newblock {\em Science}, 270(5234):255--261, 1995.

\bibitem{nielsen2002quantum}
Michael~A Nielsen and Isaac Chuang.
\newblock Quantum computation and quantum information, 2002.

\bibitem{psaroudaki2021skyrmion}
Christina Psaroudaki and Christos Panagopoulos.
\newblock Skyrmion qubits: A new class of quantum logic elements based on
  nanoscale magnetization.
\newblock {\em Physical Review Letters}, 127(6):067201, 2021.

\bibitem{xia2023universal}
Jing Xia, Xichao Zhang, Xiaoxi Liu, Yan Zhou, and Motohiko Ezawa.
\newblock Universal quantum computation based on nanoscale skyrmion helicity
  qubits in frustrated magnets.
\newblock {\em Physical Review Letters}, 130(10):106701, 2023.

\bibitem{xia2022qubits}
Jing Xia, Xichao Zhang, Xiaoxi Liu, Yan Zhou, and Motohiko Ezawa.
\newblock Qubits based on merons in magnetic nanodisks.
\newblock {\em Communications Materials}, 3(1):88, 2022.

\bibitem{leonov2015multiply}
AO~Leonov and M~Mostovoy.
\newblock Multiply periodic states and isolated skyrmions in an anisotropic
  frustrated magnet.
\newblock {\em Nature communications}, 6(1):8275, 2015.

\bibitem{lin2016ginzburg}
Shi-Zeng Lin and Satoru Hayami.
\newblock Ginzburg-landau theory for skyrmions in inversion-symmetric magnets
  with competing interactions.
\newblock {\em Physical Review B}, 93(6):064430, 2016.

\bibitem{batista2016frustration}
Cristian~D Batista, Shi-Zeng Lin, Satoru Hayami, and Yoshitomo Kamiya.
\newblock Frustration and chiral orderings in correlated electron systems.
\newblock {\em Reports on Progress in Physics}, 79(8):084504, 2016.

\bibitem{diep2019phase}
Hung~T Diep.
\newblock Phase transition in frustrated magnetic thin film—physics at phase
  boundaries.
\newblock {\em Entropy}, 21(2):175, 2019.

\bibitem{makhlin2002nonlocal}
Yuriy Makhlin.
\newblock Nonlocal properties of two-qubit gates and mixed states, and the
  optimization of quantum computations.
\newblock {\em Quantum Information Processing}, 1:243--252, 2002.

\bibitem{im2012symmetry}
Mi-Young Im, Peter Fischer, Keisuke Yamada, Tomonori Sato, Shinya Kasai,
  Yoshinobu Nakatani, and Teruo Ono.
\newblock Symmetry breaking in the formation of magnetic vortex states in a
  permalloy nanodisk.
\newblock {\em Nature communications}, 3(1):983, 2012.

\bibitem{petrovic2021skyrmion}
Alexander~Paul Petrovi{\'c}, M~Raju, XY~Tee, A~Louat, Ivan Maggio-Aprile,
  RM~Menezes, MJ~Wyszy{\'n}ski, NK~Duong, M~Reznikov, Ch~Renner, et~al.
\newblock Skyrmion-(anti) vortex coupling in a chiral magnet-superconductor
  heterostructure.
\newblock {\em Physical review letters}, 126(11):117205, 2021.

\bibitem{hals2016composite}
Kjetil~MD Hals, Michael Schecter, and Mark~S Rudner.
\newblock Composite topological excitations in ferromagnet-superconductor
  heterostructures.
\newblock {\em Physical Review Letters}, 117(1):017001, 2016.

\bibitem{diaz2021majorana}
Sebasti{\'a}n~A D{\'\i}az, Jelena Klinovaja, Daniel Loss, and Silas Hoffman.
\newblock Majorana bound states induced by antiferromagnetic skyrmion textures.
\newblock {\em Physical Review B}, 104(21):214501, 2021.

\bibitem{rex2019majorana}
Stefan Rex, Igor~V Gornyi, and Alexander~D Mirlin.
\newblock Majorana bound states in magnetic skyrmions imposed onto a
  superconductor.
\newblock {\em Physical Review B}, 100(6):064504, 2019.

\bibitem{ivanov2001non}
Dmitri~A Ivanov.
\newblock Non-abelian statistics of half-quantum vortices in p-wave
  superconductors.
\newblock {\em Physical review letters}, 86(2):268, 2001.

\bibitem{nayak2008non}
Chetan Nayak, Steven~H Simon, Ady Stern, Michael Freedman, and Sankar~Das
  Sarma.
\newblock Non-abelian anyons and topological quantum computation.
\newblock {\em Reviews of Modern Physics}, 80(3):1083, 2008.

\bibitem{alicea2011non}
Jason Alicea, Yuval Oreg, Gil Refael, Felix Von~Oppen, and Matthew~PA Fisher.
\newblock Non-abelian statistics and topological quantum information processing
  in 1d wire networks.
\newblock {\em Nature Physics}, 7(5):412--417, 2011.

\bibitem{stern2013topological}
Ady Stern and Netanel~H Lindner.
\newblock Topological quantum computation-from basic concepts to first
  experiments.
\newblock {\em Science}, 339(6124):1179--1184, 2013.

\bibitem{sarma2015majorana}
Sankar~Das Sarma, Michael Freedman, and Chetan Nayak.
\newblock Majorana zero modes and topological quantum computation.
\newblock {\em npj Quantum Information}, 1(1):1--13, 2015.

\bibitem{kitaev2001unpaired}
A~Yu Kitaev.
\newblock Unpaired majorana fermions in quantum wires.
\newblock {\em Physics-uspekhi}, 44(10S):131, 2001.

\bibitem{lutchyn2010majorana}
Roman~M Lutchyn, Jay~D Sau, and S~Das Sarma.
\newblock Majorana fermions and a topological phase transition in
  semiconductor-superconductor heterostructures.
\newblock {\em Physical review letters}, 105(7):077001, 2010.

\bibitem{oreg2010helical}
Yuval Oreg, Gil Refael, and Felix Von~Oppen.
\newblock Helical liquids and majorana bound states in quantum wires.
\newblock {\em Physical review letters}, 105(17):177002, 2010.

\bibitem{mourik2012signatures}
Vincent Mourik, Kun Zuo, Sergey~M Frolov, SR~Plissard, Erik~PAM Bakkers, and
  Leo~P Kouwenhoven.
\newblock Signatures of majorana fermions in hybrid
  superconductor-semiconductor nanowire devices.
\newblock {\em Science}, 336(6084):1003--1007, 2012.

\bibitem{das2012zero}
Anindya Das, Yuval Ronen, Yonatan Most, Yuval Oreg, Moty Heiblum, and Hadas
  Shtrikman.
\newblock Zero-bias peaks and splitting in an al--inas nanowire topological
  superconductor as a signature of majorana fermions.
\newblock {\em Nature Physics}, 8(12):887--895, 2012.

\bibitem{deng2016majorana}
MT~Deng, S~Vaitiek{\.e}nas, Esben~Bork Hansen, Jeroen Danon, M~Leijnse, Karsten
  Flensberg, Jesper Nyg{\aa}rd, P~Krogstrup, and Charles~M Marcus.
\newblock Majorana bound state in a coupled quantum-dot hybrid-nanowire system.
\newblock {\em Science}, 354(6319):1557--1562, 2016.

\bibitem{suominen2017zero}
Henri~J Suominen, Morten Kjaergaard, Alexander~R Hamilton, Javad Shabani,
  Chris~J Palmstr{\o}m, Charles~M Marcus, and Fabrizio Nichele.
\newblock Zero-energy modes from coalescing andreev states in a two-dimensional
  semiconductor-superconductor hybrid platform.
\newblock {\em Physical review letters}, 119(17):176805, 2017.

\bibitem{vaitiekenas2021zero}
S~Vaitiek{\.e}nas, Y~Liu, P~Krogstrup, and CM~Marcus.
\newblock Zero-bias peaks at zero magnetic field in ferromagnetic hybrid
  nanowires.
\newblock {\em Nature Physics}, 17(1):43--47, 2021.

\bibitem{hals2016supercurrent}
Kjetil~MD Hals.
\newblock Supercurrent-induced spin-orbit torques.
\newblock {\em Physical Review B}, 93(11):115431, 2016.

\bibitem{hals2017magnetoelectric}
Kjetil~MD Hals.
\newblock Magnetoelectric coupling in superconductor-helimagnet
  heterostructures.
\newblock {\em Physical Review B}, 95(13):134504, 2017.

\bibitem{nadj2014observation}
Stevan Nadj-Perge, Ilya~K Drozdov, Jian Li, Hua Chen, Sangjun Jeon, Jungpil
  Seo, Allan~H MacDonald, B~Andrei Bernevig, and Ali Yazdani.
\newblock Observation of majorana fermions in ferromagnetic atomic chains on a
  superconductor.
\newblock {\em Science}, 346(6209):602--607, 2014.

\bibitem{ruby2015end}
Michael Ruby, Falko Pientka, Yang Peng, Felix Von~Oppen, Benjamin~W Heinrich,
  and Katharina~J Franke.
\newblock End states and subgap structure in proximity-coupled chains of
  magnetic adatoms.
\newblock {\em Physical review letters}, 115(19):197204, 2015.

\bibitem{pawlak2016probing}
R{\'e}my Pawlak, Marcin Kisiel, Jelena Klinovaja, Tobias Meier, Shigeki Kawai,
  Thilo Glatzel, Daniel Loss, and Ernst Meyer.
\newblock Probing atomic structure and majorana wavefunctions in mono-atomic fe
  chains on superconducting pb surface.
\newblock {\em npj Quantum Information}, 2(1):1--5, 2016.

\bibitem{kim2018toward}
Howon Kim, Alexandra Palacio-Morales, Thore Posske, Levente R{\'o}zsa,
  Kriszti{\'a}n Palot{\'a}s, L{\'a}szl{\'o} Szunyogh, Michael Thorwart, and
  Roland Wiesendanger.
\newblock Toward tailoring majorana bound states in artificially constructed
  magnetic atom chains on elemental superconductors.
\newblock {\em Science Advances}, 4(5):eaar5251, 2018.

\bibitem{klinovaja2013topological}
Jelena Klinovaja, Peter Stano, Ali Yazdani, and Daniel Loss.
\newblock Topological superconductivity and majorana fermions in rkky systems.
\newblock {\em Physical review letters}, 111(18):186805, 2013.

\bibitem{vazifeh2013self}
MM~Vazifeh and Marcel Franz.
\newblock Self-organized topological state with majorana fermions.
\newblock {\em Physical review letters}, 111(20):206802, 2013.

\bibitem{braunecker2013interplay}
Bernd Braunecker and Pascal Simon.
\newblock Interplay between classical magnetic moments and superconductivity in
  quantum one-dimensional conductors: toward a self-sustained topological
  majorana phase.
\newblock {\em Physical review letters}, 111(14):147202, 2013.

\bibitem{pientka2013topological}
Falko Pientka, Leonid~I Glazman, and Felix Von~Oppen.
\newblock Topological superconducting phase in helical shiba chains.
\newblock {\em Physical Review B}, 88(15):155420, 2013.

\bibitem{alicea2012new}
Jason Alicea.
\newblock New directions in the pursuit of majorana fermions in solid state
  systems.
\newblock {\em Reports on progress in physics}, 75(7):076501, 2012.

\bibitem{pawlak2019majorana}
R{\'e}my Pawlak, Silas Hoffman, Jelena Klinovaja, Daniel Loss, and Ernst Meyer.
\newblock Majorana fermions in magnetic chains.
\newblock {\em Progress in Particle and Nuclear Physics}, 107:1--19, 2019.

\bibitem{prada2020andreev}
Elsa Prada, Pablo San-Jose, Michiel~WA de~Moor, Attila Geresdi, Eduardo~JH Lee,
  Jelena Klinovaja, Daniel Loss, Jesper Nyg{\aa}rd, Ram{\'o}n Aguado, and Leo~P
  Kouwenhoven.
\newblock From andreev to majorana bound states in hybrid
  superconductor--semiconductor nanowires.
\newblock {\em Nature Reviews Physics}, 2(10):575--594, 2020.

\bibitem{desjardins2019synthetic}
MM~Desjardins, LC~Contamin, MR~Delbecq, MC~Dartiailh, LE~Bruhat, T~Cubaynes,
  JJ~Viennot, F~Mallet, S~Rohart, A~Thiaville, et~al.
\newblock Synthetic spin--orbit interaction for majorana devices.
\newblock {\em Nature materials}, 18(10):1060--1064, 2019.

\bibitem{matos2017tunable}
Alex Matos-Abiague, Javad Shabani, Andrew~D Kent, Geoffrey~L Fatin, Benedikt
  Scharf, and Igor {\v{Z}}uti{\'c}.
\newblock Tunable magnetic textures: From majorana bound states to braiding.
\newblock {\em Solid State Communications}, 262:1--6, 2017.

\bibitem{poyhonen2016skyrmion}
Kim P{\"o}yh{\"o}nen, Alex Weststr{\"o}m, Sergey~S Pershoguba, Teemu Ojanen,
  and Alexander~V Balatsky.
\newblock Skyrmion-induced bound states in a p-wave superconductor.
\newblock {\em Physical Review B}, 94(21):214509, 2016.

\bibitem{gungordu2018stabilization}
Utkan G{\"u}ng{\"o}rd{\"u}, Shane Sandhoefner, and Alexey~A Kovalev.
\newblock Stabilization and control of majorana bound states with elongated
  skyrmions.
\newblock {\em Physical Review B}, 97(11):115136, 2018.

\bibitem{garnier2019topological}
Maxime Garnier, Andrej Mesaros, and Pascal Simon.
\newblock Topological superconductivity with deformable magnetic skyrmions.
\newblock {\em Communications Physics}, 2(1):126, 2019.

\bibitem{kubetzka2020towards}
Andr{\'e} Kubetzka, Jan~M B{\"u}rger, Roland Wiesendanger, and Kirsten von
  Bergmann.
\newblock Towards skyrmion-superconductor hybrid systems.
\newblock {\em Physical Review Materials}, 4(8):081401, 2020.

\bibitem{yang2016majorana}
Guang Yang, Peter Stano, Jelena Klinovaja, and Daniel Loss.
\newblock Majorana bound states in magnetic skyrmions.
\newblock {\em Physical Review B}, 93(22):224505, 2016.

\bibitem{nakosai2013two}
Sho Nakosai, Yukio Tanaka, and Naoto Nagaosa.
\newblock Two-dimensional p-wave superconducting states with magnetic moments
  on a conventional s-wave superconductor.
\newblock {\em Physical Review B}, 88(18):180503, 2013.

\bibitem{mohanta2021skyrmion}
Narayan Mohanta, Satoshi Okamoto, and Elbio Dagotto.
\newblock Skyrmion control of majorana states in planar josephson junctions.
\newblock {\em Communications Physics}, 4(1):163, 2021.

\bibitem{beenakker2015random}
CWJ Beenakker.
\newblock Random-matrix theory of majorana fermions and topological
  superconductors.
\newblock {\em Reviews of Modern Physics}, 87(3):1037, 2015.

\bibitem{edelstein1995magnetoelectric}
Victor~M Edelstein.
\newblock Magnetoelectric effect in polar superconductors.
\newblock {\em Physical review letters}, 75(10):2004, 1995.

\bibitem{baumard2019generation}
J~Baumard, J~Cayssol, FS~Bergeret, and A~Buzdin.
\newblock Generation of a superconducting vortex via n{\'e}el skyrmions.
\newblock {\em Physical Review B}, 99(1):014511, 2019.

\bibitem{dahir2019interaction}
Samme~M Dahir, Anatoly~F Volkov, and Ilya~M Eremin.
\newblock Interaction of skyrmions and pearl vortices in superconductor-chiral
  ferromagnet heterostructures.
\newblock {\em Physical Review Letters}, 122(9):097001, 2019.

\bibitem{dahir2020meissner}
Samme~M Dahir, Anatoly~F Volkov, and Ilya~M Eremin.
\newblock Meissner currents induced by topological magnetic textures in hybrid
  superconductor/ferromagnet structures.
\newblock {\em Physical Review B}, 102(1):014503, 2020.

\bibitem{luh1965bound}
Yu~Luh.
\newblock Bound state in superconductors with paramagnetic impurities.
\newblock {\em Acta Physica Sinica}, 21(1):75, 1965.

\bibitem{shiba1968classical}
Hiroyuki Shiba.
\newblock Classical spins in superconductors.
\newblock {\em Progress of theoretical Physics}, 40(3):435--451, 1968.

\bibitem{rusinov1969superconductivity}
AI~Rusinov.
\newblock Superconductivity near a paramagnetic impurity.
\newblock {\em JETP Lett.(USSR)(Engl. Transl.);(United States)}, 9, 1969.

\bibitem{balatsky2006impurity}
Alexander~V Balatsky, Ilya Vekhter, and Jian-Xin Zhu.
\newblock Impurity-induced states in conventional and unconventional
  superconductors.
\newblock {\em Reviews of Modern Physics}, 78(2):373, 2006.

\bibitem{yao2014enhanced}
Norman~Y Yao, Leonid~I Glazman, Eugene~A Demler, Mikhail~D Lukin, and Jay~D
  Sau.
\newblock Enhanced antiferromagnetic exchange between magnetic impurities in a
  superconducting host.
\newblock {\em Physical Review Letters}, 113(8):087202, 2014.

\end{thebibliography}
\end{document}